\documentclass[iop]{emulateapj}

\usepackage{rotating, graphicx,dsfont}
\usepackage{color}
\usepackage{amssymb, amsmath}

\usepackage[utf8]{inputenc}
\usepackage{lipsum}
\usepackage{pifont}
\usepackage{amsmath, amssymb, bm}
\usepackage{balance}
\usepackage{multirow}
\usepackage{enumitem}
\usepackage{booktabs}

\shorttitle{Ingredients for 21cm intensity mapping}
\shortauthors{Francisco Villaescusa-Navarro et al.}

\usepackage[usenames,dvipsnames]{xcolor}
\usepackage{hyperref}
\hypersetup{
    colorlinks = true,
    citecolor = {MidnightBlue},
    linkcolor=blue,
    urlcolor =blue		
}

\newcommand{\be}{\begin{equation}}
\newcommand{\ee}{\end{equation}}
\newcommand{\ba}{\begin{eqnarray}}
\newcommand{\ea}{\end{eqnarray}}

\begin{document}

\title{Ingredients for 21cm intensity mapping}

\author{Francisco Villaescusa-Navarro$^{1,\dagger}$, Shy Genel$^{1,2}$, Emanuele Castorina$^{3,4}$, Andrej Obuljen$^{5,6}$, David N. Spergel$^{7,1}$, Lars Hernquist$^8$, Dylan Nelson$^9$, Isabella P. Carucci$^{10}$, Annalisa Pillepich$^{11}$, Federico Marinacci$^{12,8}$, Benedikt Diemer$^8$, Mark Vogelsberger$^{12}$, Rainer Weinberger$^{13}$, R{\"u}diger Pakmor$^{13}$\\}
\affil{$^{1}${Center for Computational Astrophysics, Flatiron Institute, 162 5th Avenue, 10010, New York, NY, USA}}
\affil{$^{2}${Columbia Astrophysics Laboratory, Columbia University, 550 West 120th Street, New York, NY 10027, USA}}
\affil{$^{2}${Berkeley Center for Cosmological Physics, University of California, Berkeley, CA 94720, USA}} 
\affil{$^{4}${Lawrence Berkeley National Laboratory, 1 Cyclotron Road, Berkeley, CA 93720, USA}}
\affil{$^5${SISSA- International School for Advanced Studies, Via Bonomea 265, 34136 Trieste, Italy}}
\affil{$^6${INFN - National Institute for Nuclear Physics, Via Valerio 2, I-34127 Trieste, Italy}}
\affil{$^7${Department of Astrophysical Sciences, Princeton University, Peyton Hall, Princeton NJ 08544-0010, USA}}
\affil{$^8$Harvard-Smithsonian Center for Astrophysics, 60 Garden Street, Cambridge, MA 02138, USA}
\affil{$^9${Max-Planck-Institut fur Astrophysik, Karl-Schwarzschild-Strasse 1, 85740 Garching bei Munchen, Germany}}
\affil{$^{10}${Department of Physics and Astronomy, University College London, London WC1E 6BT, UK}}
\affil{$^{11}${Max-Planck-Institut fur Astronomie, Konigstuhl 17, 69117 Heidelberg, Germany}}
\affil{$^{12}${Department of Physics, Kavli Institute for Astrophysics and Space Research, MIT, Cambridge, MA 02139, USA}}
\affil{$^{13}${Heidelberg Institute for Theoretical Studies, Schloss-Wolfsbrunnenweg 35, 69118 Heidelberg, Germany}}
\altaffiltext{$\dagger$}{fvillaescusa@flatironinstitute.org}

\begin{abstract}

Current and upcoming radio telescopes will map the spatial
distribution of cosmic neutral hydrogen (HI) through its 21cm
emission. In order to extract the maximum information from these
surveys, accurate theoretical predictions are needed. We study the
abundance and clustering properties of HI at redshifts $z\leqslant5$
using TNG100, a large state-of-the-art magneto-hydrodynamic simulation
of a $75~h^{-1}{\rm Mpc}$ box size, which is part of the IllustrisTNG
Project. We show that most of the HI lies within dark matter halos and
we provide fits for the halo HI mass function, i.e.~the mean HI mass
hosted by a halo of mass $M$ at redshift $z$. We find that only halos
with circular velocities larger than $\simeq30~{\rm km/s}$ contain
HI. While the density profiles of HI exhibit a large halo-to-halo scatter, the mean profiles are universal across mass and redshift.
The HI in low-mass halos is mostly located in the central galaxy,
while in massive halos HI is concentrated in the satellites. Our simulation reproduces the DLAs bias value from observations. We show
that the HI and matter density probability distribution functions
differ significantly. Our results point out that for small halos the HI bulk velocity goes in the same direction and has the same magnitude as the halo peculiar velocity, while in large halos differences show up. We find that halo HI velocity dispersion follows a
power-law with halo mass. We find a complicated HI bias, with HI becoming non-linear already at $k=0.3~h{\rm Mpc}^{-1}$ 
at $z\gtrsim3$. The clustering of HI can however be accurately reproduced by perturbative methods.
We find a new secondary bias, by showing
that the clustering of halos depends not only on mass but also on HI
content. We compute the amplitude of the HI shot-noise and find that
it is small at all redshifts, verifying the robustness of BAO
measurements with 21cm intensity mapping. We study the clustering of
HI in redshift-space, and show that linear theory can explain the ratio
between the monopoles in redshift- and real-space down to 0.3, 0.5 and
1 $h{\rm Mpc}^{-1}$ at redshifts 3, 4 and 5, respectively. We find
that the amplitude of the Fingers-of-God effect is larger for HI than for
matter, since HI is found only in halos above a certain mass. We
point out that 21 cm maps can be created from N-body simulations
rather than full hydrodynamic simulations. Modeling the 1-halo term is
crucial for achieving percent accuracy with respect to a full hydro
treatment.

\end{abstract} 


\section{Introduction}
\label{sec:introduction}

The $\Lambda$CDM model describes how the initial quantum perturbations
in the primordial Universe grow and give rise to the cosmic web: large
accumulations of matter in the form of dark matter halos accrete
material through filaments and sheets that surround enormous diffuse
regions in space. This model has been successful in explaining a very
diverse set of cosmological observables, including, among others, the
anisotropies in the cosmic microwave background (CMB), the spatial
distribution of galaxies, the statistical properties of the
Ly$\alpha$-forest, the abundance of galaxy clusters, and correlations
in the shapes of galaxies induced by gravitational lensing.

The $\Lambda$CDM model has free parameters that describe physical
quantities such as the geometry of the Universe, the amount of cold
dark matter (CDM) and baryons, the sum of the neutrino masses, the
expansion rate of the Universe, the nature of dark energy, and the
initial conditions of the Universe.  The current quest in cosmology is
to determine the values of these parameters as precisely as possible,
by exploiting the fact that they influence the spatial distribution of
matter.  Thus, by examining the statistical properties of matter
tracers such as galaxies and cosmic neutral hydrogen, the spatial
distribution of matter can be inferred and the value of the
cosmological parameters can be constrained.

The amount of information that can be extracted from cosmological
surveys depends on several factors, such as the volume being covered
or the range in scales where theoretical predictions are reliable. For
example, in the case of the CMB, theoretical predictions are extremely
precise, because the radiation we observe was produced when the
fluctuations were in the linear regime.  Tracing the large-scale
structure of the Universe at low redshifts, through spectroscopic
galaxy surveys, represents a complementary approach for extracting
cosmological information, where much larger volumes can be surveyed
but the theoretical predictions are more uncertain. For galaxy surveys
the volume that can be probed also limits the method, because at
high redshifts galaxies are fainter and their spectroscopic detection
is challenging.

A different way to trace the matter field is through 21cm intensity
mapping
\citep{Bharadwaj_2001A,Bharadwaj_2001B,Battye:2004re,McQuinn_2006,
  Chang_2008,Loeb_Wyithe_2008,Bull2015, Santos_2015,
  Villaescusa-Navarro_2015a}.  The method consists of carrying out a
low angular resolution survey where the total 21 cm flux from
unresolved sources is measured on large areas of the sky at different
frequencies. The emission arises from the hyperfine splitting of the
ground state of neutral hydrogen into two levels because of the
spin-spin interaction between the electron and proton.  An electron
located in the upper energy level can decay into the lower energy
state by emitting a photon with a rest wavelength of 21 cm. This method
has several advantages over traditional approaches. First, given that the
observable is the 21 cm line, the method is spectroscopic in
nature. Second, very large cosmological volumes can be surveyed in a
fast and efficient manner. Third, the amplitude of the signal
depends only on the abundance and clustering of neutral hydrogen, so
cosmic HI can be traced from $z=0$ to $z\simeq50$\footnote{At higher
  redshifts the atmosphere becomes opaque at the relevant wavelengths.}.

Current, upcoming and future surveys such as the Giant Meterwave Radio
Telescope (GMRT)\footnote{http://gmrt.ncra.tifr.res.in/}, the Ooty
Radio Telescope (ORT)\footnote{http://rac.ncra.tifr.res.in/}, the
Canadian Hydrogen Intensity Mapping Experiment
(CHIME)\footnote{http://chime.phas.ubc.ca/}, the Five hundred meter
Aperture Spherical Telescope
(FAST)\footnote{http://fast.bao.ac.cn/en/},
Tianlai\footnote{http://tianlai.bao.ac.cn}, BINGO (Baryon acoustic
oscillations In Neutral Gas
Observations)\footnote{http://www.jb.man.ac.uk/research/BINGO/}, ASKAP
(The Australian Square Kilometer Array
Pathfinder)\footnote{http://www.atnf.csiro.au/projects/askap/index.html},
MeerKAT (The South African Square Kilometer Array
Pathfinder)\footnote{http://www.ska.ac.za/meerkat/}, HIRAX (The
Hydrogen Intensity and Real-time Analysis
eXperiment)\footnote{https://www.acru.ukzn.ac.za/$\sim$hirax/} and the
SKA (The Square Kilometer
Array)\footnote{https://www.skatelescope.org/} will sample the
large-scale structure of the Universe in the post-reionization era by
detecting 21 cm emission from cosmic neutral hydrogen \citep{Sarkar_2018b, Carucci_2017a, Sarkar_2018c,Marthi_2017,Sarkar_2016k,Choudhuri_2016}.

In order to extract information from those surveys, the observational
data has to be compared with theoretical predictions. To linear
order, the amplitude and shape of the 21 cm power spectrum is given by
\be 
P_{\rm 21cm}(k,\mu)=\bar{T}_b^2\left[(b_{\rm HI}+f\mu^2)^2P_{\rm m}(k)+P_{\rm SN}\right]~,
\label{Eq:P_21cm}
\ee
where $\bar{T}_b$ is the mean brightness temperature, $b_{\rm HI}$ is
the HI bias, $f$ is the linear growth rate, $\mu=k_z/k$, $P_{\rm
  m}(k)$ is the linear matter power spectrum and $P_{\rm SN}$ is the
HI shot-noise. Here, $k_z$ is the projection of $k$ along the
line-of-sight, which we take to be the z-axis.

At redshifts $z\in[0,5]$ we have relatively good knowledge of the
abundance of cosmic neutral hydrogen, and therefore, of
$\bar{T}_b$. On the other hand, little is known about the value of the
HI bias and HI shot-noise in that redshift interval. It is important
to determine their values since the signal-to-noise ratio and range of
scales where information can be extracted critically depends on
them. One of the purposes of our work is to measure these quantities
at different redshifts. Moreover, it is important to determine the
the regime where linear theory is accurate. In this work we investigate in detail at which redshifts and 
scales the clustering of HI in real-space (i.e. the HI bias) and in redshift-space (Kaiser factor)
becomes non-linear.

In order to optimize what can be learned from the surveys mentioned above,
theoretical predictions in the mildly and fully non-linear regimes are
needed. The halo model provides a reasonably accurate
framework for predicting the abundance and clustering of HI from linear to
fully non-linear scales.  To apply this method, several ingredients
are needed for a given cosmological model: 1) the linear matter power
spectrum, $P_{\rm lin}(k,z)$, 2) the halo mass function, $n(M,z)$, 3)
the halo bias, $b(M,z)$, 4) the average HI mass that a halo of mass
$M$ hosts at redshift $z$, $M_{\rm HI}(M,z)$, which we refer to as the
\textit{halo HI mass function}\footnote{note that the term ``HI mass
  function'' is commonly used to model the abundance of galaxies with
  different HI masses. Thus, in order to distinguish the two concepts
  we use ``halo HI mass function'' to refer to the function that returns the
  average HI mass inside a halo of mass $M$ at redshift $z$.} and 5)
the mean density profile of neutral hydrogen within halos of mass $M$
at redshift $z$, $\rho_{\rm HI}(r|M,z)$.  In addition, the halo model is
formulated under the assumption that all HI is confined to dark matter
halos. With the above ingredients at hand one can write the fully
non-linear HI power spectrum as the sum of 1-halo and 2-halo terms

\begin{widetext}
\begin{eqnarray}
\label{eq:HI1h_HI2h}
P_{\rm HI}(k,z)&=&P_{\rm HI,1h}(k)+P_{\rm HI,2h}(k)\\
\label{eq:HI1h}
P_{\rm HI,1h}(k,z)&=&\frac{1}{(\rho_{\rm c}^0\Omega_{\rm HI}(z))^2}\int_0^\infty dM n(M,z) M_{\rm HI}^2(M,z)\left|u_{\rm HI}(k|M,z)\right|^2\\
P_{\rm HI,2h}(k,z)&=&\frac{P_{\rm lin}(k,z)}{(\rho_{\rm c}^0\Omega_{\rm HI}(z))^2}\left[\int_0^\infty dM n(M,z)b(M,z)M_{\rm HI}(M,z)|u_{\rm HI}(k|M,z)| \right]^2
\label{eq:HI1h2h}
\end{eqnarray}
\end{widetext}
where $\rho_{\rm c}^0$ is the critical density of the Universe today and
$\Omega_{\rm HI}(z)=\bar{\rho}_{\rm HI}(z)/\rho_{\rm c}^0$, with
$\bar{\rho}_{\rm HI}(z)$ being the mean HI density at redshift
$z$. $u(k|M,z)$ is the Fourier transform of the normalized HI density
profile: $u(x|M,z)=\rho_{\rm HI}(x|M,z)/M_{\rm HI}(M,z)$.

Some of the goals of our work are to quantify: 1) the amount of HI
outside of halos, 2) the form of the halo HI mass function, and 3)
the density profiles of HI within halos.

While the halo model is a powerful analytic framework, it does not model accurately a number of things, e.g.  the transition between the 1-h and 2-h terms \citep{Massara_2014}. Thus, its accuracy can be severely limited by that. A more precise modeling can be achieved by \textit{painting} HI on top of dark matter halos according the HI halo model ingredients \citep{Villaescusa-Navarro_2014a}, i.e. more like an HI Halo Occupation Distribution (HOD) modeling.

Since 21cm intensity mapping observations are carried out in redshift-space, modeling the abundance and spatial distribution of HI in halos is not enough. A complete description also requires to know the distribution of HI velocities. In this work we investigate the HI bulk velocities, the velocity dispersion of HI inside halos and the amplitude of the Fingers-of-God in the power spectrum.

The standard halo model does not account for various complexities
expected in the real Universe, e.g. whether the HI density profiles
depend not only on mass but on the galaxy population (blue/red) of the
halo, whether the clustering of halos depends not only on mass but
also on environment, and so forth. These questions can however be
addressed with hydrodynamic simulations, and in this paper we investigate them in detail.

We also study some quantities that can help us to improve our knowledge on the spatial distribution of HI: the probability distribution function of HI, the relation between the overdensities of matter and HI, the contribution of central and satellites galaxies to the total HI mass content in halos, the HI column density distribution function and the DLAs cross-section. 

We carry out our analysis using the IllustrisTNG Project,
state-of-the-art hydrodynamic simulations that follow the evolution of
billions of resolution elements representing CDM, gas, black holes and
stars in the largest volumes ever explored at such mass and spatial
resolution. Given the realism of our hydrodynamic simulations, we
always aim to connect our results to the underlying physical
processes. We note that previous works have studied the HI content of
simulated galaxies in detail \citep{Crain_2017, Bahe_2016, DaveR_13a,
  Bird_2014,Lagos_2014,FG_2016, MarinacciF_17a, Zoldan_2017,Xie_2017}.

We also show that once the most important ingredients for modeling the
abundance and clustering properties of HI have been calibrated using
full hydrodynamic simulations, less costly dark mater-only simulations, or approximate methods such as {\sc COLA} \citep{COLA}, {\sc peak-patch} or {\sc Pinocchio} \citep{Pinocchio,Munari2017}, can be used to generate accurate 21 cm maps. Those maps can then be
used to study other properties of HI in the fully non-linear regime,
such as the 21 cm bispectrum or the properties of HI voids. In this work we investigate the accuracy achieved by creating 21cm maps from N-body with respect to full hydrodynamic simulations.

This paper is organized as follows. In section \ref{sec:simulations}
we describe the characteristics of the IllustrisTNG simulations and the
method we use to estimate the mass of neutral hydrogen associated with
each gas cell. We consider different properties of the abundance of HI in
sections \ref{sec:Omega_HI} to \ref{sec:HI_stochasticity}:
\begin{itemize}[labelindent=1pt,labelwidth=0.5em,leftmargin=!]
\item In section \ref{sec:Omega_HI} we compare the overall HI abundance in our simulations to observations.
\item In section \ref{subsec:HI_in_halos_galaxies} we quantify the fraction of HI within halos and galaxies.
\item In section \ref{subsec:M_HI} we study the halo HI mass function.
\item In section \ref{sec:HI_profile} we investigate the density profiles of HI inside halos.
\item In section \ref{sec:HI_centrals_satellites} we quantify the fraction of the HI mass in halos that is in the central and satellites galaxies.
\item In section \ref{sec:HI_pdf} we examine the probability distribution function (pdf) of the HI density and compare it 
with the total matter density pdf.  
\item In section \ref{sec:DLAs} we compute the HI column density distribution function for the absorbers with high column density and quantify the DLAs cross-section and bias.
\item In section \ref{sec:HI_bulk} we consider the bulk velocities of HI inside halos.
\item In section \ref{sec:sigma_HI} we investigate the velocity dispersion of HI inside halos and compare it against matter. 
\item In section \ref{sec:HI_stochasticity} we quantify the relation between the overdensity of matter and HI.
\end{itemize}
We investigate HI clustering in sections \ref{subsec:HI_bias} to \ref{subsec:RSD}:
\begin{itemize}[labelindent=1pt,labelwidth=0.5em,leftmargin=!]
\item In section \ref{subsec:HI_bias} we present the amplitude and shape of the HI bias and investigate how well perturbation theory can reproduce the HI clustering in real-space.
\item In section \ref{sec:assembly_bias} we show that the clustering of dark matter halos in general depends on their HI masses for fixed halo mass.
\item In section \ref{subsec:SN} we quantify the amplitude of the HI shot-noise.
\item In section \ref{subsec:RSD} we study the clustering of HI in redshift-space.
\end{itemize}
In section \ref{sec:21cm} we estimate the accuracy that can be achieved by simulating HI
through a combination of N-body simulations with the results
derived in the previous sections rather than through ful hydrodynamic
simulations. Finally, we provide the main conclusions of our work in
section \ref{sec:conclusions}. During the course of the discussion,
we provide fitting formulae that can be used to reproduce our results.


\section{Methods}
\label{sec:simulations}
\subsection{The IllustrisTNG simulations}
\label{subsec:tng}

The simulations used in this work are part of the IllustrisTNG
Project \citep{SpringelV_17a,PillepichA_17a,NelsonD_17a,NaimanJ_17a,MarinacciF_17a}. We
employ two cosmological boxes that have been evolved to $z=0$, TNG100
(which is the same volume as the Illustris
simulation; \citealp{VogelsbergerM_14a,VogelsbergerM_14b,GenelS_14a})
and TNG300, with $75~{h^{-1}{\rm Mpc}}$ and $205~h^{-1}{\rm Mpc}$
comoving on a side, respectively. In particular, we use their
high-resolution realizations that evolve baryonic resolution elements
with mean masses of $1.4\times10^6~{\rm\thinspace M_{\odot}}$ and
$1.1\times10^7~{\rm\thinspace M_{\odot}}$, respectively.

These simulations have been run with the {\small AREPO}
code \citep{Arepo}, which calculates gravity using a tree-PM method,
magneto-hydrodynamics with a Godunov method on a moving Voronoi mesh,
and a range of astrophysical processes described by sub-grid
models. These processes include primordial and metal-line cooling
assuming a time-dependent uniform UV background radiation, star and
supermassive black hole formation, stellar population evolution that
enriches surrounding gas with heavy elements or metals, galactic winds, and
several modes of black hole feedback. Importantly, where uncertainty
and freedom exist for the implementation of these models, they are
parametrized and tuned to obtain a reasonable match to a small set of
observational results. These include the galaxy stellar mass function
and the baryon content of group-scale dark matter halos, both at
$z=0$. The numerical methods and subgrid physics models build
upon \citet{VogelsbergerM_13a}, and are specified in full
in \citet{WeinbergerR_16a, Weinberger_2017}
and \citet{Pillepich_2017b}. Accounts of the match between the
simulations and observations in a number of diverse aspects, such as
galaxy and halo sizes, colors, metallicities, magnetic fields and
clustering, are presented in the references above as well as
in \citet{GenelS_17a}, \citet{VogelsbergerM_17a}
and \citet{TorreyP_18a}.

In this paper we work mainly with halos identified by the
Friends-of-Friends (FoF) algorithm with a linking length of
$b=0.2$ \citep{DavisM_85a}. We take the halo center as the position of
the most bound particle in the halo. For each FoF halo, we also use
the halo's ``virial'' radius, defined using
$M=4\pi/3\Delta_c\rho_cR^3$, where $\rho_c$ is the Universe critical
density at the halo's redshift $z$ and $\Delta_c=18\pi^2+82x-39x^2$,
with
$x=\Omega_m(1+z)^3/(\Omega_m(1+z)^3+\Omega_\Lambda)-1$ \citep{Bryan_Norman_97}. We
refer to these objects as ``FoF-SO'' halos, for `spherical
overdensity,' since a single SO (spherical overdensity) halo is identified for each FoF
halo\footnote{Notice that a pure SO algorithm may identify several SO
halos inside a single FoF halo (see
appendix \ref{sec:FoF_vs_SO}).}. Unless stated explicitly, we refer
for FoF halos when talking generally about dark matter halos. The
{\small SUBFIND} algorithm \citep{SpringelV_01} has been run to
identify self-bound substructures in each FoF halos, and those objects
are referred to as `galaxies' in what follows. This class of objects
includes both the satellites and the central {\small SUBFIND} subhalos
of each FoF halo.

\subsection{Modeling the hydrogen phases}
\label{subsec:HI}

We now describe the method we use to quantify the fraction of hydrogen
that is in each phase (neutral, ionized or molecular) for each Voronoi
cell in the simulation.

For non-star-forming gas, we use the division between neutral and
ionized mass fractions that is calculated in the IllustrisTNG runs
on-the-fly and is included in the simulation outputs. This breakdown
assumes primordial chemistry in photo-ionization equilibrium with the
cosmic background radiation \citep{FaucherGiguereC_09a}, including a
density-dependent attenuation thereof to account for self-shielding
following \cite{Rahmati_2013}.

For star-forming gas, we post-process the outputs of the simulations to
account for the multi-phase interstellar medium, including the
presence of molecular hydrogen, H$_2$. The values stored in the
simulation output are based on the mass-weighted temperature between
cold and hot phases according to the
\cite{Springel-Hernquist_2003} model, and are thus expected to
underestimate the neutral hydrogen fraction. Instead, we set the
temperature of star-forming cells to $T=10^4~{\rm K}$ and re-calculate
the equilibrium neutral hydrogen fraction, also including the
self-shielding correction.

The above procedure gives the fraction of hydrogen which is neutral:
$M_{\rm NH}/M_{\rm H}$, with $M_{\rm NH}=M_{\rm HI}+M_{{\rm
    H}_2}$. We then compute the H$_2$ fraction, $f_{{\rm H}_2}=M_{{\rm
    H}_2}/M_{\rm NH}$ employing the KMT model (\citealp{Krumholz_2008,
  Krumholz_2009, McKee_2010}; see also \citealp{SternbergA_14a}), which
we briefly review here.

The molecular hydrogen fraction, $f_{{\rm H}_2}$, which we 
assume non-zero only for star-forming gas, is estimated through
\be
f_{\rm H_2} = \left\{ 
  \begin{array}{l l}	
  
    1-\frac{0.75s}{1+0.25s} & \quad \text{if $s<2$}
    \\
    0 & \quad \text{if $s\geqslant2$}\\
  \end{array} \right.
\label{eq:f_H2}
\ee
where $s$ is given by
\be
s=\frac{\log(1+0.6\chi+0.01\chi^2)}{0.6\tau_c}~,
\ee
and
\begin{eqnarray}
\chi&=&0.756(1+3.1Z^{0.365})\\
\tau_c&=&\Sigma\sigma_d/\mu_{\rm H}~.
\end{eqnarray}
In the above equations $Z$ represents the gas metallicity in units of
solar metallicity \citep{Allende_Prieto_2001}, $\sigma_d$ is the
cross-section of dust, which we estimate from
$\sigma_d=Z\times10^{-21}~{\rm cm^2}$, $\mu_{\rm H}$ is the mean mass
per hydrogen nucleus, $\mu_{\rm H}=2.3\times10^{-24}~{\rm
  g}$, and $\Sigma$ is the surface density of the gas, which we compute as
$\Sigma=\rho R$, where $\rho$ is the gas density and
$R=(3V/4\pi)^{1/3}$ with $V$ being the volume of the Voronoi cell.

\begin{figure*}
\begin{center}
\includegraphics[width=0.497\textwidth]{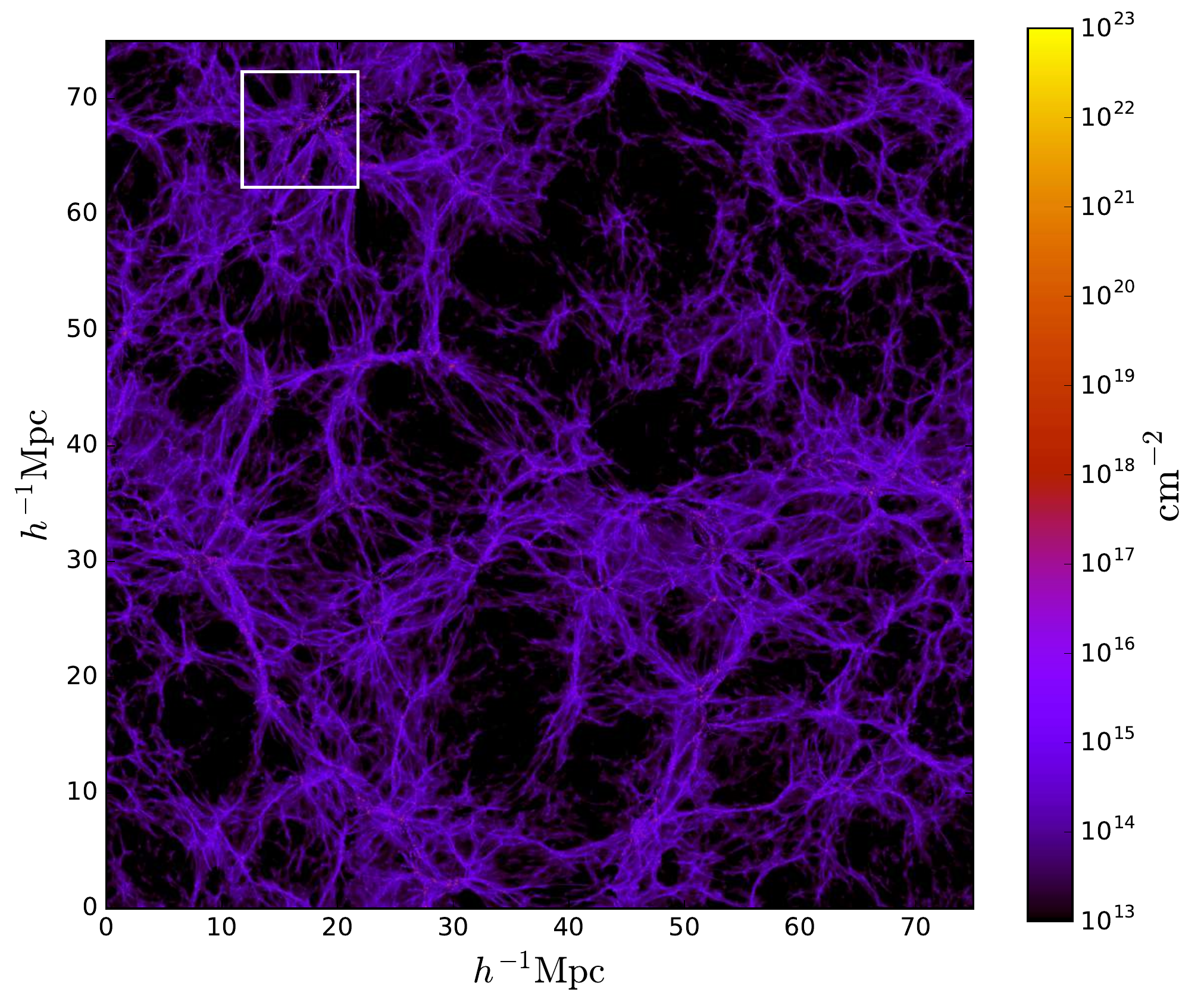}
\includegraphics[width=0.497\textwidth]{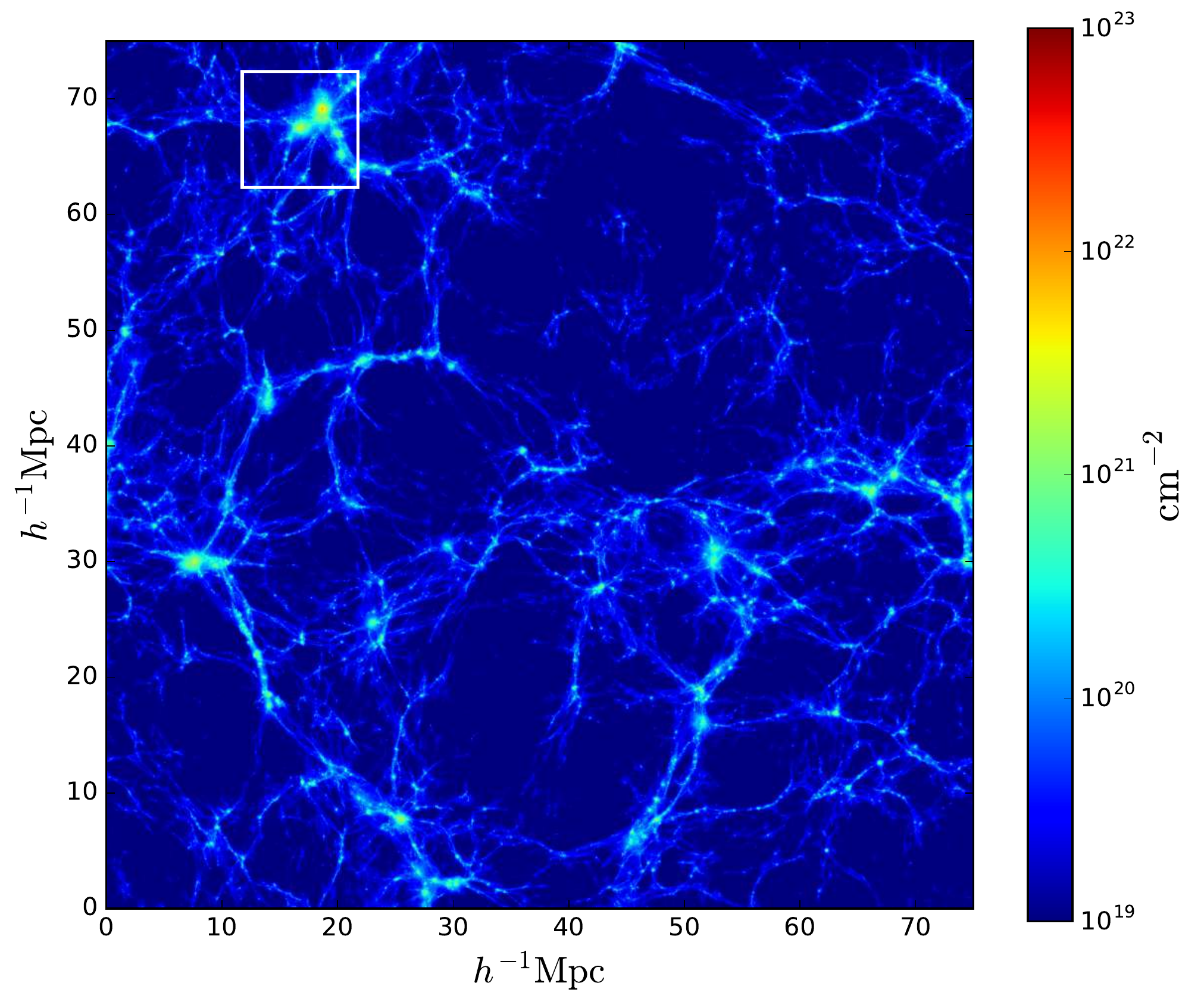}\\
\includegraphics[width=0.497\textwidth]{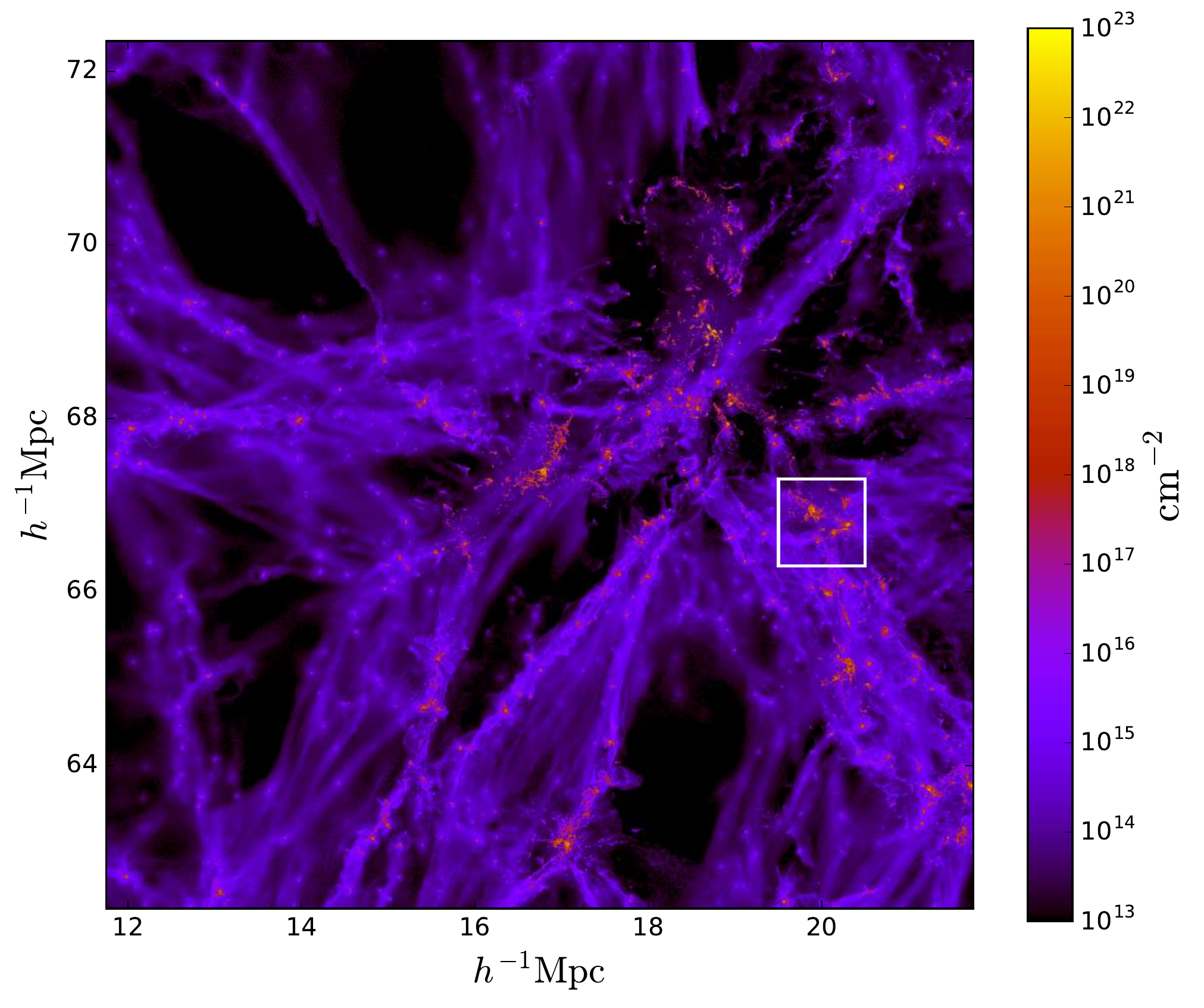}
\includegraphics[width=0.497\textwidth]{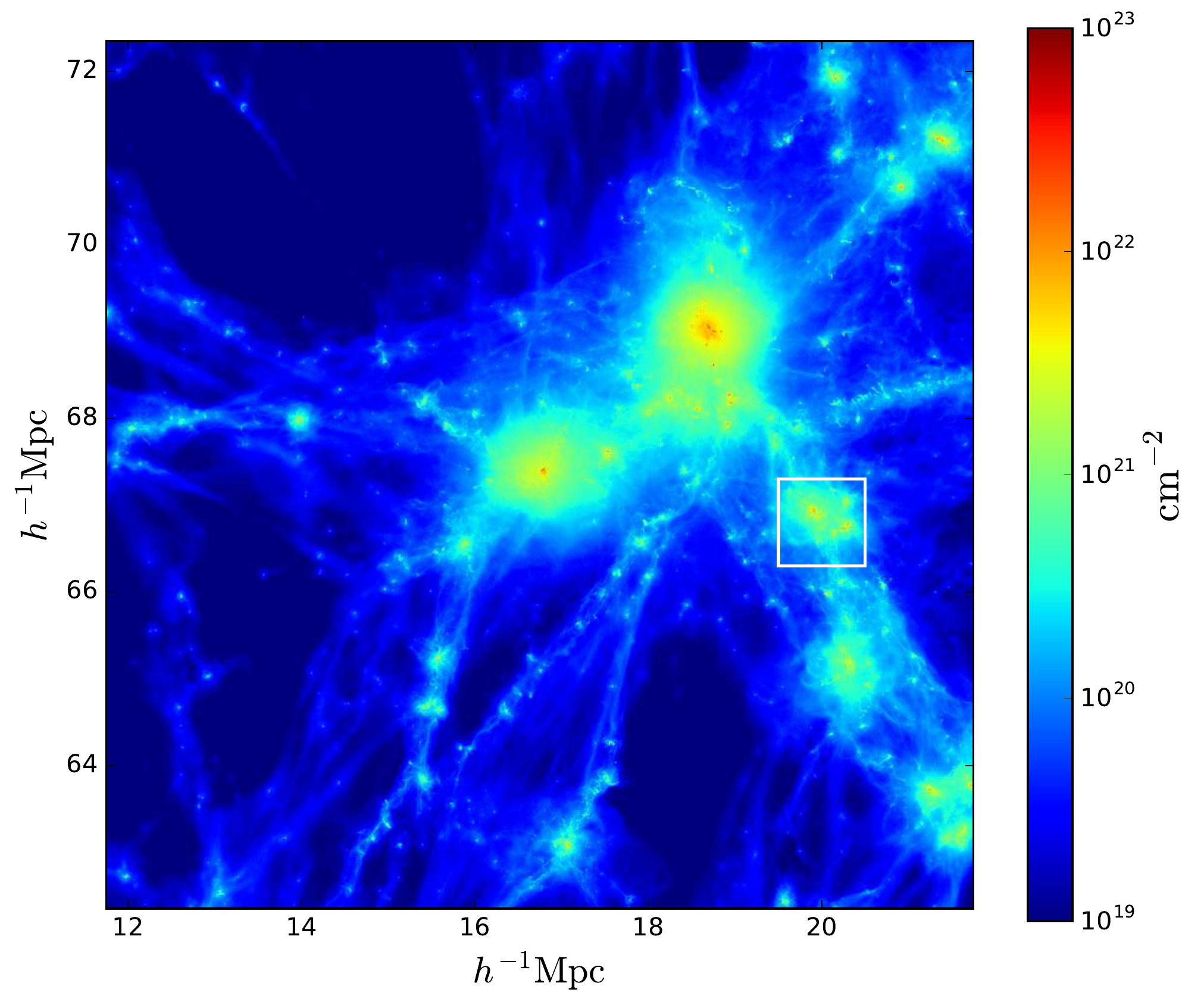}\\
\includegraphics[width=0.497\textwidth]{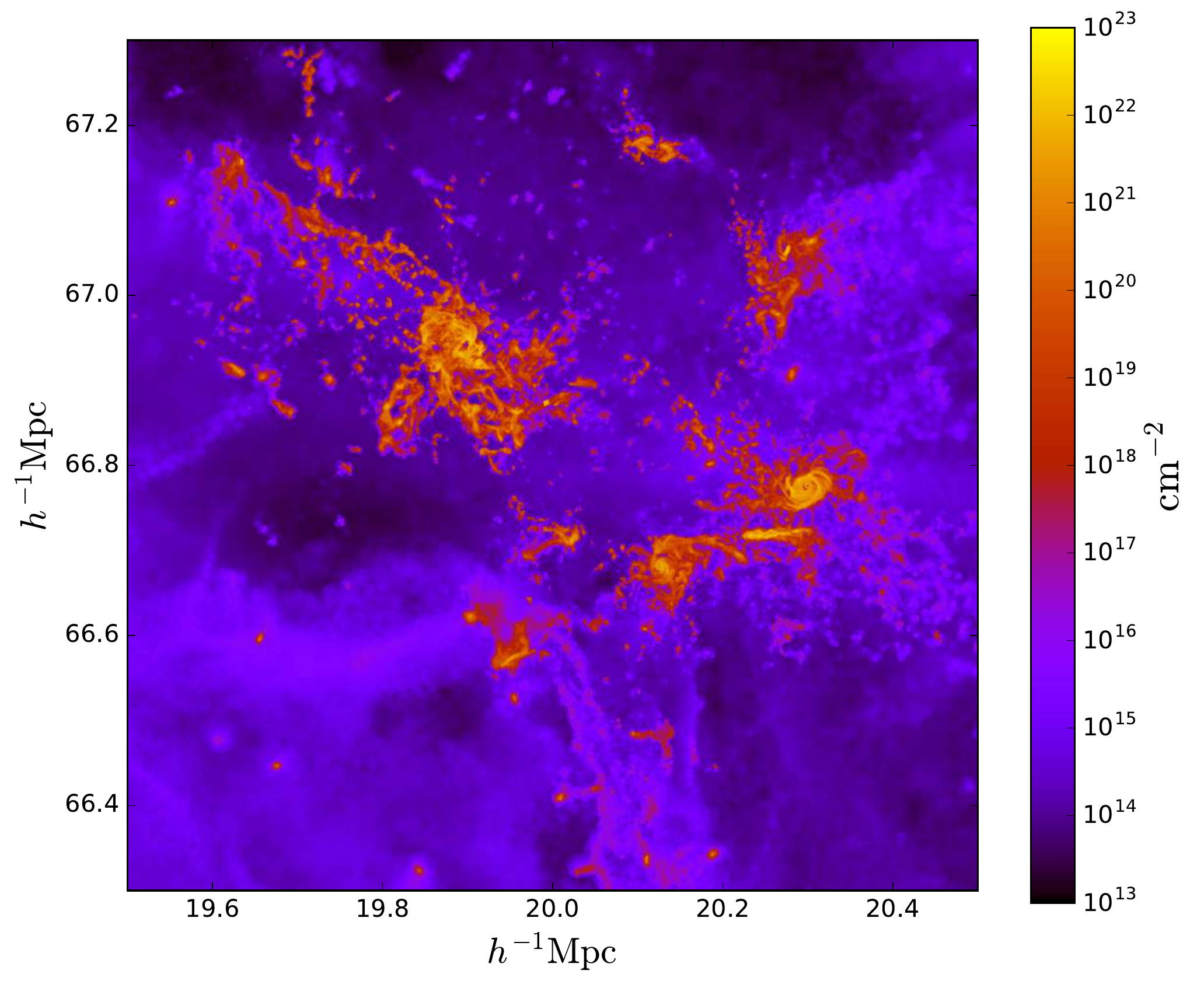}
\includegraphics[width=0.497\textwidth]{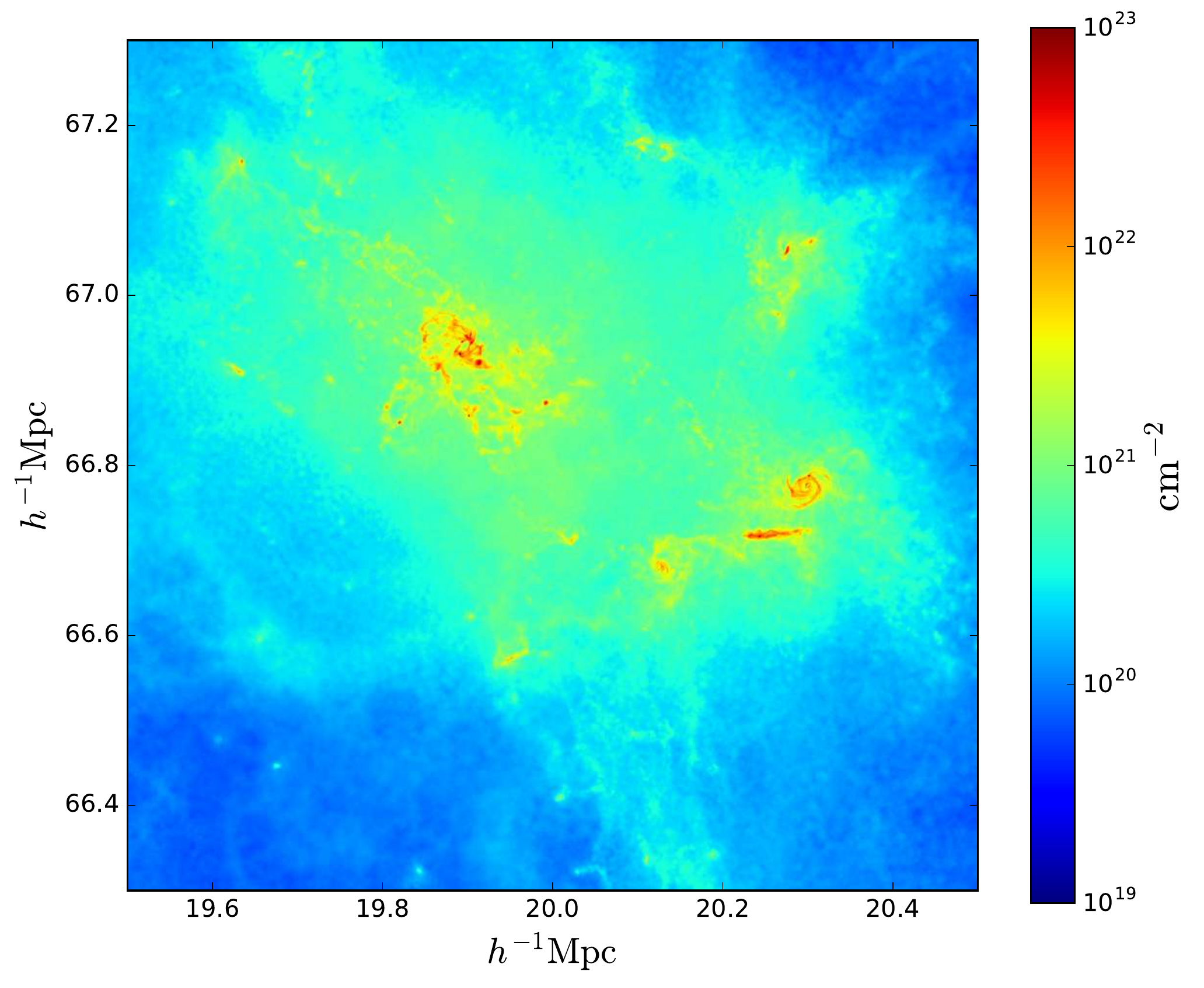}
\caption{Spatial distribution of neutral hydrogen (left) and gas (right) in slices of $5~h^{-1}{\rm Mpc}$ depth at redshift $z=1$. The upper panels show the distribution in the entire simulation volume of TNG100, while the middle and bottom panels display a zoom-in into the regions marked with a white square in the upper and middle panels, respectively. While the HI in the Ly$\alpha$-forest occupies most of the volume, the HI in galaxies represents the majority of its total mass.}
\label{fig:HI_image}
\end{center}
\end{figure*}

It is possible that our treatment may underestimate the $H_2$ fractions
since: 1) the molecular hydrogen fractions go to zero at low densities
in the KMT model, 2) we assign molecular hydrogen only to star-forming
cells, and 3) it is pessimistic to estimate the surface density from
the cell radii. However, we believe that a more precise treatment of
H$_2$ will not affect our results, as its overall
abundance is small and therefore not much HI will be transformed
into H$_2$. In order to test this more explicitly we have considered
two extreme cases in which: 1) no H$_2$ is modeled and 2) all hydrogen in
star-forming cells is in molecular form. We have computed the value
$\Omega_{\rm HI}(z)$ (see section \ref{sec:Omega_HI}) and did not find
significant changes. We thus believe that our conclusions are robust against
our H$_2$ treatment.

We note that in our approach we have considered
ionization only from the UV background. In other words, we are neglecting
the contribution of ionizing photons from, e.g., local
sources \citep{Miralda-Escude_2005, Schaye_2006, Rahmati_sources} or
X-ray heating from the intracluster medium \citep{Kannan_2015}.

Fig. \ref{fig:HI_image} shows the $z=1$ spatial distribution of HI and
gas in slices of 5 $h^{-1}{\rm Mpc}$ depth through the entire TNG100
simulation box as well as in zoomed-in regions thereof. We see that
the Ly$\alpha$-forest dominates the abundance of HI in terms of
volume, but the HI inside galaxies dominates in terms of mass.

\section{Overall HI abundance: $\Omega_{\rm HI}(\lowercase{z})$}
\label{sec:Omega_HI}

\begin{figure}
\begin{center}
\includegraphics[width=0.49\textwidth]{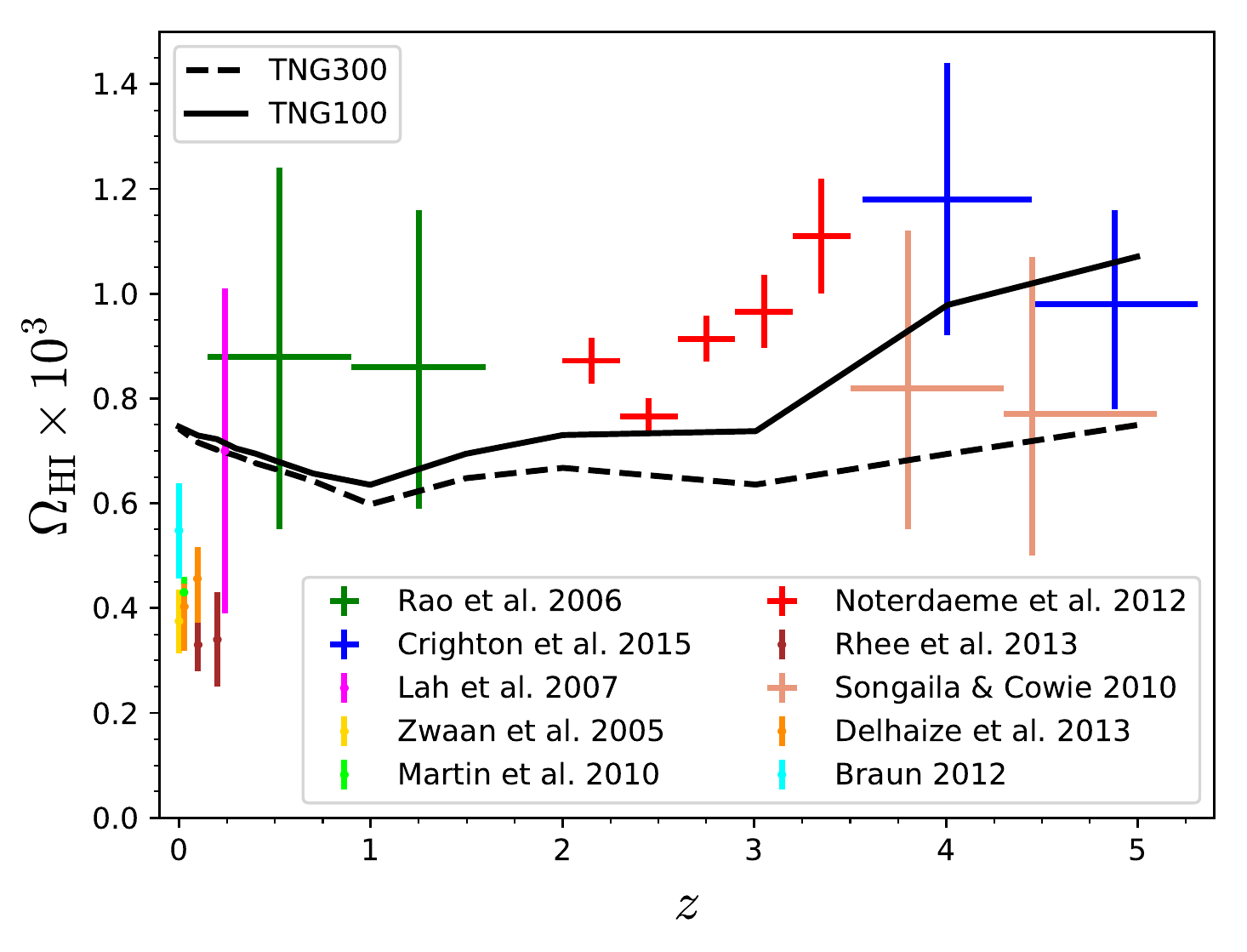}
\caption{$\Omega_{\rm HI}(z)=\rho_{\rm HI}(z)/\rho_{\rm c}^0$ from observations (colored points) and from the simulations (black lines) as a function of redshift. Our highest resolution simulation, TNG100, reproduces well the abundance of cosmic HI in the post-reionization era, although it slightly overpredicts/underpredicts the HI abundance at $z=0$/$z\in[2-3.5]$. 
A simulation with even higher mass resolution 
would likely yield an increased HI abundance.}
\label{fig:Omega_HI}
\end{center}
\end{figure}

Here, we study the overall abundance of neutral hydrogen in
the IllustrisTNG simulations. In Fig. \ref{fig:Omega_HI} we show the
value of $\Omega_{\rm HI}(z)$ from 
TNG300 (solid black) and TNG100 (dashed black). In this plot we also
indicate measurements from different observations \citep{Zwaan_2005,
  Rao_2006, Lah_2007, Songaila_2010, Martin_2010, Noterdaeme_2012,
  Braun_2012, Rhee_2013, Delhaize_2013, Crighton_2015}.

The agreement between the results from our simulations and
observations is good, although the simulations tend to overpredict the
amount of HI at redshifts $z<0.5$ and underpredict the HI abundance at
$2<z<3.5$. Compared to earlier studies with hydrodynamic simulations
\citep[e.g.][]{DaveR_13a, Bird_2014} and semi-analytic models
\citep{Lagos_2014}, however, our results agree better with
observations at $z>1$ and comparable to the agreement found by
\citep{Rahmati_2015}.
 
The overall HI mass in the simulations depends on 
resolution, such that the simulation with higher resolution, TNG100,
contains between $\simeq2.5\%$ and $\simeq40\%$ more HI in the
redshift interval $z=0 - 5$ than TNG300. This is a consequence of
the fact that the stellar mass function is not yet converged
\citep{PillepichA_17a}.
For example, in TNG100 there is HI in halos with
masses that TNG300 cannot resolve (see section
\ref{subsec:M_HI}). This can be seen in Fig. \ref{fig:M_HI},
where we show the HI mass within halos versus total halo mass. In
TNG300, halos with masses only above $2\times10^9~h^{-1}M_\odot$ can
be resolved, assuming a minimum of 50 CDM particles in
a halo.  From Fig. \ref{fig:M_HI} we see that the amount of HI in
halos below that mass is not negligible at high-redshift, thus we
would expect that $\Omega_{\rm HI}$ will be lower at high-redshift in
TNG300 in comparison with TNG100, as we find.

In this paper we examine the most important properties of
cosmic neutral hydrogen over a wide range of redshifts. Not being able
to resolve the HI that is contained within the smallest halos impacts our results
in several ways. For example, the values of the HI bias, HI shot-noise,
the HI halo mass function or the amplitude of the HI Fingers-of-God effect
will be affected by this. For this reason, from now on we
focus our analysis to the TNG100 simulation.

\section{HI fraction in halos and galaxies}
\label{subsec:HI_in_halos_galaxies}

The fraction of the total HI mass that resides within halos is an
important ingredient for theoretical frameworks that aim at modeling
the abundance and clustering properties of cosmic neutral hydrogen,
such as the halo model \citep{Cooray_Sheth_2002, Barnes_Haehnelt_2014, Villaescusa-Navarro_2014a, Padmanabhan_2015, 
Padmanabhan_2016, Padmanabhan_2017}. In particular, these methods make the
assumption that all HI is confined within halos, whose properties,
such as spatial distribution or abundance, are well-described by
analytic models and/or numerical simulations.

\begin{figure}
\begin{center}
\includegraphics[width=0.49\textwidth]{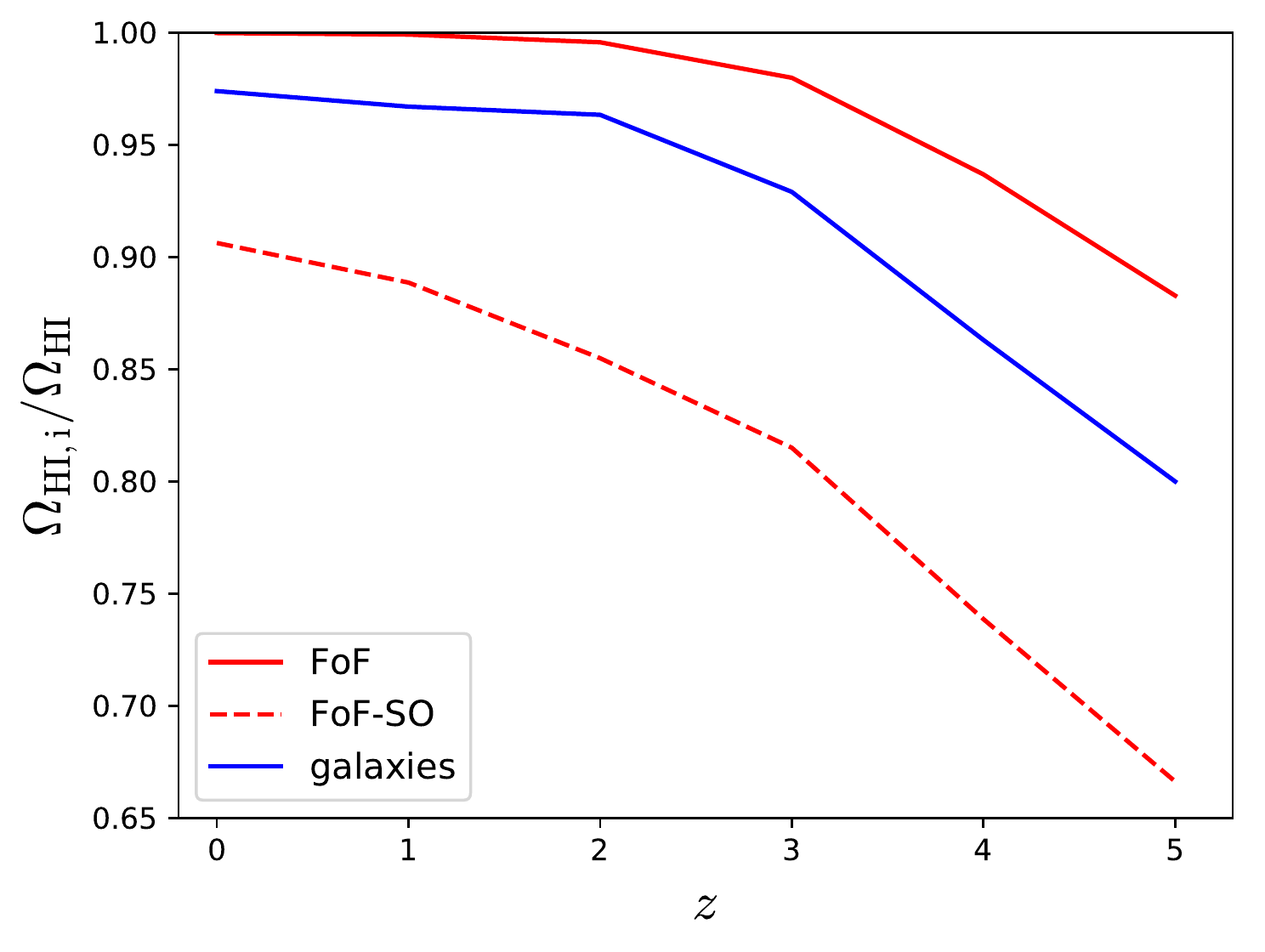}
\caption{The fraction of total HI mass that is inside FoF halos (solid red lines), FoF-SO halos (dashed red lines) and galaxies (blue lines) as a function of redshift. At low redshift nearly all HI is located within FoF halos and galaxies, while at high-redshift the amount of HI outside FoF halos/galaxies can be 10\%/20\%. There is a significant amount of HI in the outskirts of FoF halos: the fraction of HI inside FoF-SO halos ranges from $90\%$ at $z=0$ to $67\%$ at z=5.}
\label{fig:HI_halos_galaxies}
\end{center}
\end{figure}

In contrast to the gas in halos, the properties of the gas in the
intergalactic medium (IGM) are more difficult to model analytically
\cite[see however][]{Vid_2018}. The standard approach has been to
characterize the gas in the IGM using numerical
simulations. If a significant amount of HI is found outside halos, any
standard HI halo model will need to be complemented with either
simulations or further analytic ingredients. Below, we 
determine the amount of HI that is outside of halos, to
quantify the limitations of standard HI halo models.

We have computed the HI mass inside each FoF, FoF-SO and galaxy in the
simulation at several redshifts. In Fig. \ref{fig:HI_halos_galaxies}
we show, for each of these object types, the fraction of the total HI
mass in the simulation that resides inside all objects combined.

We find that at redshifts $z\leqslant2$ more than $99\%$ of all HI is
contained within FoF halos. While a significant fraction of the baryons
lie outside these regions, the IGM is highly ionized at these times.
At these redshifts, the fraction of HI
within galaxies is larger than $95\%$. We note that \textsc{subfind}
may not identify any subhalo/galaxy within a FoF halo. This could
happen for several reasons, like low-density or virialization not
having been reached. Thus, we conclude that at these redshifts $\simeq5\%$ of the
cosmic neutral hydrogen is outside galaxies\footnote{We emphasize that
  the term ``galaxy'' should be considered in our framework, not in
  the traditional observational definition. For instance, gas far away
  from the center of a halo but gravitationally bound to it will still
  be considered as belonging to that galaxy.}  but inside halos.

The fraction of HI within FoF halos and galaxies decreases
monotonically with redshift. At redshift $z=5$ the HI inside FoF halos
only accounts for $88\%$ of the total HI, while the mass within
galaxies is $80\%$. These results are in qualitative agreement with
\cite{Villaescusa-Navarro_2014a}, who studied the HI outside halos
using a different set of hydrodynamic simulations. Morever, we find
that our results are not significantly affected by mass resolution,
since the same analysis carried out for the TNG300
simulation gives similar results.

We consider our finding that the fraction of HI outside halos
increases with redshift to be reasonable. At high-redshift the gas in the
IGM is denser and the amplitude of the UV background is lower, and so
it is easier for that gas to host higher fractions of neutral hydrogen
(see Appendix \ref{sec:HI_Lya} for further details).

The fraction of HI outside FoF-SO halos is not negligible, varying from
$10\%$ at $z=0$ to $33\%$ at $z=5$. On average, the ratio
of HI mass in FoF-SO halos to that in FoF halos is
similar to the ratio between their total masses.
Thus, FoFs host more HI that FoF-SO halos simply because
they are larger and more massive. This also tells
us that the regions beyond the virial radius of typical halos are
neither HI poor nor HI rich, while when this is examined specifically
in massive halos we find these regions to be HI rich (see appendix
\ref{sec:FoF_vs_SO} for further details).

We thus conclude that while the standard assumption
that all HI lies within halos is reasonable at $z\leqslant2$, at high
redshift it begins to break down since a small fraction is
located outside halos ($\sim10\%$ at $z=5$). The numbers derived here
can be used to quantify the limitations of HI halo models that target
the distribution of HI at high redshift.

\section{Halo HI mass function}
\label{subsec:M_HI}

\begin{figure*}
\begin{center}
\includegraphics[width=0.33\textwidth]{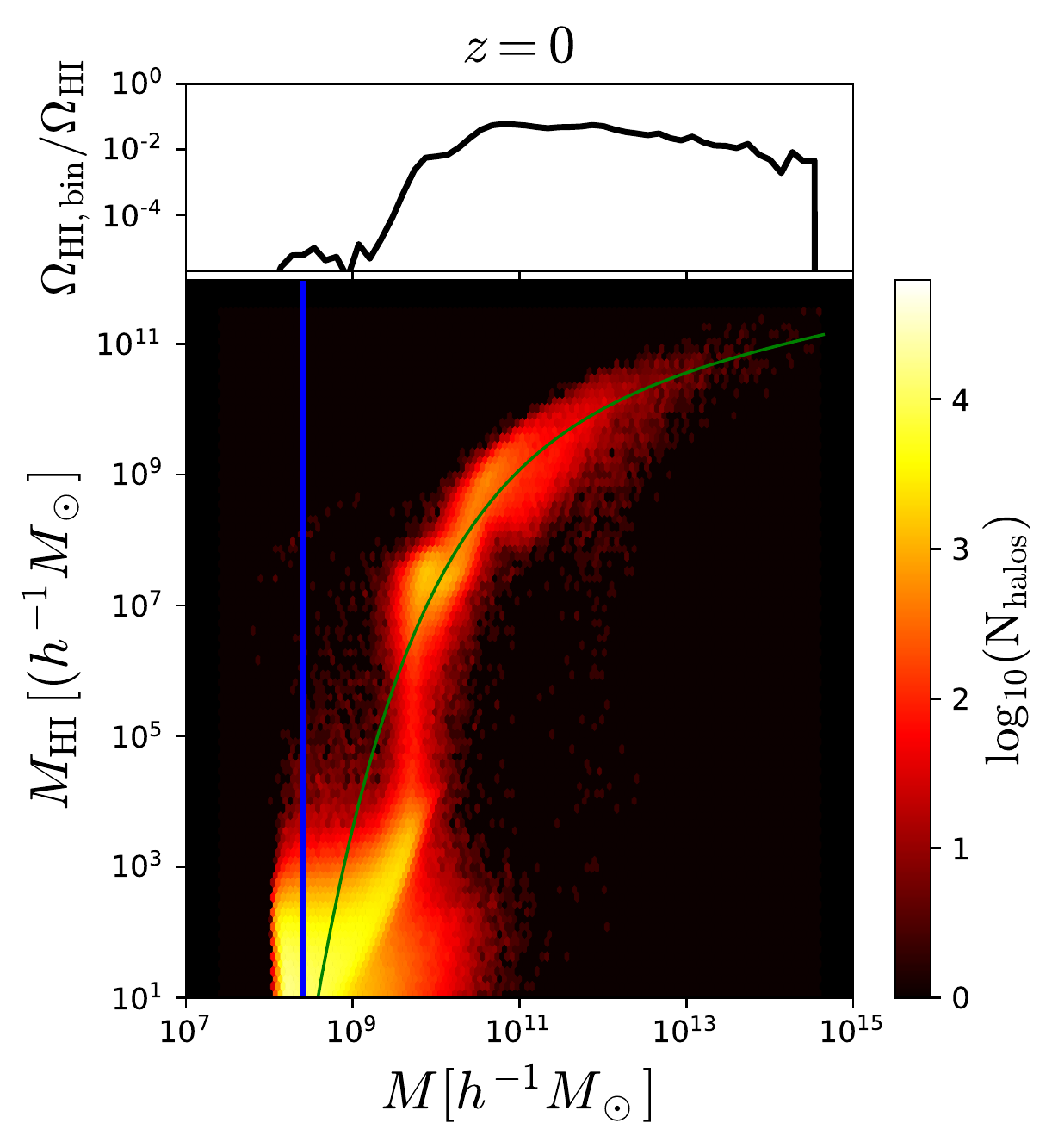}
\includegraphics[width=0.33\textwidth]{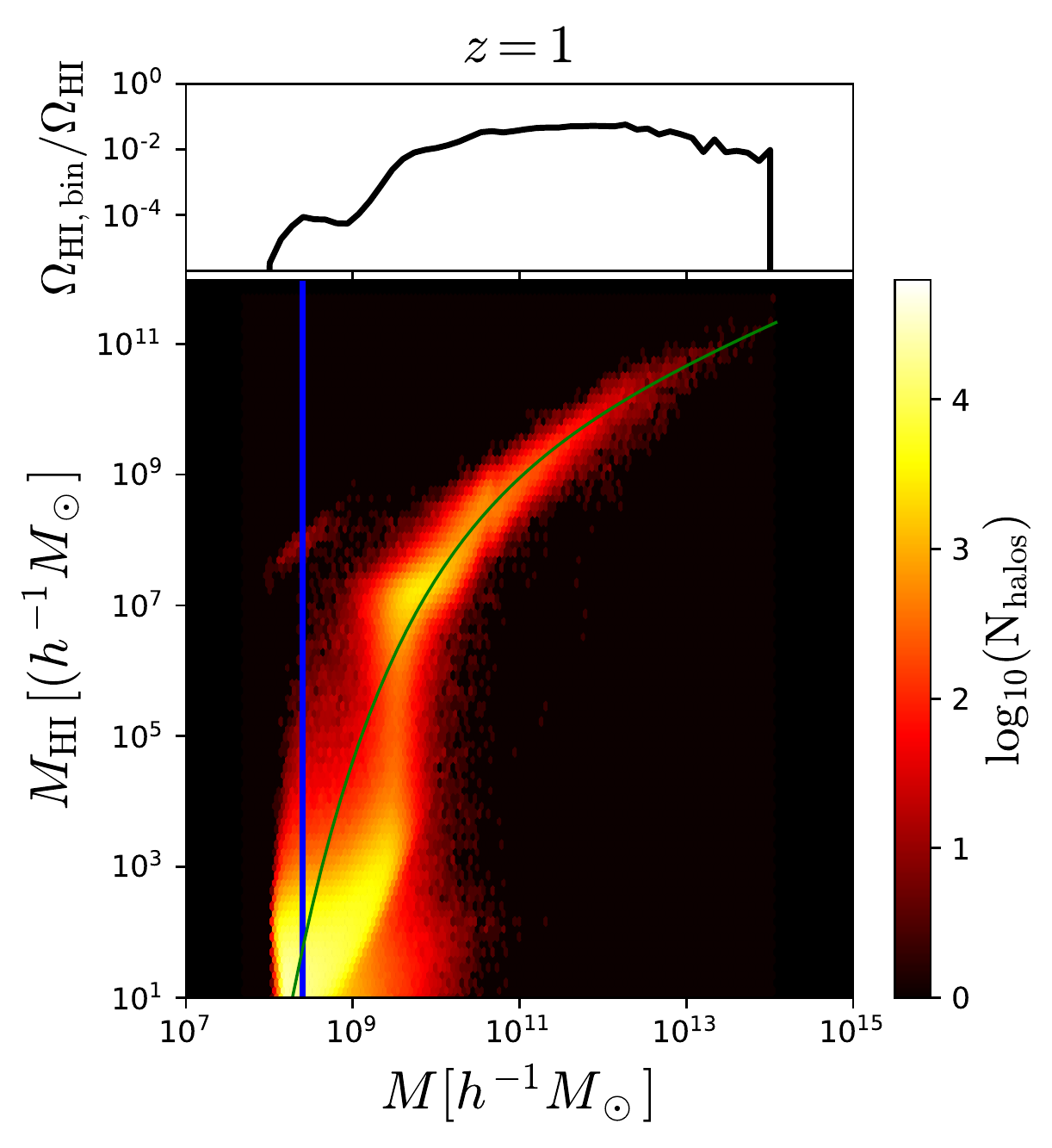}
\includegraphics[width=0.33\textwidth]{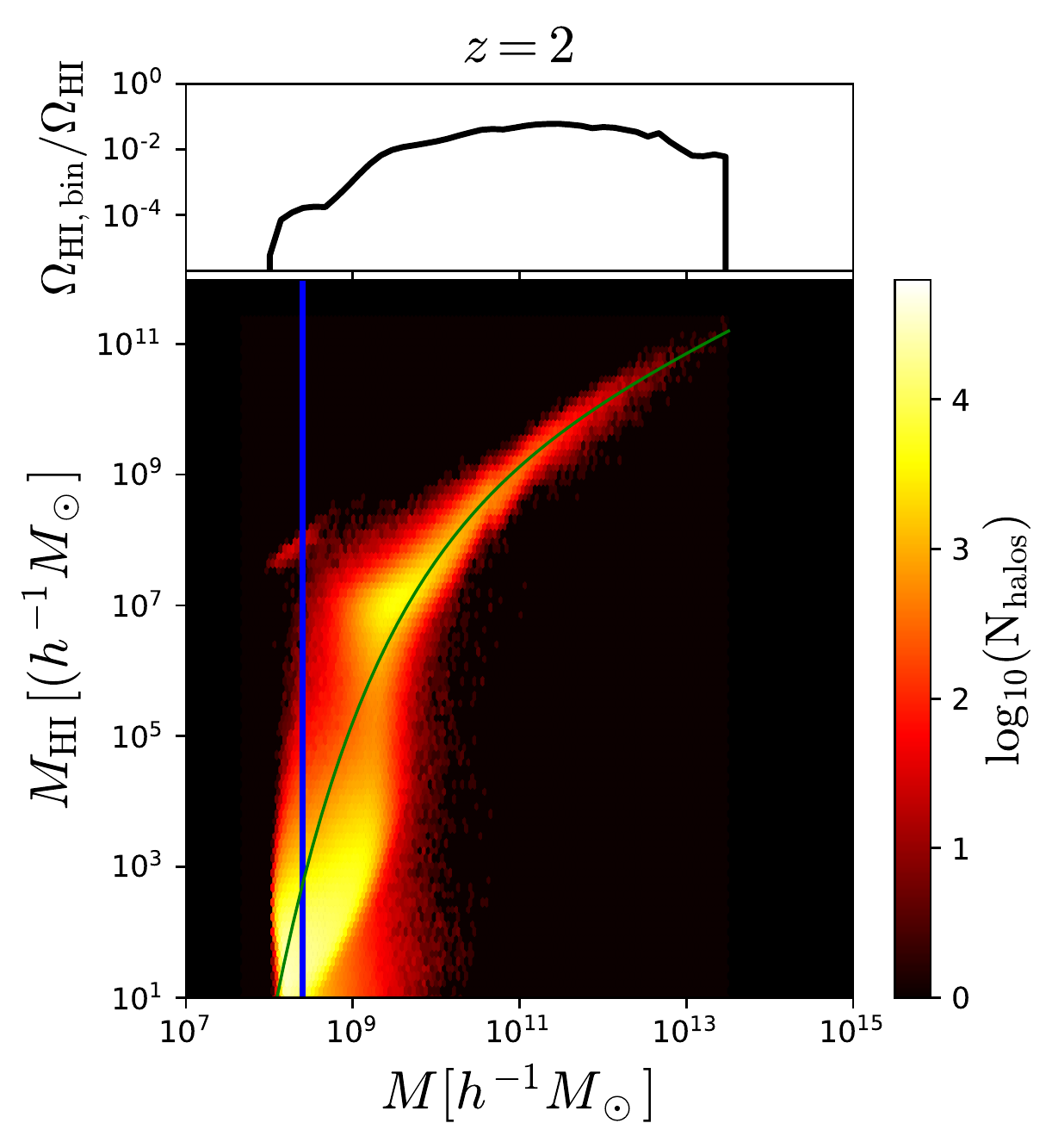}\\
\includegraphics[width=0.33\textwidth]{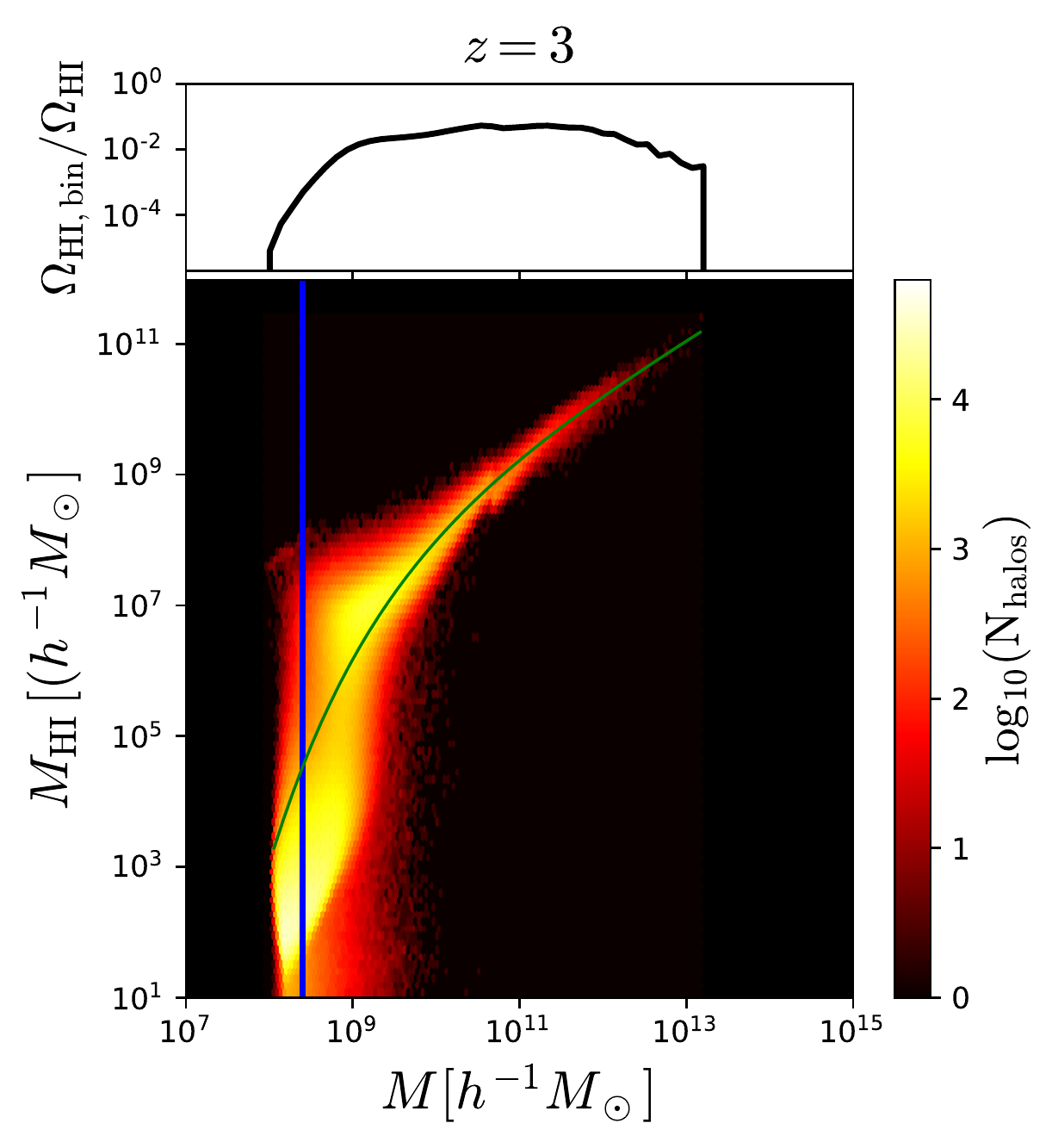}
\includegraphics[width=0.33\textwidth]{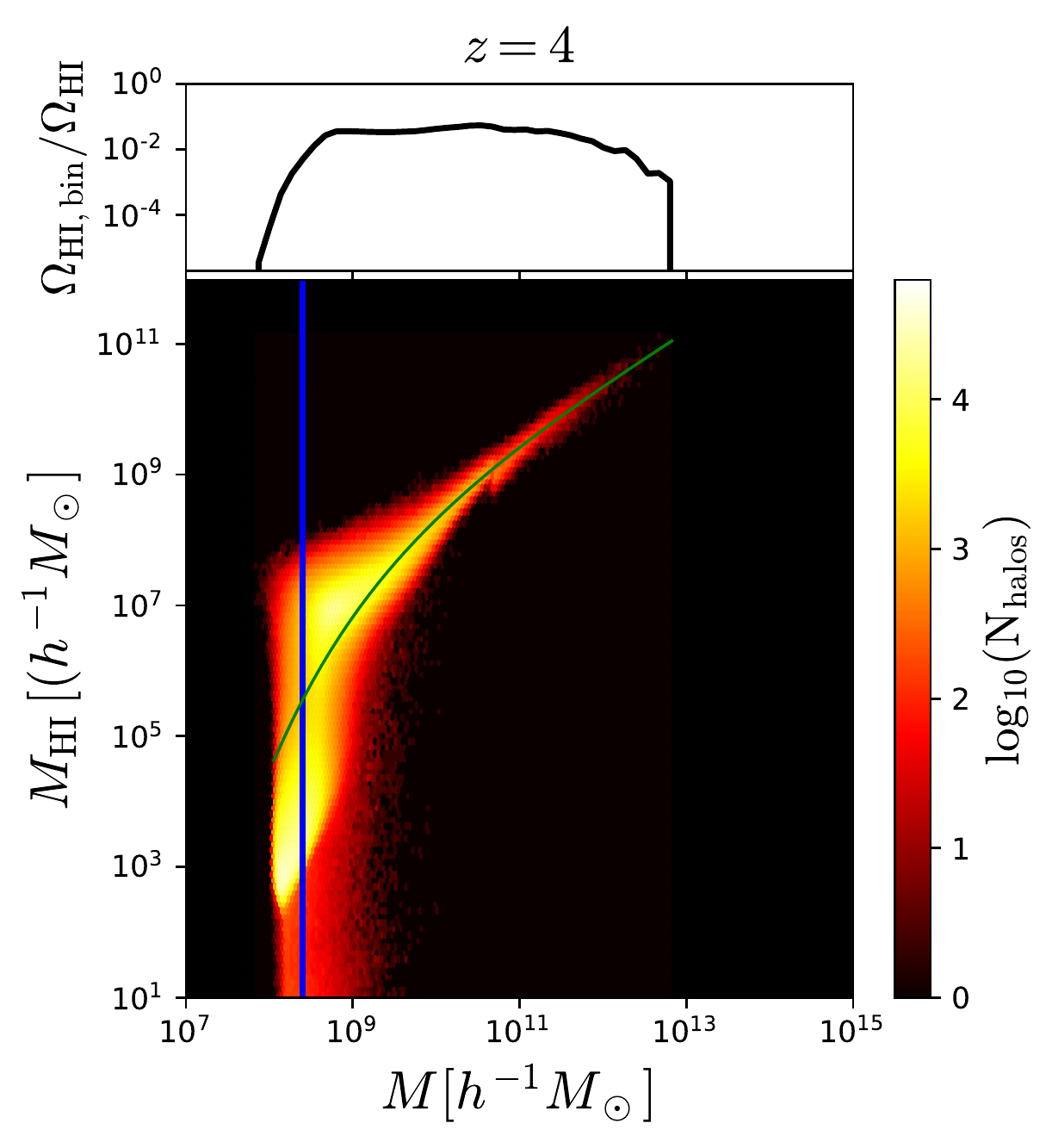}
\includegraphics[width=0.33\textwidth]{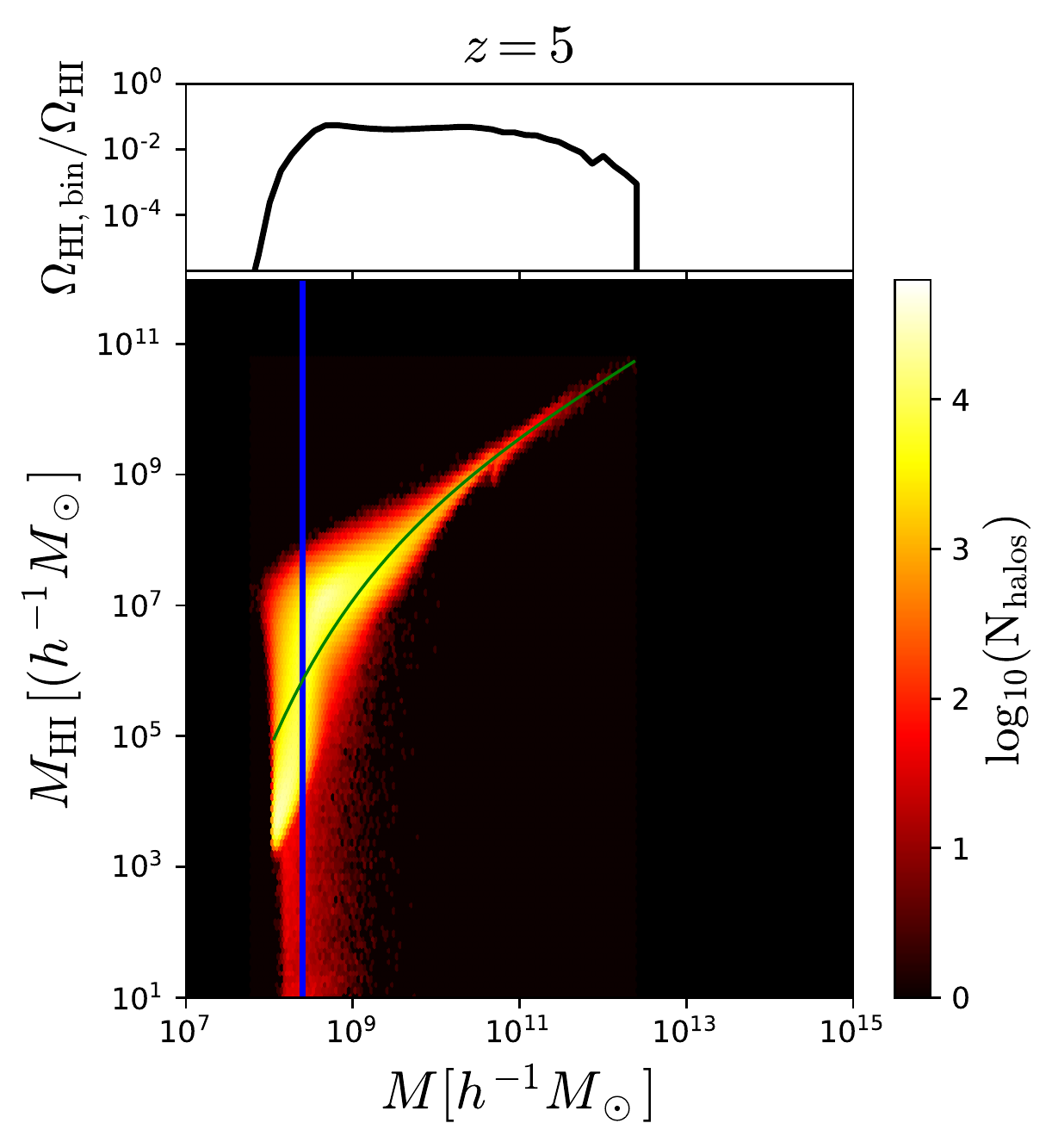}
\caption{Halo HI mass function. For each FoF halo of the simulation TNG100 we compute the total HI mass it hosts. The plots show the HI mass versus halo mass, color coded by the number of halos in each bin. The blue vertical lines show the mass corresponding to a halo that hosts 50 dark matter particles, which we adopt a rough mass resolution threshold. We take narrow bins in halo mass and compute the total HI mass in each of them. The top part of each panel shows the ratio between the total HI mass in the bin and the total HI mass in all halos. Our results can be well reproduced by the fitting formula $M_{\rm HI}(M,z)=M_0x^\alpha e^{-1/x^{0.35}}$, with $x=M/M_{\rm min}$. The best fits are shown with green lines at each redshift. }
\label{fig:M_HI}
\end{center}
\end{figure*}
\begin{table*}
\begin{center}
\resizebox{1.0\textwidth}{!}{\begin{tabular}{|c|| c| c| c|| c||| c|c|c||c|} 
 \hline
 & \multicolumn{4}{|c|||}{FoF} & \multicolumn{4}{|c|}{FoF-SO}\\[0.5ex]
 \hline
 $z$ & $\alpha$ & $M_0$ & $M_{\rm min}$ & $M_{\rm hard}$ & $\alpha$ & $M_0$ & $M_{\rm min}$ & $M_{\rm hard}$\\ [0.5ex] 
 & & $[h^{-1}M_\odot]$ & $[h^{-1}M_\odot]$ & $[h^{-1}M_\odot]$& & $[h^{-1}M_\odot]$ & $[h^{-1}M_\odot]$ & $[h^{-1}M_\odot]$ \\[0.5ex] 
 \hline\hline
 
 0 & $0.24\pm0.05$ & $(4.3\pm1.1)\times10^{10}$ & $(2.0\pm0.6)\times10^{12}$ & $1.5\times10^{10}$&
        $0.16\pm0.05$ & $(4.1\pm1.0)\times10^{10}$ & $(2.4\pm0.7)\times10^{12}$ & $1.3\times10^{10}$\\ 
 
 \hline
1 & $0.53\pm0.06$ & $(1.5\pm0.7)\times10^{10}$ & $(6.0\pm2.9)\times10^{11}$ & $6.9\times10^{9}$&
      $0.43\pm0.06$ & $(1.8\pm0.8)\times10^{10}$ & $(8.6\pm4.2)\times10^{11}$ & $6.1\times10^{9}$\\
 
 \hline
2 & $0.60\pm0.05$ & $(1.3\pm0.6)\times10^{10}$ & $(3.6\pm1.6)\times10^{11}$ &$3.1\times10^{9}$ &
       $0.51\pm0.05$ & $(1.5\pm0.7)\times10^{10}$ & $(4.6\pm2.1)\times10^{11}$ & $2.5\times10^{9}$\\
 
 \hline
 3 & $0.76\pm0.05$ & $(2.9\pm2.0)\times10^9$ &  $(6.7\pm4.0)\times10^{10}$ & $9.9\times10^{8}$&
        $0.69\pm0.06$ & $(3.7\pm2.6)\times10^9$ &  $(9.6\pm6.0)\times10^{10}$ & $7.6\times10^{8}$\\
 
\hline
4 & $0.79\pm0.04$ & $(1.4\pm1.0)\times10^9$ &  $(2.1\pm1.3)\times10^{10}$ & $3.9\times10^{8}$&
      $0.61\pm0.06$ & $(4.5\pm2.7)\times10^9$ &  $(7.6\pm4.4)\times10^{10}$ & $2.3\times10^{8}$\\ 
 
 \hline
5 & $0.74\pm0.04$ & $(1.9\pm1.2)\times10^9$ &  $(2.0\pm1.2)\times10^{10}$ & $2.7\times10^{8}$&
      $0.59\pm0.07$ & $(4.1\pm2.8)\times10^9$ &  $(5.4\pm3.6)\times10^{10}$ & $1.7\times10^{8}$\\
 
 \hline
\end{tabular}}
\caption{We fit our results for the $M_{\rm HI}(M,z)$ function to the form $M_0x^\alpha\exp(-1/x^{0.35})$, where $x=M/M_{\rm min}$. This table shows the best-fit value of the free parameters, for FoF and FoF-SO halos, at different redshifts. The column $M_{\rm hard}$ indicates the value of our hard cutoff mass, which is defined so that halos with masses $M\geqslant M_{\rm hard}$ host $98\%$ of all HI in halos. For FoF-SO halos we can express $M_{\rm hard}$ in terms of circular velocities, giving 34, 35, 31, 24, 19 and 18 km/s at redshifts 0, 1, 2, 3, 4 and 5, respectively.}
\label{table:M_HI_fit}
\end{center}
\end{table*}

In the previous section we have shown that most of the HI is inside
halos, justifying the use of HI halo models to characterize the
spatial distribution of HI.  As discussed in the introduction, besides
the linear matter power spectrum, halo mass function and halo bias, we
need to know the halo HI mass function (i.e.~the average HI mass
hosted by a halo of mass $M$ at redshift $z$) and the spatial
distribution of HI inside halos. Below, we investigate the
former: $M_{\rm HI}(M,z)$.

We emphasize the paramount importance of this function by noting that
knowing it is sufficient for predicting the amplitude and shape of the
21 cm power spectrum to linear order (see Eq. \ref{Eq:P_21cm}):
\be
P_{\rm 21cm}(k,\mu,z)=\bar{T}_b(z)^2\left[(b_{\rm HI}(z)+f(z)\mu^2)^2P_{\rm m}(k,z)+P_{\rm SN}(z)\right]~,\nonumber\\
\ee
where $\bar{T}_b(z)$, $b_{\rm HI}(z)$ and $P_{\rm SN}(z)$ can all be derived from $M_{\rm HI}(M,z)$ as
\begin{align}
\bar{T}_b(z)&=189h\left(\frac{H_0(1+z)^2}{H(z)}\right)\Omega_{\rm HI}(z) ~{\rm mK}\\
\Omega_{\rm HI}(z)&=\frac{1}{\rho_{\rm c}^0}\int_0^\infty n(M,z)M_{\rm HI}(M,z)dM\\
b_{\rm HI}(z)&=\frac{1}{\rho_{\rm c}^0\Omega_{\rm HI}(z)}\int_0^\infty n(M,z)b(M,z)M_{\rm HI}(M,z)dM\\
P_{\rm SN}(z)&=\frac{1}{(\rho_{\rm c}^0\Omega_{\rm HI}(z))^2}\int_0^\infty n(M,z)M_{\rm HI}^2(M,z)dM~,
\label{eq:M_HI_terms}
\end{align}
where $n(M,z)$ and $b(M,z)$ are the halo mass function and halo bias, respectively. Knowledge of this function can be used to understand the impact of different phenomena on the amplitude and shape of the 21cm power spectrum such as neutrino masses \citep{Villaescusa-Navarro_2015a}, warm dark matter \citep{Carucci_2015} or modified gravity \citep{Carucci_2017}.

For each dark matter halo in the simulation we have computed its
enclosed HI mass. In Fig. \ref{fig:M_HI} we show the HI mass versus
halo mass for each single FoF halo in the simulation at redshifts 0,
1, 2, 3, 4 and 5. The color indicates the number of halos in each
bin. We show this map rather than the $M_{\rm HI}(M,z)$ function
since the former contains more information, such as the scatter in
$M_{\rm HI}(M,z)$.

The halo HI mass function increases monotonically with
halo mass. Two trends can be identified: 1) in the high-mass end
$M_{\rm HI}(M,z)$ can be approximated by a power law, and 2) in the
low-mass end it has a sharp cutoff. A good fit to our
results is given by
\be
M_{\rm HI}(M,z)=M_0\left(\frac{M}{M_{\rm min}}\right)^\alpha \exp(-(M_{\rm min}/M)^{0.35})~.
\label{Eq:M_HI}
\ee
The free parameters are $M_{\rm min}$, which sets the cutoff mass in
$M_{\rm HI}(M,z)$, $\alpha$, which controls the slope of the function at
the high-mass end, and $M_0$, which determines the overall normalization
and represents $\simeq40\%$ of the HI mass of a halo of mass $M_{\rm
  min}$. We have fitted our results to this function and give the best fitting 
values for both FoF and FoF-SO halos in Table
\ref{table:M_HI_fit}. The green lines in Fig. \ref{fig:M_HI} indicate the best
fits at each redshift.

At redshifts $z\geqslant3$ for FoF halos 
$\alpha$ is $\simeq0.75$, while it declines at lower
redshifts: $\alpha=0.60$ at $z=2$, $\alpha=0.53$ at $z=1$ and
$\alpha=0.24$ at $z=0$. We interpret this as a result of several
physical processes such as AGN feedback, ram pressure and tidal
stripping being more efficient at removing gas from galaxies at low
redshift than at higher redshifts.

The value of $M_{\rm min}$ decreases monotonically with redshift, from
$\simeq2\times10^{12}~h^{-1}M_\odot$ at $z=0$ to
$2\times10^{10}~h^{-1}M_\odot$ at $z=5$. This indicates that as the
redshift increases, lower mass halos host HI. 
In the appendix \ref{sec:UVB_cutoff} we discuss the
physical origin of the cutoff in the halo HI mass function and the
relative importance of supernova feedback, gas stripping and the UV
background for this.

Similar conclusions can be reached when computing $M_{\rm
  HI}(M,z)$ using FoF-SO halos (see Table \ref{table:M_HI_fit}). The
derived values of $\alpha$, $M_0$ and $M_{\rm min}$ are roughly
compatible, within the errors, between FoF and FoF-SO halos. There are,
however, systematic differences between the best fit
values from FoF and SO, which is because the total
amount of HI in a given halo is sensitive to its definition as we
will see below.

For FoF-SO halos we can relate halo masses to circular velocities
through $V_{\rm circ}=\sqrt{GM/R}$, where $G$ is the gravitational
constant and $M$ and $R$ are the halo mass and radius. Expressing
$M_{\rm min}$ in terms of circular velocities we obtain: $V_{\rm
  circ}(M_{\rm min})\simeq180\pm20$ km/s for $z\in[0,2]$ and $V_{\rm
  circ}(M_{\rm min})\simeq120\pm20$ km/s for $z\in[3,5]$. This
suggests that the minimum halo mass that can host HI depends primarily
on the depth of its gravitational potential and that at lower
redshifts the potential has to be deeper since astrophysical processes
such as AGN feedback, tidal stripping, and so forth are more effective
at removing gas from small halos.

Since our parametrization of the halo HI mass function
does not have a ``hard'' cutoff, the value of $M_{\rm min}$ only
represents a mass scale at which the halo HI mass function changes its
trend. In other words, halos with masses around $M_{\rm min}$ host a
significant amount of HI. It is also very interesting to quantify the
cutoff in the halo HI mass function more rigidly, i.e. so that
halos below a certain mass contain a negligible amount of HI.

We have calculated the halo mass at which $98\%$ of all HI in halos is
above that mass, and only $2\%$ is in smaller halos. We term this halo
mass as a ``hard cutoff mass'', $M_{\rm hard}$,
\be
\frac{\int_0^{M_{\rm hard}} n(M,z)M_{\rm HI}(M,z)dM}{\int_0^\infty n(M,z)M_{\rm HI}(M,z)dM}=0.02
\label{Eq:Hard_cutoff}
\ee
and we show corresponding values in Table \ref{table:M_HI_fit}. For FoF-SO halos
we obtain: $1.3\times10^{10}$, $6.1\times10^{9}$, $2.5\times10^9$,
$7.6\times^8$, $2.3\times10^8$ and $1.7\times10^8~h^{-1}M_\odot$ at
redshifts 0, 1, 2, 3, 4 and 5, respectively. These can be transformed to 
circular velocities, giving: 34, 35, 31, 24, 19 and 18 km/s
at redshifts 0, 1, 2, 3, 4 and 5, correspondingly. The values we infer 
do not change much if we use a threshold equal to $99\%$. We thus
conclude that at redshifts $z\leqslant2$ only halos with circular
velocities above about 30 km/s host HI, while at redshifts $z\geqslant3$
the HI is only in halos with circular velocities above $\sim20$ km/s.

A more conventional parametrization of the halo HI mass function
\be
M_{\rm HI}(M,z)=M_0\left(\frac{M}{M_{\rm min}}\right)^\alpha \exp(-M_{\rm min}/M)~.
\ee
also reproduces our results. The fit in Eq. \ref{Eq:M_HI} is,
however, preferred for our results since at high-redshift $M_{\rm
  HI}(M,z)$ falls more slowly for very low halo masses than the 
standard profile. In order to facilitate the comparison with works using the
above parametrization \citep[e.g.][]{Bagla_2010, EmaPaco,
  Obuljen_17,Paco_Alonso,Penin_17,Padmanabhan_2017} we also provide
the best-fit values for the more conventional fit in the Appendix
\ref{sec:HHIMF}.

\section{HI density profile}
\label{sec:HI_profile}

\begin{figure*}
\begin{center}
\includegraphics[width=1\textwidth]{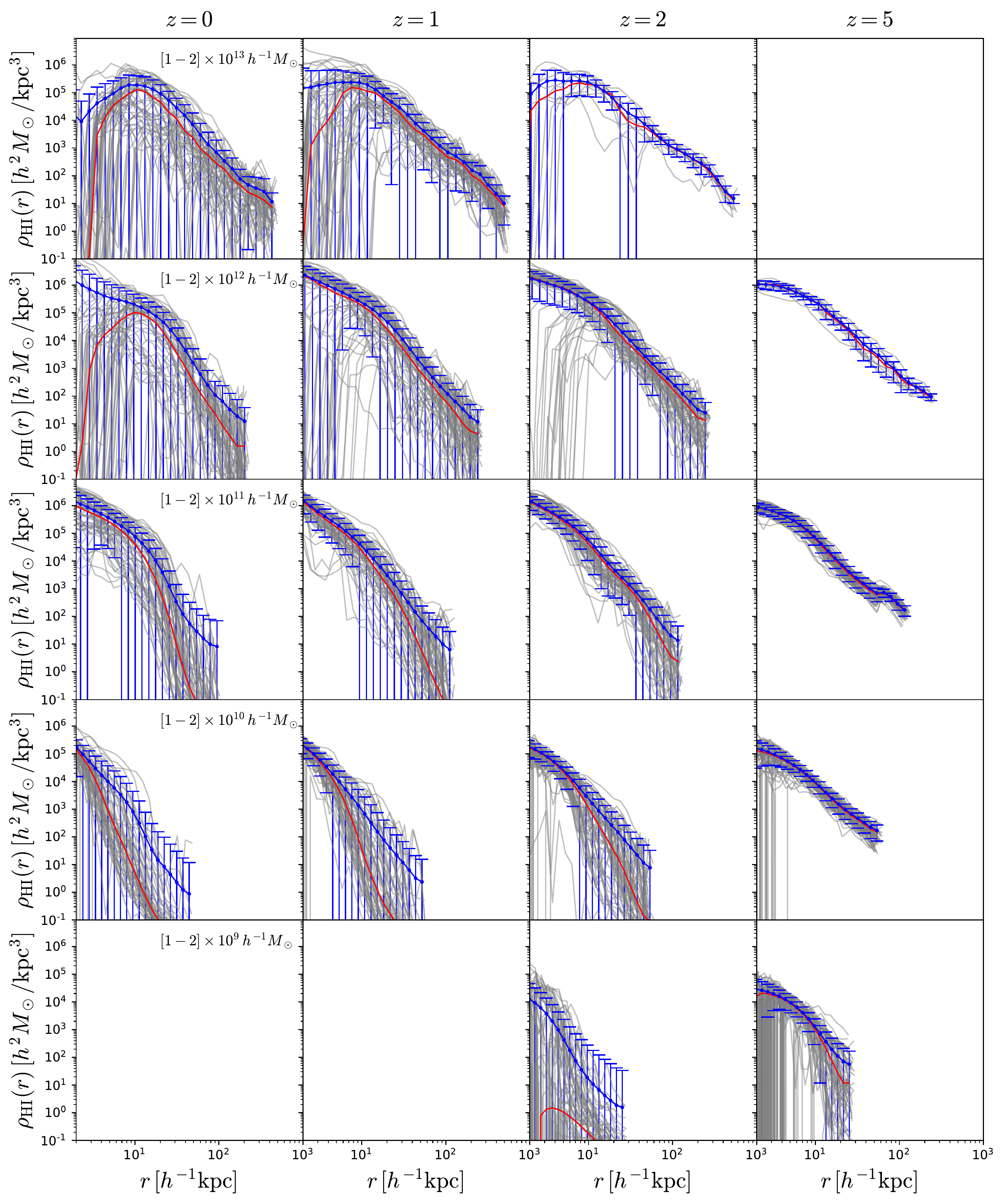}
\caption{Density profiles of HI for halos of different masses in different rows (see labels in the left column) at redshifts $z=0$ (left), $z=1$ (middle-left), $z=2$ (middle-right) and $z=5$ (right). In each panel we display up to 50 individual profiles (grey lines), the mean profile and the standard deviation (blue lines) and the median profile (red lines). Empty panels correspond to situations with either no halos (top-right) or with halos far below the cutoff mass $M_{\rm min}$. In contrast to dark matter, HI density profiles are not universal, and they exhibit, in most of the cases, a very large scatter. The HI-H$_2$ transition saturates the amplitude of the profiles in the core, while processes such as AGN feedback  create HI holes in the core of the most massive halos. The mean and the median can be quite different, indicating that the distribution is asymmetric. In some cases, that asymmetry is due to the presence of two different populations such as blue and red galaxies.}
\label{fig:HI_profiles}
\end{center}
\end{figure*}

Another important ingredient in describing the spatial distribution of
cosmic neutral hydrogen using HI halo models is the density profile
of HI inside halos (see Eqs. \ref{eq:HI1h_HI2h}, \ref{eq:HI1h},
\ref{eq:HI1h2h}). In this section we investigate the spatial
distribution of HI inside simulated dark matter halos.

\begin{figure*}[ht]
\begin{center}
\includegraphics[width=1\textwidth]{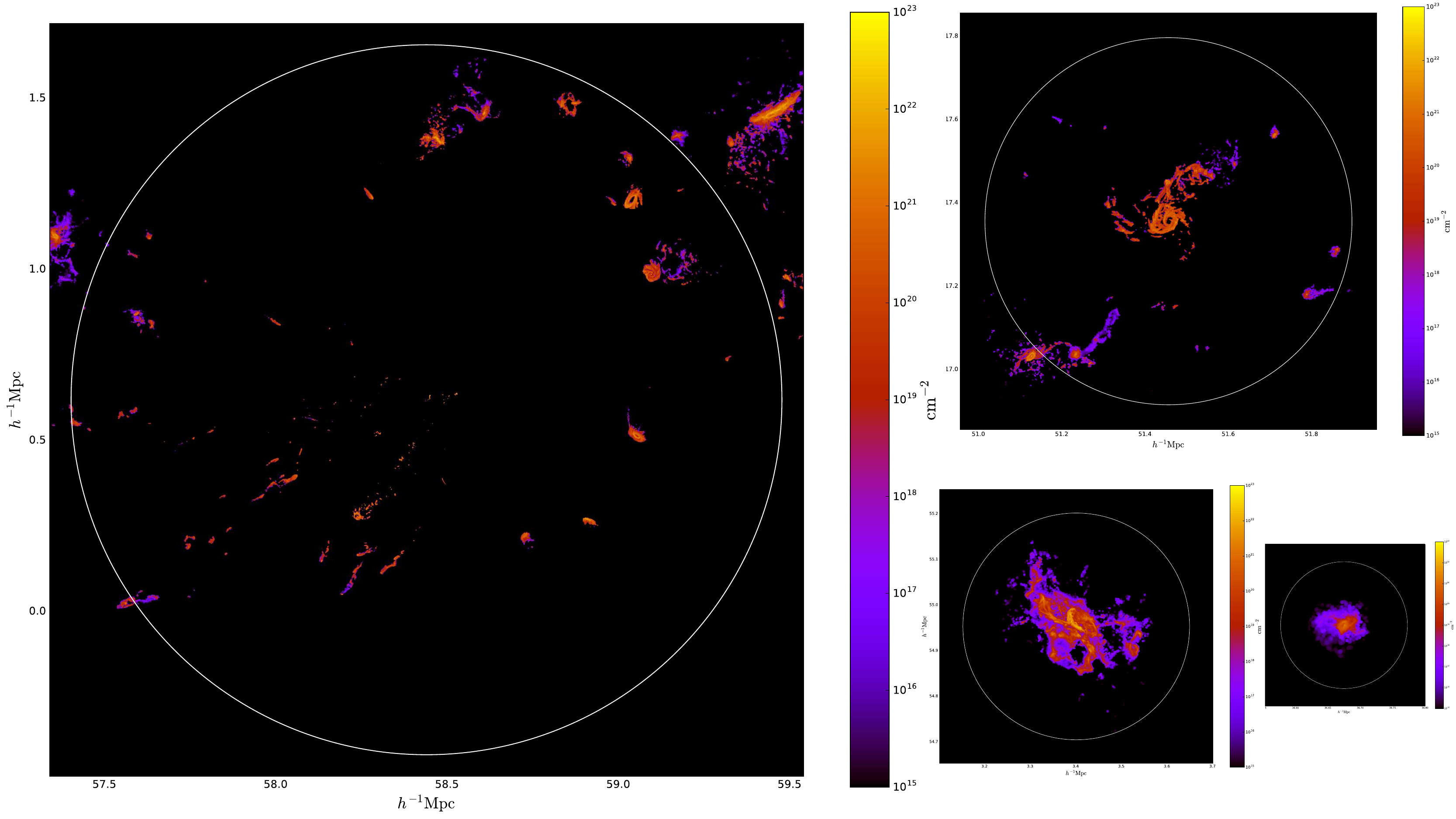}
\caption{To better understand the features in the HI profiles of Fig. \ref{fig:HI_profiles}, we have chosen halos with HI profiles close to the mean. The images show the HI column density for halos of mass $\sim10^{14}~h^{-1}M_\odot$ at $z=0$ (left), $\sim10^{13}~h^{-1}M_\odot$ at $z=0$ (top-right), $\sim10^{12}~h^{-1}M_\odot$ at $z=1$ (bottom-middle) and $\sim10^{11}~h^{-1}M_\odot$ at $z=0$ (bottom-right). The center of galaxy clusters is typically occupied by HI poor ellipticals, whereas HI rich spirals reside in the centers of lower mass halos. Processes such as tidal-stripping and ram pressure efficiently remove gas from galaxies near the centers of galaxy clusters. In small halos ($\lesssim10^{12}~h^{-1}M_\odot$) gas can cool and accumulate in the center while in groups AGN feedback produces holes in the core of the HI profile (see top-right panel). The ``cuspyness'' of the HI profiles increases with decreasing halo mass, but saturates due to the formation of H$_2$.}
\label{fig:HI_profile_images}
\end{center}
\end{figure*}

Since FoF halos can have very irregular shapes, we have computed the
HI profiles inside FoF-SO halos. For each FoF-SO halo we have computed
the HI mass within narrow spherical shells up to the virial radius,
and from them the HI profile. Fig. \ref{fig:HI_profiles} shows
individual HI profiles for halos in a narrow mass bin at different
redshifts with grey lines. The large halo-to-halo scatter is
surprising, and highlights that individual HI profiles, as opposed to dark
matter ones, are far from universal.

The scatter is particularly large towards the centers of massive
halos, as discussed below, as well as for halos with masses around or
below the cutoff we observe in Fig. \ref{fig:M_HI}. This is expected
as the halo HI mass function also exhibits large scatter in that
range. As we will see later in section \ref{sec:assembly_bias}, the
clustering of halos in that mass range depends significantly on
their HI mass. Thus, it is likely that the HI content of these halos
is influenced by their environment, so small halos around more
massive ones may have lose or gain a significant fraction of their
HI mass due to related effects.

The scatter generally tends to be lower at higher redshifts, and, in
particular, is small in halos with masses above
$10^{10}~h^{-1}M_\odot$ at redshift $z=5$. This is related to the
lower scatter we find at high redshift in the halo HI mass function,
$M_{\rm HI}(M,z)$ (see Fig. \ref{fig:M_HI}). We speculate that this
originates from a reduced role that AGN feedback and environmental gas
stripping play at earlier times.

The blue lines in Fig. \ref{fig:HI_profiles} show the mean and the
standard deviation of the HI profiles from all halos that lie in each
mass bin and redshift, while the red lines display the
median. Clearly, in some cases they differ substantially. This
behavior can be partially attributed to the HI profiles arising from
two distinct populations: i.e.~HI-rich blue galaxies versus HI-poor
red ones \citep{NelsonD_17a}. This can clearly be seen in the panel in
Fig. \ref{fig:HI_profiles} corresponding to halos in the mass range
$M\in[1-2]\times10^{12}~h^{-1}M_\odot$ at $z=0$. In this range,
some halos have a core in their HI profiles while others do not. The
reason is that the central galaxy of some halos is experiencing AGN
feedback (those with holes in the profile) and are therefore becoming
red, while the galaxies in the other halos are not yet being affected
by AGN feedback, remaining blue \citep{NelsonD_17a}. 

\begin{table*}
\begin{center}
\renewcommand{\arraystretch}{2}
\resizebox{1.0\textwidth}{!}{ \begin{tabular}{|c|| cc| cc| cc| cc| cc| cc|} 
 \hline
 & \multicolumn{12}{c|}{\large{Model 1 --- power law + exponential cutoff: $\alpha_\star$, $\log_{10}r_{\rm 0}\,[h^{-1}{\rm Mpc}]$}}\\[0.7ex]
 \hline
  $z$ & \multicolumn{2}{c|}{$M_h=10^9\,[h^{-1} M_\odot]$} & \multicolumn{2}{c|}{$M_h=10^{10}\,[h^{-1} M_\odot]$} & \multicolumn{2}{c|}{$M_h=10^{11}\,[h^{-1} M_\odot]$} & \multicolumn{2}{c|}{$M_h=10^{12}\,[h^{-1} M_\odot]$} & \multicolumn{2}{c|}{$M_h=10^{13}\,[h^{-1} M_\odot]$} & \multicolumn{2}{c|}{$M_h=10^{14}\,[h^{-1} M_\odot]$}\\[0.7ex] 
 \hline\hline
 0 & --- & --- & $3.04^{+0.04}_{-0.03},$ & $-3.59^{+0.85}_{-0.92}$ & $3.03^{+0.03}_{-0.02},$ & $-2.8^{+0.5}_{-1.2}$ & $3.02^{+0.03}_{-0.03},$ & $-2.32^{+0.33}_{-1.15}$ & $3.00^{+0.04}_{-0.04},$ & $-1.71^{+0.09}_{-0.12}$ & $2.92^{+0.03}_{-0.03},$ & $-1.91^{+0.11}_{-0.14}$\\[0.7ex]
 \hline
1 & $3.3^{+1.3}_{-0.7},$ & $-2.5^{+1.1}_{-1.6}$ &  $3.05^{+0.02}_{-0.02},$ & $-3.72^{+0.77}_{-0.84}$ & $3.02^{+0.02}_{-0.02},$ & $-3.3^{+0.7}_{-1.1}$ & $3.00^{+0.03}_{-0.02},$ & $-2.32^{+0.16}_{-0.28}$ & $2.99^{+0.03}_{-0.03},$ & $-1.77^{+0.09}_{-0.11}$& --- & ---\\[0.7ex] 
 \hline
2 &  $3.07^{+0.10}_{-0.08},$ & $-3.2^{+0.9}_{-1.2}$ &  $3.03^{+0.01}_{-0.02},$ & $-3.64^{+0.78}_{-0.89}$ & $3.01^{+0.01}_{-0.01},$ & $-2.75^{+0.26}_{-0.68}$ & $3.00^{+0.02}_{-0.02},$ & $-2.18^{+0.09}_{-0.12}$ & $2.98^{+0.02}_{-0.01},$ & $-1.74^{+0.04}_{-0.05}$& --- & ---\\[0.7ex]
 \hline
3 &  $3.05^{+0.02}_{-0.02},$ & $-3.63^{+0.85}_{-0.93}$ &  $3.02^{+0.02}_{-0.02},$ & $-3.1^{+0.5}_{-1.1}$ & $3.00^{+0.01}_{-0.01},$ & $-2.52^{+0.13}_{-0.20}$ & $3.00^{+0.02}_{-0.02},$ & $-2.09^{+0.06}_{-0.07}$ & --- & --- & --- & ---\\[0.7ex]
 \hline
4 &  $3.04^{+0.02}_{-0.02},$ & $-3.3^{+0.7}_{-1.0}$ &  $3.00^{+0.01}_{-0.01},$ & $-2.46^{+0.15}_{-0.24}$ & $3.00^{+0.01}_{-0.01},$ & $-2.32^{+0.07}_{-0.08}$ & $2.99^{+0.01}_{-0.01},$ & $-2.04^{+0.03}_{-0.04}$ & --- & --- & --- & ---\\[0.7ex]
 \hline
5 & $3.03^{+0.02}_{-0.02},$ & $-2.9^{+0.5}_{-1.2}$ &  $3.00^{+0.01}_{-0.01},$ & $-2.28^{+0.09}_{-0.12}$ & $3.00^{+0.01}_{-0.01},$ & $-2.18^{+0.04}_{-0.05}$ & $3.00^{+0.01}_{-0.01},$ & $-2.02^{+0.03}_{-0.03}$ & --- & --- & --- & ---\\[0.7ex] 
 \hline \hline

 & \multicolumn{12}{c|}{{\large Model 2 --- altered NFW + exponential cutoff: $\log_{10}r_{\rm s}\,[h^{-1}{\rm Mpc}]$, $\log_{10}r_{\rm 0}\,[h^{-1}{\rm Mpc}]$}}\\[0.5ex]
 \hline
0 & --- & --- &  $-4.0^{+0.7}_{-0.7},$ & $-3.8^{+0.8}_{-0.8}$ & $-3.7^{+0.6}_{-0.9},$ & $-3.4^{+0.7}_{-1.0}$ & $-3.2^{+0.7}_{-1.1},$ & $-3.1^{+0.8}_{-1.3}$ & $-3.0^{+1.0}_{-1.3},$ & $-1.8^{+0.1}_{-1.0}$ & $-2.3^{+0.5}_{-1.7},$ & $-2.6^{+0.7}_{-1.6}$\\[0.7ex]
 \hline
1 & $-2.8^{+1.8}_{-1.5},$ & $-3.3^{+1.2}_{-1.1}$ & $-4.0^{+0.5}_{-0.6},$ & $-3.7^{+0.6}_{-0.8}$ & $-3.9^{+0.6}_{-0.7},$ & $-3.6^{+0.7}_{-0.9}$ & $-3.0^{+0.4}_{-1.2},$ & $-2.9^{+0.6}_{-1.4}$ & $-2.2^{+0.3}_{-1.6},$ & $-2.4^{+0.6}_{-1.7}$ & --- & --- \\[0.7ex] 
 \hline
2 & $-3.7^{+0.9}_{-0.8},$ & $-3.7^{+0.9}_{-0.9}$  &  $-3.8^{+0.5}_{-0.7},$ & $-3.5^{+0.5}_{-0.9}$ & $-3.6^{+0.5}_{-0.9},$ & $-3.3^{+0.6}_{-1.1}$ & $-2.8^{+0.3}_{-1.3},$ & $-2.6^{+0.4}_{-1.5}$ & $-1.8^{+0.1}_{-0.2},$ & $-3.0^{+0.8}_{-1.3}$& --- & ---\\[0.7ex]
 \hline
3 & $-3.8^{+0.6}_{-0.8},$ & $-3.5^{+0.6}_{-0.9}$  &  $-3.6^{+0.5}_{-0.8},$ & $-3.3^{+0.5}_{-1.1}$ & $-3.3^{+0.4}_{-1.0},$ & $-3.0^{+0.5}_{-1.3}$ & $-2.8^{+0.4}_{-1.3},$ & $-2.4^{+0.3}_{-1.4}$ & --- & --- & --- & ---\\[0.7ex]
 \hline
4 & $-3.6^{+0.5}_{-0.9},$ & $-3.2^{+0.5}_{-1.1}$  &  $-3.2^{+0.4}_{-1.1},$ & $-3.0^{+0.5}_{-1.3}$ & $-2.9^{+0.3}_{-1.3},$ & $-2.7^{+0.4}_{-1.4}$ & $-2.6^{+0.3}_{-1.4},$ & $-2.4^{+0.3}_{-1.2}$ & --- & --- & --- & ---\\[0.7ex]
 \hline
5 & $-3.4^{+0.5}_{-1.0},$ & $-3.1^{+0.5}_{-1.1}$  &  $-2.7^{+0.2}_{-1.1},$ & $-3.1^{+0.7}_{-1.3}$ & $-2.5^{+0.1}_{-1.1},$ & $-3.3^{+1.0}_{-1.2}$ & $-2.2^{+0.1}_{-0.4},$ & $-3.1^{+0.8}_{-1.2}$ & --- & --- & --- & ---\\[0.7ex]
 \hline
\end{tabular}}
\caption{\label{table:HI_profiles} Best-fit values of the parameters determining the HI density profiles. We show the resulting parameters for the two different models considered (see text): an altered NFW profile with an exponential cutoff on small scales (top) and a simple power law with an exponential cutoff on small scales (bottom), as a dependence on the halo mass (columns) and redshift (rows).}
\end{center}
\end{table*}

We find that the HI density profiles of small halos
($M\lesssim10^{12}~h^{-1}M_\odot$) increase towards their halo
center. We note, however, that the amplitude of the HI profile tends to
saturate; i.e.~the slope of the profiles declines significantly towards
the halo center. For example, at $z=0$ and $z=1$ and for halos with
masses larger than $10^{11}~h^{-1}M_\odot$, the mean HI profiles
change slope around $\sim20~h^{-1}{\rm kpc}$.  This is expected since
neutral hydrogen at high densities will turn into molecular hydrogen
and stars on short time scales. For higher halo masses
($M\simeq10^{13}~h^{-1}M_\odot$) the HI density profile exhibits a
hole in the center. This is caused by AGN feedback in the central
galaxy of those halos. We notice that higher densities in the center of halos
can give rise to the formation of molecular hydrogen, that can produce a 
similar effect \citep{MarinacciF_17a}.
Holes, which extend even further than in
groups, are also found in the HI profiles of galaxy clusters, which we
however do not show here since there are only a few of them and only at
low redshift.

We illustrate these features of the HI profiles in
Fig. \ref{fig:HI_profile_images}, where we show the spatial
distribution of HI in and around four individual halos with masses
$10^{11}$, $10^{12}$, $10^{13}$ and $10^{14}$ $h^{-1}M_\odot$ at
redshifts 0 or 1. We have selected these halos by requiring that their
HI density profiles are close to the mean. It can be seen
that HI is localized in the inner regions of small halos, while for
groups, the central galaxy exhibits a hole produced by AGN feedback.
For galaxy clusters the
central regions have little HI. This happens because the central
galaxy is an HI poor elliptical, and 
ram-pressure and tidal stripping are very efficient in removing the
gas content of galaxies passing near the center.
The analysis of this section suggests that analytical approaches to the distribution of Damped Lyman-$\alpha$ systems employing a universal HI profile, \textit{e.g.} \citep{Padmanabhan_2017}, will not be able to reproduce observations. 

In order to quantitatively investigate what is the effective average HI density profile across different halo masses and redshift, we use the mean measured HI density profile and test two models of HI density that both include an exponential cutoff on small scales. 

First we consider a simple power law with an exponential cutoff on small scales --- Model 1:
\begin{equation}
\rho_{\rm HI}(r)=\frac{\rho_0}{r^{\alpha_\star}}\exp(-r_0/r),
\end{equation}
where $\rho_0$ is the overall normalisation, $\alpha_\star$ is the slope parameter and $r_0$ is the inner radius at which the density drops and the profile changes its slope. 

Second, we consider an altered NFW profile \citep{Ari_2004, Barnes_Haehnelt_2014}, found to be a good fit to the multiphase gas distribution at high redshifts in hydrodynamical simulations, with an exponential cutoff on small scales --- Model 2:
 \begin{equation}
\rho_{\rm HI}(r)=\frac{\rho_0 r_s^3}{(r+3/4r_s)(r+r_s)^2}\exp(-r_0/r),
\end{equation}
where $\rho_0$ is the overall normalisation and $r_s$ is the scale radius of the HI cloud. In both cases the overall normalisation --- $\rho_0$, is fixed such that the volume integral of the model density profile integrated up to the virial radius of a given halo matches the mean total HI mass obtained from the density profile found in simulations (blue lines in Fig. 5). We are then left with two free parameters for each model: $\{\alpha_\star, r_0\}$ and $\{r_s, r_0\}$. 
We fit these models to the measured mean HI density profiles limiting our analysis only to the scales above $r\ge2 h^{-1}{\rm kpc}$. For the uncertainties in the density profiles we use the scatter among different galaxies (blue error-bars in Fig. 5) and assume that these uncertainties are uncorrelated between different scales. 

The best-fit values along with the 68\% confidence intervals are presented in table \ref{table:HI_profiles}, while in Fig. \ref{fig:rho_HI_fit} in Appendix \ref{sec:rho_HI_fit} we show the best-fit results for the Model 1. Based on the resulting best-fit $\chi^2$, we find that both Model 1 and 2 are good fits for all the considered redshifts and halo masses, except for the most massive halo bin $M_h=10^{14}\,[h^{-1} M_\odot]$ at $z=0$. We find that the difference in the best-fit $\chi^2$ between the two models to be negligible. This is to be expected since the models are rather similar and have the same slope on large scales. In the case of Model 1, we find the HI density profile slope to be consistent with a value of $\alpha_\star=3$ for all the halo masses and redshifts. The inner radius $r_0$ depends on the halo mass and is larger for larger halo masses at a fixed redshift, while at a fixed halo mass, it increases with increasing redshift. For example, for halos with $M_h\le10^{11} h^{1-}{\rm Mpc}$ and $z\le2$, $r_0$ is below the minimum scale considered and the uncertainties are rather large. In the case of Model 2, we find a similar behaviour. The inferred values of $r_0$ are consistent between two models, with Model 2 having larger uncertainties which is due to the degeneracy between parameters $r_0$ and $r_s$.
	
We note that other observational and simulation studies have found that the HI surface density profile of galaxies can be reproduced by an exponential profile \citep{Wang_2014,Obreschkow_2009}. Based on these studies, other spherically averaged density models have been used in the literature, e.g. an exponential profile \citep{Padmanabhan_2017}. We find that using an exponential profile for the spherically averaged profile does not reproduce our mean data very well.

\begin{figure*}
\begin{center}
\includegraphics[width=0.33\textwidth]{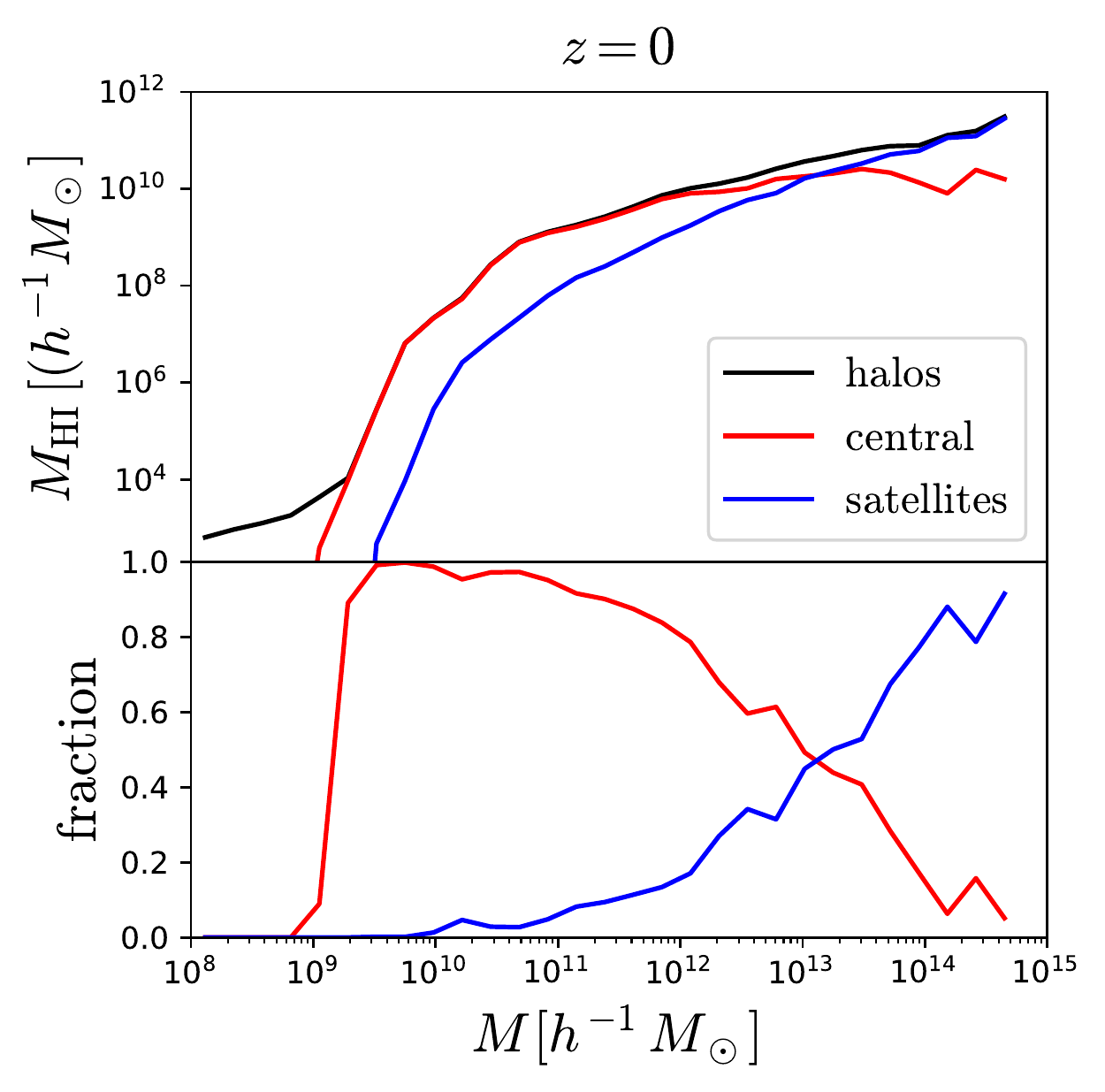}
\includegraphics[width=0.33\textwidth]{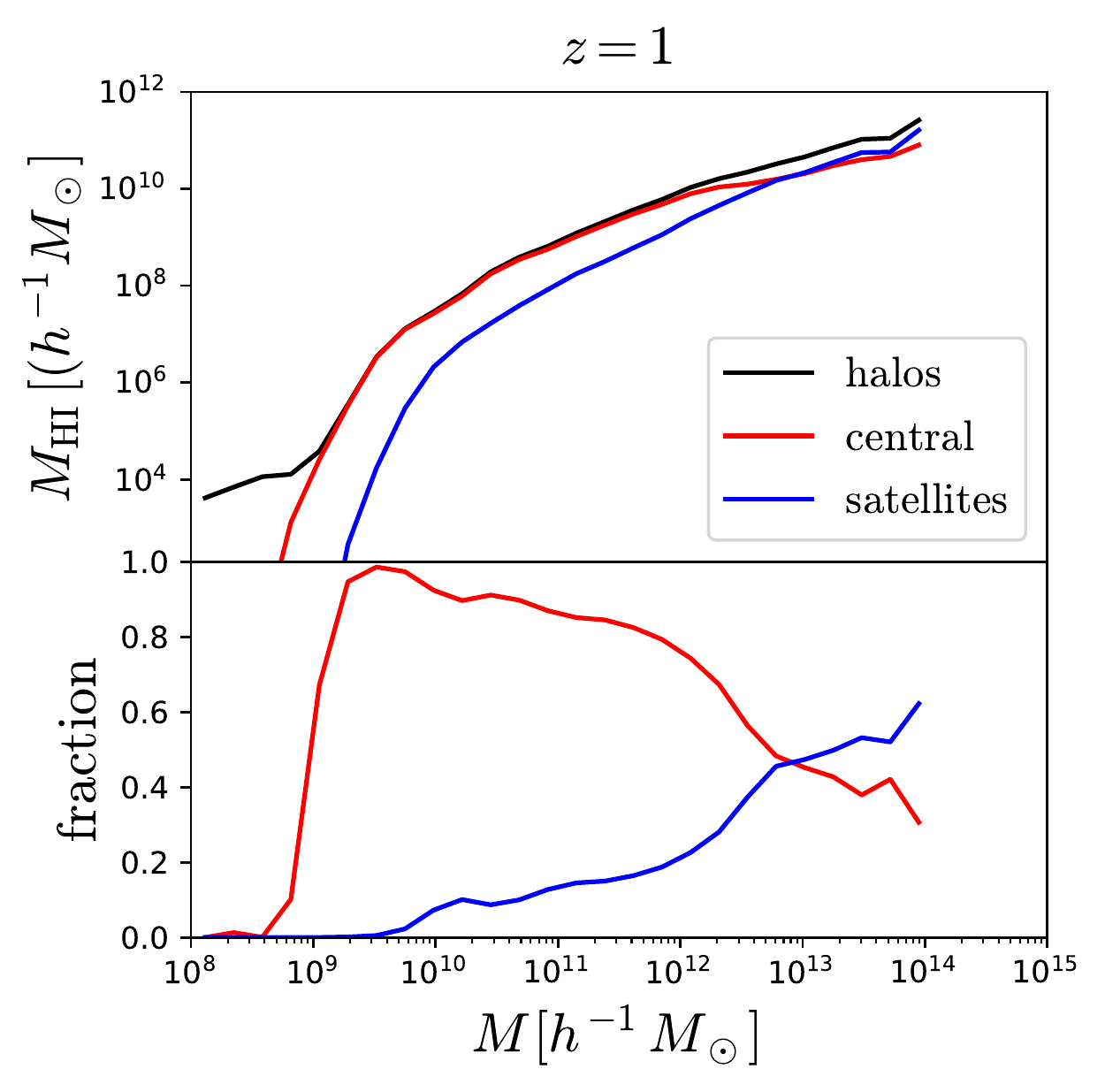}
\includegraphics[width=0.33\textwidth]{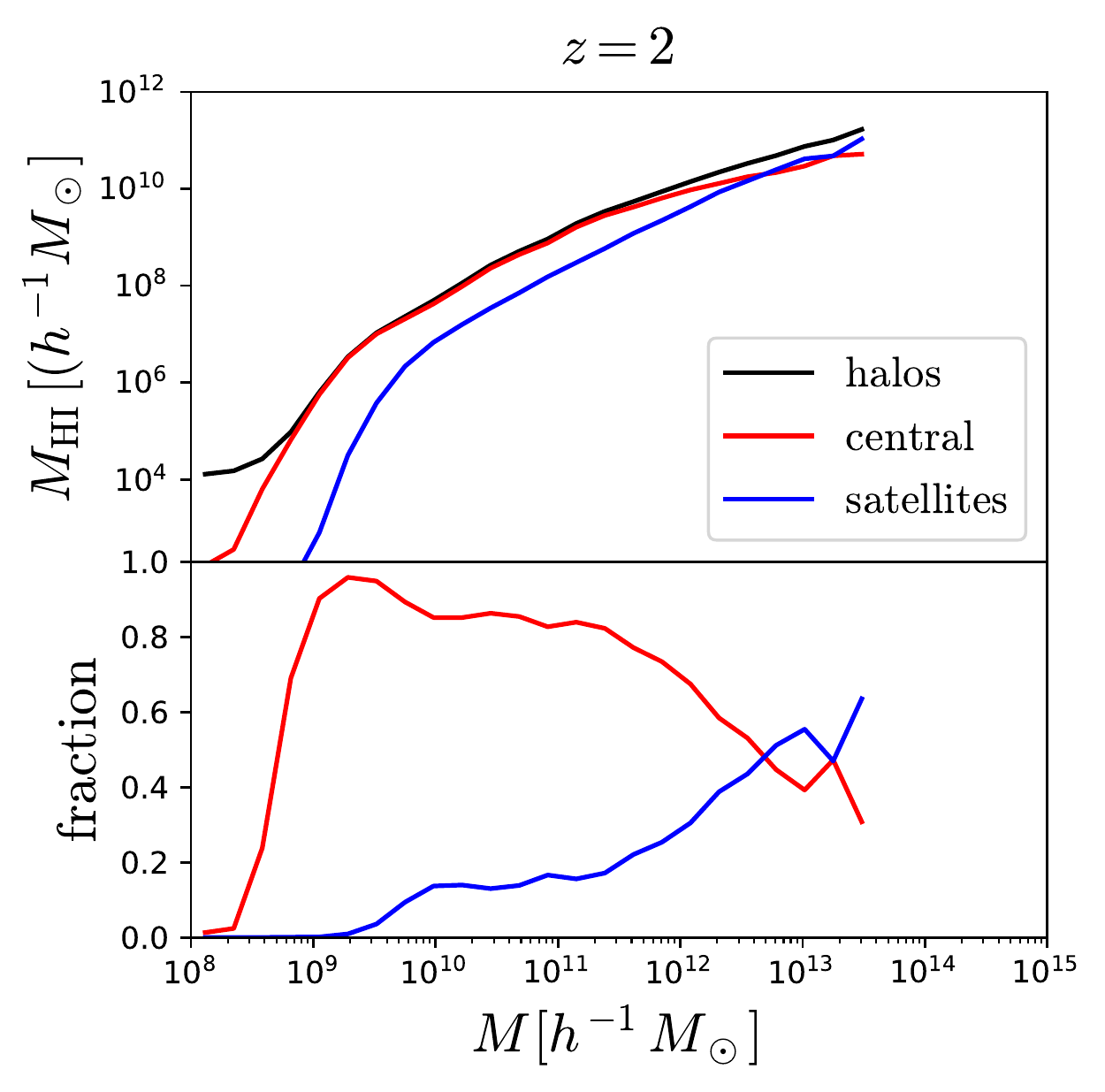}\\
\includegraphics[width=0.33\textwidth]{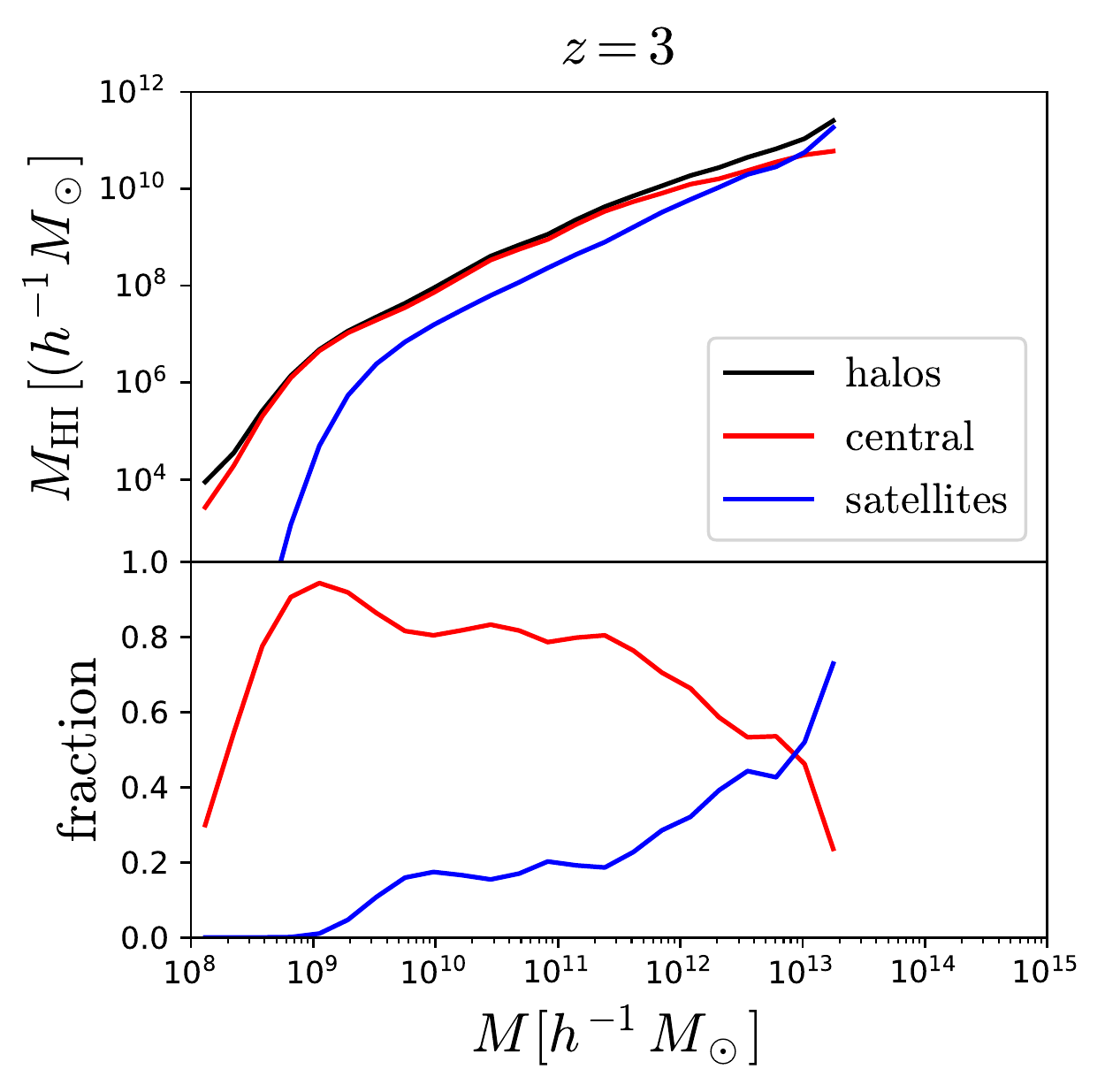}
\includegraphics[width=0.33\textwidth]{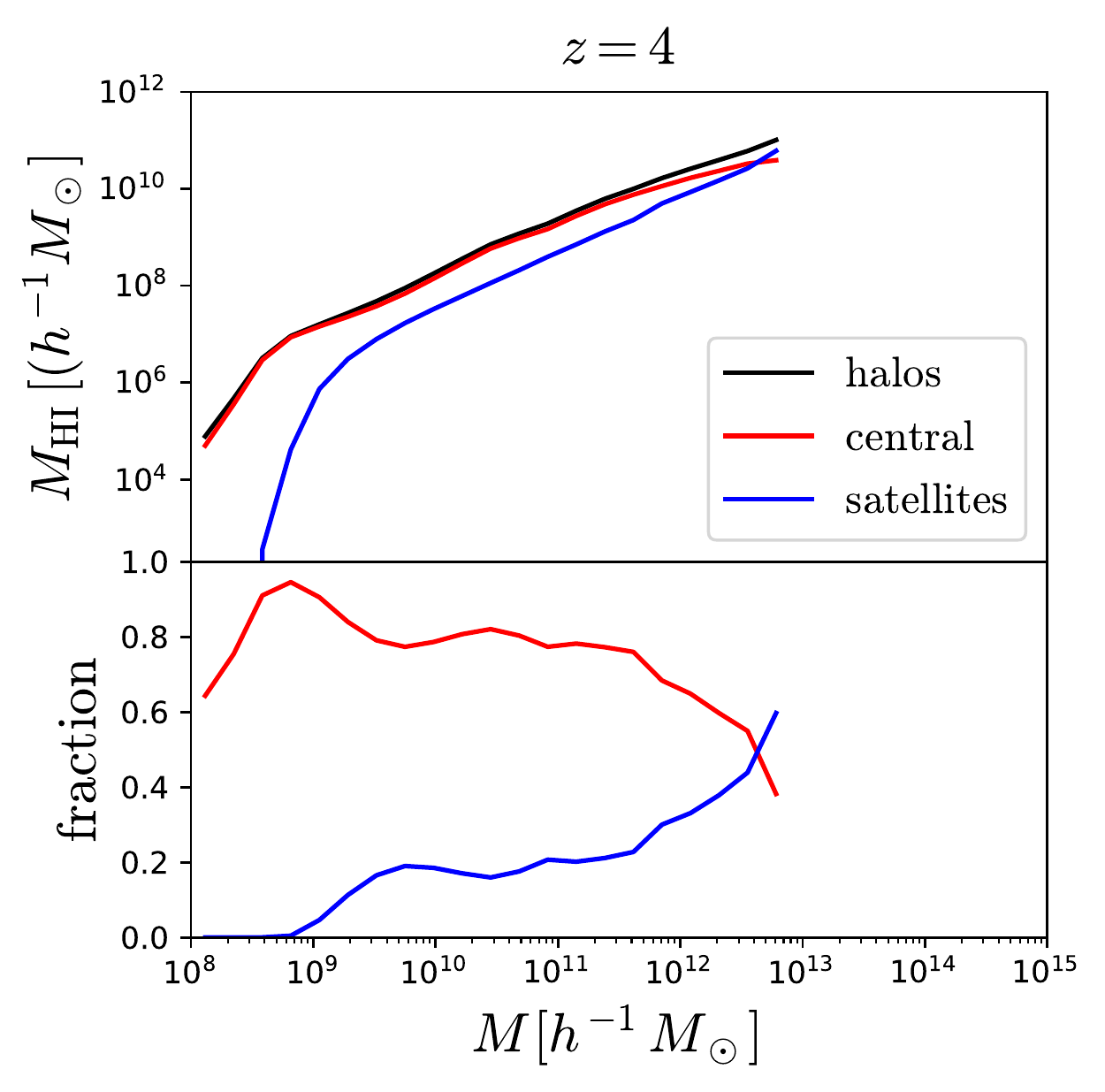}
\includegraphics[width=0.33\textwidth]{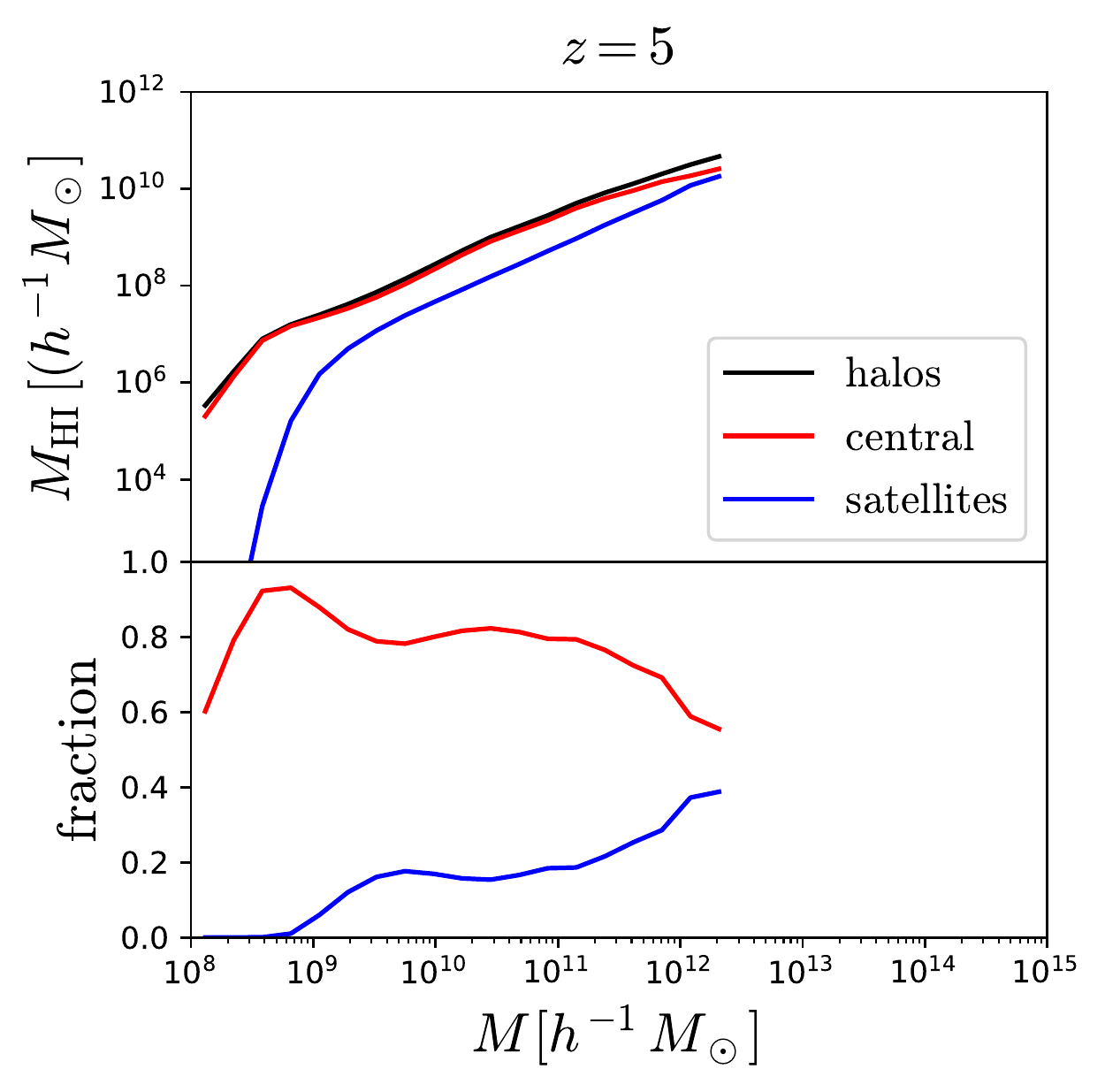}
\caption{The black lines show the average HI mass inside halos as a function of mass; i.e. the halo HI mass function, at redshifts 0 (top-left), 1 (top-middle), 2 (top-right), 3 (bottom-left), 4 (bottom-middle), 5 (bottom-right). The red and blue lines represent the average HI mass within central and satellites galaxies as a function of halo mass. The bottom panels display the fraction of the HI mass inside halos that is embedded in centrals and satellites. The HI mass of halos below $\simeq5\times10^{12}~h^{-1}M_\odot$ is dominated by HI in the central galaxy, while the HI in satellites dominates the HI content of more massive halos. For small halos, the HI mass in the central and satellites galaxies is less than the total HI mass in halos. This happens because the HI in those halos is small in mass and ``diffuse''.}
\label{fig:HI_centrals}
\end{center}
\end{figure*}

\section{HI in centrals and satellites galaxies}
\label{sec:HI_centrals_satellites}

It is interesting to quantify what fraction of HI mass inside
halos comes from their central and satellites galaxies. This will help
us to better understand the HI density profiles (see section
\ref{sec:HI_profile}) and improves our intuition for the amplitude of
the HI Fingers-of-God effect (see section \ref{subsec:RSD}).

For each FoF halo we have computed its total HI mass, the HI
mass within its central galaxy and the HI mass inside its satellites
We emphasize that our definition of central galaxy departs
significantly from that used in observations, as we consider the
central galaxy to be the most massive subhalo. In general, this
subhalo hosts the particle at the minimum of the gravitational
potential and is therefore the classical central galaxy, but it also
has significant spatial extent and particles far away from the halo center 
can be associated to this subhalo, unless they are bound to a satellite.

We take narrow bins in halo mass and compute the average HI mass, for
each of the above quantities. The outcome is shown in
Fig. \ref{fig:HI_centrals}. The black lines in the upper panels show
the halo HI mass function, while the red and blue ones display the average
HI mass inside the central and satellites galaxies as a function of
halo mass. The bottom panels show the fraction of HI mass within
halos that comes from the central and the satellites galaxies.

Aside from very low mass halos ($M\lesssim10^9h^{-1}M_\odot$), the
fraction of HI in the central galaxy decreases with halo mass, while
the fraction of HI in satellites increases, independent of
redshift. For small halos, nearly all the HI is located in the central
galaxy, as expected. For halos of masses
$\sim5\times10^{12}~h^{-1}M_\odot$, the fraction of HI in the central
galaxy and in satellites is roughly the same, almost independent of
redshift. For more massive halos, the total HI mass is dominated by
the HI in satellites galaxies.

At high-redshift, the contribution of satellites to the total HI mass
in small ($M\in[10^{10}-10^{11}]~h^{-1}M_\odot$) halos is
non-negligible: $\simeq20\%$. At $z=0$ and for galaxy clusters
$M\geqslant10^{14}~h^{-1}M_\odot$, the contribution of the central
galaxy to the total HI mass is negligible, as expected.

The HI mass in very low mass halos ($M\lesssim10^9h^{-1}M_\odot$) is
small, in particular at low-redshift. For some of these halos, the sum
of the HI mass in the central and satellite subhalos is much less
than the total HI mass. In such cases, some of
the HI mass was determined by {\sc SUBFIND} to be unbound. In a
fraction of these halos, no bound structure was identified by {\sc
  SUBFIND} altogether, rendering the combined HI masses of the central
and satellites, which by definition do not exist, to be zero, even if
the FoF group contains some HI.

\section{HI pdf}
\label{sec:HI_pdf}

We now study other quantities that, although are not ingredients for
HI halo models, will help us better understand the spatial
distribution of neutral hydrogen. One of those quantities is the
density probability distribution function (pdf), which we investigate
in detail in this section and compare to that of matter.

\begin{figure*}
\begin{center}
\includegraphics[width=0.40\textwidth]{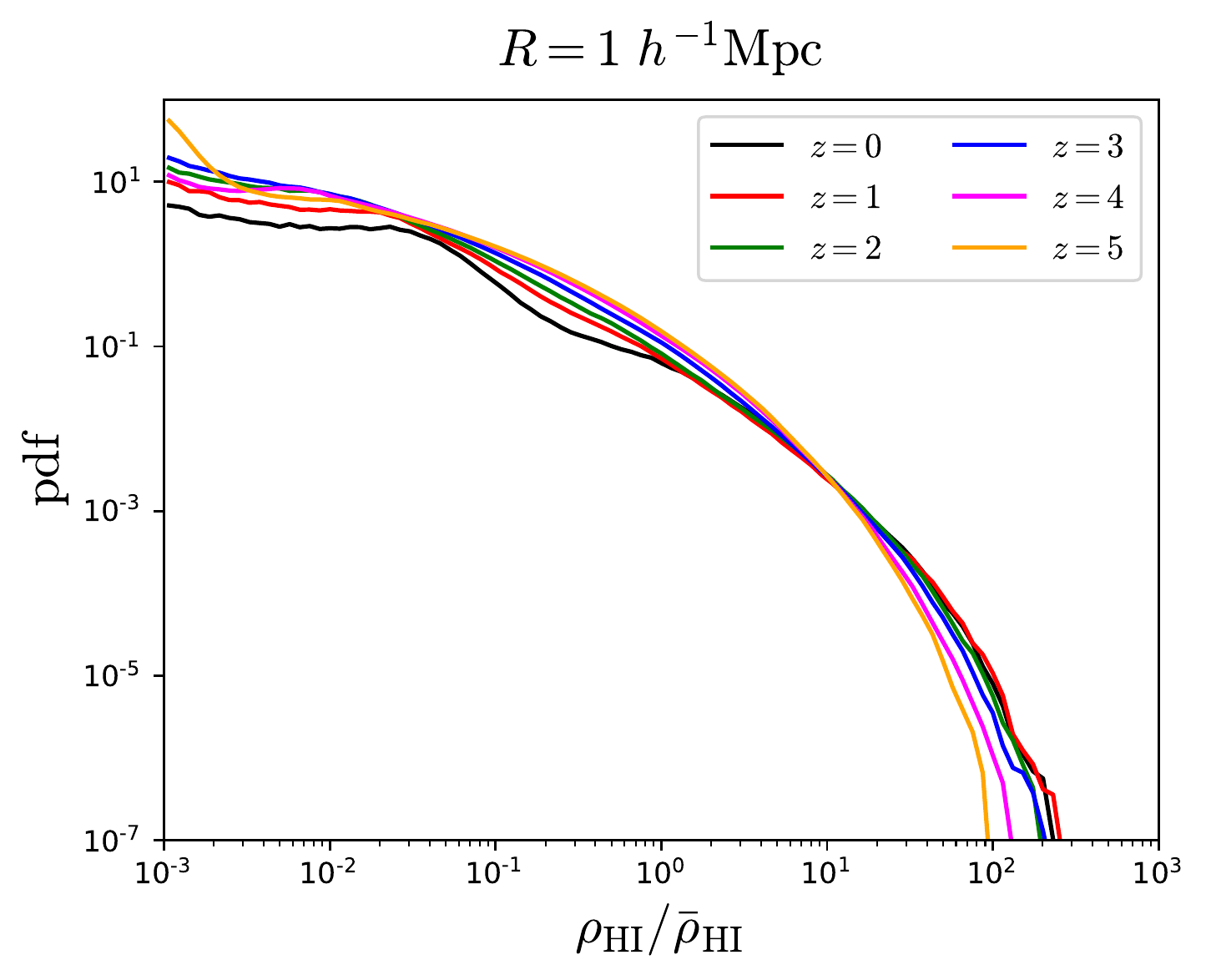}
\includegraphics[width=0.40\textwidth]{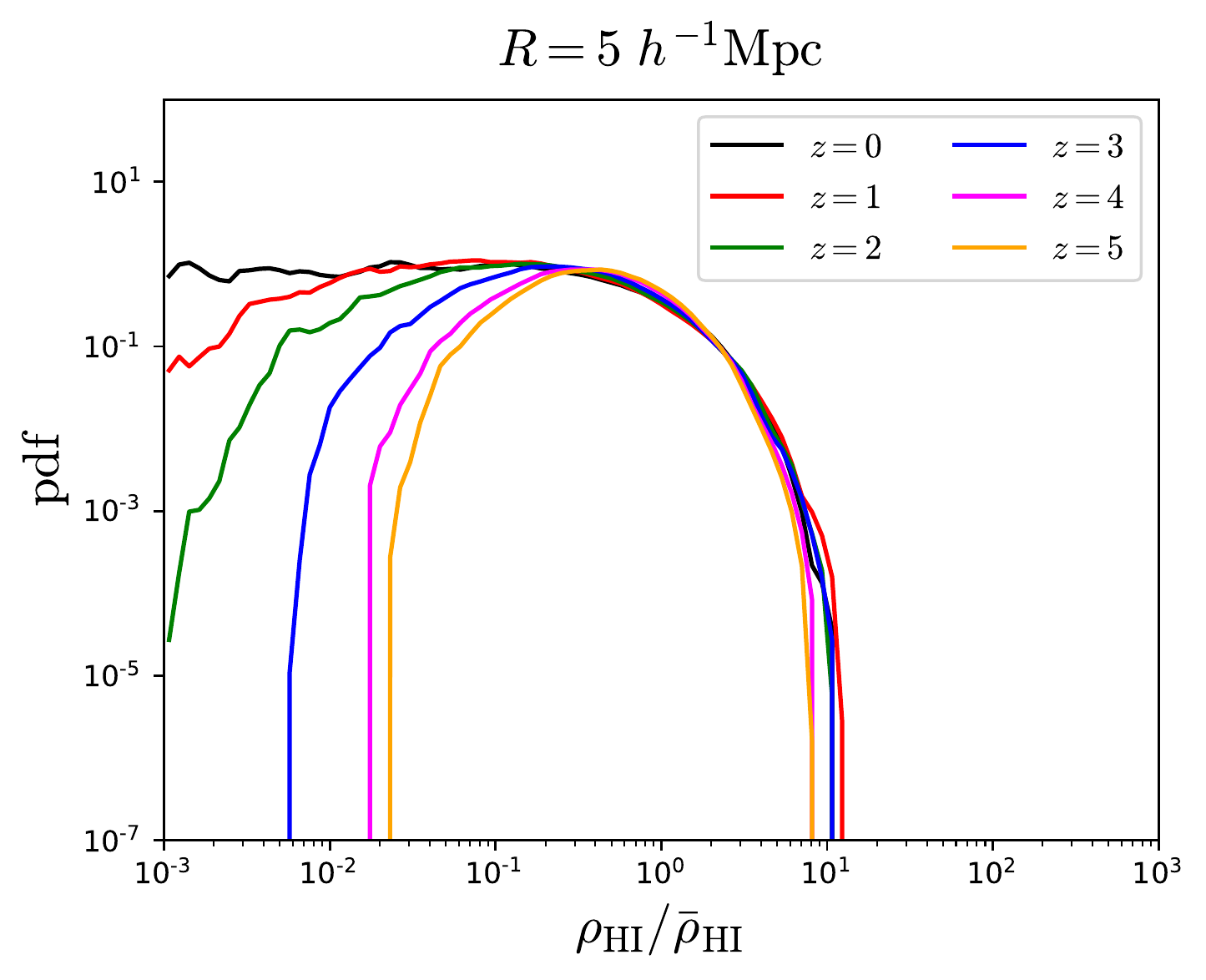}\\
\includegraphics[width=0.40\textwidth]{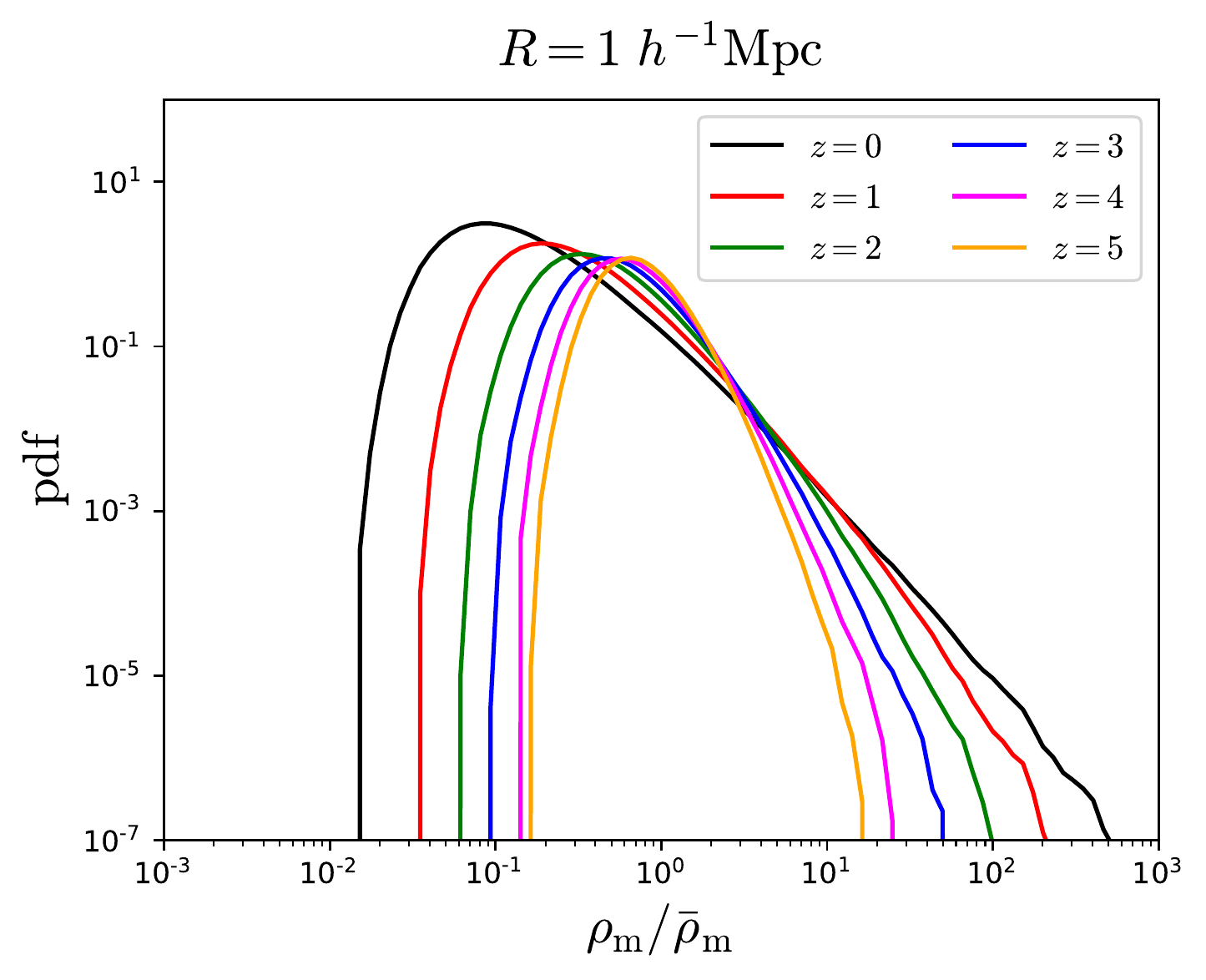}
\includegraphics[width=0.40\textwidth]{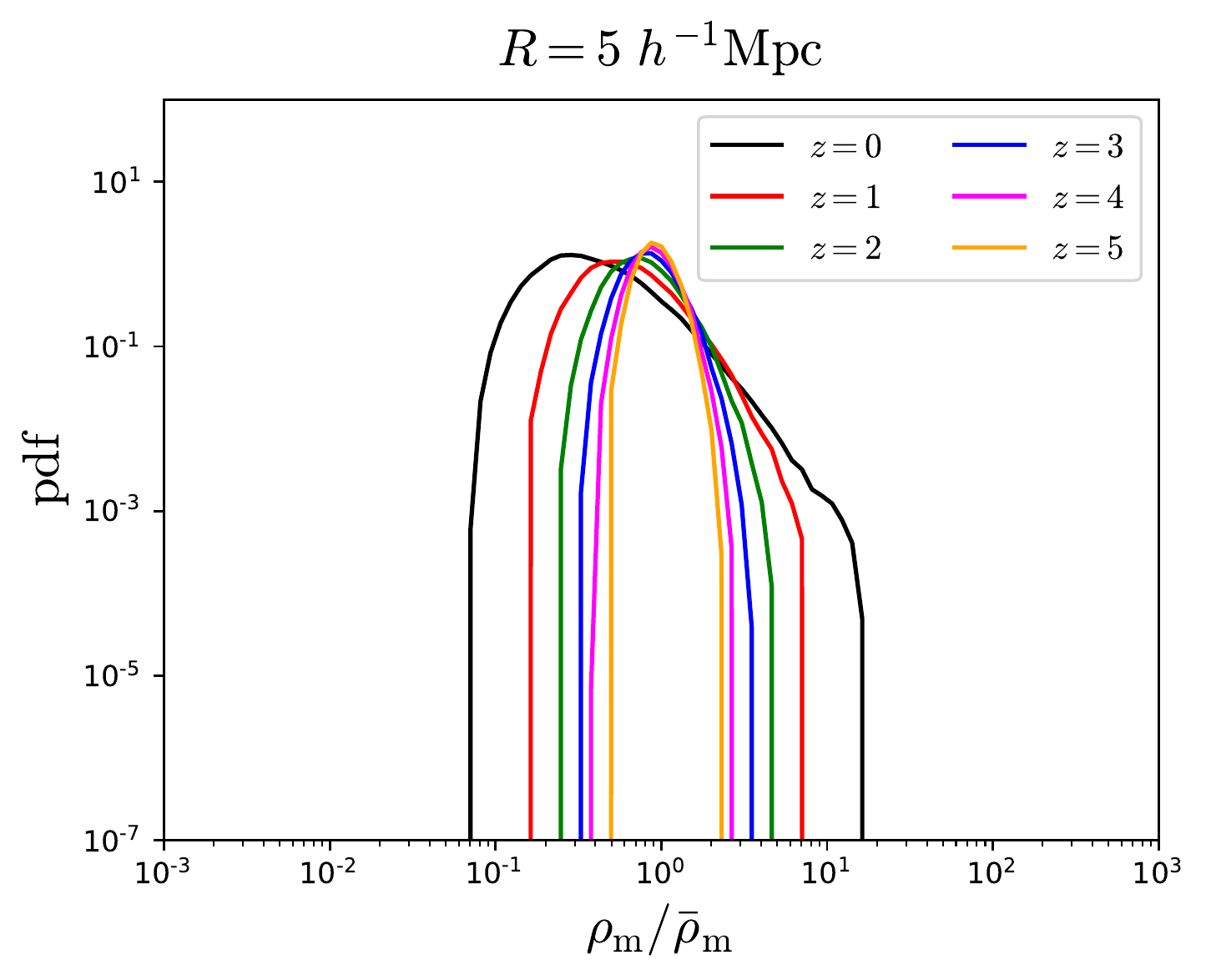}
\caption{The density fields of HI (top) and matter (bottom) smoothed with a top-hat filter of 1 (left) and 5 (right) $h^{-1}{\rm Mpc}$. This figure shows the pdf of those fields at several redshifts (see legend). Deviations from Gaussianity are larger in the HI field than in the matter field at all redshifts. For small smoothing scales the HI density pdf barely changes with redshift. At high-redshifts both the HI and the matter fields are well described by a log-normal distribution.}
\label{fig:HI_pdf}
\end{center}
\end{figure*}

We compute the density fields of neutral hydrogen and total matter in
the whole simulation volume using cloud-in-cell (CIC) interpolation on
a grid with $2048^3$ cells in real-space, namely
$\approx36.6~h^{-1}{\rm kpc}$ across each grid cell. We then smooth
those fields with top-hat filters of radii 1 and 5 $h^{-1}{\rm
  Mpc}$. We have chosen those values for the smoothing scale, $R$, as
a compromise between large and small scales. On one hand the volumes
of our simulations do not allow us to explore values much larger than
$\simeq5~h^{-1}{\rm Mpc}$, while on the other hand we take
$1~h^{-1}{\rm Mpc}$ as a representative estimate of the non-linear
regime. In Fig. \ref{fig:HI_pdf} we show the pdfs, computed as the
number of cells in a given interval in overdensity, over the total
number of cells, divided by the width of the overdensity interval.

While the density pdf of the matter field is highly
non-Gaussian at low-redshift, for either of the two smoothing scales we
consider here, at high-redshift it becomes more nearly Gaussian, as
expected. At redshifts $z\geqslant4$ and for $R=5~h^{-1}{\rm Mpc}$
the matter pdf can be well approximated by a lognormal distribution
\be
{\rm pdf}(\delta)=\frac{1}{\sqrt{2\pi\sigma^2}}\exp\left[\frac{(\log(1+\delta)+\sigma^2/2)^2}{2\sigma^2}\right]
\ee
where we consider $\sigma$ as a free parameter. We find $\sigma=0.21$
at $z=5$ and $\sigma=0.28$ at $z=4$. At lower redshifts and for smaller
values of the smoothing scale the lognormal
function does not provide a good fit to our results, as expected
\citep[see e.g.][]{Uhlemann_15}.

The HI density pdf exhibits a different behavior compared to the
matter pdf. First, the abundance of large HI overdensities
remains roughly constant with redshift, independent of the smoothing
scale considered. Second, for a smoothing scale of 1 $h^{-1}{\rm
  Mpc}$, the HI pdf hardly changes with redshift.

For redshifts $z\geqslant3$ a lognormal characterizes  our results
relatively well: for $R=1~h^{-1}{\rm Mpc}$
$\sigma\simeq1.9$ while for for $R=5~h^{-1}{\rm Mpc}$ $\sigma\simeq1$
at $z=3$, $\sigma\simeq0.9$ at $z=4$ and $\sigma\simeq0.8$ at
$z=5$. At lower redshifts, a log-normal distribution does not 
provide a good match to the simulations.

To understand the physical origin of the differences
between the pdfs of HI and matter density it is useful to relate
the width of the pdf to the amplitude of the HI power spectrum.
This is possible since the amplitude of the HI power spectrum
represents a measurement of the variance of the field at a given
scale. Low values of the HI power spectrum indicate that HI is
distributed homogeneously, while higher values mean that 
spatial variations in HI density can be
large.

One of the reasons that the HI density pdf is roughly similar across
redshifts while this is less true of the matter density pdf is that
the amplitude of the HI power spectrum depends more weakly on redshift
than does the matter power spectrum (see section
\ref{subsec:HI_bias}). Thus, the variance of the HI pdf is necessarily
smaller than that of the matter pdf. Since the amplitude of the HI
power spectrum is larger than that of the matter power spectrum at
high redshifts, the variance of the HI pdf is larger than that of the
matter pdf at those redshifts, as we find. Finally, in the
central region of a void, the matter density will be low, but
the HI density will be even lower\footnote{In order to have a
  significant amount of HI, self-shielding is required. Thus, in
  low-density regions, HI will be highly ionized.}. Thus, we
find that there will be more cells with low HI overdensity than
with low matter overdensity.

It can be seen for both HI and matter that the distributions are
broader for a smoothing scale of $1~h^{-1}{\rm Mpc}$ than for
$5~h^{-1}{\rm Mpc}$. This is expected since when smoothing over larger
scales, any field will become more homogeneous and therefore the width
of its pdf will become smaller. This can also be quantified through
the amplitude of the power spectrum, using the same reasoning as above.

\section{HI column density distribution function and DLAs cross-sections}
\label{sec:DLAs}

\begin{figure*}
\begin{center}
\includegraphics[width=0.497\textwidth]{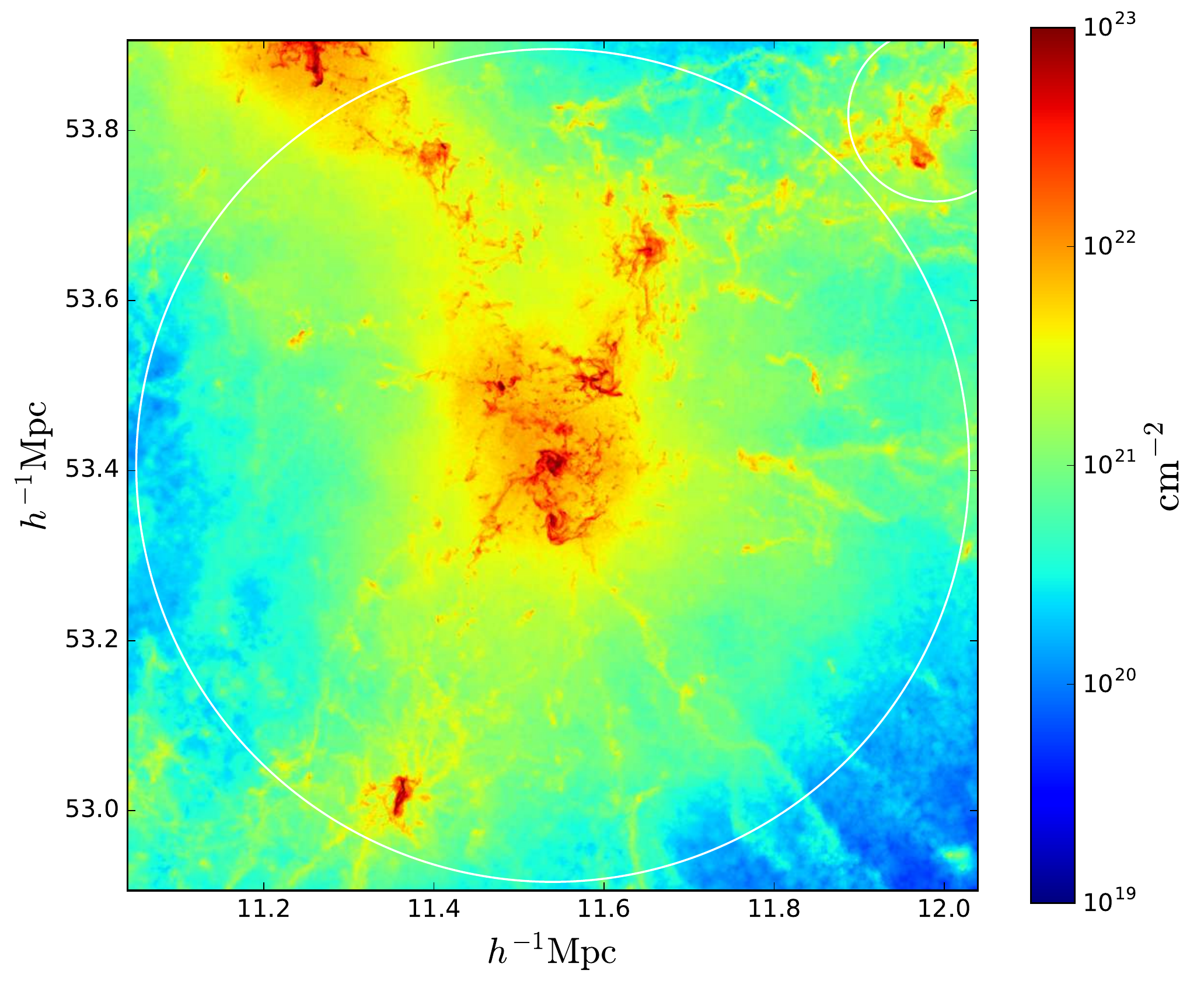}
\includegraphics[width=0.497\textwidth]{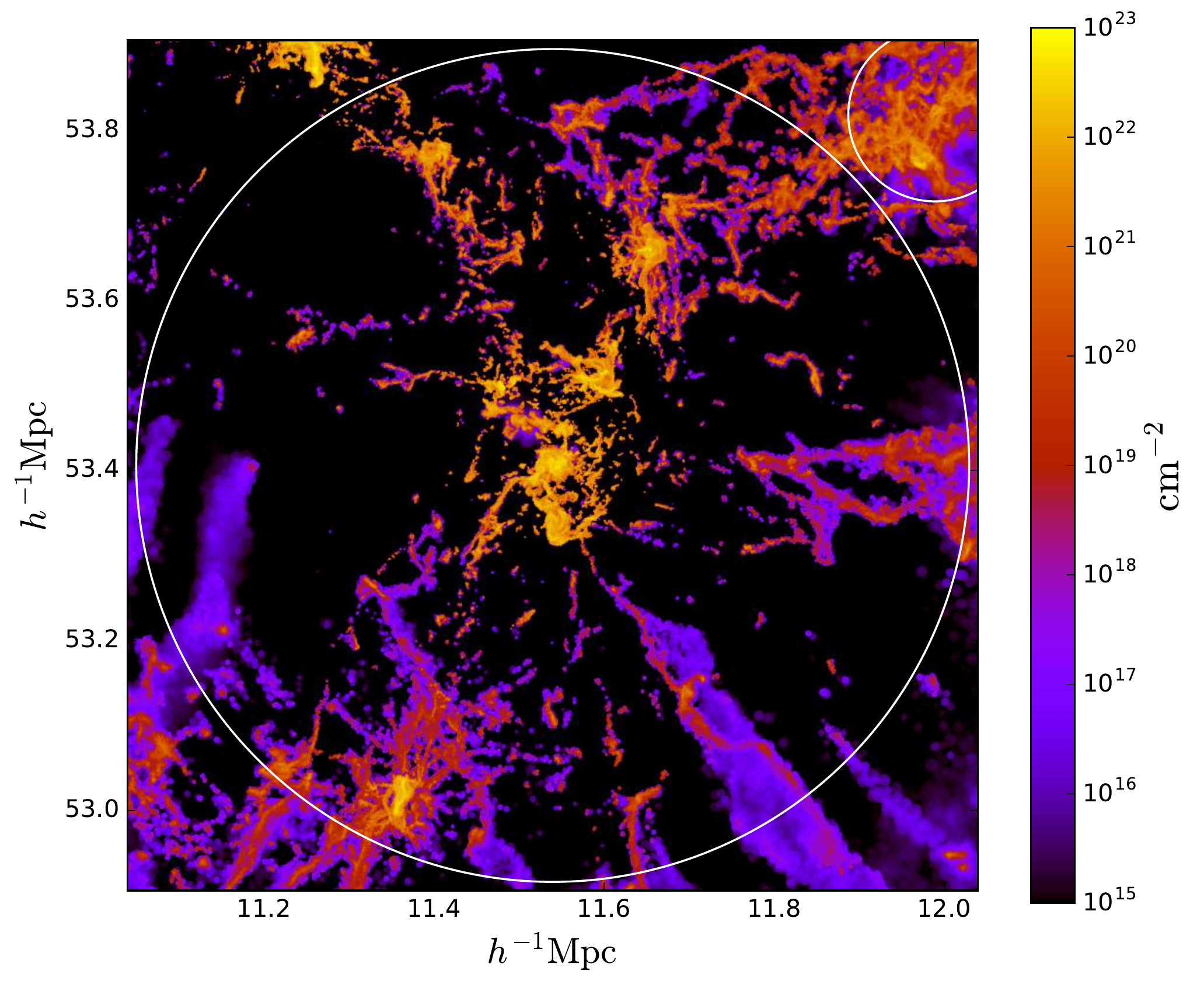}\\
\caption{Column density of gas (left panel) and HI (right panel) around a massive halo of mass $1.5\times10^{13}~h^{-1}M_\odot$ 
at $z=3$.  The white circles show the position and radius of dark matter halos.}
\label{fig:DLAs_image}
\end{center}
\end{figure*}

Another quantity commonly employed to study the abundance of neutral
hydrogen in the post-reionization era is the HI column density
distribution function (HI CDDF), defined as
\be
f_{\rm HI}(N_{\rm HI})=\frac{d^2n(N_{\rm HI})}{dN_{\rm HI}dX}~,
\ee
where $n$ is the number of lines-of-sight with column densities
between $N_{\rm HI}$ and $N_{\rm HI}+dN_{\rm HI}$, and $dX = H_0(1 +
z)^2/H(z)dz$ is the absorption distance. This quantity
can be inferred directly from
observations of the Ly$\alpha$-forest.

Here, we investigate the HI CDDF focusing on absorbers with
high column densities: damped Lyman alpha systems (DLAs), $N_{\rm
  HI}\geqslant10^{20.3}~{\rm cm}^{-2}$. We also examine the
DLA cross-sections, which are required both observationally 
\citep{Font_2012, Rafols_2018,Alonso_2017} and theoretically \citep{EmaPaco}.

In Fig. \ref{fig:DLAs_image} we show an example of the spatial
distribution of gas and HI around a massive halo at redshift
$z=3$. As in this case, DLAs correspond to gas in galaxies, 
gas recently stripped from galaxies, and gas in streams.

The HI CDDF at redshifts 0, 1, 2, 3, 4 and 5 is computed using the
following procedure \cite[we refer the reader to appendix B of][for
  further details]{Villaescusa-Navarro_2014a}. We approximate each
Voronoi cell by an uniform sphere with radius equal to
$R=(3V/4\pi)^{1/3}$, where $V$ is the volume of the cell, and
determine the HI column density of a line through it from $N_{\rm
  HI}=\rho_{\rm HI}d$, where $\rho_{\rm HI}=M_{\rm HI,cell}/V$ and $d$
is the length of the segment intersecting the sphere. The simulation
volume is projected along the z-axis and a grid with 20000x20000
points is overlaid.  Each point is considered to be a line-of-sight, and
the column density along it is estimated as the sum of
the column densities of all
Voronoi cells contributing to it. Since our box size is relatively
small, the probability of encountering more than a single absorber
with a large column density along the line of sight is
negligible. Thus, if the column density of a given line-of-sight is
larger than $\sim10^{19}~{\rm cm^{-2}}$, it can be attributed to a
single absorber. We repeated the tests carried out in
\cite{Villaescusa-Navarro_2014a} to verify that: 1) the grid is fine
enough to achieve convergence in the CDDF, and 2) the results do not
change if the CDDF is computed by slicing the box into slabs of
different widths.

\begin{figure}
\begin{center}
\includegraphics[width=0.45\textwidth]{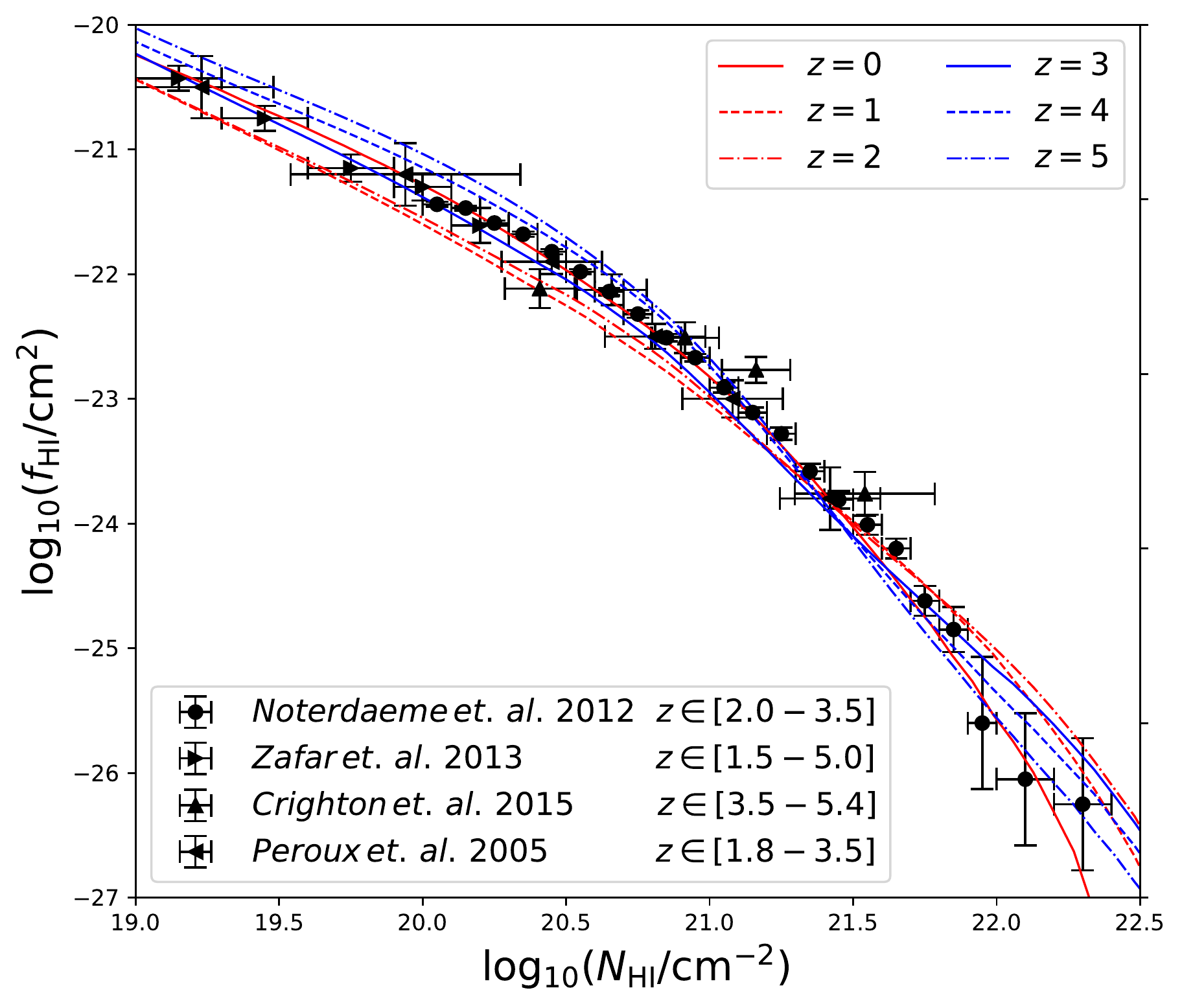}
\caption{HI column density distribution function as a function of the HI column density from the TNG100 simulation at redshifts 0 (solid red), 1 (dashed red), 2 (dot-dashed red), 3 (solid blue), 4 (dashed blue) and 5 (dot-dashed blue). Data from observations are shown as black points with errorbars.}
\label{fig:CDDF}
\end{center}
\end{figure}

We show the results in Fig. \ref{fig:CDDF}. We find excellent agreement
with the observations, which are shown as black points with errorbars, at
redshifts [1.8-3.5] \citep{Peroux_2005}, [2.0-3.5]
\citep{Noterdaeme_2012}, [3.5-5.4] \citep{Crighton_2015} and [1.5-5.0]
\citep{Zafar_2013}. The differences between the observed
and simulated CDDFs, e.g. the
amplitude of the HI CDDF around $10^{20}-10^{21}~{\rm cm}^{-2}$, are
related to the mismatch between $\Omega_{\rm HI}$ from
observations and TNG100 (see Fig. \ref{fig:Omega_HI}), since
\be
\Omega_{\rm HI}(z)=\frac{m_{\rm H}H_0}{c\rho_c^0}\int_0^\infty f_{\rm HI}(N_{\rm HI},z)N_{\rm HI}dN_{\rm HI}~,
\ee
where $m_{\rm H}$ is the mass of the hydrogen atom and $c$ is the
speed of light. In agreement with previous works, we find that the HI
CDDF exhibits a weak dependence on
redshift \cite[see e.g.][]{Rahmati_2013}. This self-similarity can be
associated with the weak redshift dependence that we observe in the
high overdensity tail of the HI density pdf for small smoothing
scales (see Fig. \ref{fig:HI_pdf}).

\begin{figure}
\begin{center}
\includegraphics[width=0.45\textwidth]{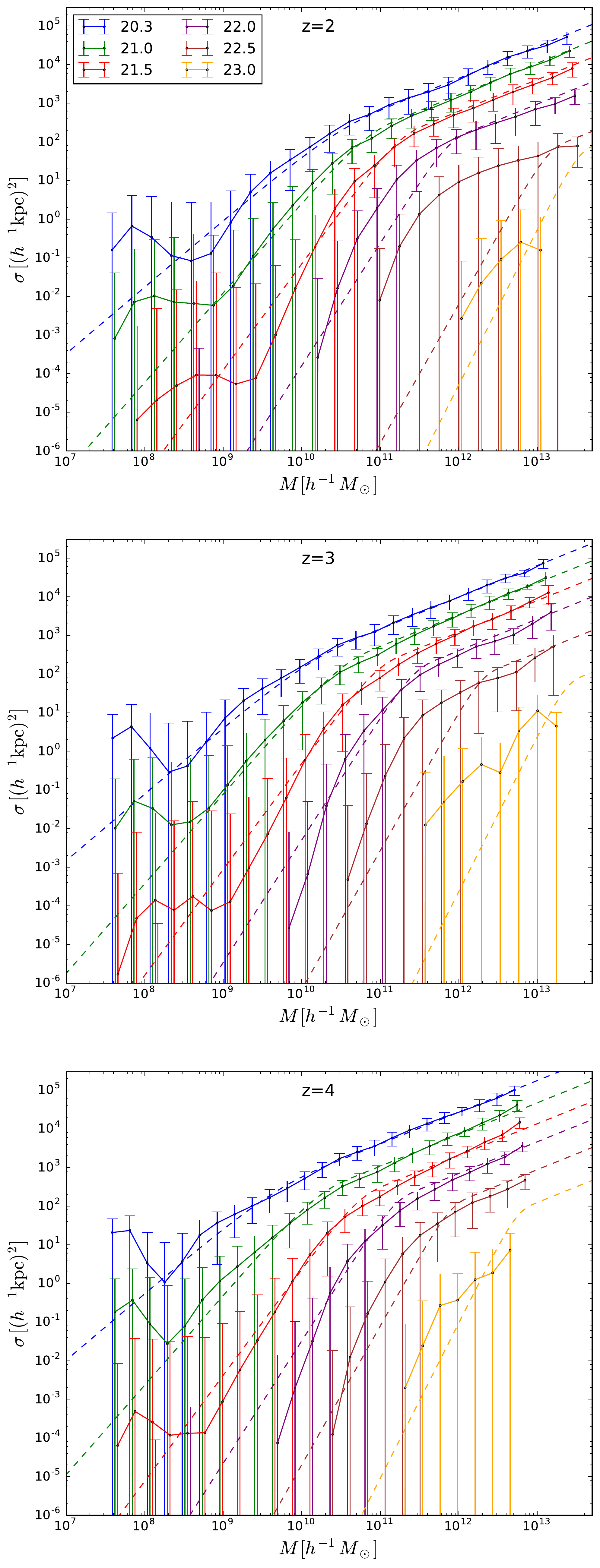}
\caption{The cross-section of DLAs (dark matter halo area covered by DLAs) with column densities above $10^{20.3}$ (blue), $10^{21}$ (green), $10^{21.5}$, $10^{22}$ (purple), $10^{22.5}$ (brown) and $10^{23}$ (orange) ${\rm cm^{-2}}$ at redshifts $z=2$ (top), $z=3$ (middle) and $z=4$ (bottom). The points with error bars are measurements from the TNG100 simulation while the dashed lines represent our fit using Eq. \ref{eq:DLAs_fitting_function}.}
\label{fig:DLAs_cross_section}
\end{center}
\end{figure}

Next, we examine the DLAs cross-section. For each dark matter halo of
the simulation the area covered by DLAs with different column
densities is computed.  Then, all halos within mass bins are selected
and the mean and standard deviation of their DLA cross-sections are
determined. As shown in Fig. \ref{fig:DLAs_cross_section} we find
that, for fixed column density, the DLA cross-section increases with
halo mass, while the cross-section decreases with column density for
halos of fixed mass.

The cross-section of the DLAs is well fitted by the following function
\be
\sigma(M|N_{\rm HI},z)=A\left(\frac{M}{h^{-1}M_\odot}\right)^\alpha\left(1-e^{-(M/M_0)^\beta}\right)~.
\label{eq:DLAs_fitting_function}
\ee
Here, $A$ is a parameter that controls the overall normalization of
function, while $\alpha$ sets the slope of the cross-section for
large halo masses, and $M_0$ determines the characteristic halo mass where
the DLA cross-section exponentially decreases at a rate controlled by
$\beta$.

We fit our results at redshifts $z\in[2,4]$ using the above form
and find $\alpha=0.82$ in the large majority of the cases,
while $\beta$ is well approximated by $\beta=0.85\cdot\log_{10}(N_{\rm
  HI}/{\rm cm}^{-2})-16.35$. There is also a strong correlation between
$A$ and $M_0$, given by $A\cdot M_0=0.0141~h^{-2}{\rm
  kpc}M_\odot$. The only redshift-dependence enters through $M_0$, the
value of which is given in Table \ref{table:M0}. We find that 
$M_0$ decreases with redshift, in agreement with 
the halo HI mass function which implies that less massive halos
can host HI at higher redshifts.

\begin{table}
\begin{center}
 \begin{tabular}{|c|| c| c| c|} 
 \hline
 & $z=2$ & $z=3$ & $z=4$ \\ [0.5ex] 
 \hline\hline
 20.0 & 10.23 & 9.89 & 9.41 \\
 \hline
 20.3 & 10.34 & 10.00 & 9.56 \\ 
 \hline
 21.0 & 10.77 & 10.45 & 10.14 \\
 \hline
 21.5 & 11.20 & 10.91 & 10.68 \\
 \hline
 22.0 & 11.83 & 11.39 & 11.14 \\
 \hline
 22.5 & 13.11 & 12.26 & 11.87 \\ 
 \hline
 23.0 & 13.49 & 13.34 & 12.72 \\
 \hline
\end{tabular}
\end{center}
\caption{ \label{table:M0} Fits to the DLA cross-section from simulations using Eq. \ref{eq:DLAs_fitting_function}. This table shows the value of $\log_{10}M_0$ at different redshifts and DLA column densities. The value of the other parameters in function \ref{eq:DLAs_fitting_function} are given by $\alpha=0.82$, $\beta=0.85\cdot\log_{10}(N_{\rm HI}/{\rm cm^{-2}})-16.35$ and $A\cdot M_0=0.0141~~h^{-2}{\rm kpc}M_\odot$.}
\end{table}

The fits to the simulation results, shown as dashed lines in
Fig. \ref{fig:DLAs_cross_section}, are a good approximation for column
densities below $10^{22}~{\rm cm^{-2}}$, but apparently less so at
higher column densities; e.g. the fit for column densities above
$10^{22.5}~{\rm cm^{-2}}$ and low halo masses is several orders of
magnitude below the mean. This is mainly an illusion of the fact that
some error bars are larger than the value. The reduced $\chi^2$
obtained from the fits in all cases is below 0.35.  The preferred
value for $\alpha$ is slightly larger at higher redshifts, but the
redshift-dependence is so weak that for simplicity we did not use it
in our fitting. The largest discrepancy between the fit and our
results occurs at $z=2$ for the DLAs with column densities larger than
$10^{22}~{\rm cm}^{-2}$.
In \cite{EmaPaco} it was suggested that a very good fit to the column density distribution and the DLA bias can be obtained assuming the differential cross section, $\rm{d} \sigma /\rm{d} N_{\rm HI}$ is roughly independent of column density. This implies that the linear bias of different absorbers will be very similar, and the measurements in the BOSS survey of \citep{Rafols_2018} confirm this simple picture, although with large errorbars.
As discussed above, our fit to Eq.~\ref{eq:DLAs_fitting_function} indicates the the slope $\alpha$ to a very good approximation is not a function of column density, and $N_{\rm HI}$ dependence of $\beta$ can be moved to the normalization constant $A$ using their tight correlation. If we then look at the expression for the linear bias of a given absorber

\begin{equation}
b_{N_{\rm HI}}(z) = \frac{\int_0^\infty b(M,z)n(M,z)\sigma(M|N_{\rm HI},z) dM}{\int_0^\infty n(M,z)  \sigma(M|N_{\rm HI},z) dM}\;,
\end{equation}
we notice that that $A$ cancels between the numerator and the denominator, and the only column density dependence is left in $\beta$ and it is rather small.
The analytical calculation in \cite{EmaPaco} therefore agrees with the measurements in IllustrisTNG, and future observations will tell us if our current understanding of the cross section is correct or not.

We have used the above expression to estimate the bias of the DLAs. We take the DLAs cross-section (for absorbers with $N_{\rm HI}>10^{20}~{\rm cm}^{-2}$) and halo mass function from our simulations and use the formula in \cite{SMT} to compute the halo bias. We obtain values of DLAs bias equal to 1.7 at $z=2$ and 2 and $z=3$. Considering that the DLAs bias follows a linear relation between $z=2$ and $z=3$ we obtain $b_{\rm DLA}(z=2.3)=1.8$, in agreement with the latest observations by \cite{Rafols_2018}: $b_{\rm DLAs}=1.99\pm0.11$. We have repeated the above calculations using our fit for the DLAs cross-section, taking the halo mass function from \cite{Sheth-Tormen} or \cite{Crocce_2010} and find that our results barely change.

We believe that the above calculation should be considered as a lower bound. In other words, the halo bias may be underestimated when calculated using \cite{SMT}. The reason is that we obtain a value of the HI bias (see section \ref{subsec:HI_bias}), computed without any assumption, of 2 at $z=2$ and 2.56 at $z=3$, i.e. $b_{\rm HI}(z=2.3)=2.17$. Following the theoretical arguments in \cite{EmaPaco} it is reasonable to expect that $b_{\rm HI}\simeq b_{\rm DLAs}$. We this conclude that both estimations of the DLAs bias are in agreement with observations.

\section{HI bulk velocity}
\label{sec:HI_bulk}

In section \ref{subsec:HI_in_halos_galaxies} we showed that nearly all
the HI at redshifts $z\leqslant5$ resides within halos. Thus, the
elements needed to describe the abundance and spatial distribution of
HI in real-space through HI halo models are the halo HI mass function
and the HI density profiles. To model the distribution of HI in
redshift space, an additional ingredient is required: the velocity
distribution of HI inside halos. This quantity can be used in both HI
halo models and HI HOD models. The accuracy that can be achieved with
the former may not be high, due to the limitations of the formalism
itself. On the other hand, HI HOD, i.e.~painting HI on top of dark
matter halos from either N-body or fast numerical simulations like
COLA \citep{COLA}, can produce highly accurate results. Hence, we
examine the velocity distribution of HI inside halos, beginning with
the HI bulk velocity in this section, and continuing with the HI
velocity dispersion in the next section, and, in both cases, comparing
with the results for all matter.

For each dark matter halo in the simulation we have computed the HI bulk velocity as 
\be
\vec{V}_{\rm HI}=\frac{\sum_i M_{\rm HI,i}\vec{V}_{\rm HI,i}}{\sum_i M_{\rm HI,i}}~,
\ee
where the sum runs over all gas cells belonging to the halo and
$M_{\rm HI,i}$ and $\vec{V}_{\rm HI,i}$ are the HI mass and peculiar
velocity of cell $i$, respectively. The peculiar velocity of halos,
$\vec{V}_h$, is computed in a similar manner, but summing over
all resolution elements in the halo (gas, CDM, stars and black holes)
and weighting their velocities by their corresponding masses.

\begin{figure*}
\begin{center}
\includegraphics[width=0.33\textwidth]{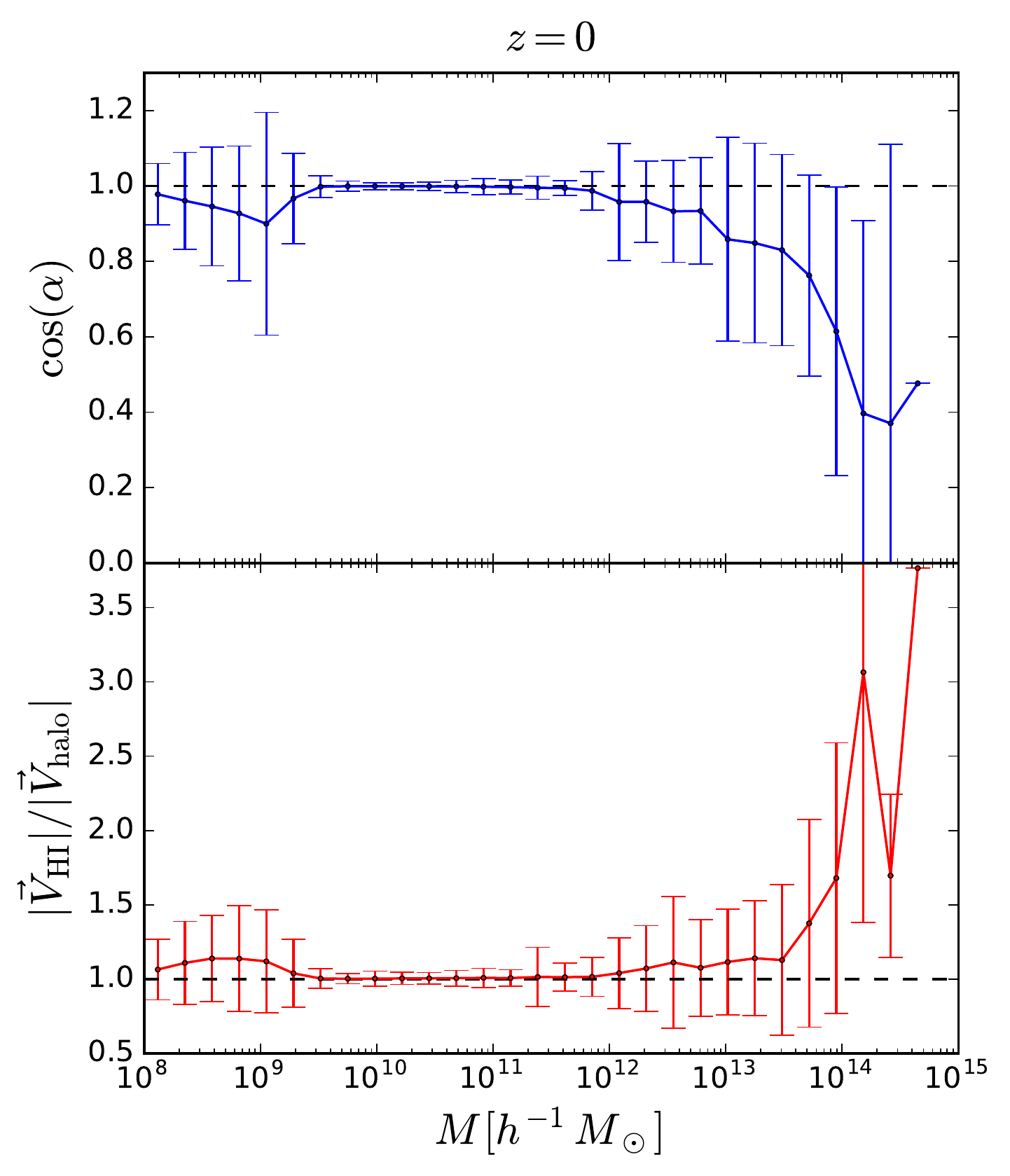}
\includegraphics[width=0.33\textwidth]{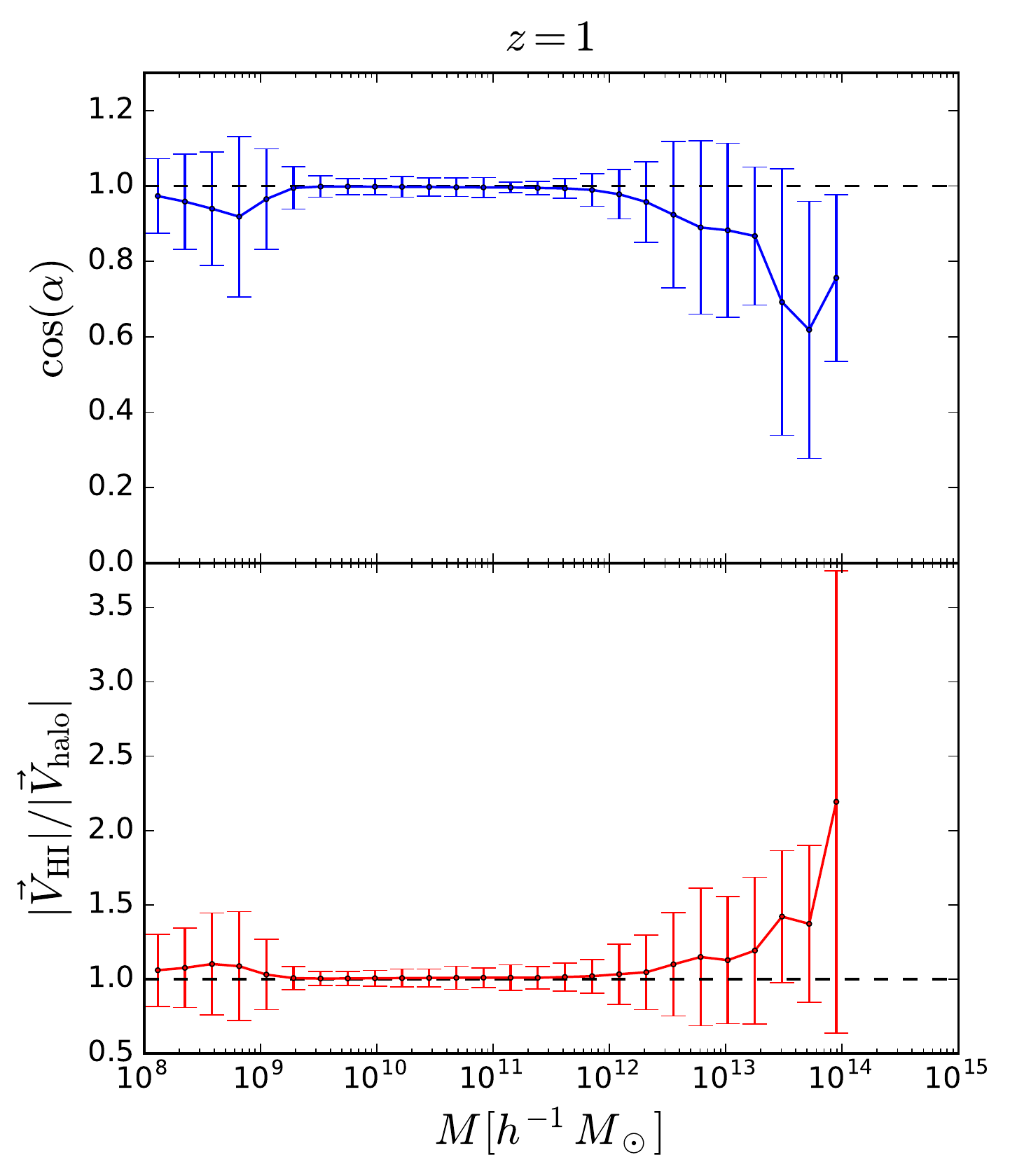}
\includegraphics[width=0.33\textwidth]{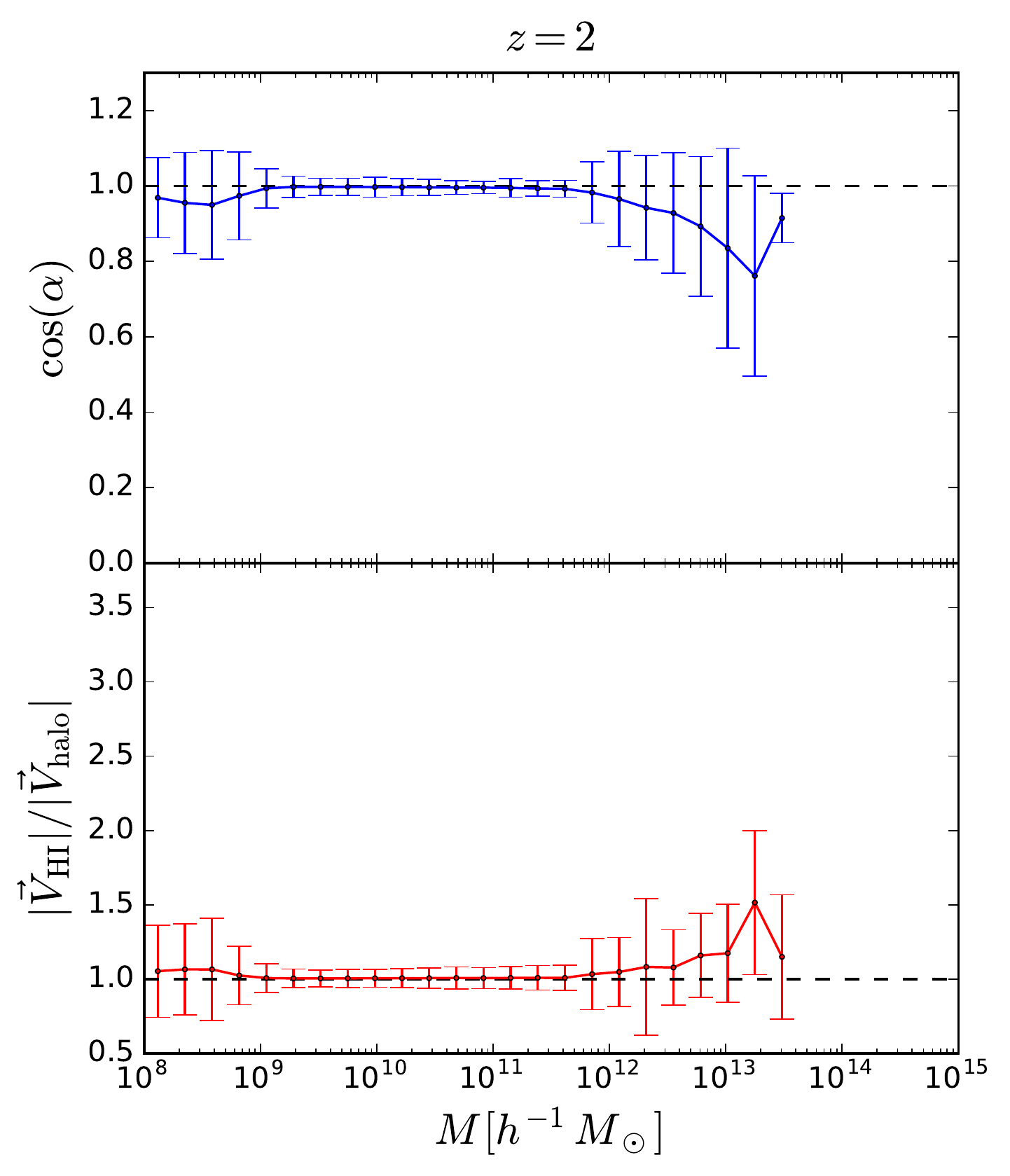}\\
\includegraphics[width=0.33\textwidth]{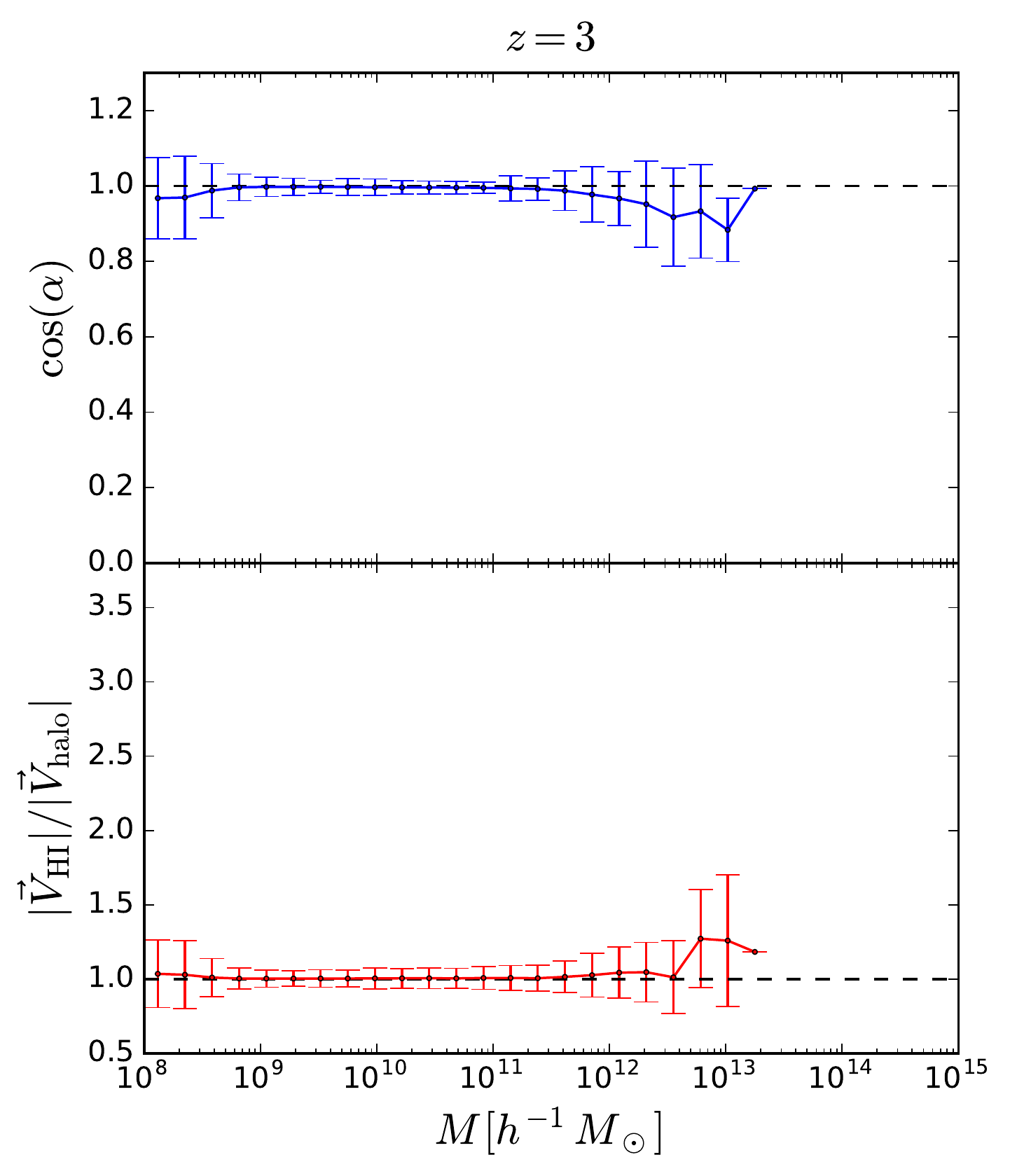}
\includegraphics[width=0.33\textwidth]{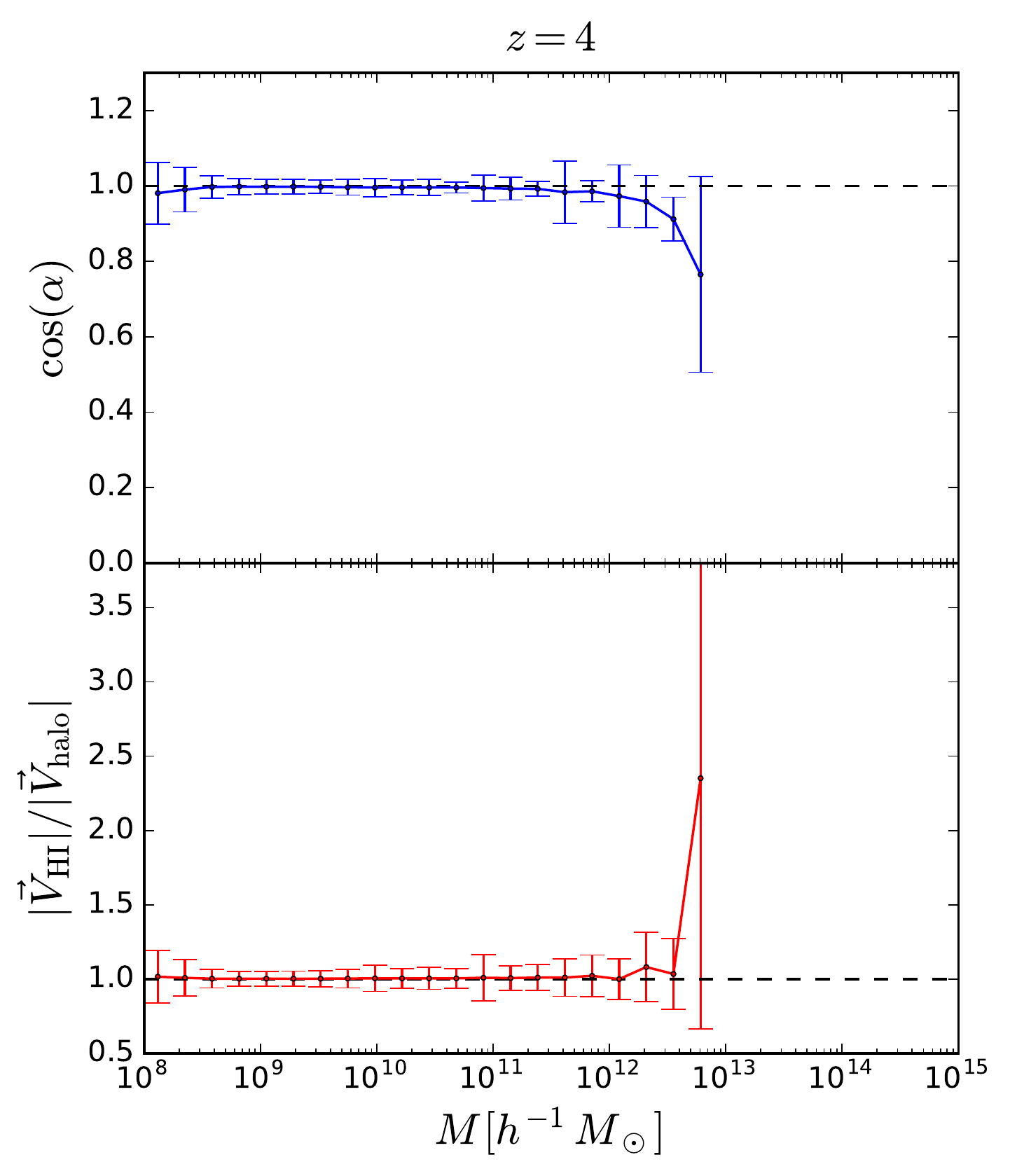}
\includegraphics[width=0.33\textwidth]{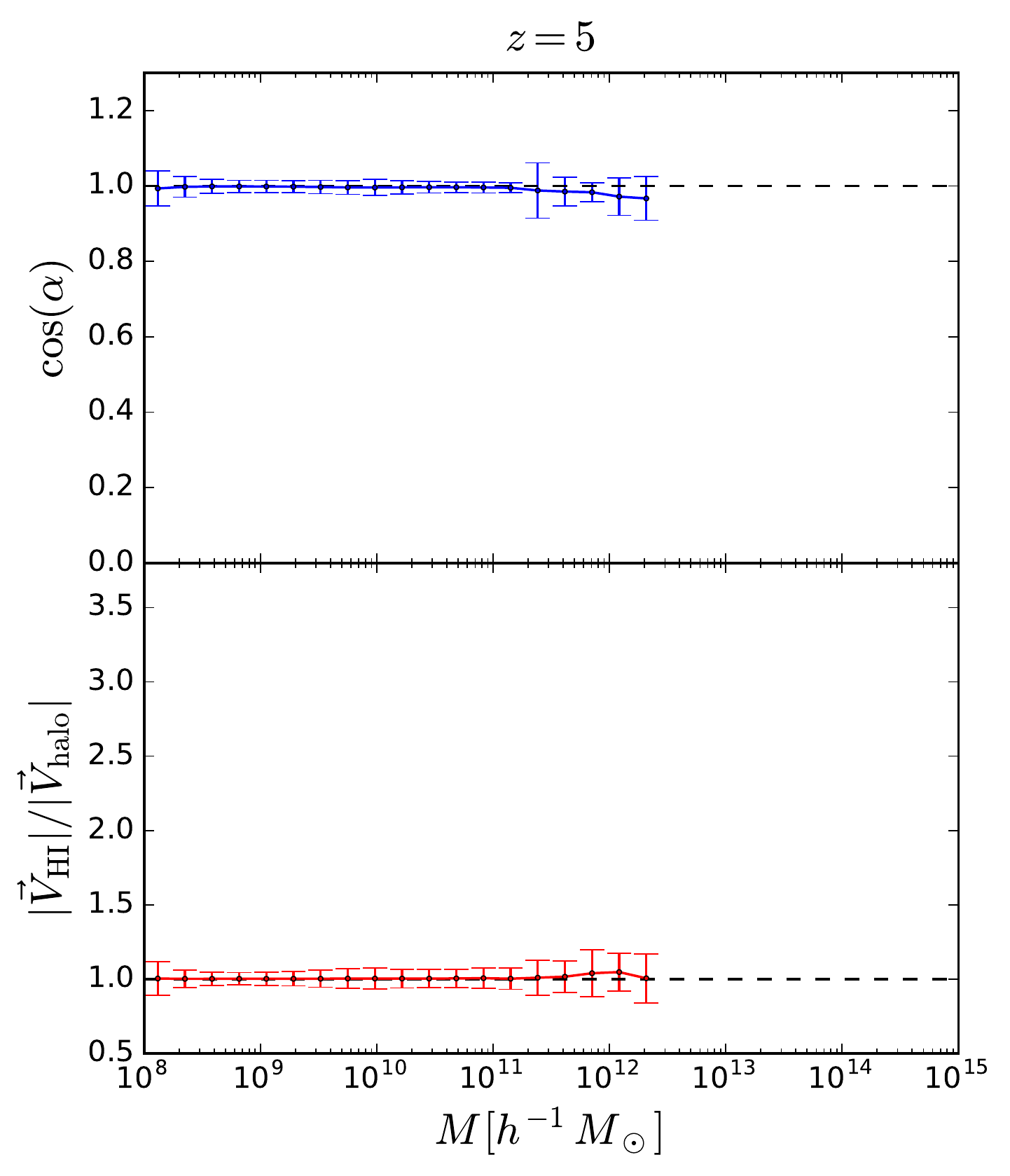}
\caption{The upper panels show the average angle between the HI and halo peculiar velocity vectors, $\cos(\alpha)=\vec{V}_{\rm HI}\cdot\vec{V}_h/(|\vec{V}_{\rm HI}||\vec{V}_h|)$, as a function of halo mass. The bottom panels display the average ratio between the moduli of the HI and halo peculiar velocity vectors as a function of halo mass. Only halos with total HI mass above $10^5~h^{-1}M_\odot$ are included. We show results at redshift 0 (top-left), 1 (top-middle), 2 (top-right), 3 (bottom-left), 4 (bottom-middle) and 5 (bottom-right). While the HI bulk velocity traces the halo peculiar velocity for small halos in both modulus and direction, there are departures for larger halos. This happens because most of the HI in small halos is in the central galaxy while in larger halos the contribution from satellites becomes more important.}
\label{fig:HI_bulk}
\end{center}
\end{figure*}

Here, we examine: 1) whether the peculiar velocity of the HI points
in the same direction as the halo peculiar velocity, and 2)
whether the modulus of the HI peculiar velocity is the same as that
of the halo peculiar velocity.  The first point is addressed by
computing the angle between the peculiar velocities of HI and the halo
from
\be
\cos(\alpha)=\frac{\vec{V}_{\rm HI}\cdot\vec{V}_h}{|\vec{V}_{\rm HI}||\vec{V}_h|}
\label{Eq:HI_bulk}
\ee
for each halo in the simulation. We do not consider halos with total
HI masses below $10^{5}~h^{-1}M_\odot$, since we expect the HI
peculiar velocities of those halos to be uncorrelated with halo
peculiar velocity. For example, the HI in such halos mass can be from
a single cell that is partially self-shielded and not bound to the
halo.  Moreover, in the limit where the HI mass is close to zero, the
HI velocity dispersion is not well defined. Thus, in order to avoid
such circumstances, we adopt the above threshold, which
corresponds to the mass of $\simeq1/5$ of a completely self-shielded
gas cell.  However, we find that this threshold does not have a 
significant impact on our results.  Choosing a different value
hardly changes our results,
with the only consequence being that the scatter of very small halos is
affected. We then take narrow bins in halo mass and compute the mean
value of $\cos(\alpha)$ and its standard deviation. The
resuts are shown in the upper panels of Fig. \ref{fig:HI_bulk}.

For small halos, $M\lesssim10^{12}~h^{-1}M_\odot$,
$\cos(\alpha)\simeq1$, indicating that the HI and halo peculiar
velocities are aligned. This is expected because the HI is
mainly located in the inner regions in low-mass 
halos, which usually traces well the
peculiar velocity of the halo. For smaller halos the
value of $\cos(\alpha)$ deviates from 1, with increased scatter.
This happens for halo masses below the cutoff scale, $\sim
M_{\rm min}$. In at least some cases, this is likely due to halos
acquiring HI through an unusual mechanism,
e.g. by passing through an HI rich filament, so that the HI
bulk velocity will not be correlated with the halo peculiar
velocity. On the other hand, we find significant misalignments between
the HI and halo peculiar velocities for the most massive halos at any
redshift. This is because the HI content of these halos is largely
contributed by satellites, whose peculiar motions do not
necessarily trace that of the halo. We return to this point below.

Further, in the bottom panels of Fig. \ref{fig:HI_bulk}, we show the
average and standard deviation of the ratio between the moduli of the
HI and halo peculiar velocities, $|\vec{V}_{\rm
  HI}|/|\vec{V}_h|$. This quantity is again calculated in narrow bins
in halo mass, for all halos with HI mass larger than
$10^5~h^{-1}M_\odot$.

For small halos, the moduli of the HI and halo peculiar velocities are
essentially the same. For halos with masses below $\sim M_{\rm min}$,
the modulus ratio can be larger than 1 and its scatter increases. This
is for the same reason as above: the HI content of some of those halos
may not be bound to the halos and are instead part of a filament. For
massive halos, the modulus of the HI peculiar velocity can be much
larger than that of the halo peculiar velocity. As earlier, this is
because the HI peculiar velocity is dominated by the HI in satellites,
whose peculiar velocities do not perfectly trace the halo peculiar
velocity.

To corroborate the assertion that the peculiar velocities of
satellites do not trace the halo peculiar velocity in either modulus
or direction, we have performed the following test. We compute the
peculiar velocities of halo satellites and compared their mean,
weighted by the total mass of each satellite, against the peculiar
velocity of their host halo.  The velocities of the satellites do not
have the same modulus or direction as those of the host halo, showing
similar trends to those for HI, with differences increasing with halo
mass.

Thus, for small halos, where most of the HI is in the central galaxy,
the HI bulk velocity traces the halo peculiar velocity well, in both
modulus and direction. On the other hand, the contribution of
satellites to the total HI mass in halos increases with mass, and
since the bulk velocities of satellites do not trace the halo peculiar
velocity, the HI bulk velocities will depart, in modulus and
direction, from the halo peculiar velocity, with differences
increasing with halo mass.

\section{HI velocity dispersion}
\label{sec:sigma_HI}

\begin{figure*}
\begin{center}
\includegraphics[width=0.33\textwidth]{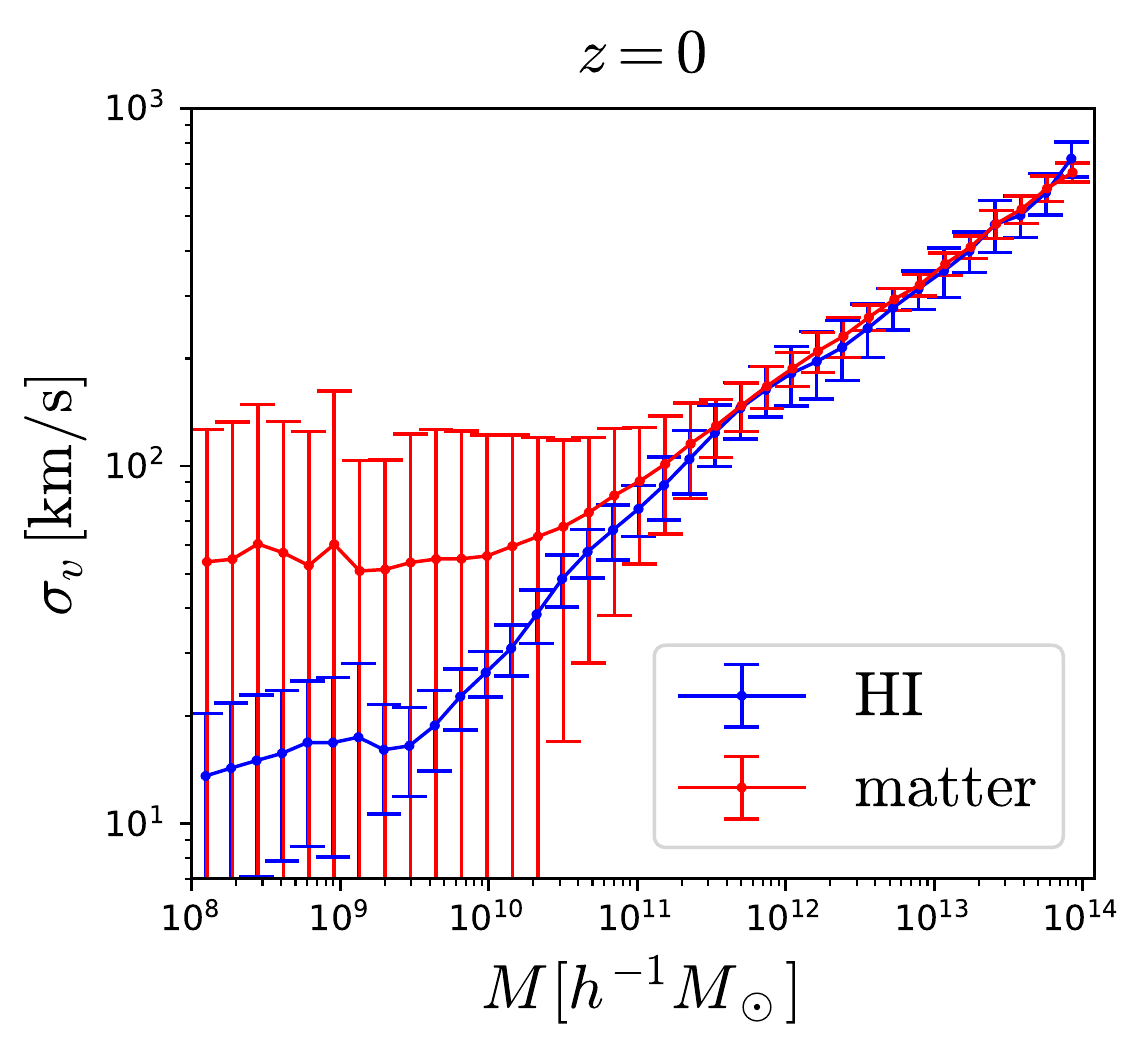}
\includegraphics[width=0.33\textwidth]{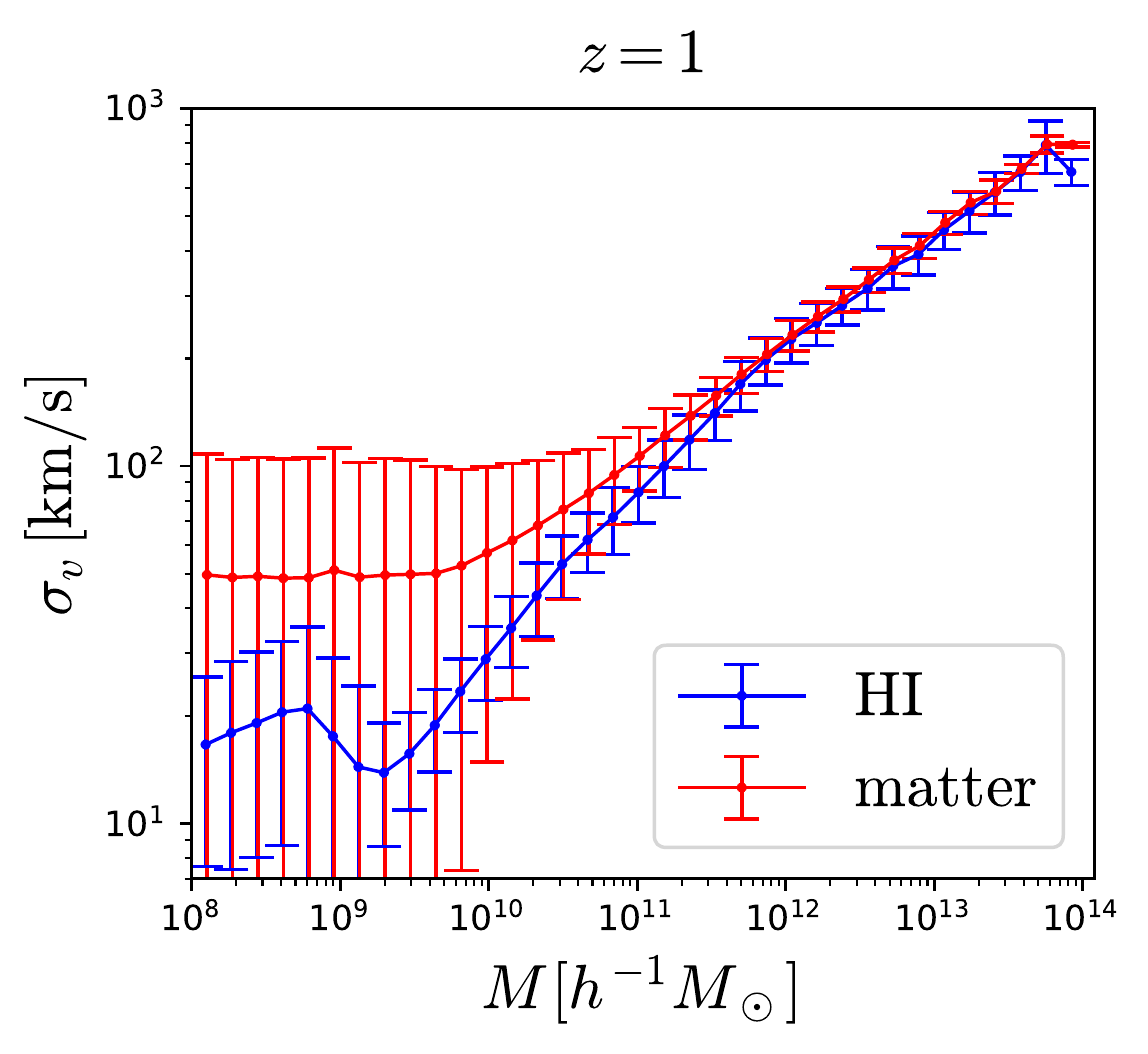}
\includegraphics[width=0.33\textwidth]{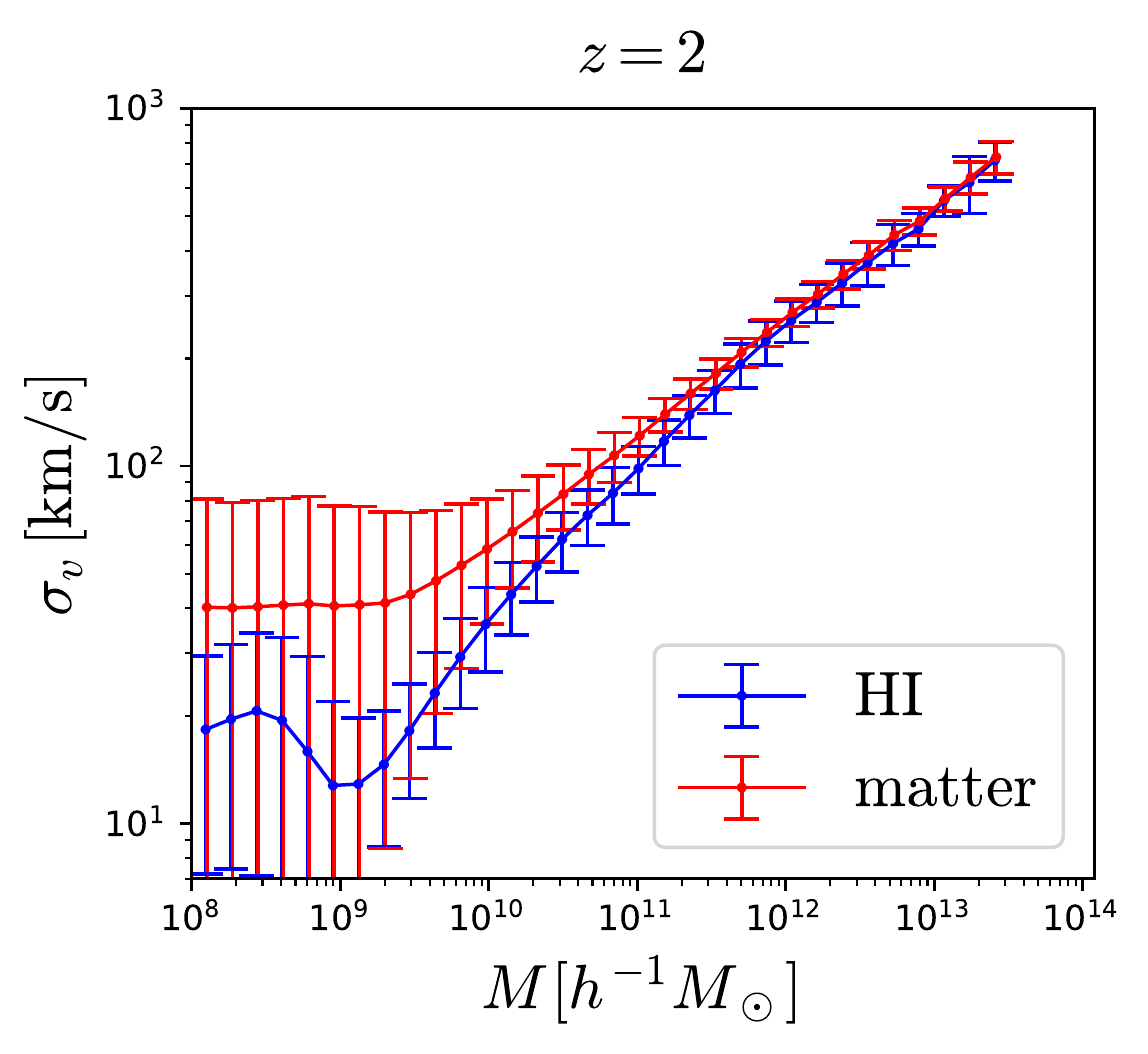}\\
\includegraphics[width=0.33\textwidth]{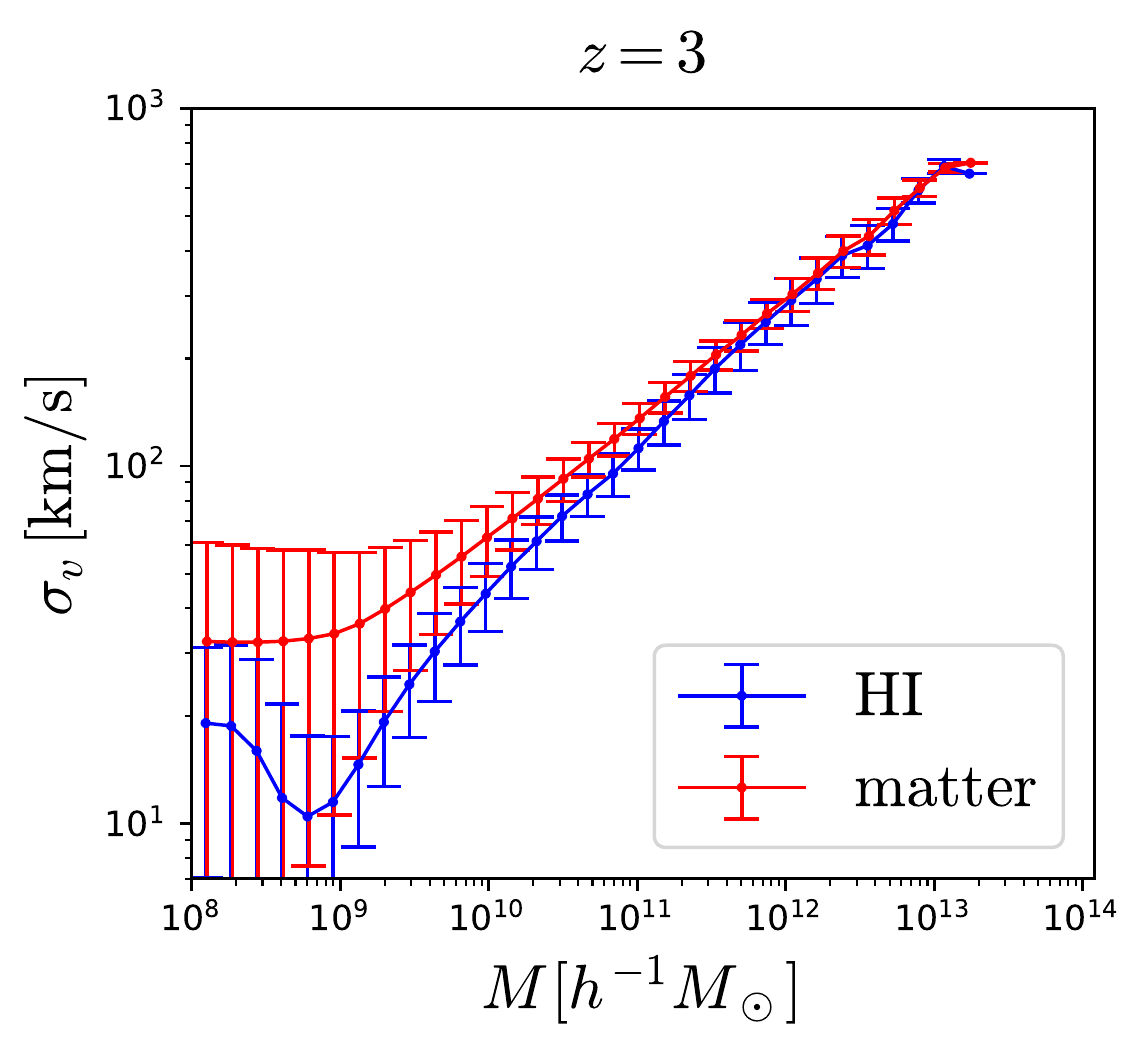}
\includegraphics[width=0.33\textwidth]{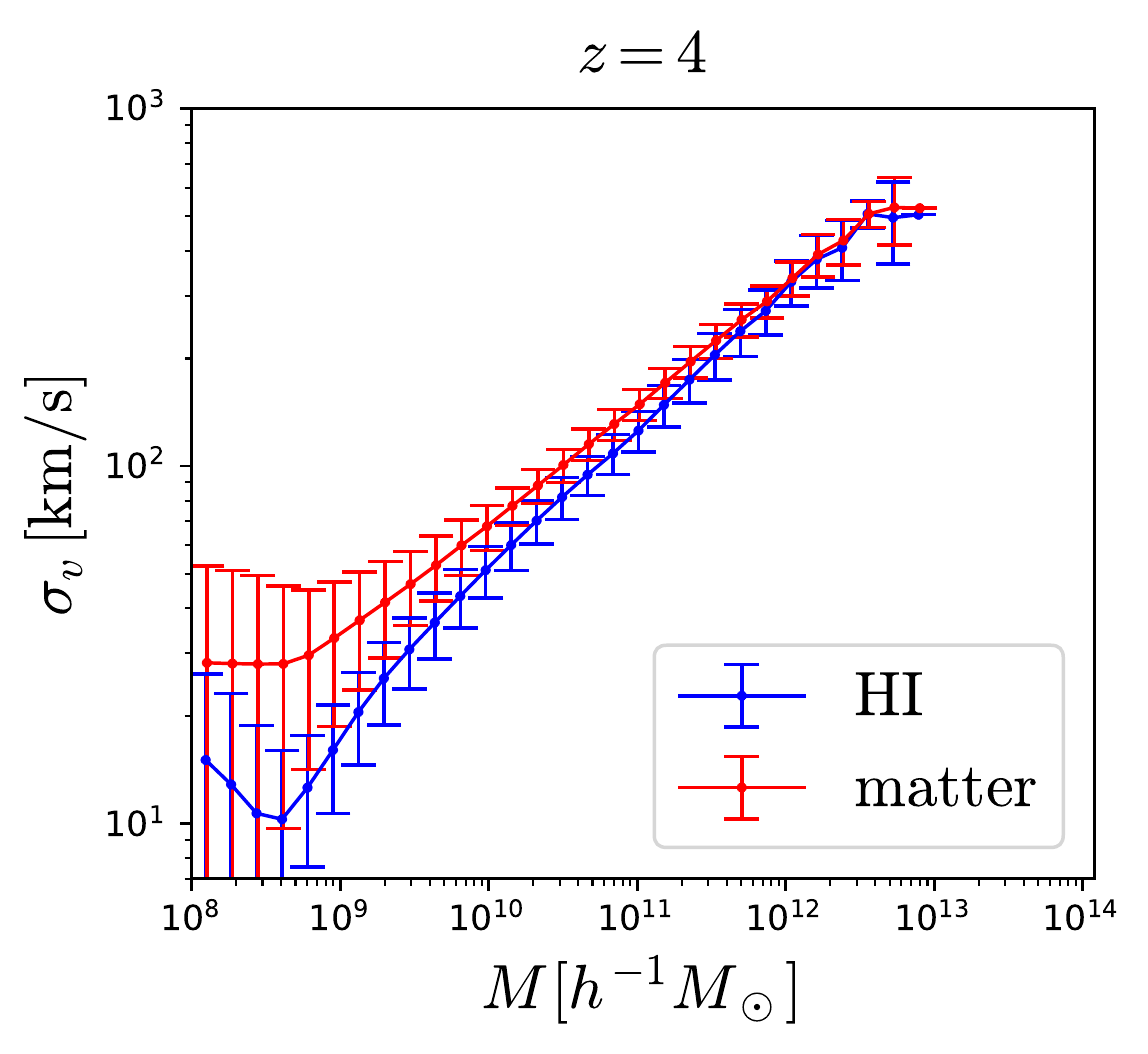}
\includegraphics[width=0.33\textwidth]{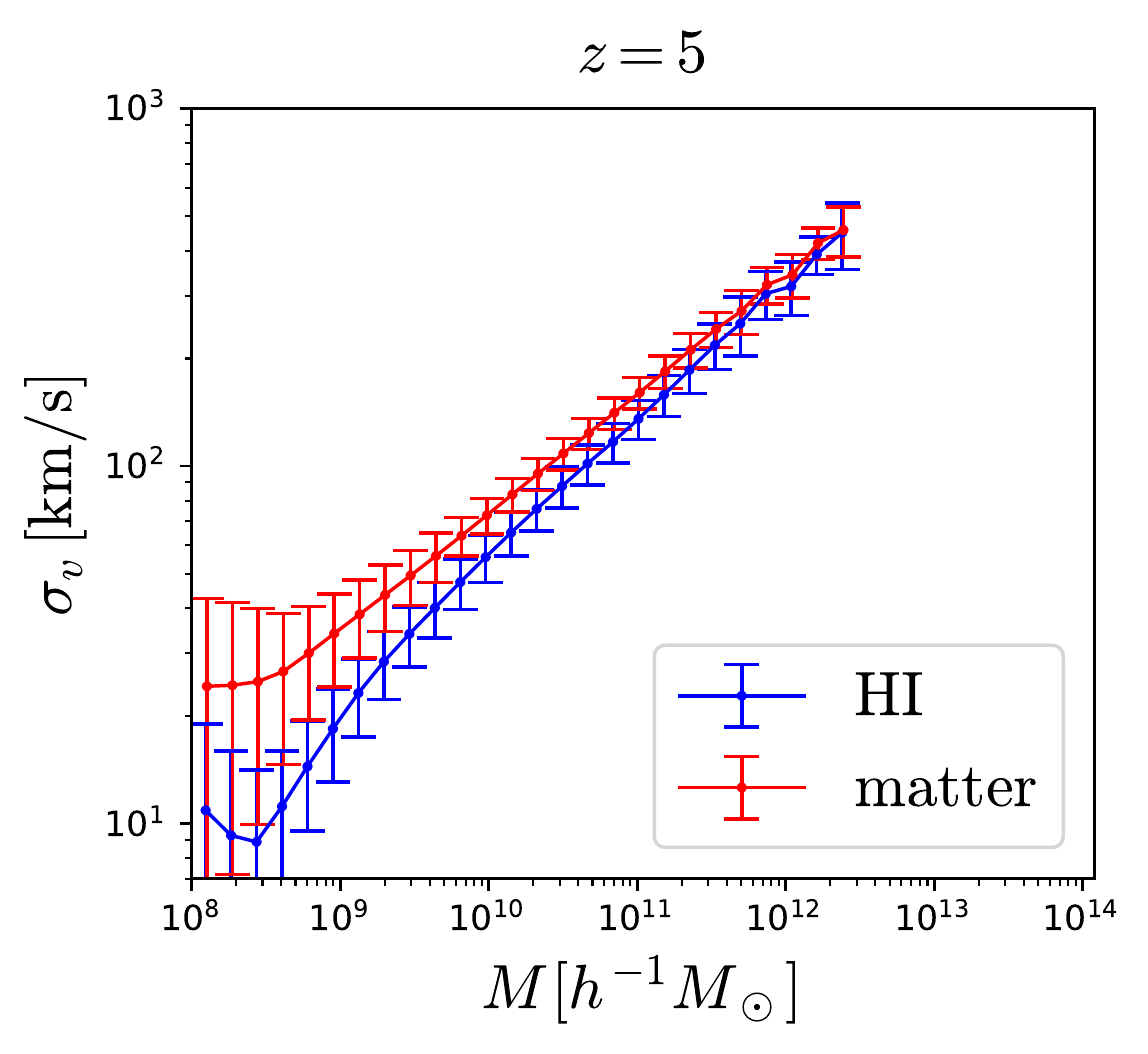}
\caption{The mean and standard deviation of the 3D velocity dispersion of HI (blue lines) and CDM (red lines) as a function of halo mass at redshifts 0 (top-left), 1 (top-middle), 2 (top-right), 3 (bottom-left), 4 (bottom-middle) and 5 (bottom-right). Our results are well-represented by a simple power law, $\sigma=\sigma_{10}(M/10^{10}~h^{-1}M_\odot)^\alpha$, with best-fit parameter given in Table \ref{table:HI_sigma}. At low redshift, the velocity dispersion of HI is very similar to that of CDM, while at high-redshift CDM exhibits larger values than HI.}
\label{fig:HI_sigma}
\end{center}
\end{figure*}

For each halo in the simulation we have computed the 3D velocity dispersion of its HI from
\be
\sigma_{\rm HI}^2=\frac{\sum_i M_{{\rm HI},i}|\vec{V}_{\rm HI,i}-\vec{V}_{\rm HI}|^2}{\sum_i M_{{\rm HI},i}}~,
\ee
where the sums run over all gas cells belonging to the halo. $M_{{\rm
    HI},i}$ and $\vec{V}_{\rm HI,i}$ are the HI mass and velocity of
gas cell $i$, and $\vec{V}_{\rm HI}$ is the HI bulk velocity,
computed as in Eq. \ref{Eq:HI_bulk}.

We then take narrow bins in halo mass and compute the mean and
standard deviation of the HI velocity
dispersion. Fig. \ref{fig:HI_sigma} shows the results, as well as a
comparison with matter, whose properties are calculated analogously,
but considering all mass elements within halos; i.e. gas, CDM, stars
and black holes.

As expected, the velocity dispersion of both HI and CDM increases with
halo mass, independent of redshift. The results can be
represented by a simple power-law
\be
\sigma_v(M)=\sigma_{10}\left(\frac{M}{10^{10}~h^{-1}M_\odot}\right)^\alpha~,
\ee
where $\sigma_{10}$ and $\alpha$ are free parameters with best-fit
values provided in Table \ref{table:HI_sigma}.

The mean HI velocity dispersions are always equal to or smaller than
the matter velocity dispersions. For large halo masses both exhibit
the same amplitude, but for low mass halos the HI velocity dispersion
is less than that of matter. The typical halo masses where the
velocity dispersions diverge is around $10^{12}~h^{-1}M_\odot$, with
higher redshifts exhibiting departures at larger masses. This behavior
is embedded in the slope of the $\sigma_{\rm HI}(M)$ relation, whose
value, $\alpha$, is larger than that of $\sigma_{\rm m}(M)$ at all
redshifts, and more so at higher redshifts.

For very small halos, and particularly at low redshift, the velocity
dispersion of HI is much smaller than that of matter. This in a
consequence of several factors. First, $\sigma_{\rm m}(M)$ is
artificially high, as the relation flattens out towards low masses. By
comparing to a version of Fig.~\ref{fig:HI_sigma} (not shown)
generated from a lower-resolution analogue of the same cosmological
volume (TNG100-2; see \citealp{NelsonD_17a}), we conclude that this is
due to finite numerical resolution, driven by particles in the
outskirts of the halos, which does not apply to the HI, which is
centrally-concentrated. In addition, we have examined a few individual
low-mass halos and found that in some case the HI arises from just a
few cells, or even a single one. In those cases, the HI bulk velocity
will be set by these few cells and the HI velocity dispersion will be
artificially suppressed due to sampling, again from finite resolution.
As we move to more massive halos, the contribution of HI from
satellites increases, and those satellites trace the underlying matter
distribution more closely.

\begin{table}
\begin{center}
 \begin{tabular}{|c|| c| c|| c| c|} 
 \hline
  & \multicolumn{2}{|c||}{matter} & \multicolumn{2}{|c|}{HI}\\[0.5ex]
 \hline
 z & $\sigma_{10}$ [km/s] & $\alpha$ & $\sigma_{10}$ [km/s] & $\alpha$ \\ [0.5ex] 
 \hline\hline
 0 & $49\pm5$ & $0.28\pm0.01$ & $31\pm1$ & $0.35\pm0.01$ \\ 
 \hline
 1 & $56\pm4$ & $0.30\pm0.01$ & $34\pm1$ & $0.37\pm0.01$\\
 \hline
 2 & $59\pm3$ & $0.32\pm0.01$ & $39\pm2$ & $0.38\pm0.01$\\
 \hline
 3 & $64\pm3$ & $0.33\pm0.01$ & $44\pm2$ & $0.39\pm0.01$\\
 \hline
 4 & $70\pm2$ & $0.33\pm0.01$ & $51\pm2$ & $0.39\pm0.01$\\ 
 \hline
 5 & $75\pm2$ & $0.33\pm0.01$ & $54\pm2$ & $0.40\pm0.01$\\
 \hline
\end{tabular}
\caption{\label{table:HI_sigma} The mean 3D velocity dispersion of both matter and HI inside halos can be represented by the relation $\sigma=\sigma_{10}(M/10^{10}~h^{-1}M_\odot)^\alpha$. The table gives the best-fits for $\sigma_{10}$ and $\alpha$ for matter and HI at different redshifts.}
\end{center}
\end{table}

\begin{figure*}
\begin{center}
\includegraphics[width=0.32\textwidth]{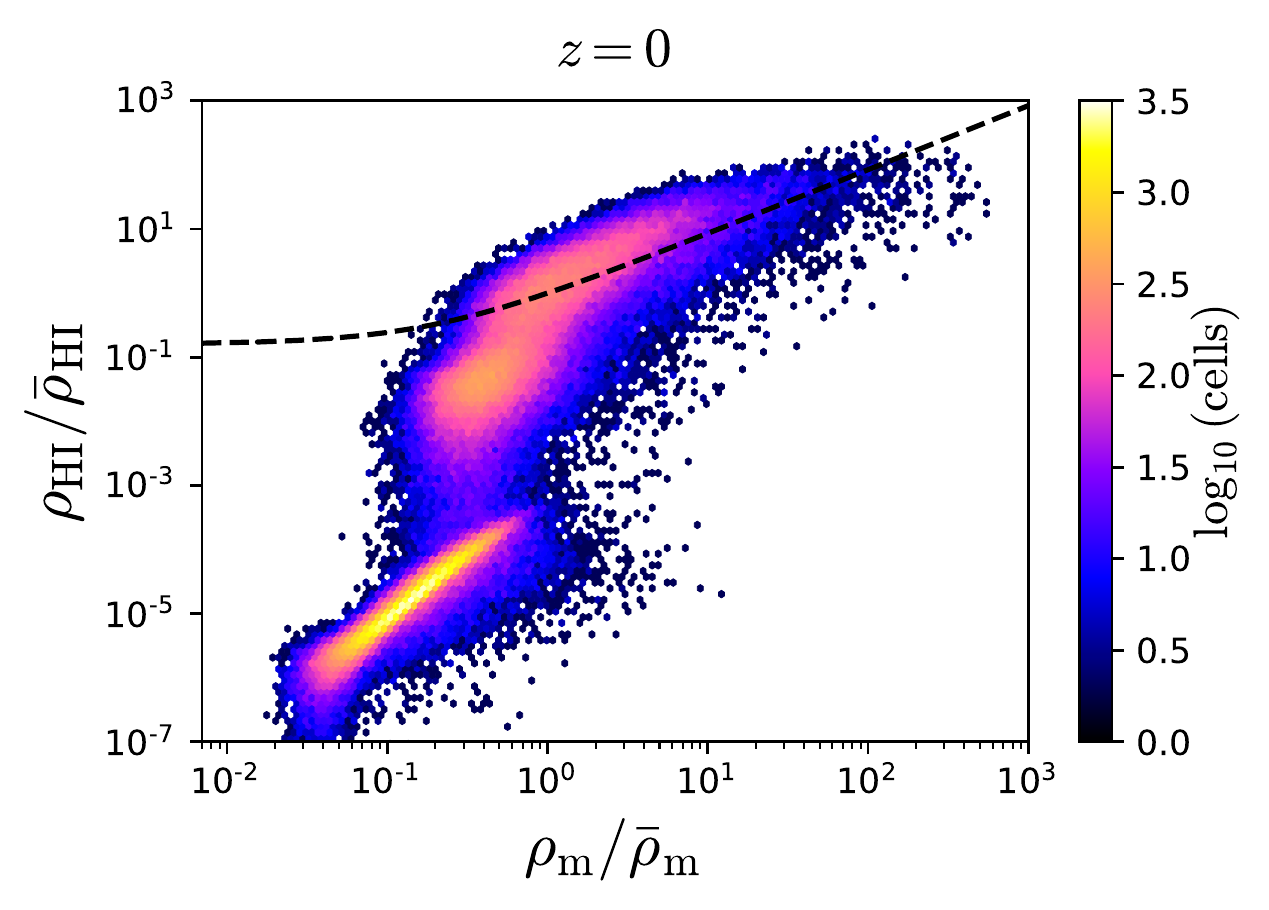}
\includegraphics[width=0.32\textwidth]{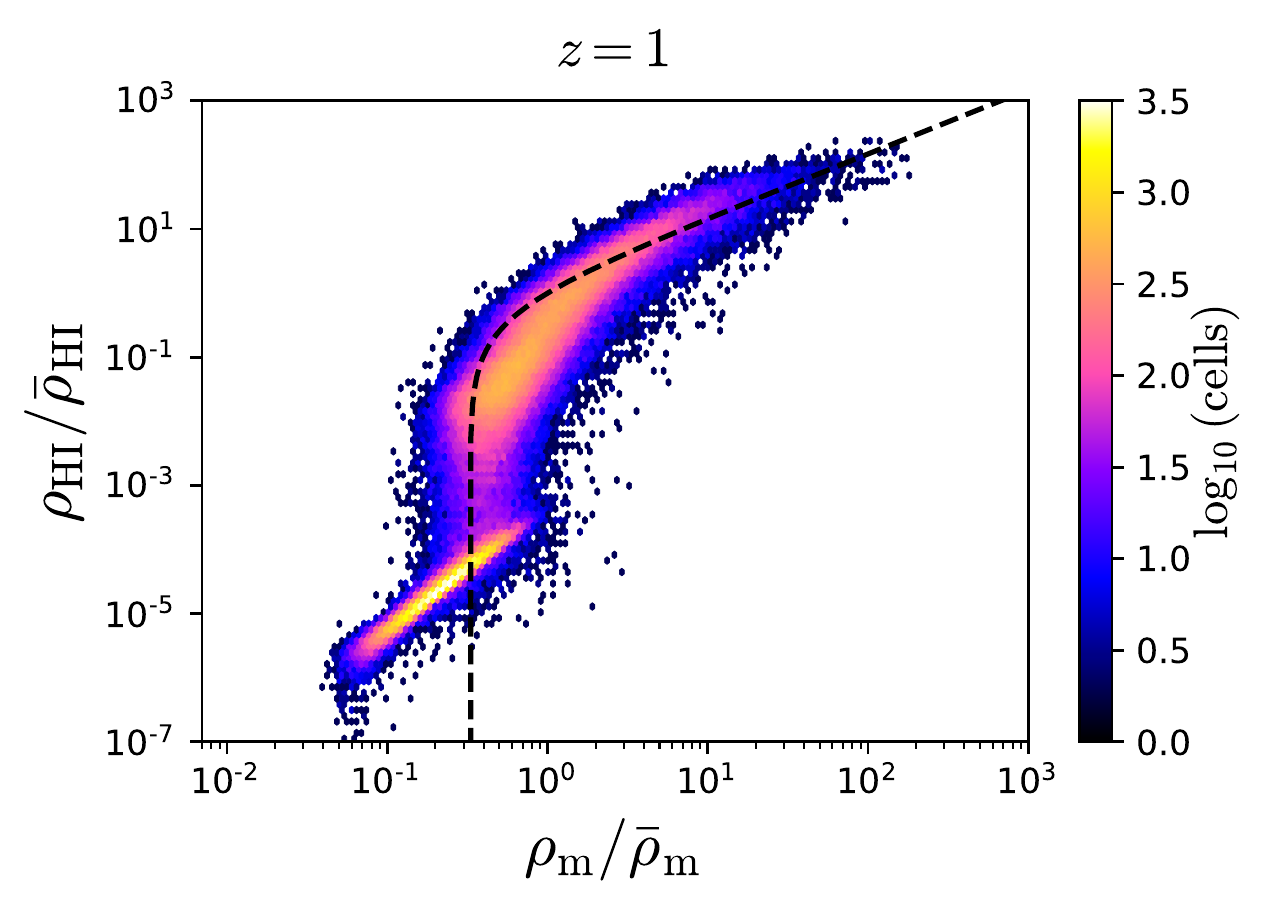}
\includegraphics[width=0.32\textwidth]{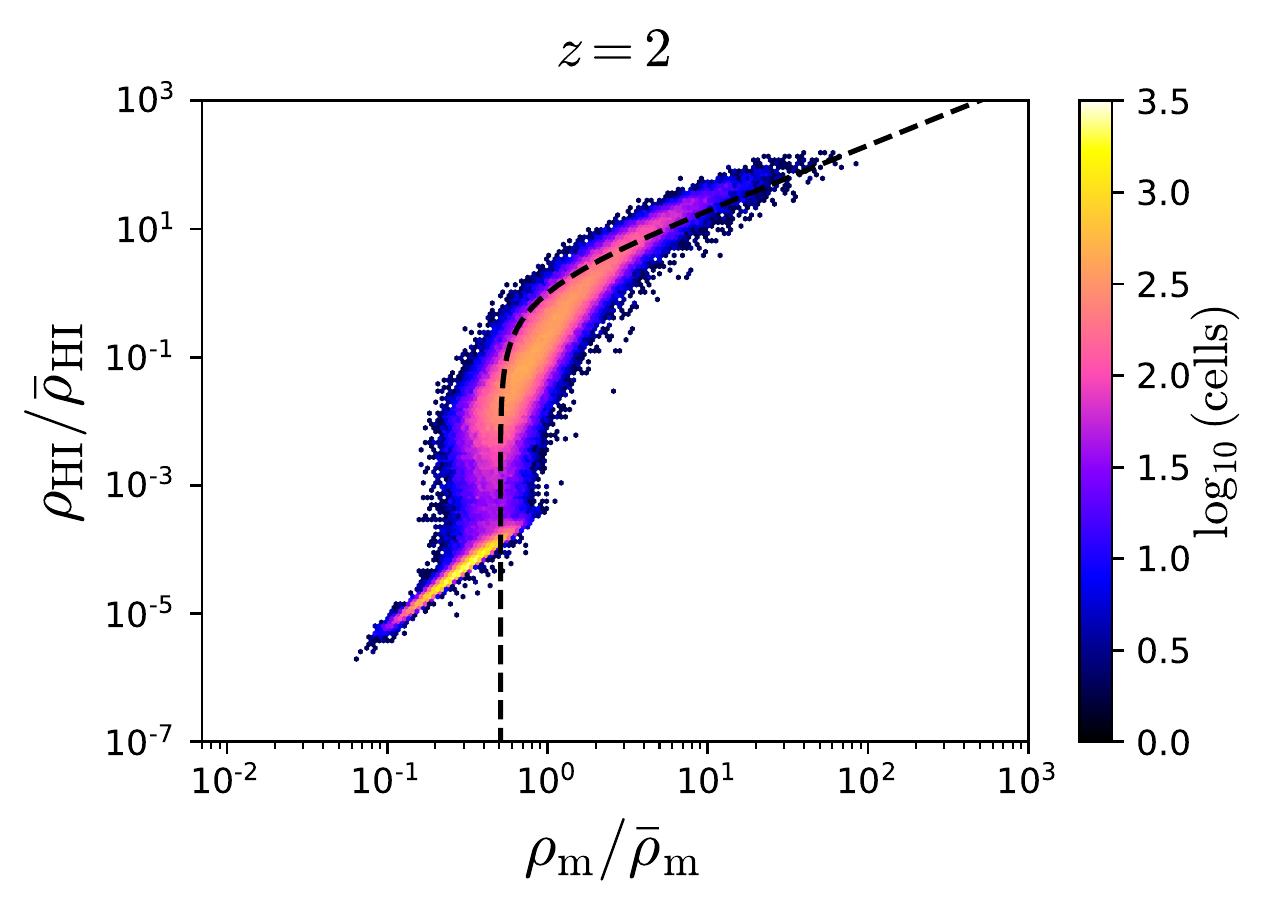}\\
\includegraphics[width=0.33\textwidth]{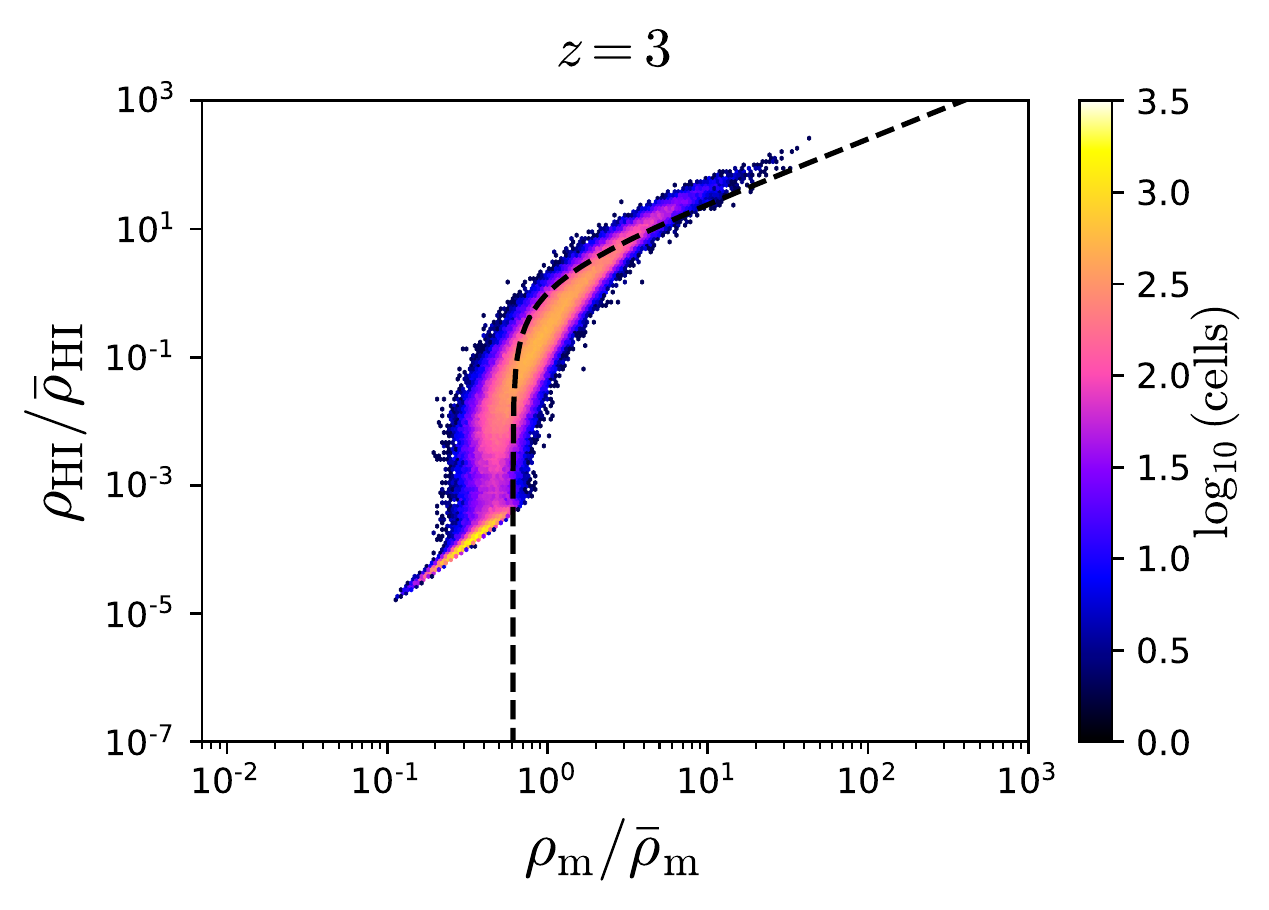}
\includegraphics[width=0.33\textwidth]{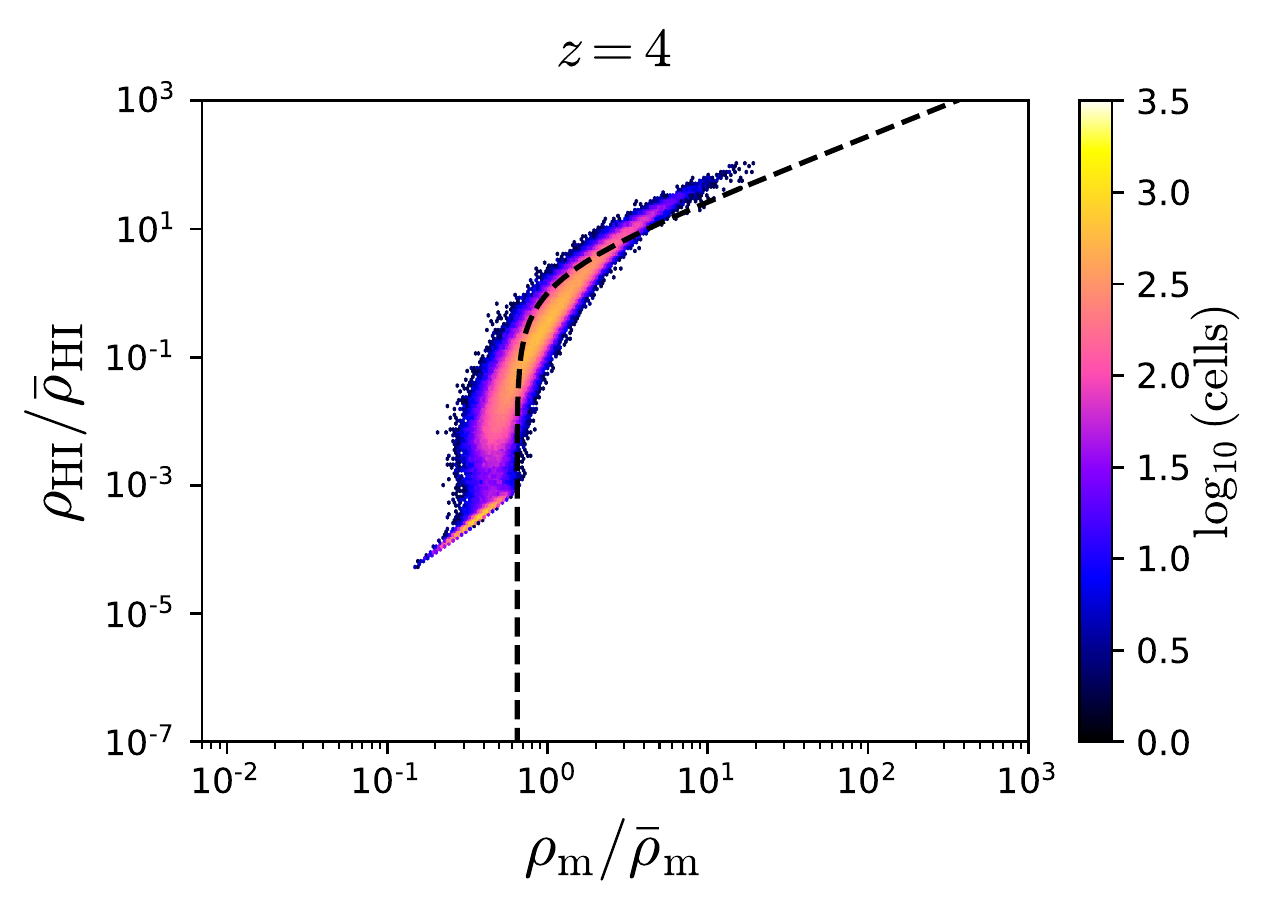}
\includegraphics[width=0.33\textwidth]{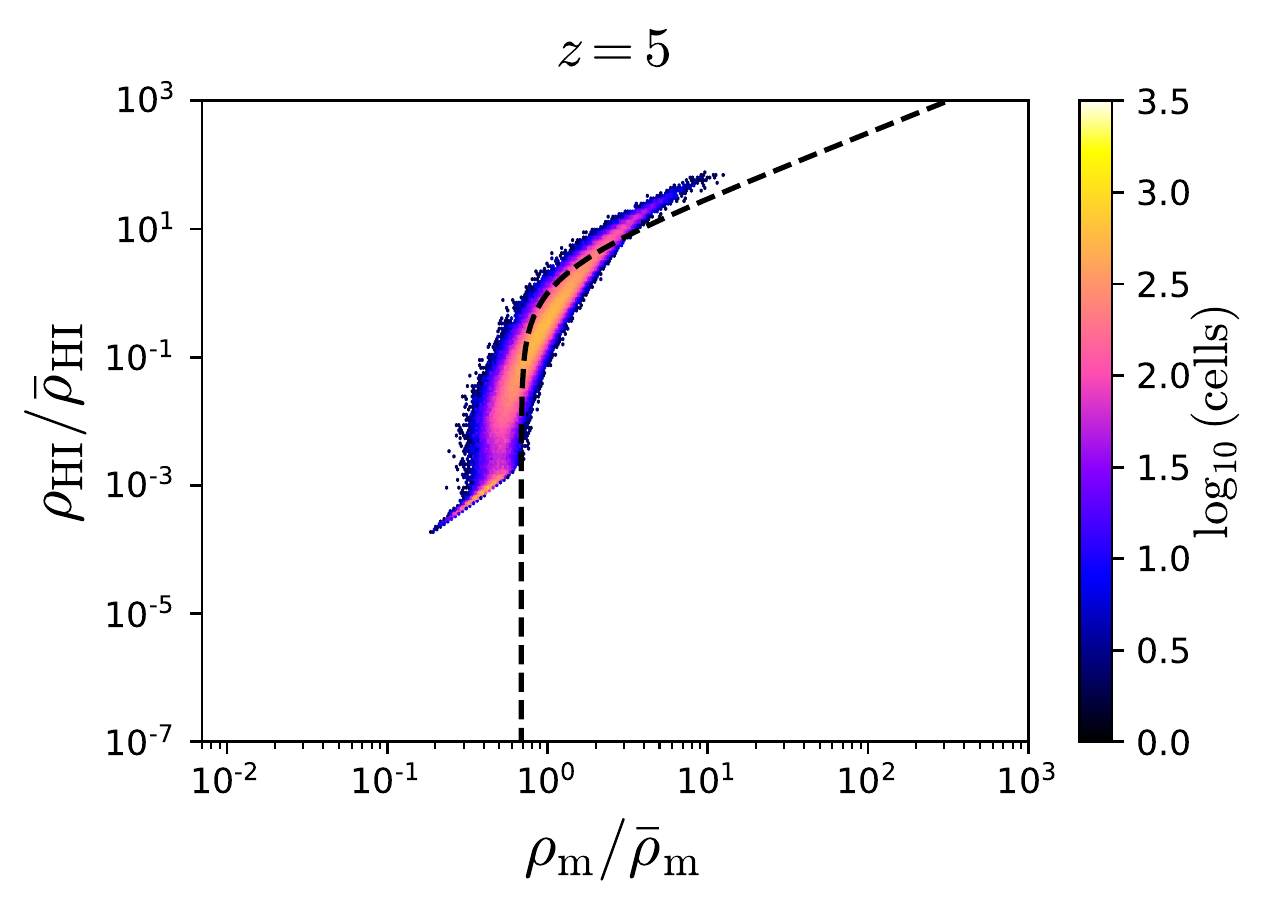}
\caption{Relations between the HI and matter density fields smoothed with a top-hat filter of radius $R=1~h^{-1}{\rm Mpc}$, shown as scatter plots between the respective overdensities at redshifts $z=0$ (top-left), $z=1$ (top-middle), $z=2$ (top-right), $z=3$ (bottom-left) $z=4$ (bottom-middle) and $z=5$ (bottom-right). The colors indicate the number cells in each hexabin. The presence of the Ly$\alpha$-forest can be seen at low matter and HI overdensities, while the HI inside halos dominates the behavior of the relation for $\rho_{\rm HI}/\bar{\rho}_{\rm HI}\gtrsim0.1$. The HI-matter relation is tighter at high-redshift than at low-redshift. The  dashed black lines show the expectation from linear theory, $\delta_{\rm HI}=b_{\rm HI}\delta_{\rm m}$, where $b_{\rm HI}$ is the large-scale HI bias, taken from table \ref{table:SN}.}
\label{fig:delta_HI_vs_delta_m_1}
\end{center}
\end{figure*}

\begin{figure*}
\begin{center}
\includegraphics[width=0.33\textwidth]{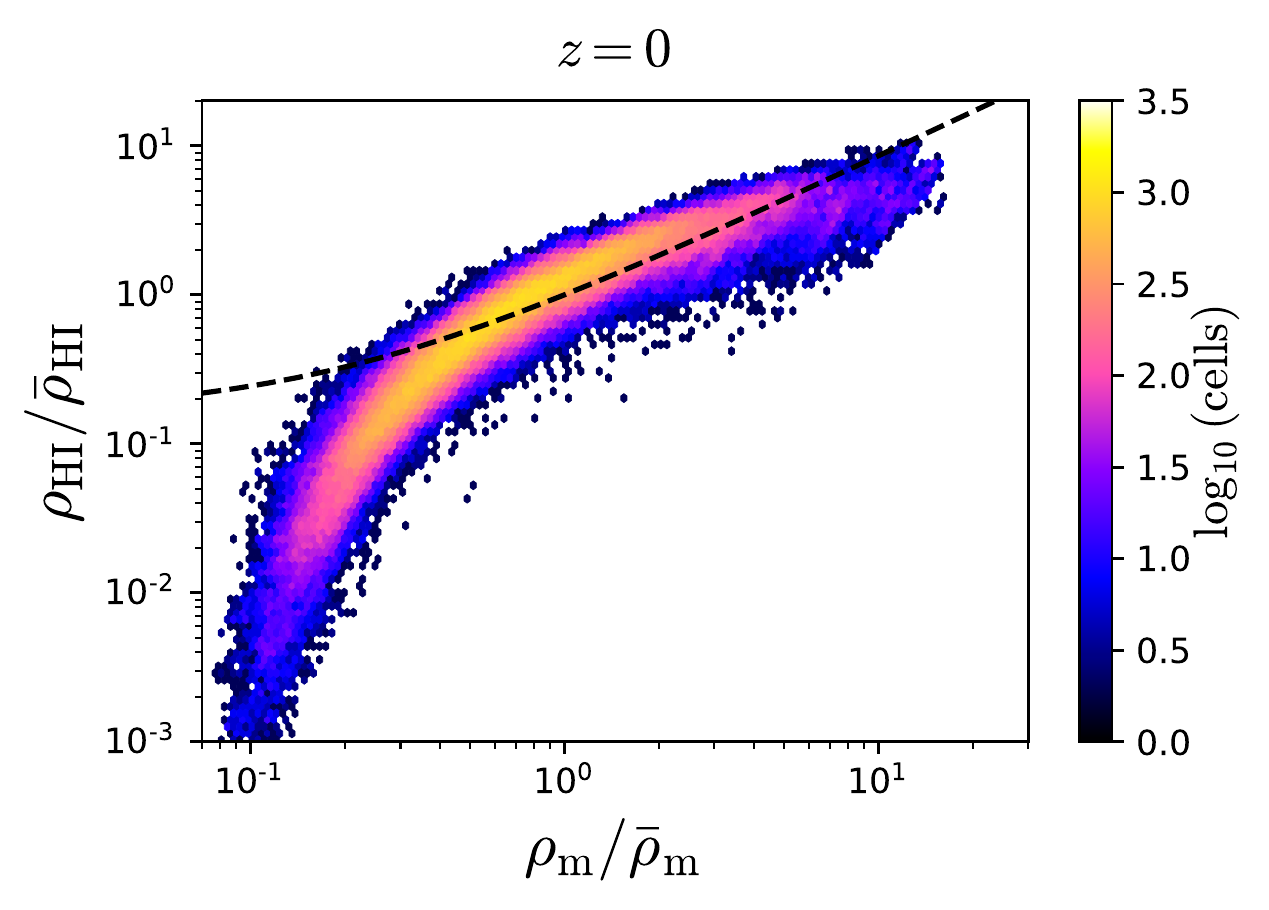}
\includegraphics[width=0.33\textwidth]{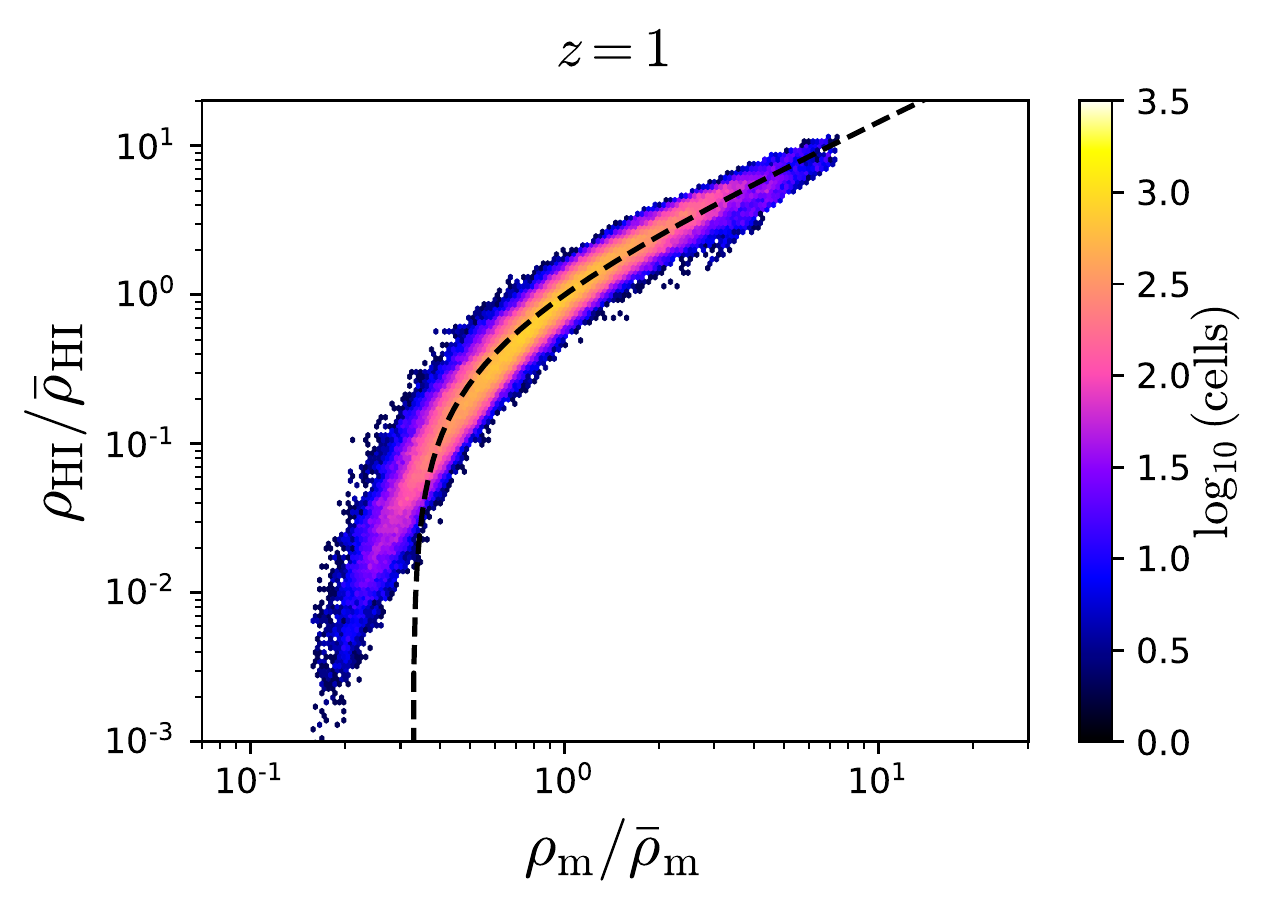}
\includegraphics[width=0.33\textwidth]{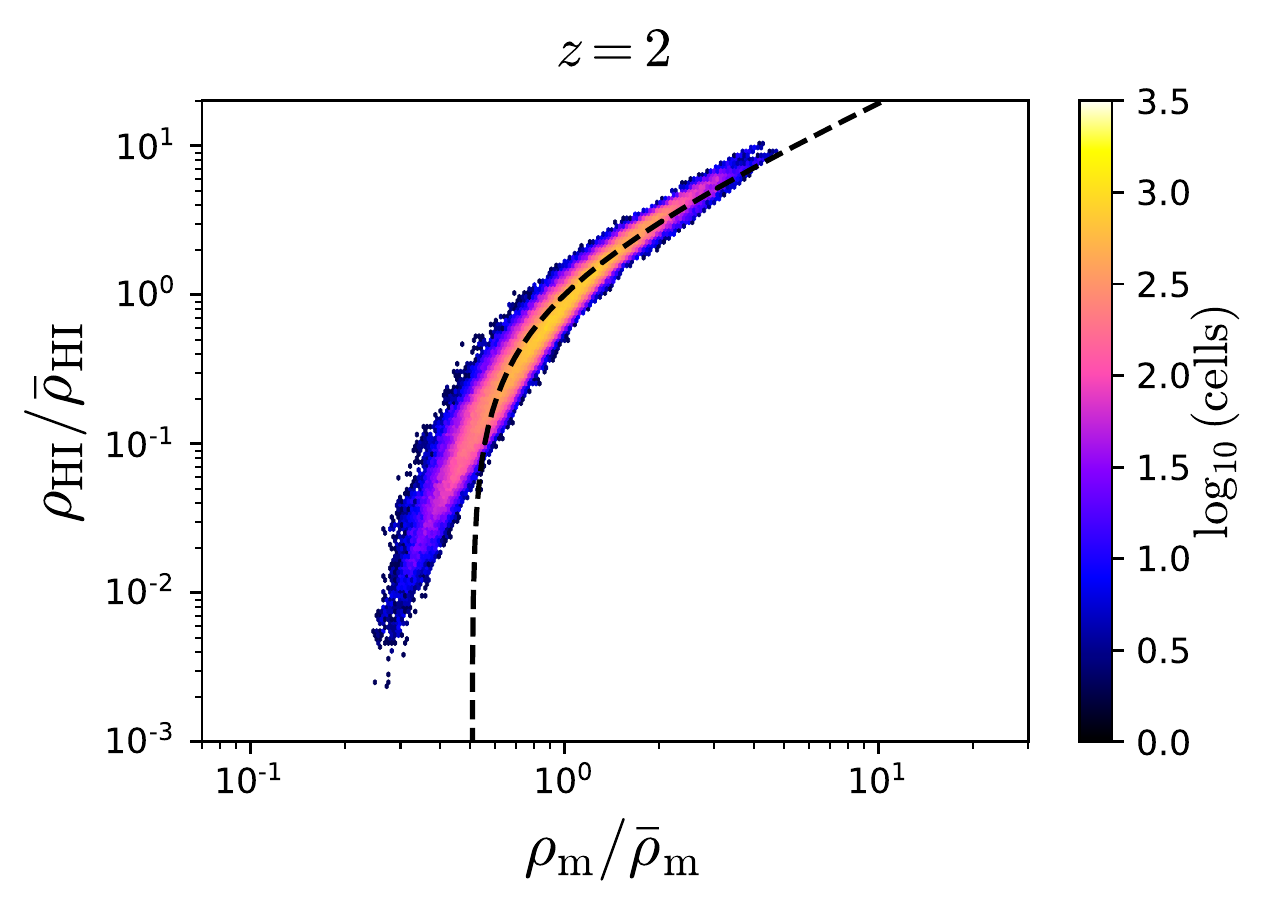}\\
\includegraphics[width=0.33\textwidth]{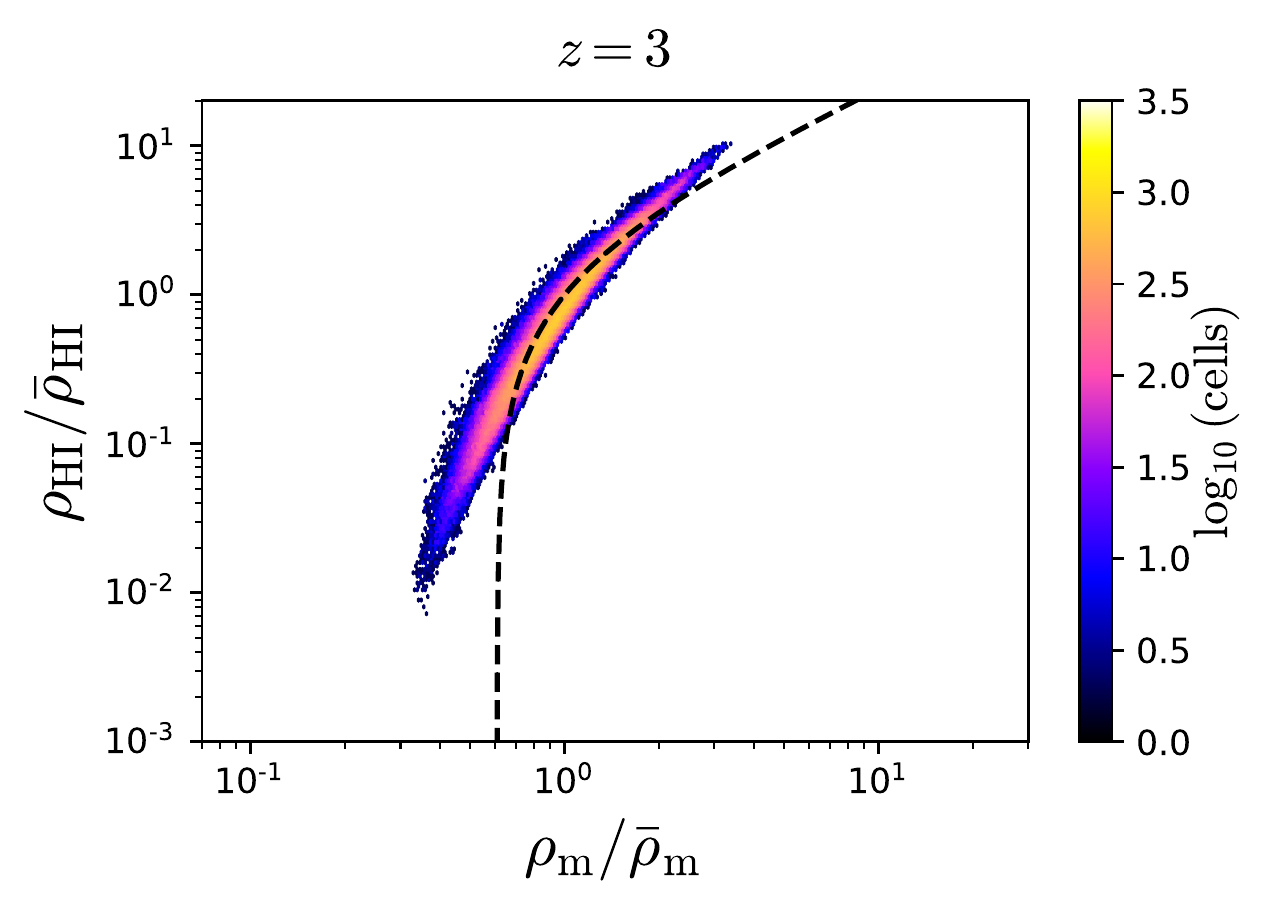}
\includegraphics[width=0.33\textwidth]{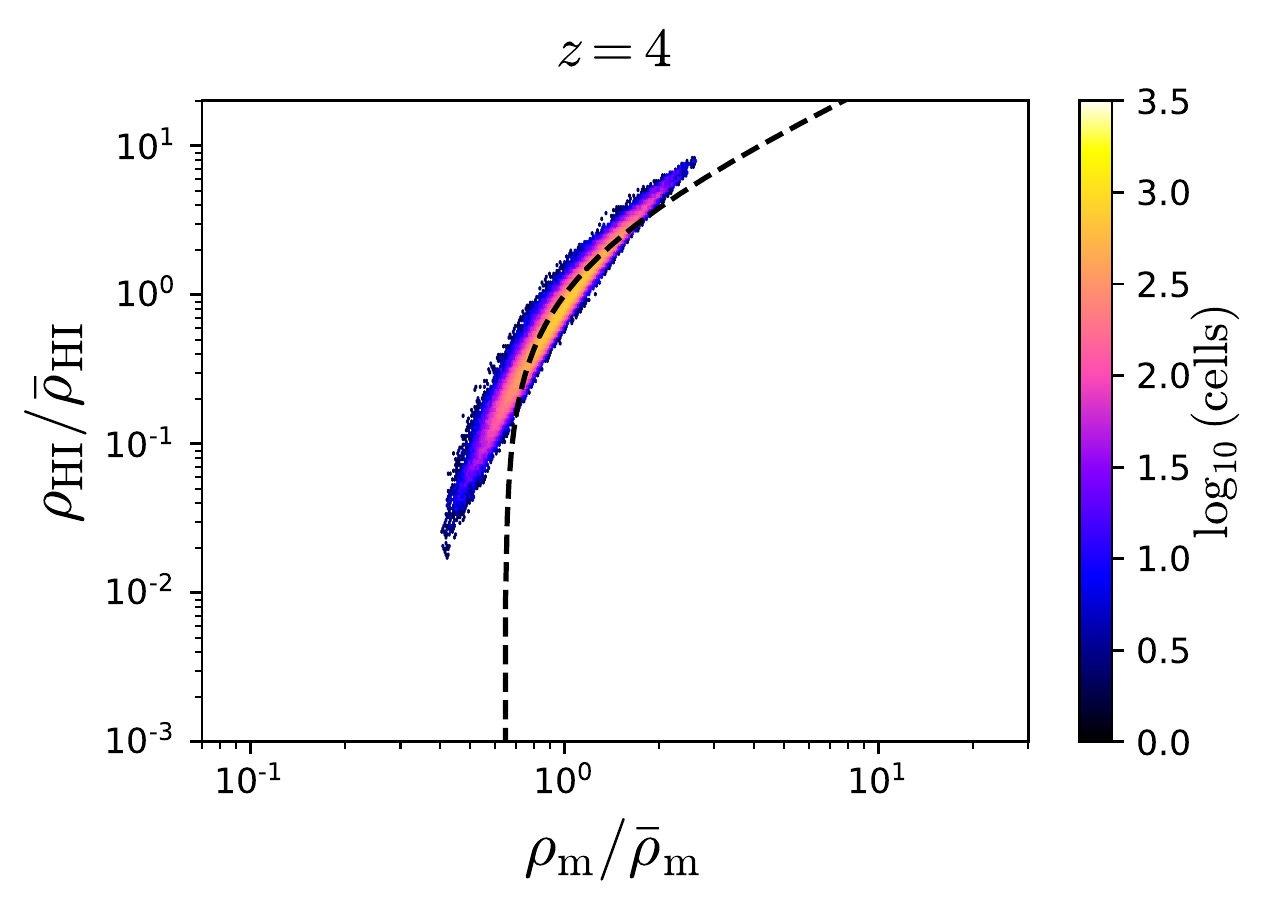}
\includegraphics[width=0.33\textwidth]{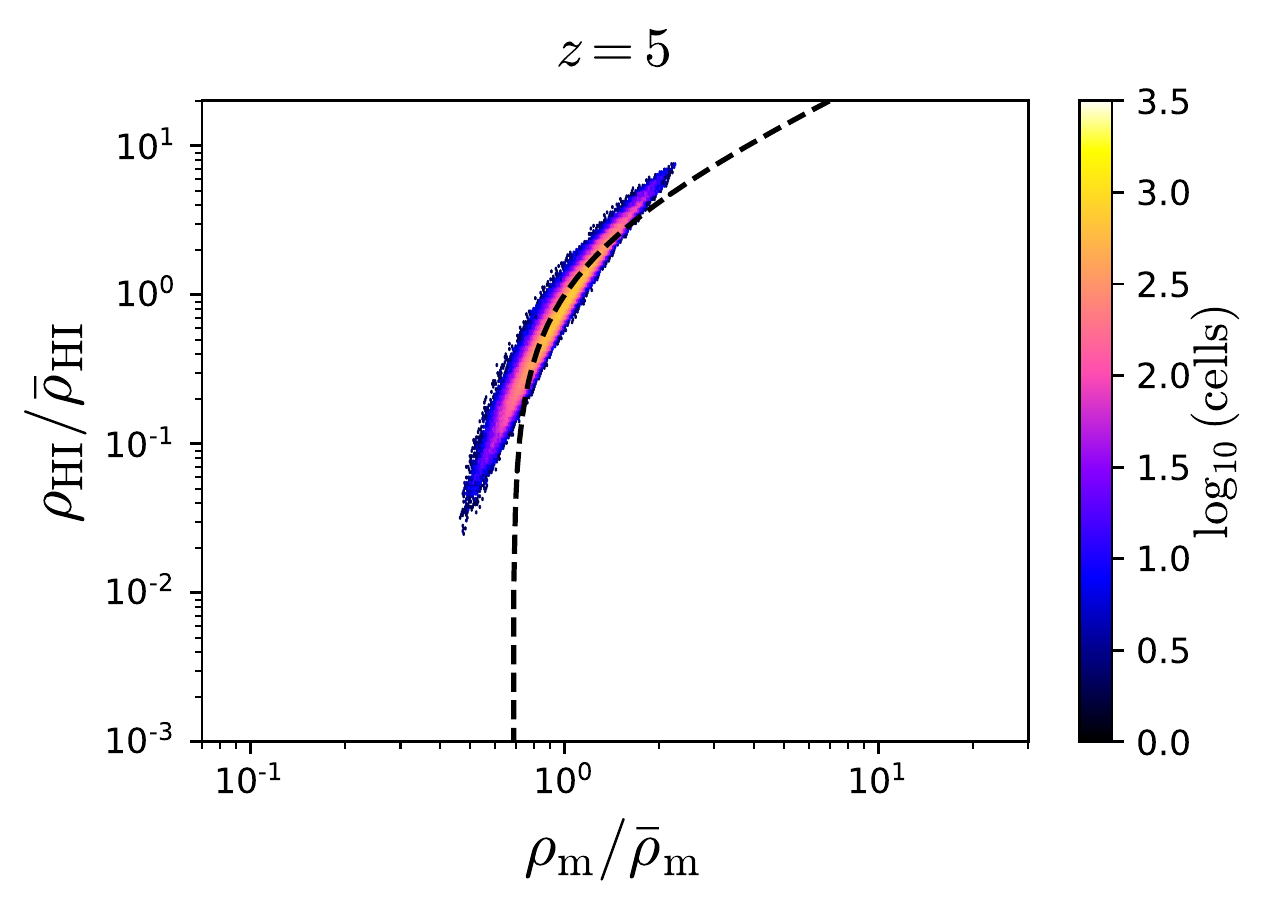}
\caption{Same as Fig. \ref{fig:delta_HI_vs_delta_m_1} but for a smoothing scale of $5~h^{-1}{\rm Mpc}$.}
\label{fig:delta_HI_vs_delta_m_5}
\end{center}
\end{figure*}

The scatter in the HI velocity dispersion for very low-mass halos is
typically much smaller than the scatter in the matter velocity
dispersion. One reason for this is that the HI velocity dispersion has
been computed only for halos with total HI masses above
$10^5~h^{-1}M_\odot$.  Without such a threshold, the scatter in the HI
velocity dispersion would be much larger. That is because if the HI
mass in a halo is very low, it will often not be bound to the halo,
e.g. the halo is crossing a filament that hosts a small amount of
HI. In that case, HI velocity dispersions can be large. For example,
several highly ionized unbound gas cells can produce a large,
unphysical, velocity dispersion.


\section{HI stochasticity}
\label{sec:HI_stochasticity}

The relation between HI and matter is given, to linear order, by
$\delta_{\rm HI}=b_{\rm HI}\delta_{\rm m}+\epsilon$, where $\epsilon$
is the stochasticity. Below, we examine whether or not this relation
reproduces our results in real-space and the amplitude of the
stochasticity.

\begin{figure*}
\begin{center}
\includegraphics[width=0.49\textwidth]{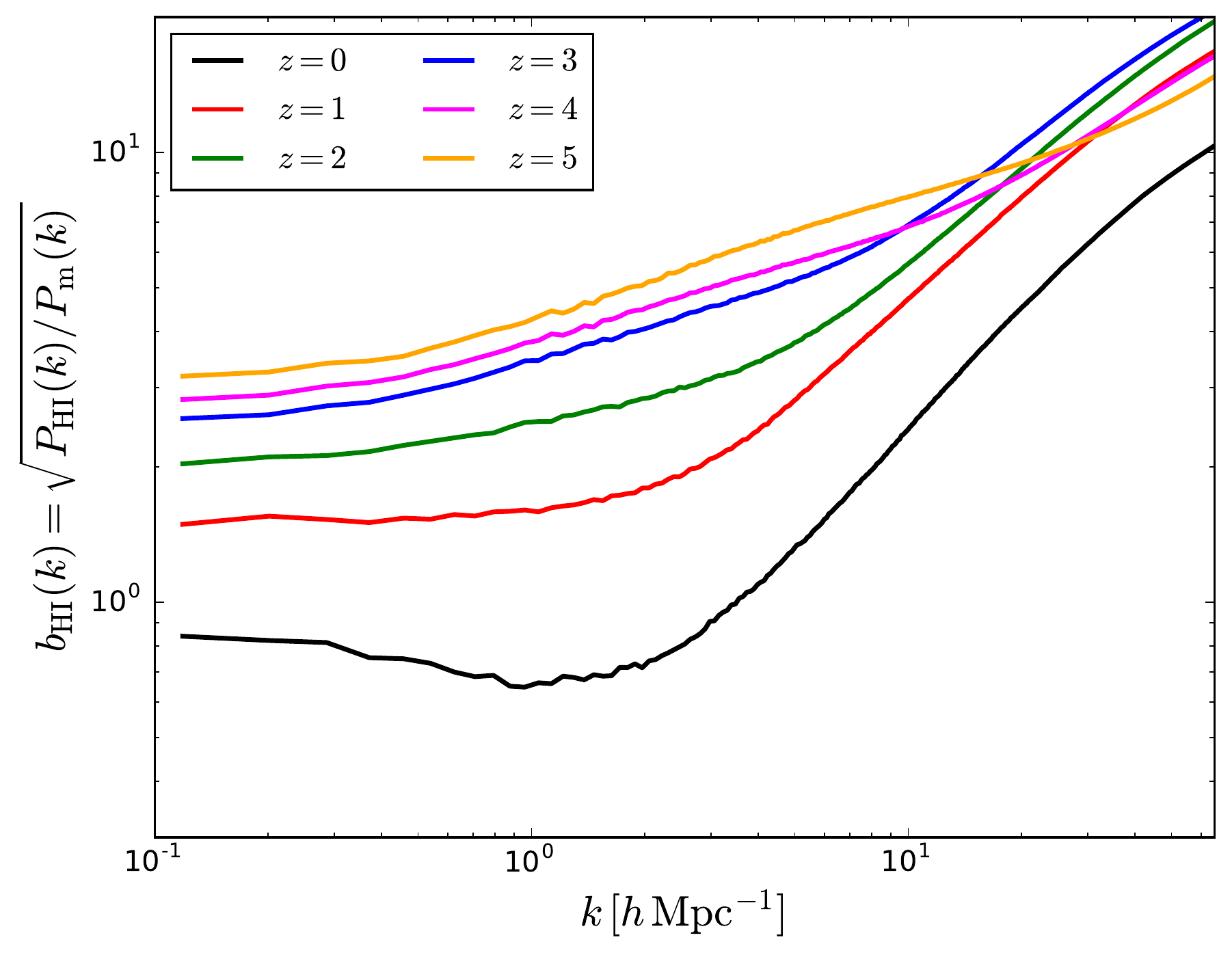}
\includegraphics[width=0.49\textwidth]{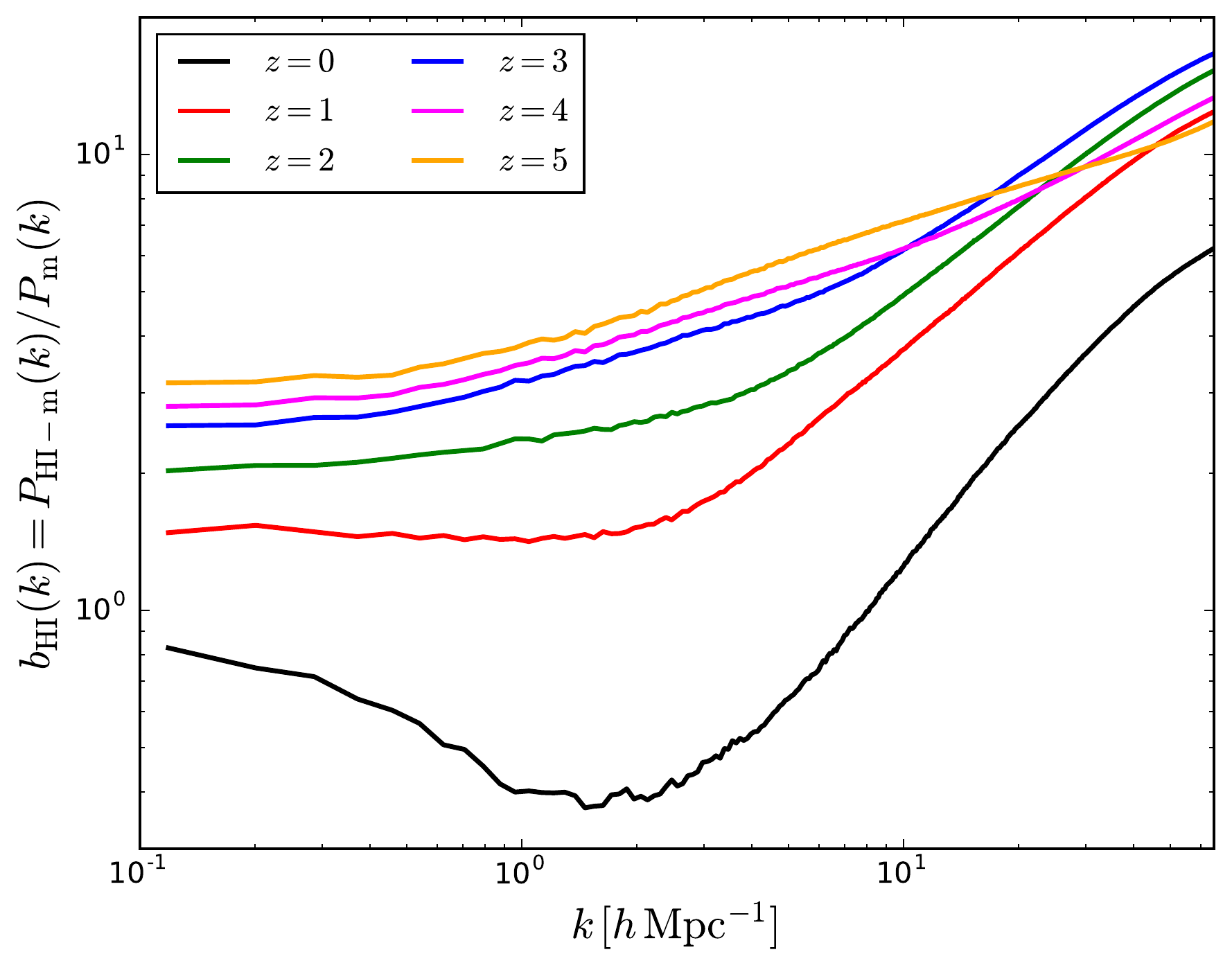}
\caption{HI bias at redshifts 0 (black), 1 (red), 2 (green), 3 (blue), 4 (purple) and 5 (orange) computed as the square root of the ratio between the HI and matter power spectra (left), and as the ratio between the HI-m cross-power spectrum and the matter power spectrum. The value of the HI bias increases with redshift. At high redshift, the HI bias becomes non-linear on scales $k\gtrsim0.3~h{\rm Mpc}^{-1}$.}
\label{fig:HI_bias}
\end{center}
\end{figure*}

As for the density pdfs (see section \ref{sec:HI_pdf}) we compute the
density fields of HI and matter on a grid with $2048^3$ cells
employing the CIC mass-assignment scheme. We then compute the
overdensity of each field and smoothed them with a top-hat filter of
radius 1 or 5 $h^{-1}{\rm Mpc}$. Next, we randomly select a subset of
cells and make a scatter plot between the overdensities of HI and
matter for each chosen cell. The results are shown in
Figs. \ref{fig:delta_HI_vs_delta_m_1} ($R=1~h^{-1}{\rm Mpc}$) and
\ref{fig:delta_HI_vs_delta_m_5} ($R=5~h^{-1}{\rm Mpc}$).

For $R=1~h^{-1}{\rm Mpc}$ two trends can be distinguished.  The
Ly$\alpha$-forest shows up as cells with matter overdensities below
the mean and very low HI overdensities because the gas there is mostly
ionized.  For large matter overdensities, the HI within halos is
self-shielded.  The density $\rho_{\rm m}/\bar{\rho}_{\rm m}\simeq0.4$
marks a transition from one regime to the other, indicating that HI
self-shielding does not take place for lower matter overdensities. In
all cases, the HI overdensity increases with matter overdensity.

At higher redshifts the range occupied by matter and HI overdensities
is smaller.  As the Universe becomes more homogeneous, fluctuations
are smaller. This behavior can also be seen in the pdfs of
Fig. \ref{fig:HI_pdf}. The scatter in the overdensity relations also
decreases towards higher redshift.

The dashed black lines show the predictions from linear theory,
$\delta_{\rm HI}=b_{\rm HI}\delta_{\rm m}$, where $b_{\rm HI}$ is the
linear HI bias measured from the simulation (see section
\ref{subsec:HI_bias} and Table \ref{table:SN}). As expected, linear
theory is not accurate in this regime because the bias is not linear
on the smoothing scale considered.  An exception if for $z=1$ where
linear HI bias reproduces the results reasonably well. This is because
the HI bias is relatively flat at that redshift (see
Fig. \ref{fig:HI_bias}).

As we move to a larger smoothing radius, the morphology of the results
changes, as can be seen in Fig. \ref{fig:delta_HI_vs_delta_m_5}, where
$R=5~h^{-1}{\rm Mpc}$.  Now, the HI and matter overdensities extend
over a smaller range, because the smoothing is over a larger scale,
making the field more homogeneous. In addition, the Ly$\alpha$ forest
is no longer visible because in the neighborhood of the highly ionized
HI in filaments, i.e.~the Ly$\alpha$ forest, there will always be some
halo within $5~h^{-1}{\rm Mpc}$ that contains self-shielded HI gas.

As above, we find that HI overdensities increase with matter
overdensities at all redshifts. However, at high-redshift the slope of
the relation becomes more pronounced. This behavior can be partly
explained by linear HI bias, shows as dashed black lines. As with
$R=1~h^{-1}{\rm Mpc}$, linear bias can explain the results relatively
well at $z=1$. At other redshifts, linear bias is more accurate than
for smaller smoothing scales, as expected, but the agreement is not
good for both large and small matter overdensities. Again, the scatter
reduces with redshift.

\section{HI bias}
\label{subsec:HI_bias}

We now examine different aspects of HI clustering in detail. In this
section we focus on the amplitude and shape of the HI bias.

The relation between the clustering of HI and that of dark matter
involves the HI bias through $P_{\rm HI}(k)=b^2_{\rm HI}(k)P_{\rm
  m}(k)$. The matter power spectrum, the quantity that contains the
information on the values of the cosmological parameters, can thus be
inferred only if bias is understood \citep[see][for a detailed discussion on HI scale-dependence bias]{Penin_2018}.  On linear scales the HI bias is constant, but on small scales we expect to see  scale-dependence. It
is important to determine the scales on which the HI bias is
scale-dependent and whether or not analytic models can reproduce that
behavior.

We have computed the HI and matter auto-power spectrum and the
HI-matter cross-power spectrum of the simulation at redshifts 0, 1, 2,
3, 4 and 5. The HI bias is then obtained using two different
definitions: $b_{\rm HI}(k)=\sqrt{P_{\rm HI}(k)/P_{\rm m}(k)}$ and
$b_{\rm HI}(k)=P_{\rm HI-m}(k)/P_{\rm m}(k)$. While the latter is
``preferred'', as it does not suffer from stochasticity, the former is
closer to observations. The results are shown in Fig. \ref{fig:HI_bias}.

The amplitude of the HI bias on large scales increases
with redshift, from $\simeq0.85$ at $z=0$ to $\simeq3.20$ at $z=5$. On
the largest scales that can be probed with TNG100, the amplitude of the HI
bias is independent of the method used to estimate it. We
assume that the values on large scales are equal to the linear HI bias, 
but note that there could be small corrections to those because of 
box-size.  The linear HI bias at different redshifts 
is given in Table \ref{table:SN}. These values
can be reproduced from the halo HI mass function as
\be
b_{\rm HI}(z)=\frac{\int_0^\infty n(M,z)b(M,z)M_{\rm HI}(M,z)dM}{\int_0^\infty n(M,z)M_{\rm HI}(M,z)dM}~,
\ee
and therefore the amplitude of the HI bias is sensitive to the
astrophysical parameters $\alpha$ and $M_{\rm min}$ (see section
\ref{subsec:M_HI}).  Note, however, that the agreement between the
above expression and the simulation results is not perfect 
because, among other things, our models for the halo mass function and 
halo bias do not include corrections for baryonic effects.

\begin{figure}
\begin{center}
\includegraphics[width=0.49\textwidth]{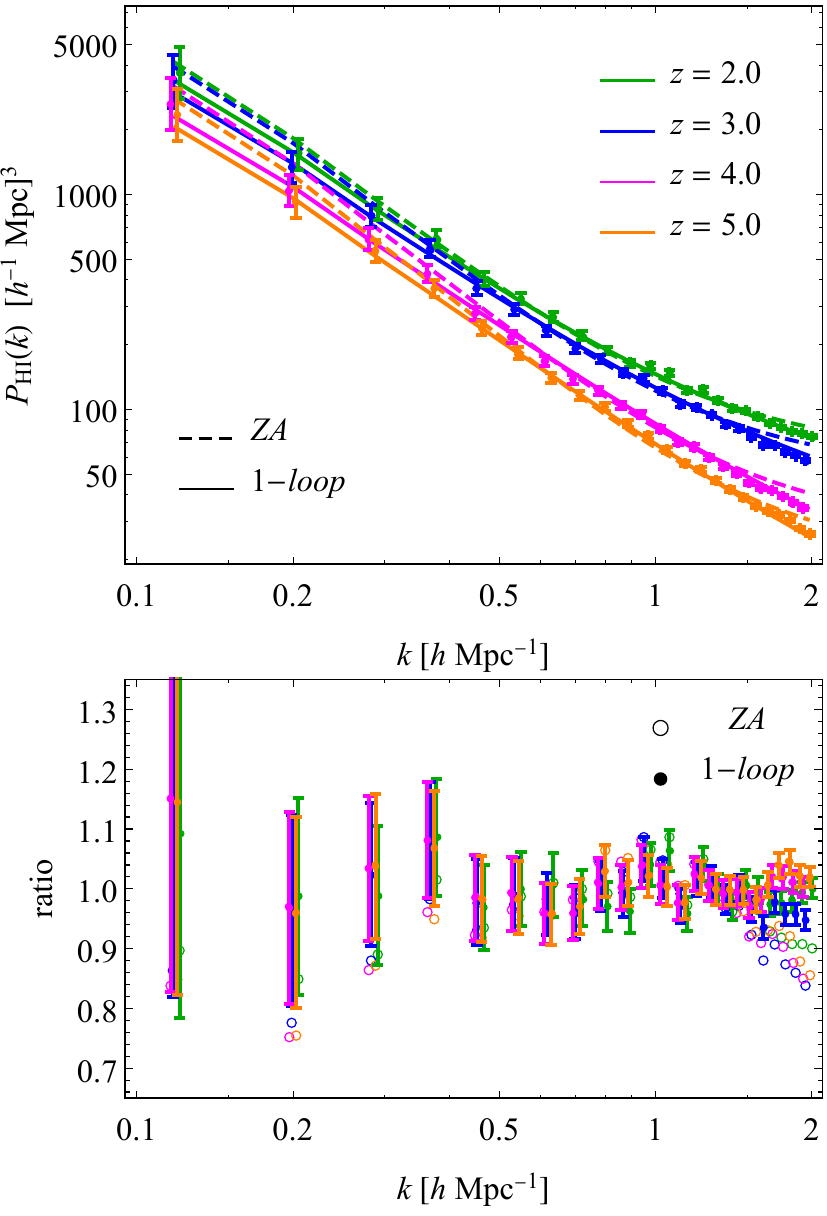}
\caption{Top panel: HI power spectrum as a function of redshift measured in TNG100 (points with error bars). For visualization purposes only, the data at different redshift have been shifted horizontally to avoid overlap. The dashed lines show the analytical calculation assuming the ZA, whereas the continuous lines correspond to the 1-loop calculation. Both models have been fitted to $k = 1\,h{\rm Mpc}^{-1}$. See text for details. Bottom panel: Ratio of the measured power to the analytical models. Filled points show the 1-loop result, whilst the empty ones the comparison to the ZA.}
\label{fig:PkHI}
\end{center}
\end{figure}

At $z=0$ the HI bias exhibits a scale-dependence even on the largest
scales we can probe, due to the fact that the matter power spectrum at the scales probed by TNG100 is not
in the linear regime at such low-redshift. It is interesting to notice the dip in the HI bias at $k\simeq1~h{\rm Mpc}^{-1}$, that has been also found in observations  \citep{Anderson_2018}. At $z=1$ the bias remains almost constant down
to rather small scales, $k\simeq1~h{\rm Mpc}^{-1}$. These trends agree
with the findings of \cite{SpringelV_17a}, who studied galaxy bias for
different galaxy populations at different redshifts. At
high-redshifts, $z\geqslant2$, the HI bias exhibits a dependence on
scale already at $k=0.3~h{\rm Mpc}^{-1}$, even though these
scales are close to linear at those redshifts. Our results are also in qualitative 
agreement with \cite{Sarkar_2016}, who studied the HI bias by painting HI on top of dark matter halos. 
The scale-dependence of the bias is not necessarily a bad thing, as long we can use perturbative methods to predict the shape of the HI power spectrum.
For this purpose we have compared the measurements of the HI power spectrum in TNG100 to analytical calculations using Lagrangian Perturbation Theory (LPT). 
The first order LPT solution is the well known Zeldovich approximation (ZA) \citep{Zeldovich,White2014}, for which we can simply write
\begin{align}
\label{eq:PkZA}
P_{\rm HI } = b_{\rm HI }^2P_{\rm ZA} (k)+N
\end{align}
and then fit for the two free parameters in the above equation. The constant piece takes care of the shot-noise and any other term which is scale independent and uncorrelated with the HI field and therefore can be treated as noise in a cosmological analysis \citep{SV2015}.
Given the small volume of TNG100, a perturbative analysis makes sense only at high redshifts, where linear and mildly non-linear modes are contained in the box, thus we restrict the comparison of Eq.~\ref{eq:PkZA} with the measurements in the simulation to $z\ge2$.
The upper panel in Fig~\ref{fig:PkHI} shows the measurements of the HI power spectrum at different redshift, using the same color scheme of the previous figures.
The points with error bars have been shifted horizontally to avoid overlap and facilitate the visual comparison with the theoretical models.
The dashed lines display the fit to Eq.~\ref{eq:PkZA} including all the modes up to $k_{\rm max} = 1\,h\text{Mpc}^{-1}$. The fit is quite accurate, despite its simple functional form, and it confirms the HI distribution as an ideal tracer for cosmological studies. The continuous lines show the next to leading order, \textit{i.e.} 1-loop, calculation in LPT \citep{Modi2017,Vlah2016}, which includes an improved treatment of non-linearities in the matter fields as well as several non-linear bias parameters. Up to the scale we include in the fit there is no difference between the two approaches, with the 1-loop calculation also working on smaller scales not included in the analysis.
The fact that the ZA works so well in describing the simulation measurements could vastly simplify the cosmological analysis and interpretation of 21cm surveys observing at high redshift. For instance, interferometric surveys with large instantaneous field of view like CHIME will be forced to include all the complications arising from the curved-sky, that are very easy to handle in the ZA \citep{CW2017,CW2018}.

\begin{table}
\begin{center}
 \begin{tabular}{|c|| c| c| c| c| c| c|} 
 \hline
 $z$ & $0$ & $1$ & $2$ & $3$ & $4$ & $5$\\ [0.7ex] 
 \hline
  $b_{\rm HI}$ & 0.84 & 1.49 & 2.03 & 2.56 & 2.82 & 3.18\\[0.7ex]
 \hline
$P_{\rm HI}^{\rm SN}$ & \multirow{ 2}{*}{104} & \multirow{ 2}{*}{124} & \multirow{ 2}{*}{65} & \multirow{ 2}{*}{39} & \multirow{ 2}{*}{14} & \multirow{ 2}{*}{7}\\
$[(h^{-1}{\rm Mpc})^3]$ & & & & & &\\
 \hline
\end{tabular}
\caption{Values of the HI bias and HI shot-noise from the simulation at different redshifts.}
\label{table:SN}
\end{center}
\end{table}

\section{Secondary HI bias}
\label{sec:assembly_bias}

\begin{figure*}
\begin{center}
\includegraphics[width=0.33\textwidth]{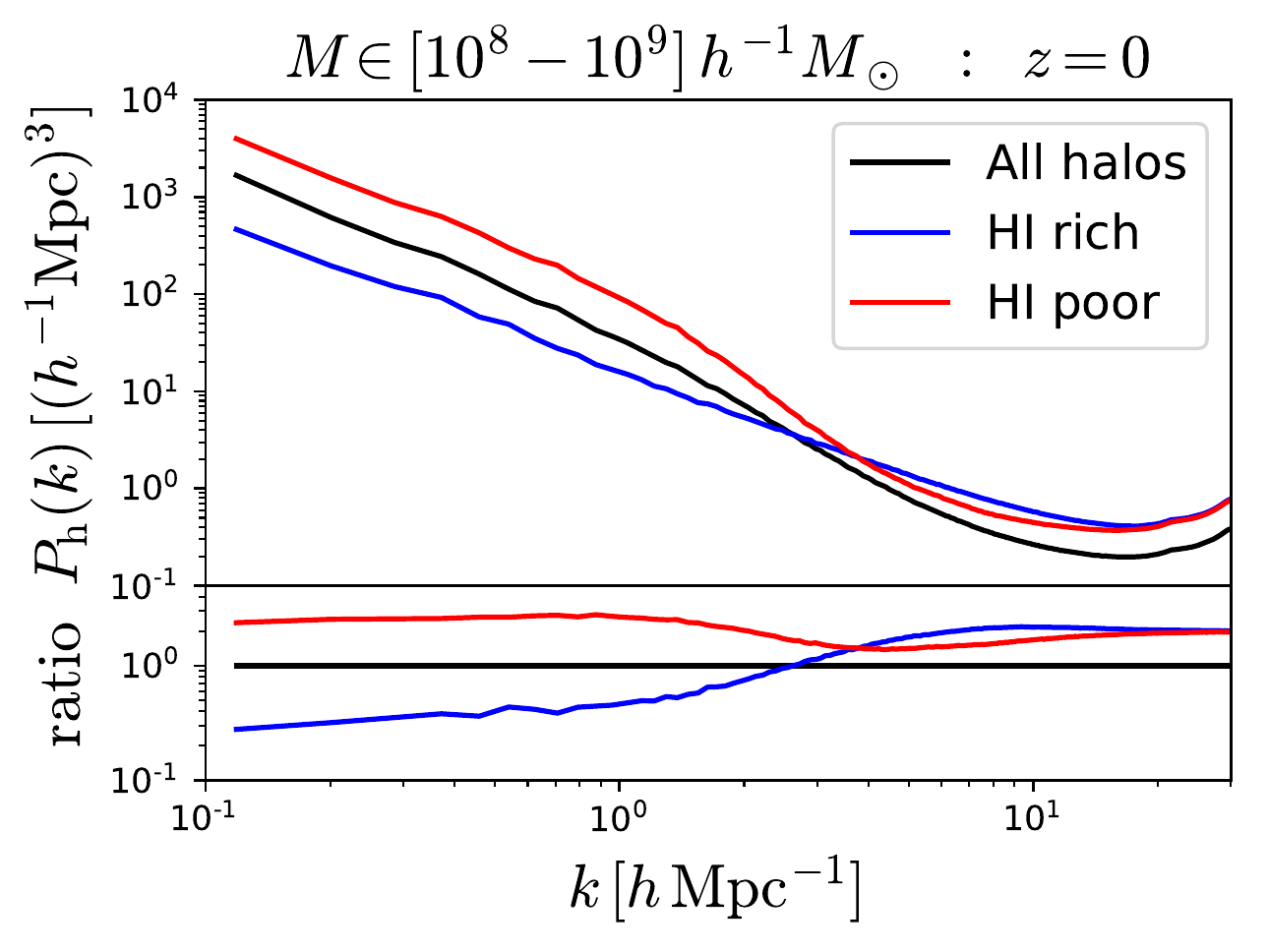}
\includegraphics[width=0.33\textwidth]{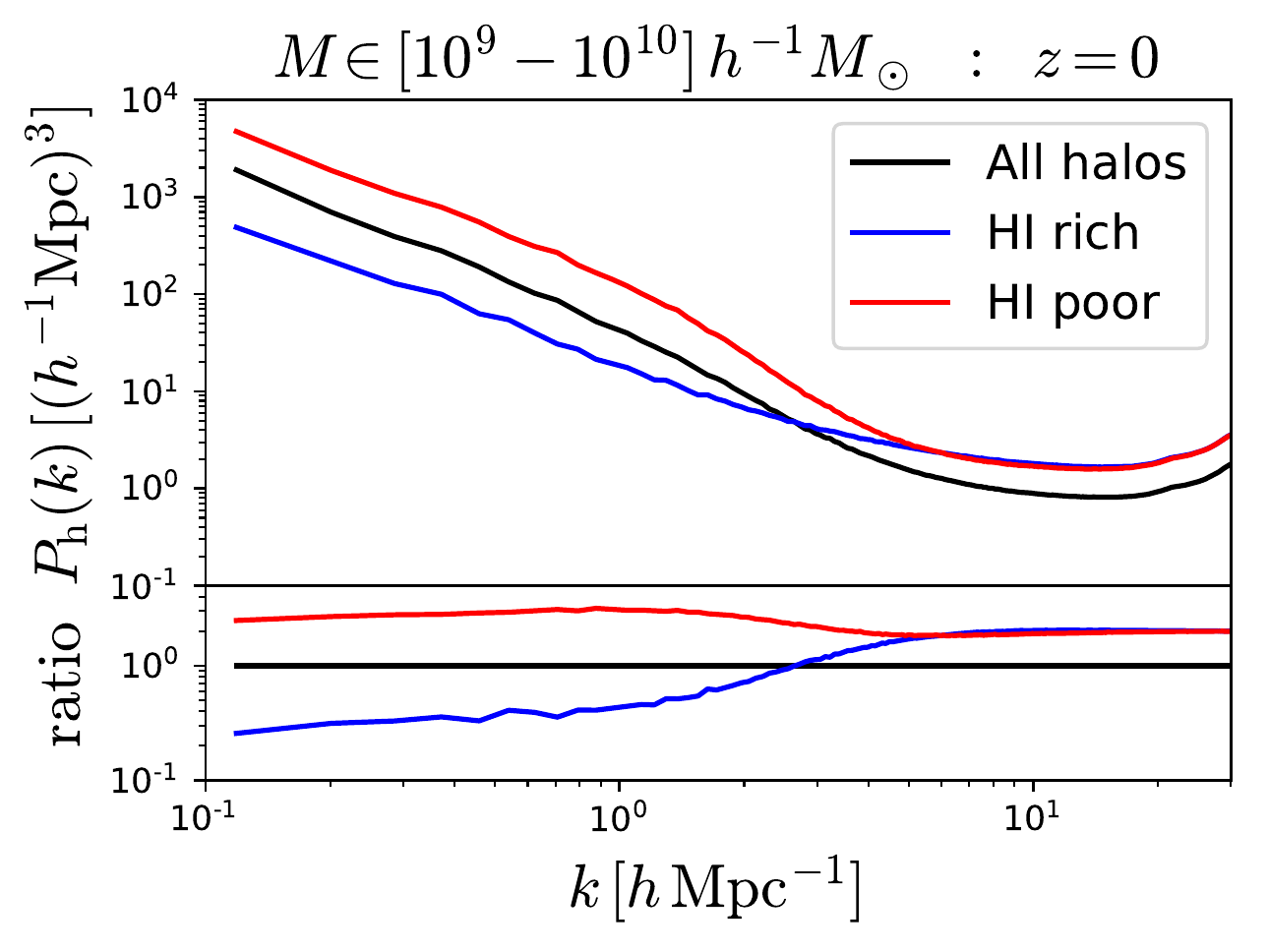}
\includegraphics[width=0.33\textwidth]{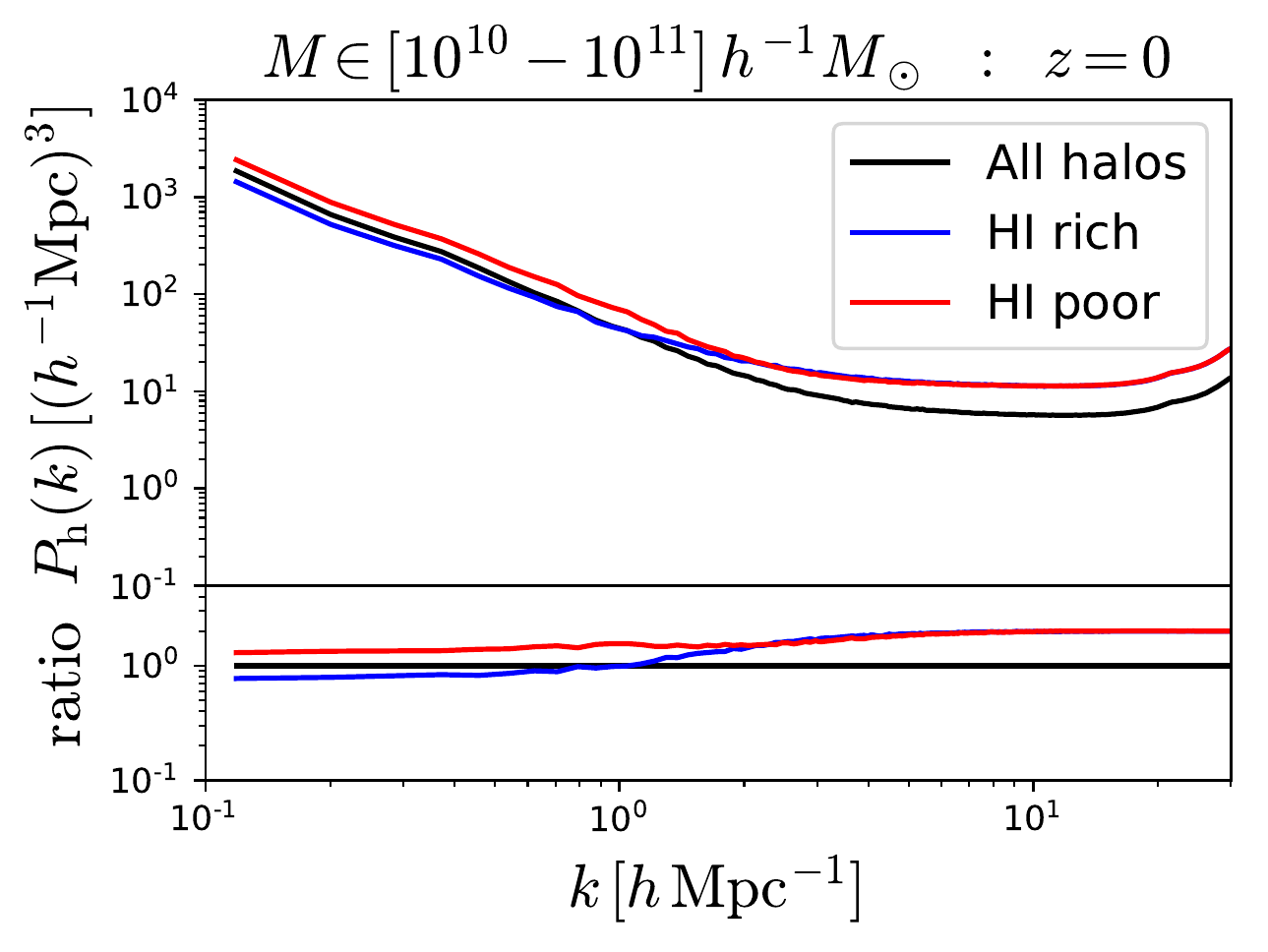}\\
\includegraphics[width=0.33\textwidth]{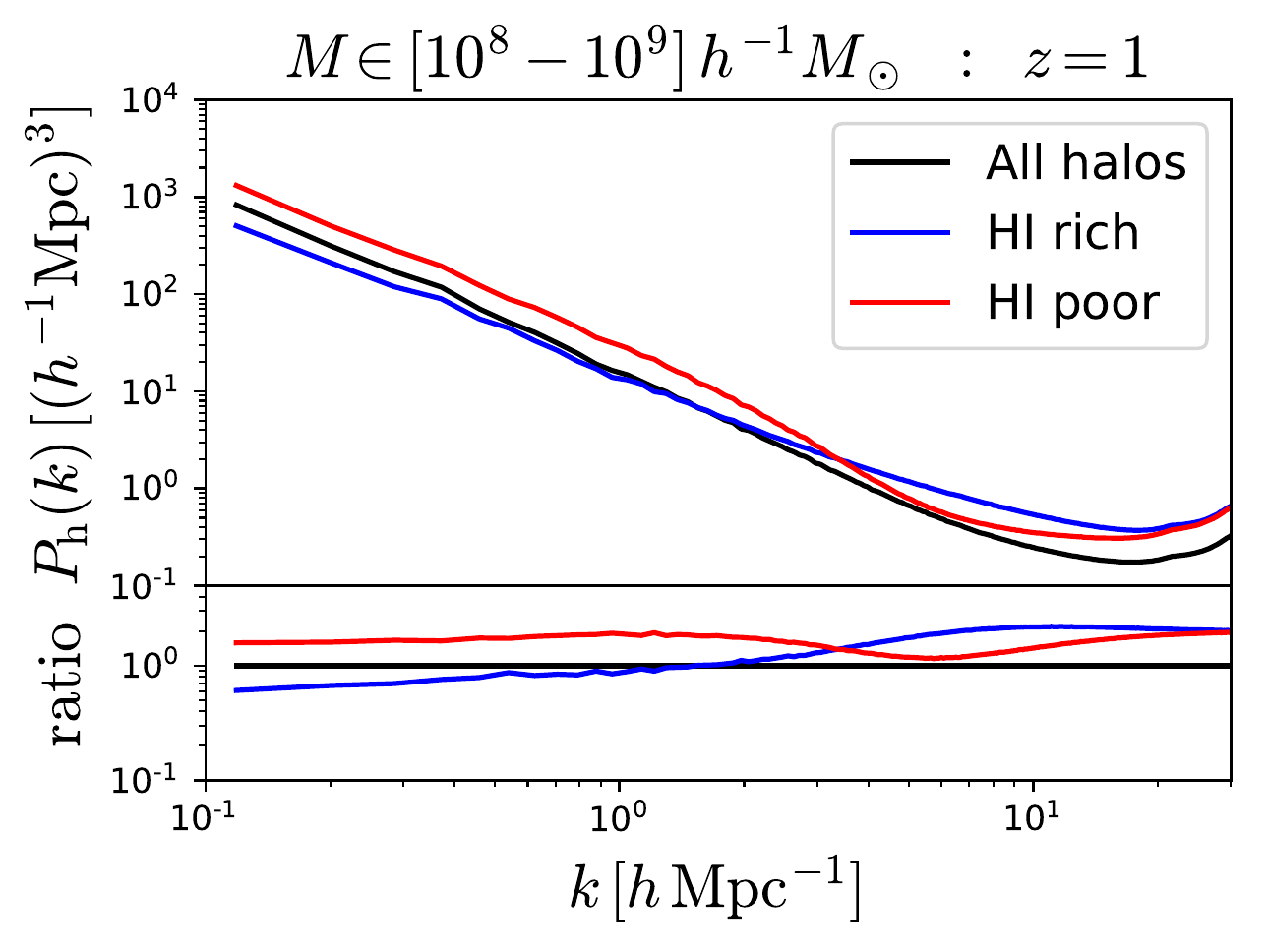}
\includegraphics[width=0.33\textwidth]{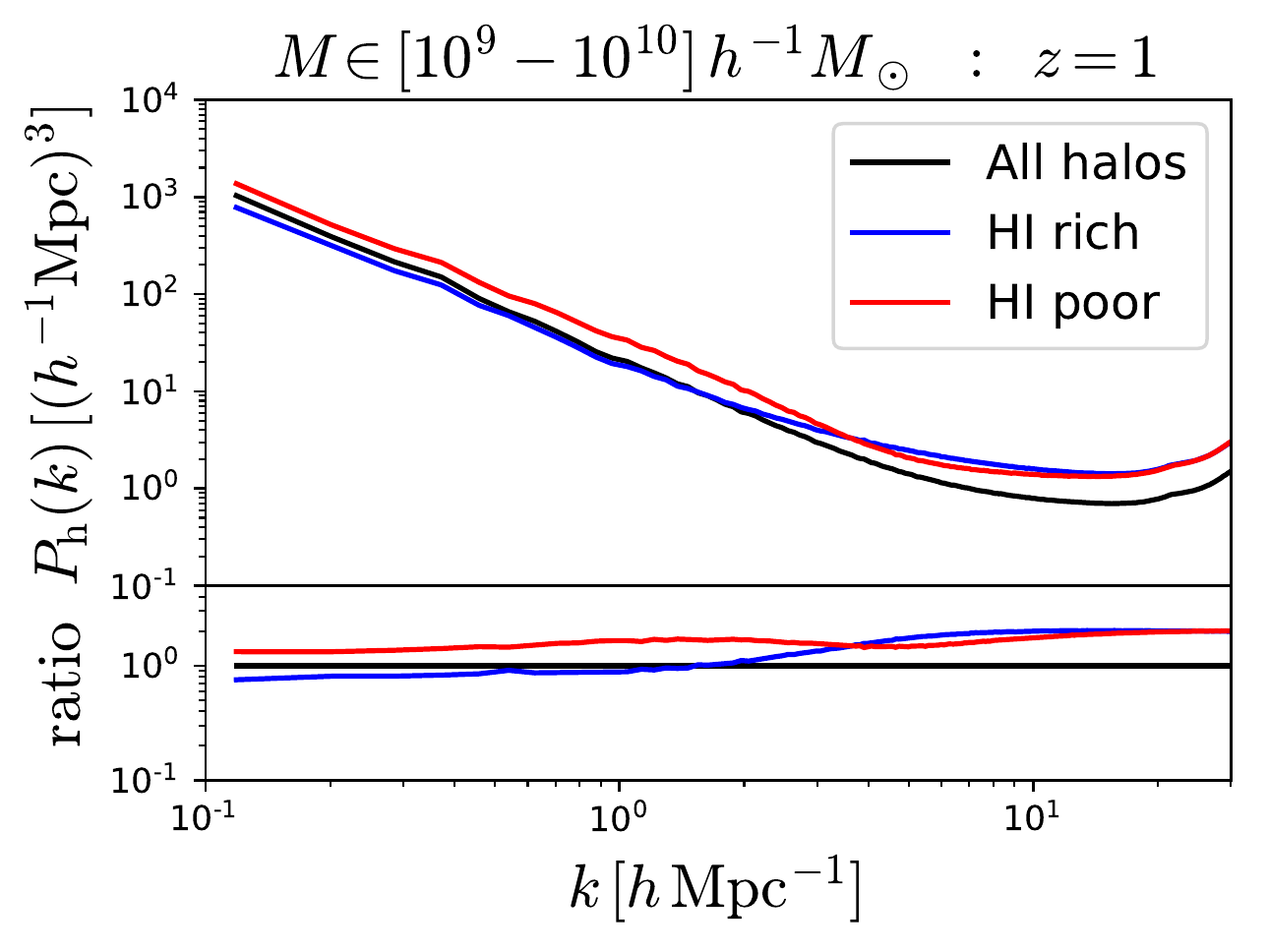}
\includegraphics[width=0.33\textwidth]{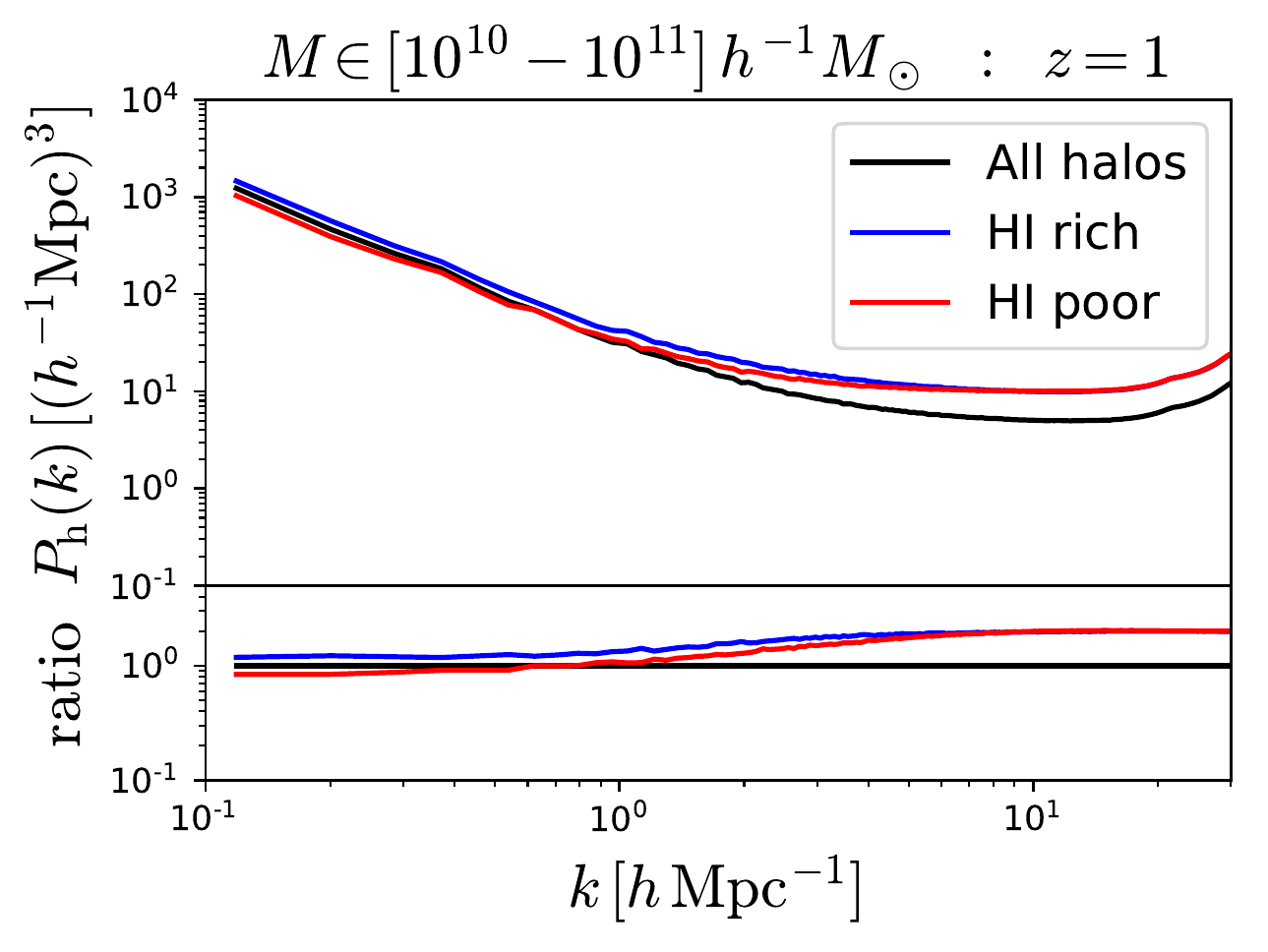}\\
\includegraphics[width=0.33\textwidth]{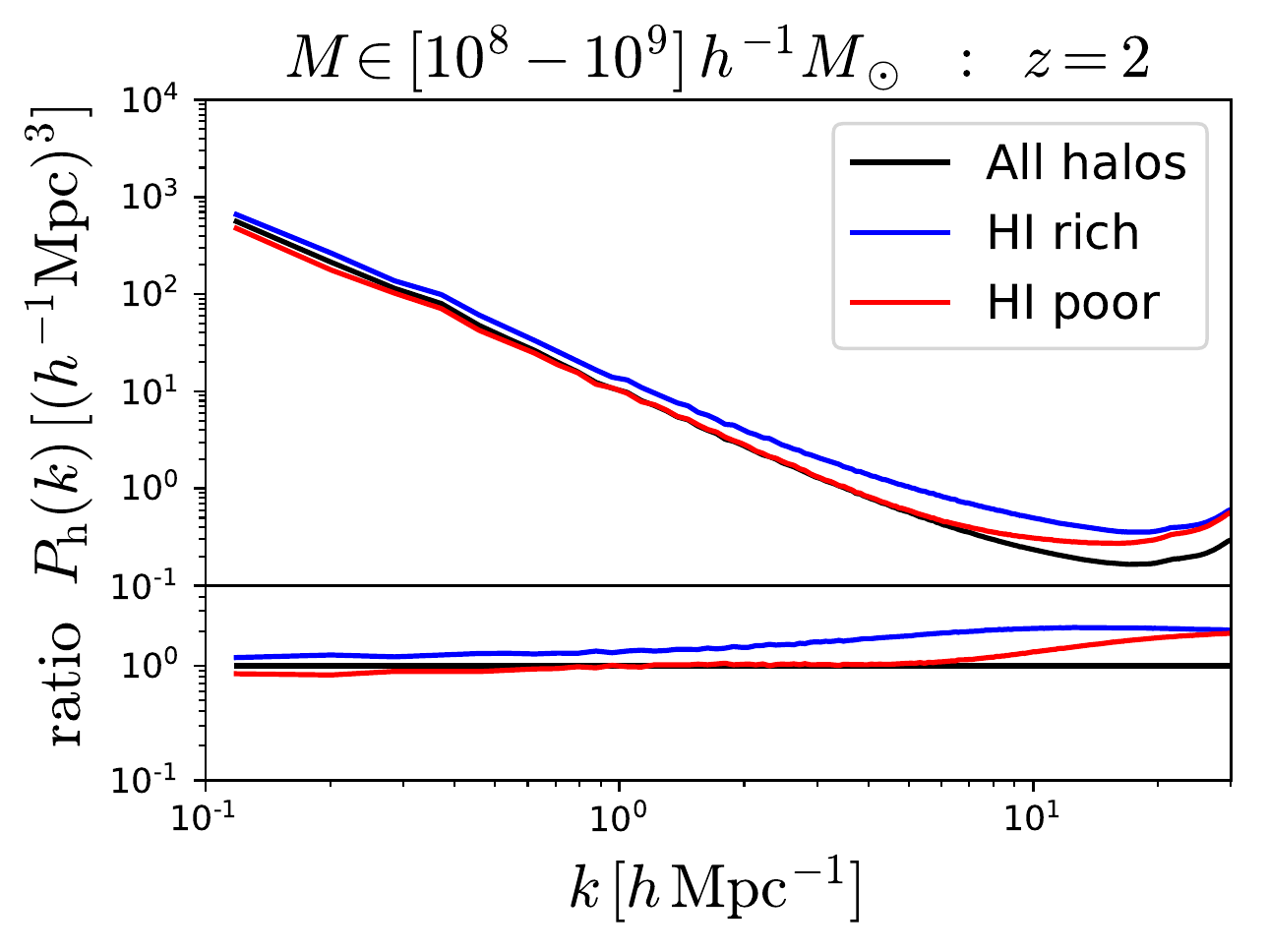}
\includegraphics[width=0.33\textwidth]{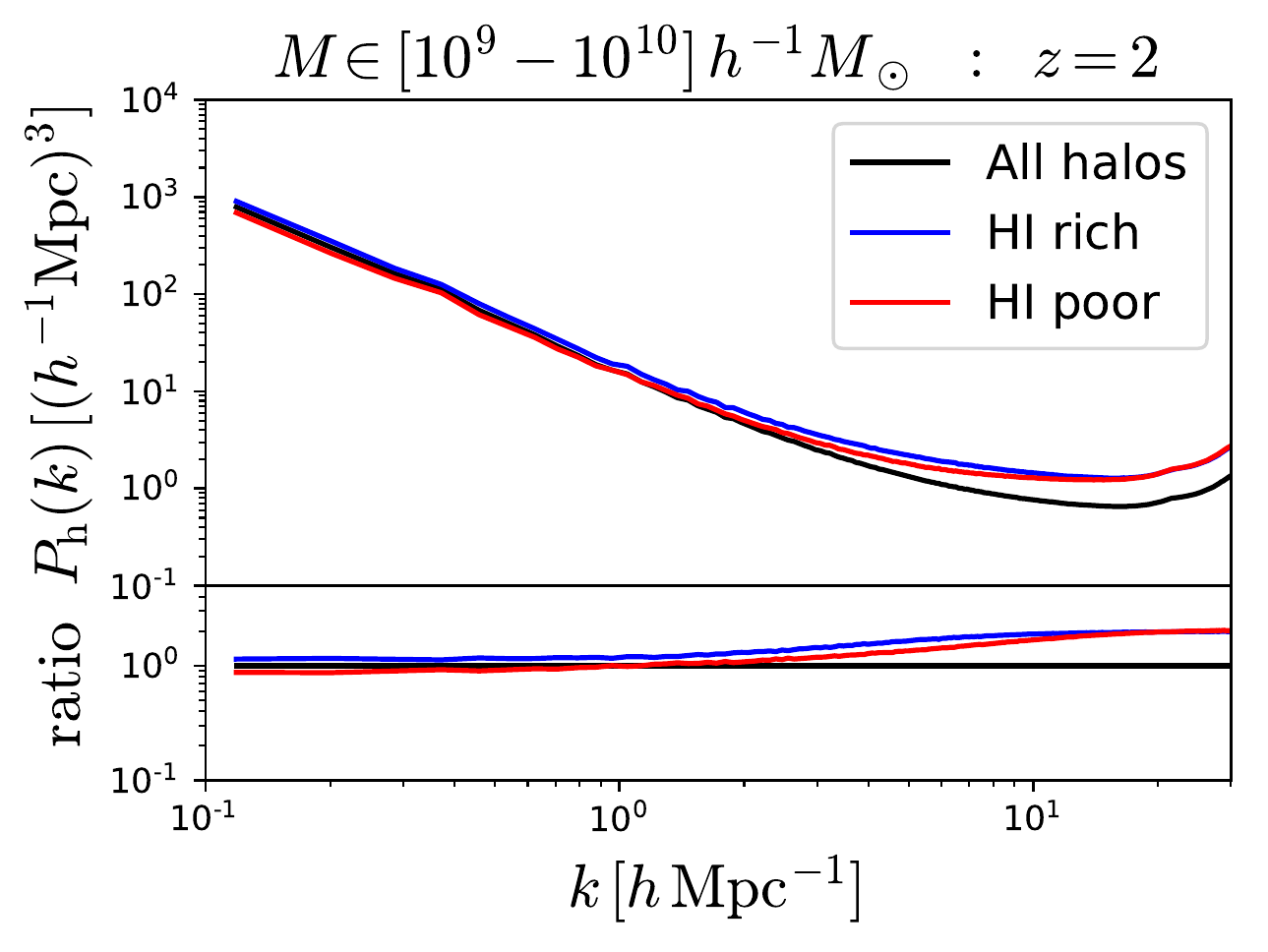}
\includegraphics[width=0.33\textwidth]{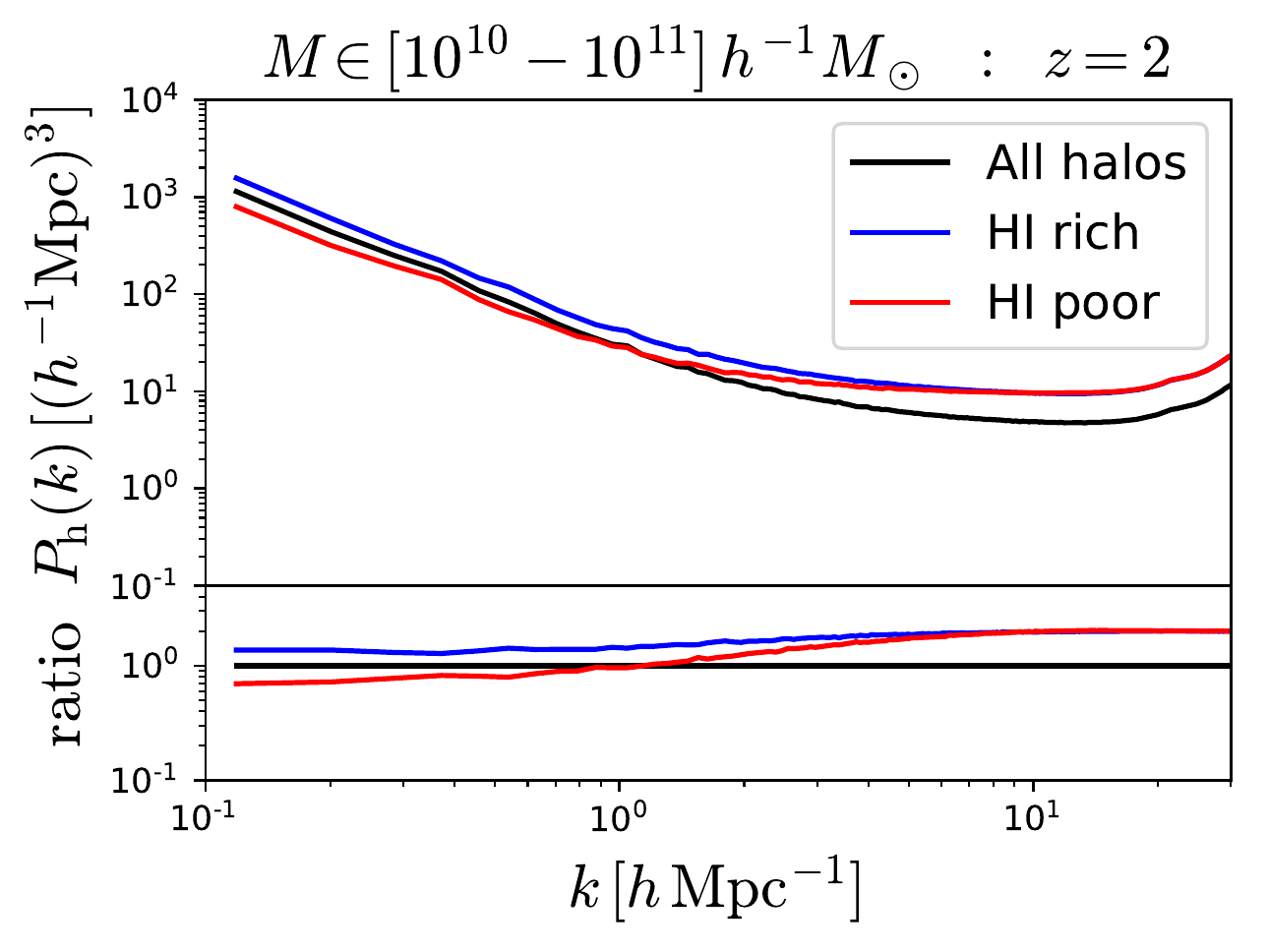}\\
\includegraphics[width=0.33\textwidth]{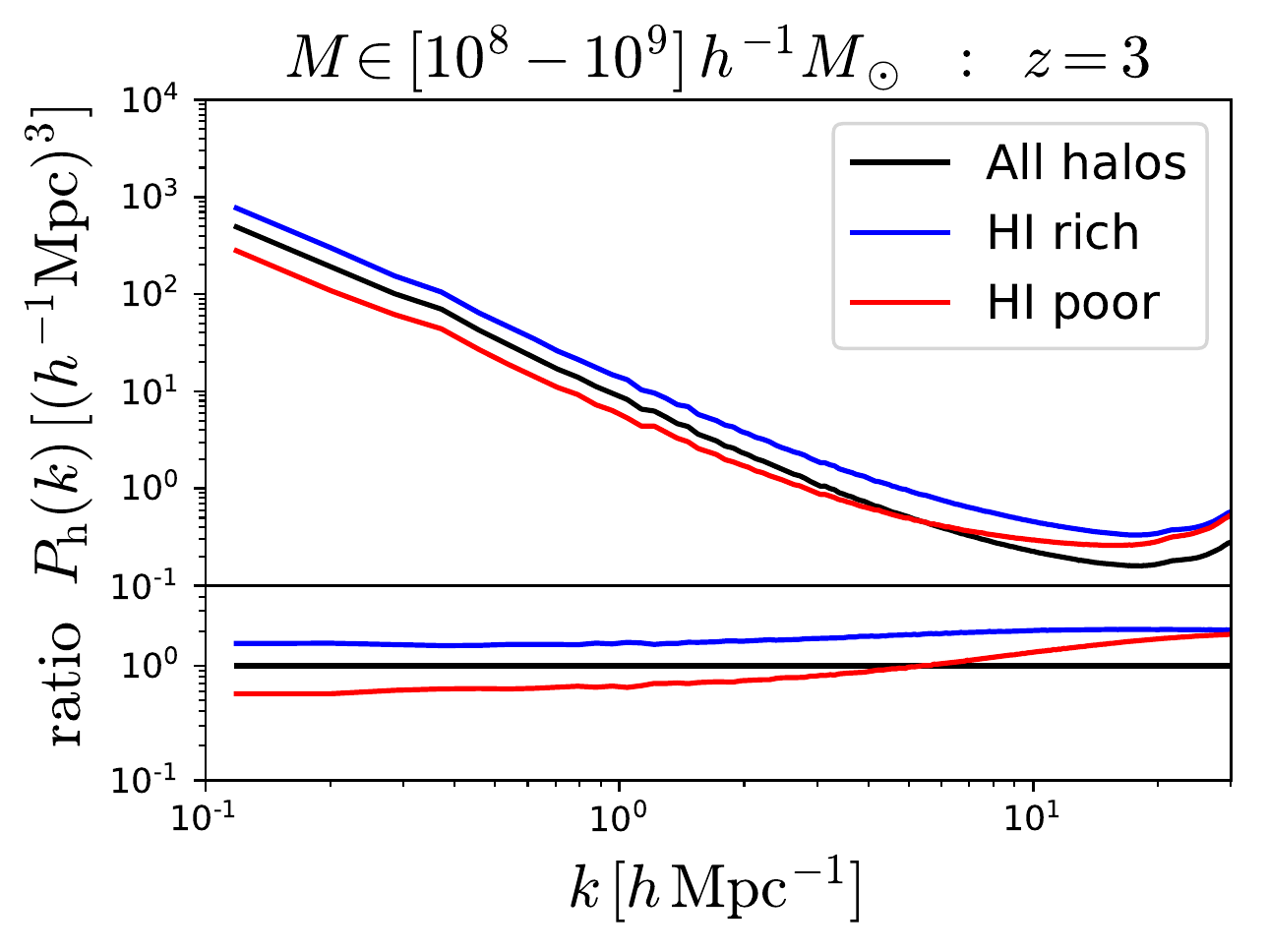}
\includegraphics[width=0.33\textwidth]{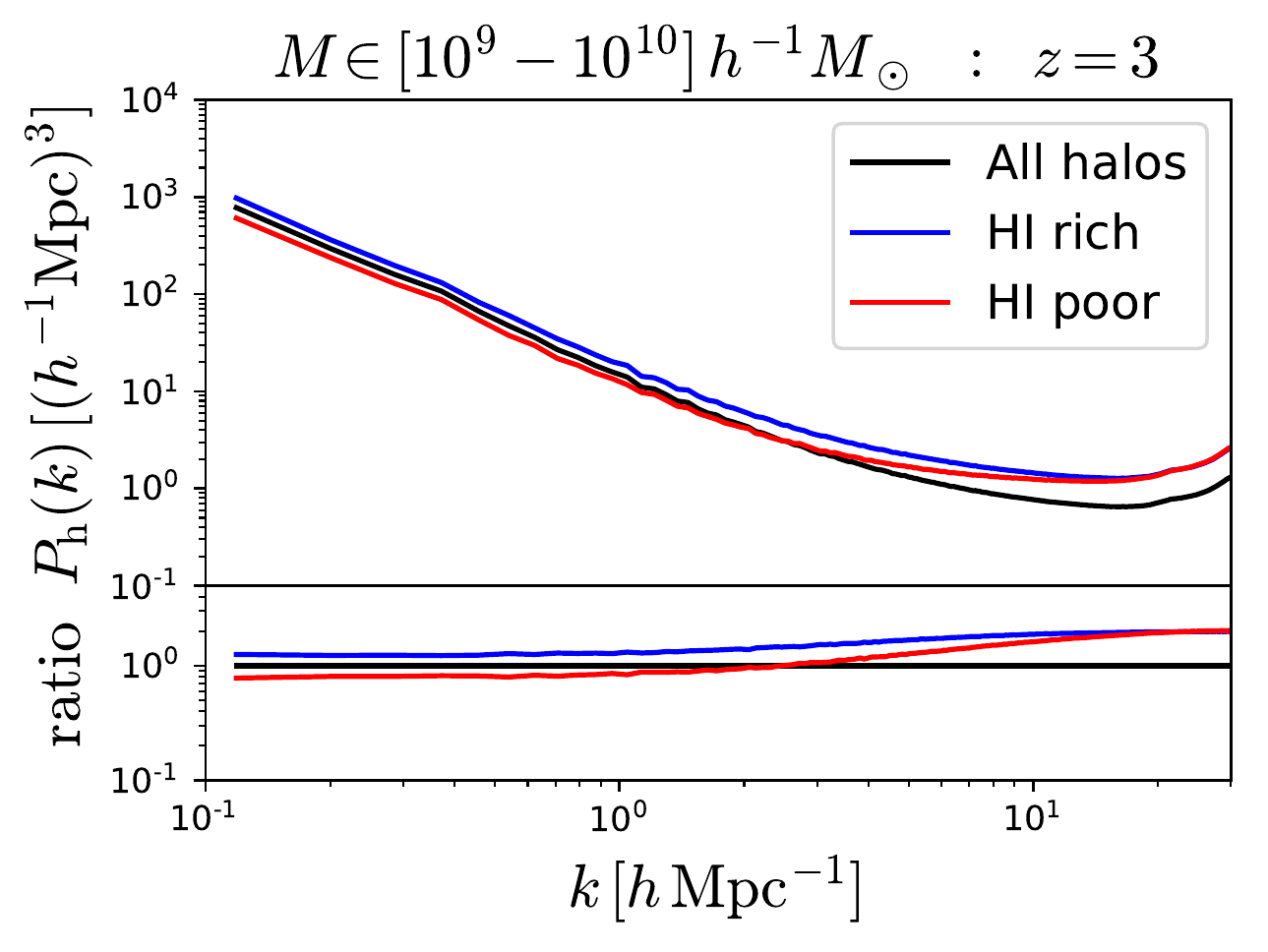}
\includegraphics[width=0.33\textwidth]{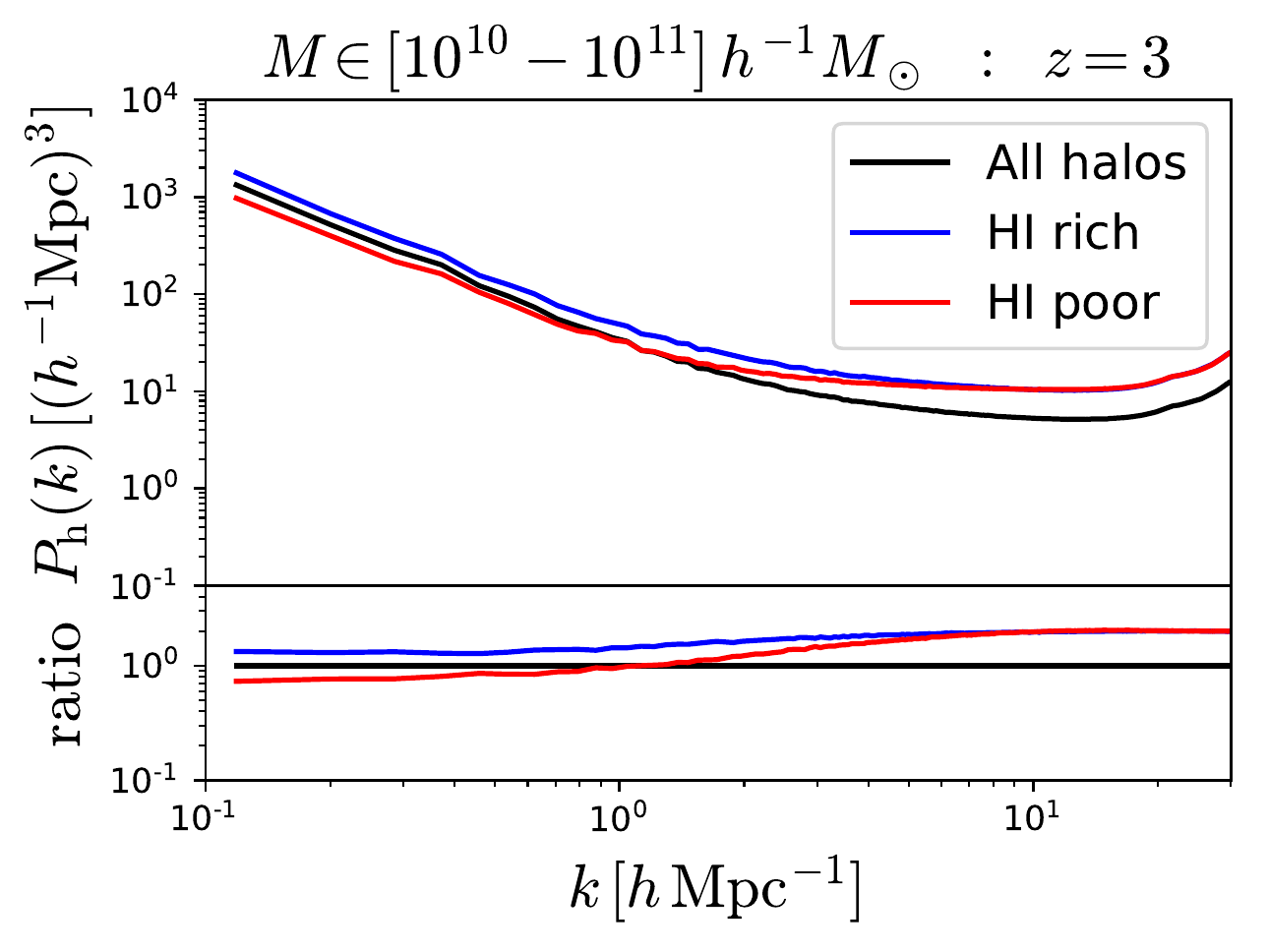}\\
\includegraphics[width=0.33\textwidth]{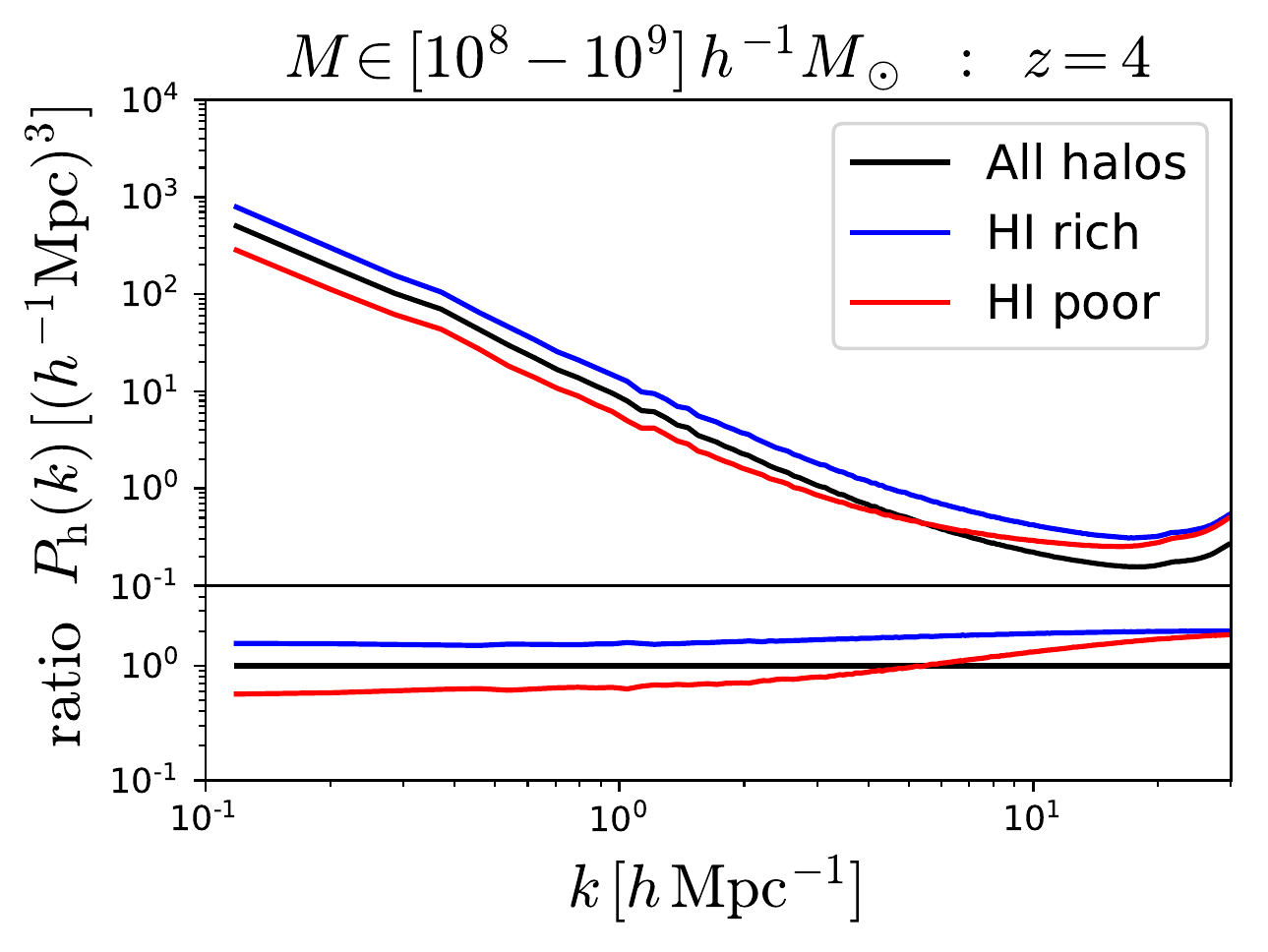}
\includegraphics[width=0.33\textwidth]{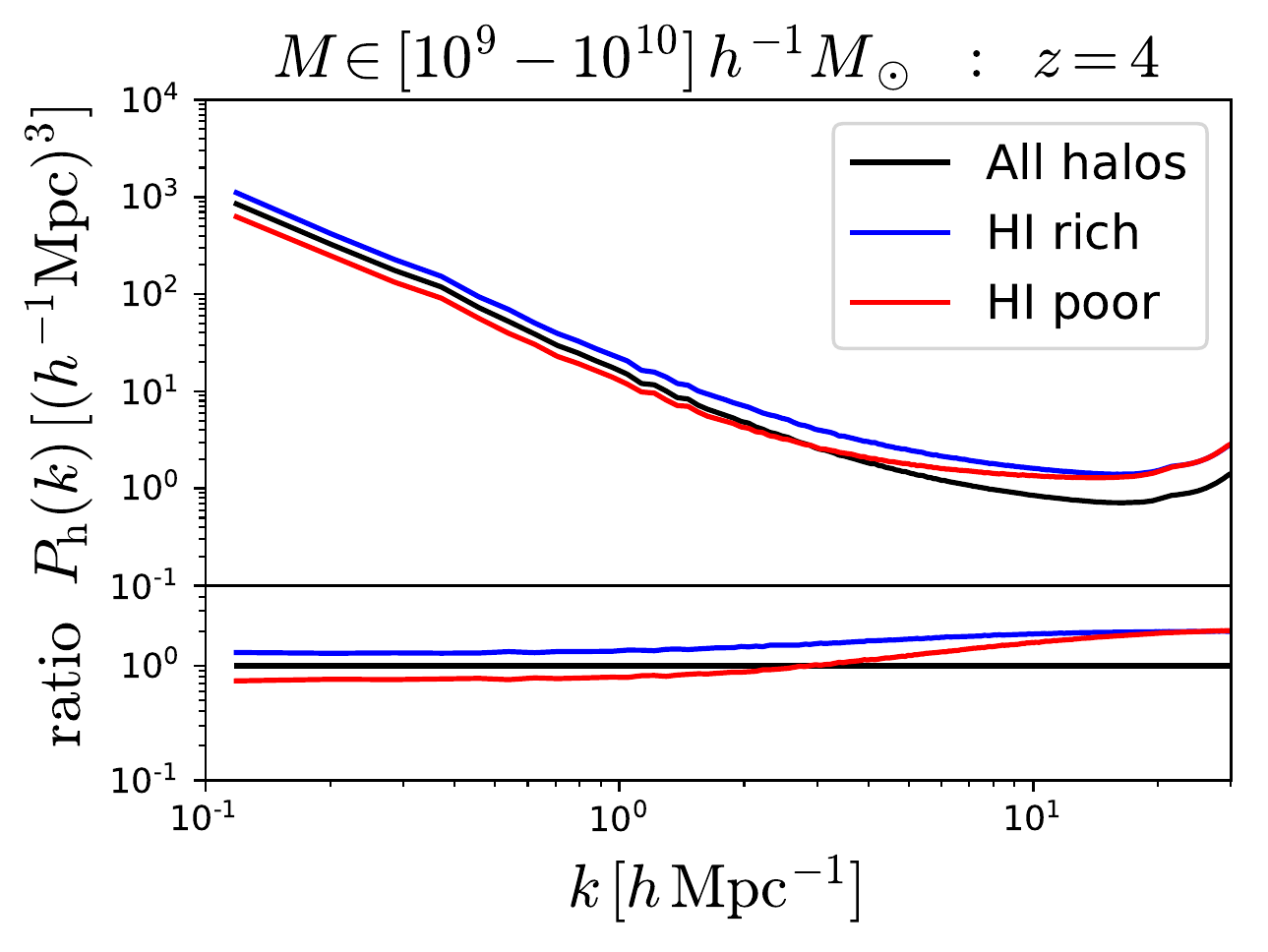}
\includegraphics[width=0.33\textwidth]{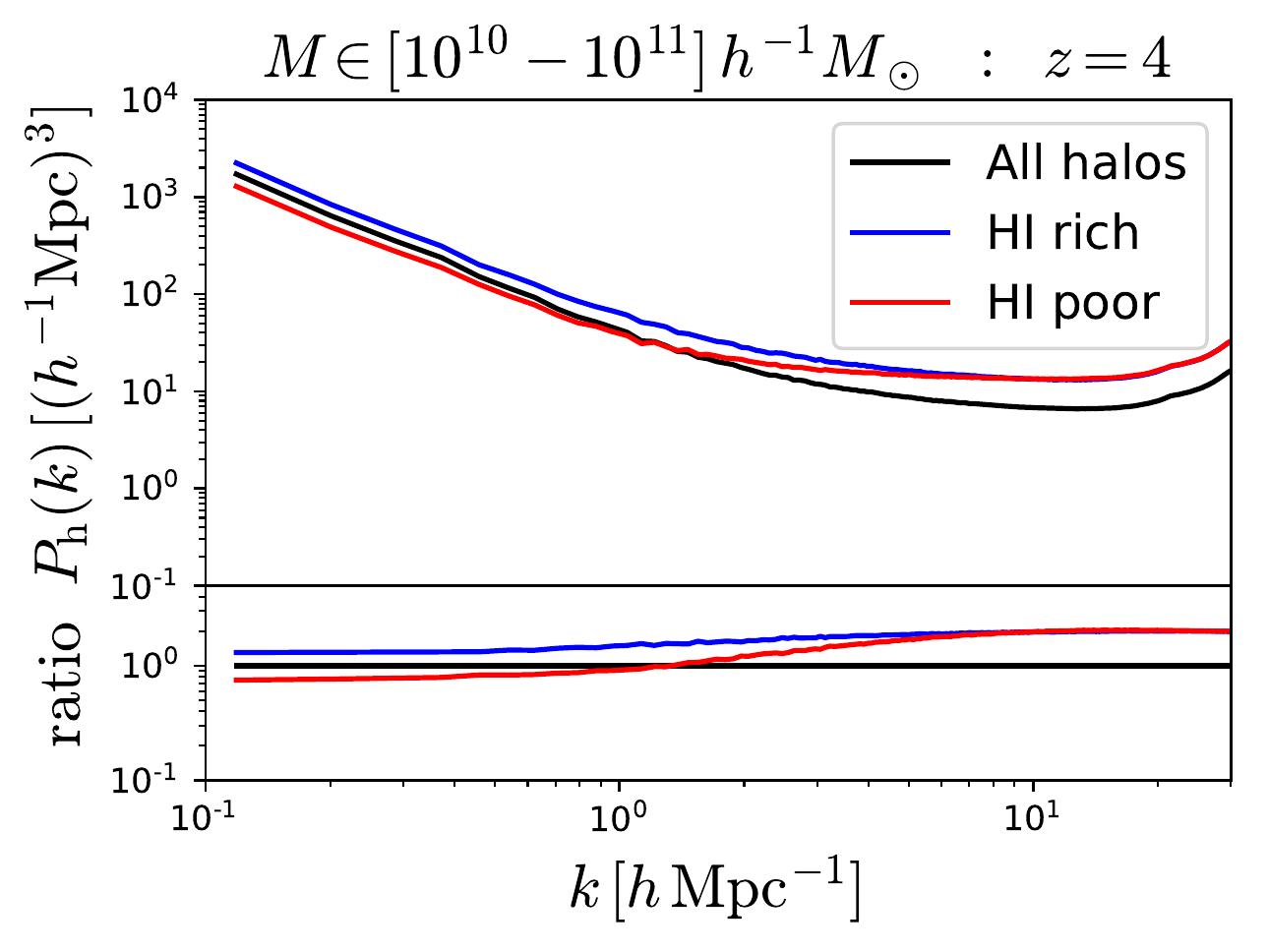}\\
\caption{We select dark matter halos in narrow mass bins and compute their power spectra, shown with solid black lines. We then compute the HI mass inside each halo and find the median value of all halos in that mass bin. Next, we split the halos in two sets, those with HI mass above (HI rich) or below (HI poor) the median. Finally, we compute the power spectrum of the halos in each set and show the results with blue (HI rich) and red (HI poor) lines. Results are shown at redshifts 0, 1, 2, 3 and 4 from top to bottom and for halo in the $10^8-10^9$ (left column), $10^9-10^{10}$ (middle column) and $10^{10}-10^{11}$ $h^{-1}M_\odot$ mass bin intervals. The upper part of each panel displays the different halo power spectra while the bottom part shows the same results normalized by the power spectrum of all halos. The clustering of halos depends not only on mass but also on HI content. The magnitude of this effect is generally larger for smaller halos. At low redshift, HI-poor halos are more clustered than HI-rich halos, but at high redshift the trend reverses.}
\label{fig:assembly_bias}
\end{center}
\end{figure*}

It is well known that the clustering of halos depends primarily on
mass. However, mass is not the only variable that determines halo
clustering; there is also a dependence on halo age \citep{ST_2004,
  Gao_2005}, concentration \citep{Wechsler_2006}, subhalo abundance
\citep{Wechsler_2006, Gao_2007}, halo shape \citep{Faltenbacher_2010},
spin \citep{Gao_2007}, and environment \citep{Salcedo_2018,Han2018}.  Here, we
identify a new secondary bias, originating from the HI content of halos.

Differently from the previous quantities, which are properties of the
dark matter halos, the HI content is more related to the properties of
the galaxies inside a halo; e.g. whether galaxies are red or
blue. Thus, a study of this kind can only be carried out using
hydrodynamic simulations.

To study this issue, we apply the following procedure. First, all
halos whose total mass is within a relatively narrow mass bin are
selected.  The HI mass inside each of those halos and the median value
are determined.  Next, the halos are split into two categories: HI
rich and HI poor, depending on whether the HI content of a particular
halo is above or below the median, respectively. Finally, we compute
the power spectrum of: 1) all halos, 2) HI rich halos and 3) HI poor
halos.

The results are shownin Fig. \ref{fig:assembly_bias} at redshifts 0,
1, 2, 3 and 4 and for three different mass bins:
$M\in[10^8-10^9]~h^{-1}M_\odot$, $M\in[10^9-10^{10}]~h^{-1}M_\odot$
and $M\in[10^{10}-10^{11}]~h^{-1}M_\odot$. The black lines indicate
the power spectrum of all halos, while the blue and red lines
represent the power spectra of the HI rich and HI poor halos,
respectively.

Going towards smaller scales, the amplitudes of the different power
spectra first flatten and then rise back up. This happens because: 1)
we approach the shot-noise limit, and 2) due to aliasing.  The
shot-noise level of the HI rich/poor halos is different from that of
all halos, as the latter contain, by definition, twice more halos than
the former. Thus, on small scales, the amplitude of the HI rich and HI
poor halos is expected to be the same but higher by a factor of 2 than
that of all halos, as is indeed seen.

For all redshifts and mass intervals considered, the clustering of HI
rich galaxies is different from that of HI poor ones, showing that
halo clustering depends not only on mass but on HI content as
well. The difference in the clustering of HI poor and HI rich halos
decreases, in general, with halo mass. At $z=0$, the amplitude of the
halo power spectrum of the HI rich and HI poor can be almost one order
of magnitude different for halos in the [$10^9$-$10^{10}$]
$h^{-1}M_\odot$ mass bin. The largest differences are seen for halos
with masses around or below $M_{\rm min}$ at that particular redshift,
namely around the mass scale where the HI content starts being
exponentially suppressed (see section \ref{subsec:M_HI}).

At $z=0$, and for the mass bin intervals considered here, the HI poor
halos are more strongly clustered than the HI rich halos. On the other
hand, at high-redshift the situation is the opposite, and HI rich
halos are more strongly clustered than HI poor halos. At $z=1$, we
find that depending on the halo mass considered, HI rich halos can be
more or less clustered than HI poor halos.

Although the halo mass bins are fairly narrow, is not unreasonable to
suspect that the most massive halos will have larger HI masses and
therefore this could introduce some natural splitting that arises just
from halo mass and not from HI secondary bias. In order to test this,
we split the halos according to their median total halo mass and
repeated the above analysis.  We find that the clustering of the two
samples is almost indistinguishable, ruling out the possibility that
halo mass is affecting our results.

At low redshift, small halos near big ones are more likely to be
stripped of their gas content. Thus, HI poor halos should be more
strongly clustered than HI rich halos, as we find. This possibility
has been recently suggested to explain the secondary bias that arises
from several halo properties in \cite{Salcedo_2018,Han2018}.

On the other hand, at high redshift, gas stripping by nearby halo
neighbors should be less effective, as the largest halos are not yet
very massive and there has been less physical time for these processes
to operate.  We speculate that at high-redshift, regions around
massive halos are richer in HI than other regions.  For example, in
regions with higher density we would expect the filaments to be
slightly more dense and therefore will host more HI. Halos connected
by those filaments may thus become HI rich. This naive picture can be
seen in Fig. \ref{fig:HI_Lya}, where at high-redshift, the filaments
in the denser regions host more HI than in less denser regions.

\section{HI shot-noise}
\label{subsec:SN}

An important consideration in any cosmological survey is shot-noise,
as its amplitude determines the maximum scale where cosmological
information can be extracted from(see Eq. \ref{Eq:P_21cm}). However, it can also be used
to learn about the galaxy population hosting the HI \citep{Wolz_2017, Wolz_2018}.
The purpose of this section is to quantify the amplitude of the HI shot-noise from
our simulations.

\begin{figure}
\begin{center}
\includegraphics[width=0.45\textwidth]{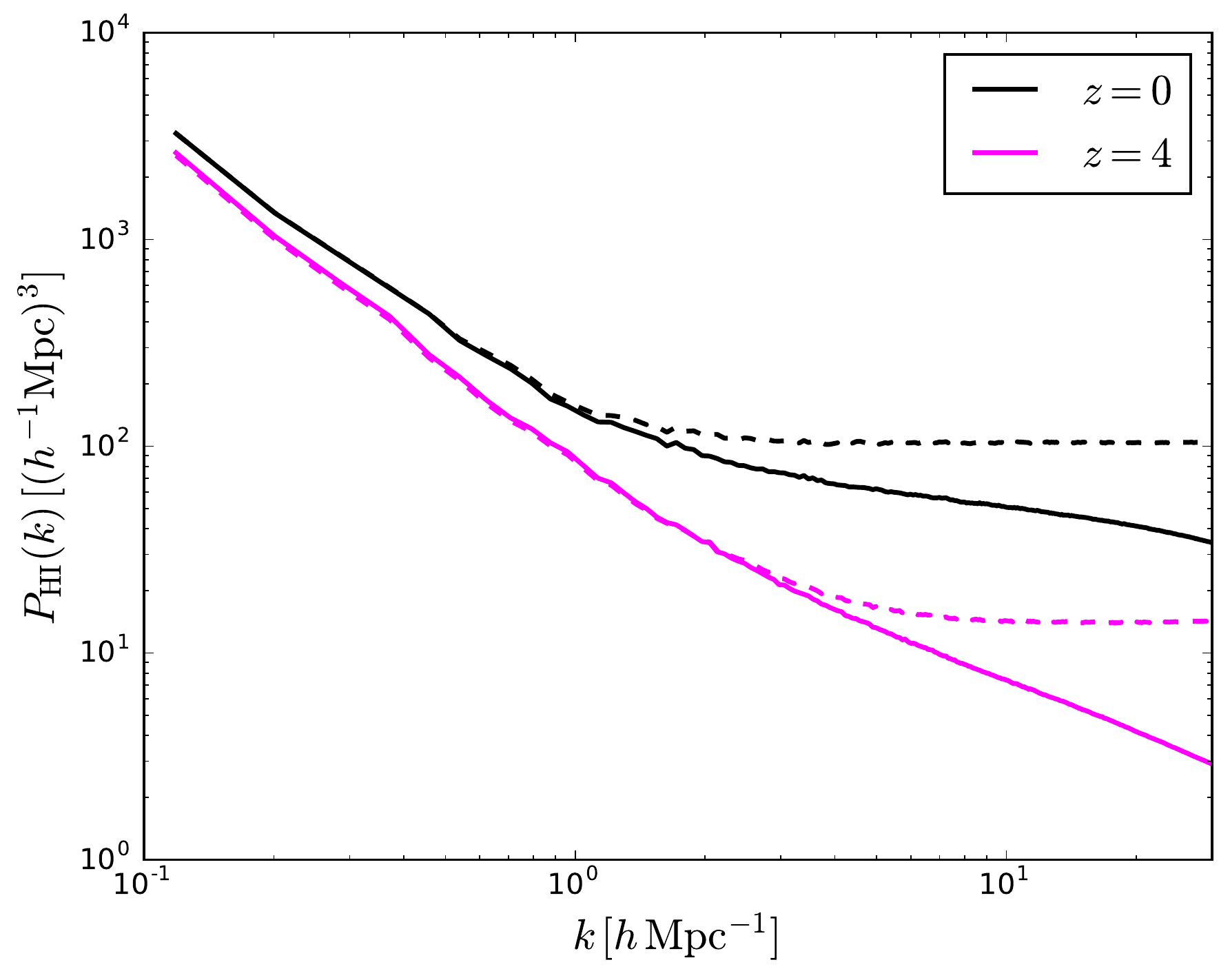}
\caption{HI shot-noise. This figure shows the standard HI power spectrum (solid lines) and the power spectrum when the HI inside halos is placed in the halo center (dashed lines) at redshifts $z=0$ (black) and $z=4$ (purple). For the former we can see both the 1- and 2-halo terms, while for the latter the 1-halo term is just the HI shot-noise.}
\label{fig:SN}
\end{center}
\end{figure}

We now illustrate why computing the HI shot-noise is slightly more
complicated than determining the shot-noise for other tracers, such as
halos, where the value of the shot-noise is simply given by the
amplitude of the power spectrum on small scales.

The solid lines of Fig. \ref{fig:SN} show the HI power spectrum at
redshifts 0 and 4. It can be seen that those power spectra receive
contributions from both the 1- and 2-halo terms; i.e.~the power
spectrum on very small scales does not become constant, in contrast
with the halo power spectrum. This happens simply because there is 
structure in HI inside halos and galaxies (see
Fig. \ref{fig:HI_profile_images}).

In order to isolate the contribution of the HI shot-noise, i.e.~to
avoid the 1-halo term contribution, we do the following. We
compute the total amount of HI inside every halo in the simulation and
place that HI mass in the halo center. We then compute the HI power
spectrum of that configuration. Since in that case there is
no HI structure inside halos there is no 1-halo term, and the
amplitude of the HI power spectrum on small scales is just the HI
shot-noise. The dashed lines in Fig. \ref{fig:SN} display the results.

It can be seen that on large scales, the amplitude of the HI power
spectrum of the two configurations is essentially identical. The small
differences at $z=4$ arise from the HI that is outside of halos (see
Fig. \ref{fig:HI_halos_galaxies}), whose contribution is not accounted
for with our procedure. That contribution should, however, have
negligible impact on the amplitude of the HI shot-noise. On small
scales, the lack of the 1-halo term when we artificially place the HI
at the halos center makes possible to isolate the value of the 
HI shot-noise. We determine the value of the HI shot-noise by
averaging the amplitude of the HI power spectrum on scales
$k\in[20-30]~h{\rm Mpc}^{-1}$ and show the results in table
\ref{table:SN}.

The HI shot-noise can also be computed using the halo model
framework (assuming all HI is in halos) as \citep{EmaPaco}
\begin{widetext}
\be
P_{\rm HI}^{\rm SN}(z)=\lim_{k\to0} P_{\rm 1h, HI}(k,z)=\frac{1}{[\rho_{\rm c}^0\Omega_{\rm HI}(z)]^2}\int_0^\infty n(M,z)M_{\rm HI}^2(M,z)dM=\frac{\int_0^\infty n(M,z)M_{\rm HI}^2(M,z)dM}{\left[\int_0^\infty n(M,z)M_{\rm HI}(M,z)dM\right]^2}~.
\ee
\end{widetext}
If we use the Sheth \& Tormen (ST) formula
\citep{Sheth-Tormen} for the halo mass function
in this expression\footnote{We find that
  the halo mass function in IllustrisTNG is well reproduced by the ST
  form.} and our fitting formula for $M_{\rm HI}(M,z)$ (see
table \ref{table:M_HI_fit}), we obtain values for the HI shot-noise in
agreement with those measured directly in the simulation.

In order to validate our results we have estimated the HI
shot-noise by measuring the stochasticity between the HI and the matter fields \citep{Seljak2009}
\be
P_{\rm HI}(k)-\frac{P_{\rm HI-m}^2(k)}{P_{\rm m}(k)}~,
\ee
which at low enough $k$ should correspond to the uncorrelated part of the HI power spectrum, \textit{i.e.}
$P_{\rm SN}$. However, the shot-noise amplitudes we obtain in this case are much
larger than those found using our fiducial procedure. By repeating the
this approach for halos we also find that we cannot recover the
standard $1/\bar{n}$ result. Our findings are in agreement with other studies, see \cite{Hand2017} and reference therein, and point towards some lack of understanding of the noise properties in low mass halos even in gravity only simulations. 
For better comparison with previous work, that always considered Poissonian shot-noise, we assume $P_{\rm SN}$
is given by the fiducial halo model procedure, and defer a study of this unexpected disagreement to future work.

We find that the HI shot-noise is low at all redshifts. This is in broad
agreement with the analytical work of \cite{EmaPaco}, whose results at high redshift had however a large theoretical error. Our shot-noise
values are slightly lower than those in that paper, mostly because 
values of $M_{\rm min}$ at high $z$ we find here are lower than those considered
in \cite{EmaPaco}.

\begin{figure}
\begin{center}
\includegraphics[width=0.45\textwidth]{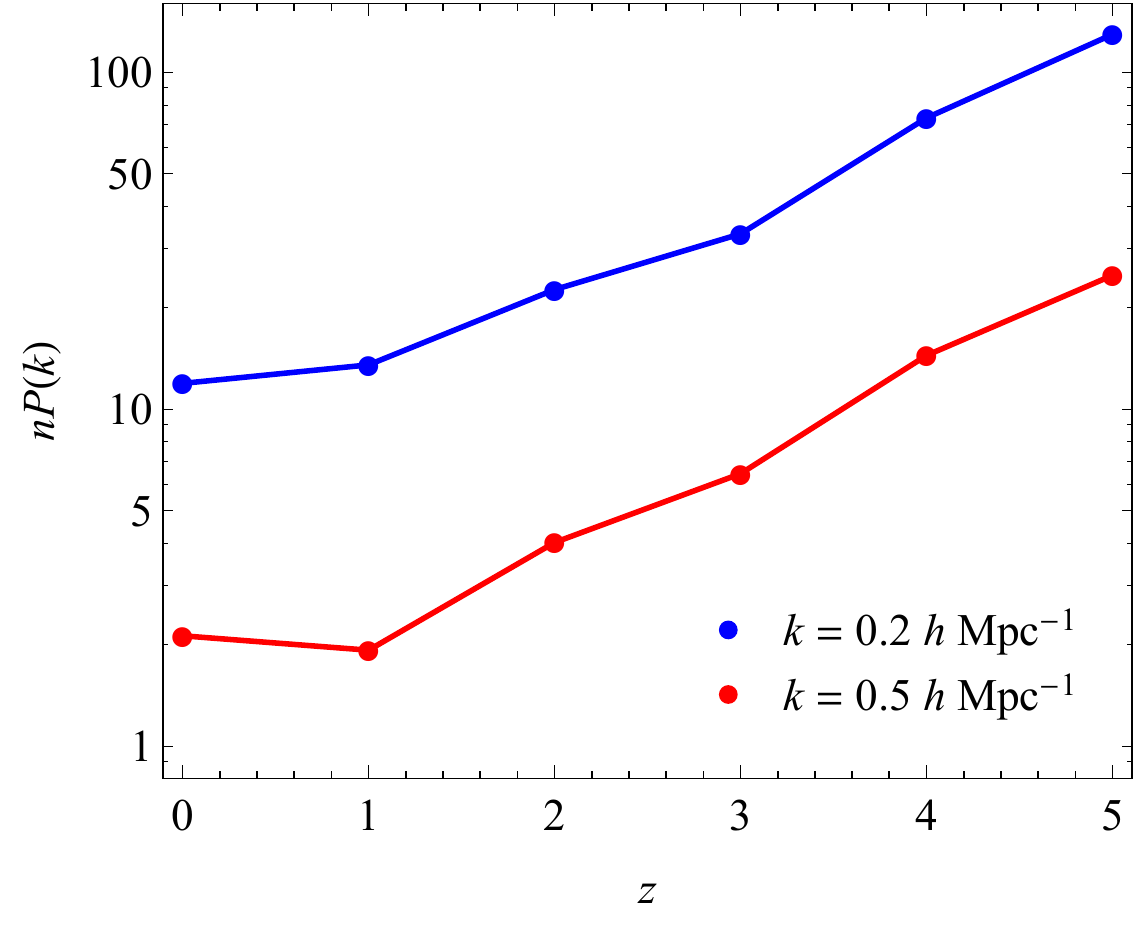}
\caption{An important quantity for BAO studies is $nP_{\rm 0.2}$, defined as the ratio between the amplitudes of the cosmological signal and the shot-noise level at $k=0.2~h{\rm Mpc}^{-1}$. In this plot we show this quantity in blue as a function of redshift using TNG100 measurements. We find that it is large at all redshifts, indicating that shot-noise is not important for BAO studies with 21 cm intensity mapping. In order to extract information from small scales it is also crucial to have large $nP$ at high-$k$. The red line shows that this is indeed the case at $k = 0.5 ~h\text{Mpc}^-{1}$ for all redshifts considered in our work.
}
\label{fig:nP}
\end{center}
\end{figure}

\begin{figure*}
\begin{center}
\includegraphics[width=0.33\textwidth]{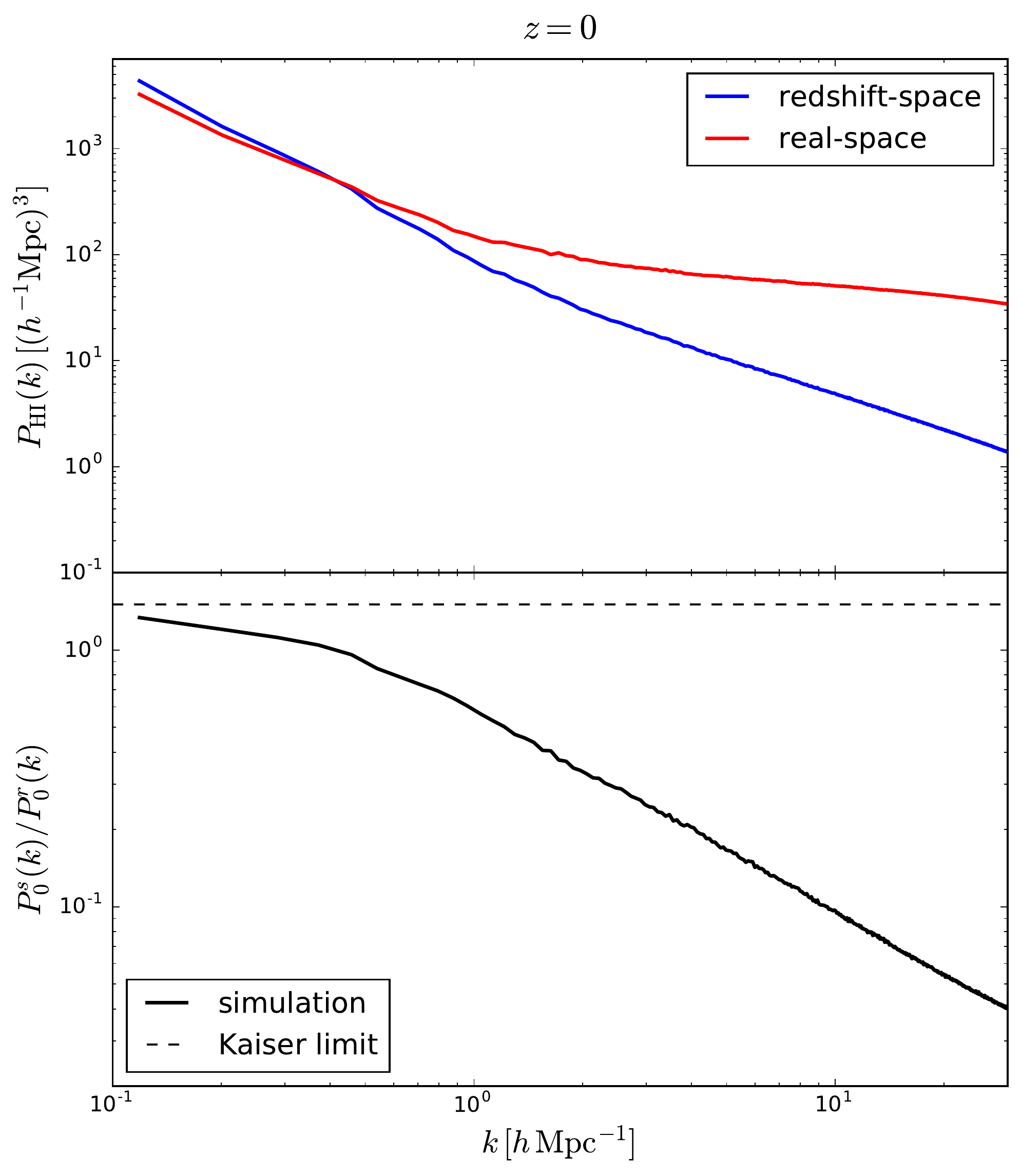}
\includegraphics[width=0.33\textwidth]{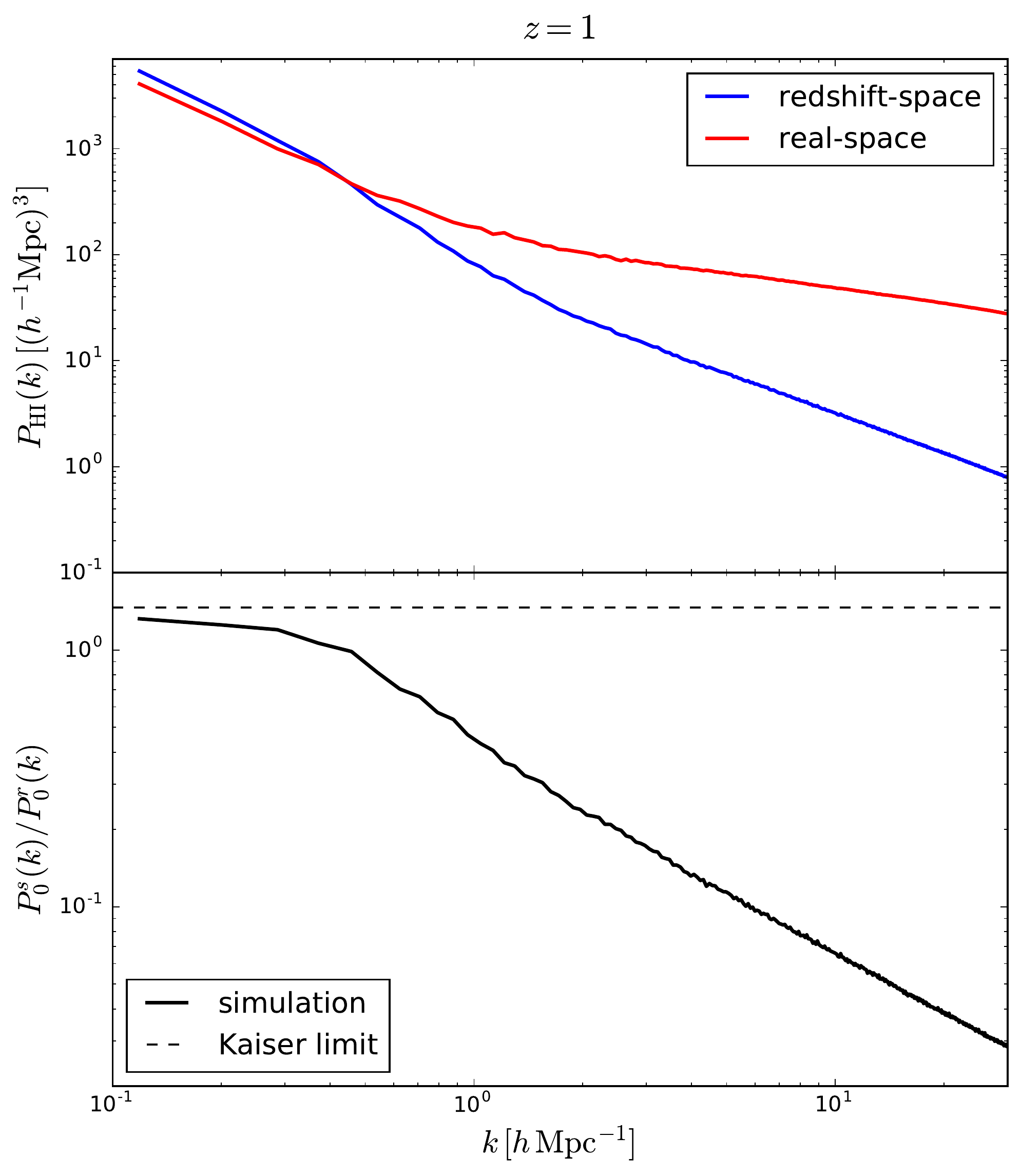}
\includegraphics[width=0.33\textwidth]{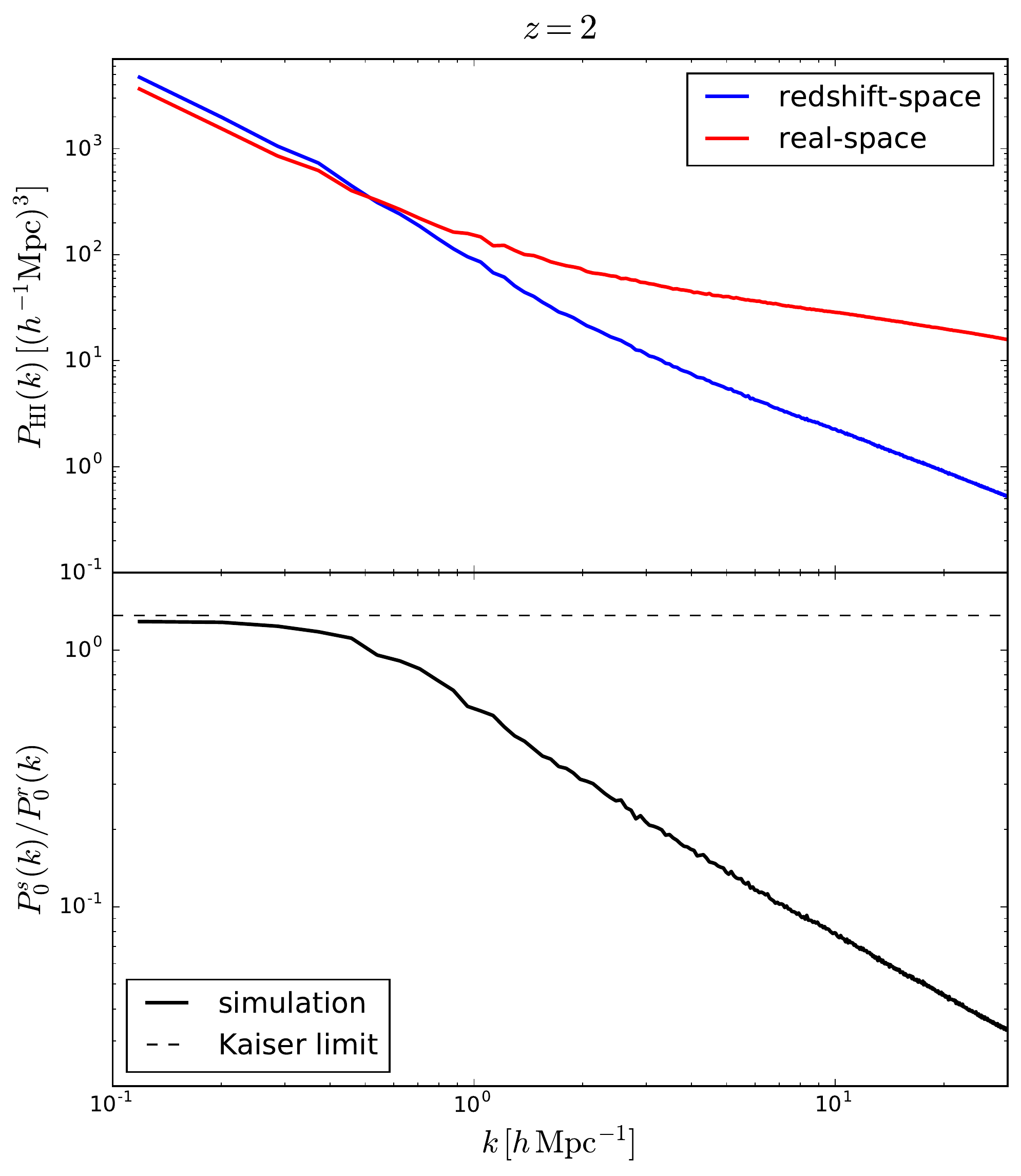}\\
\includegraphics[width=0.33\textwidth]{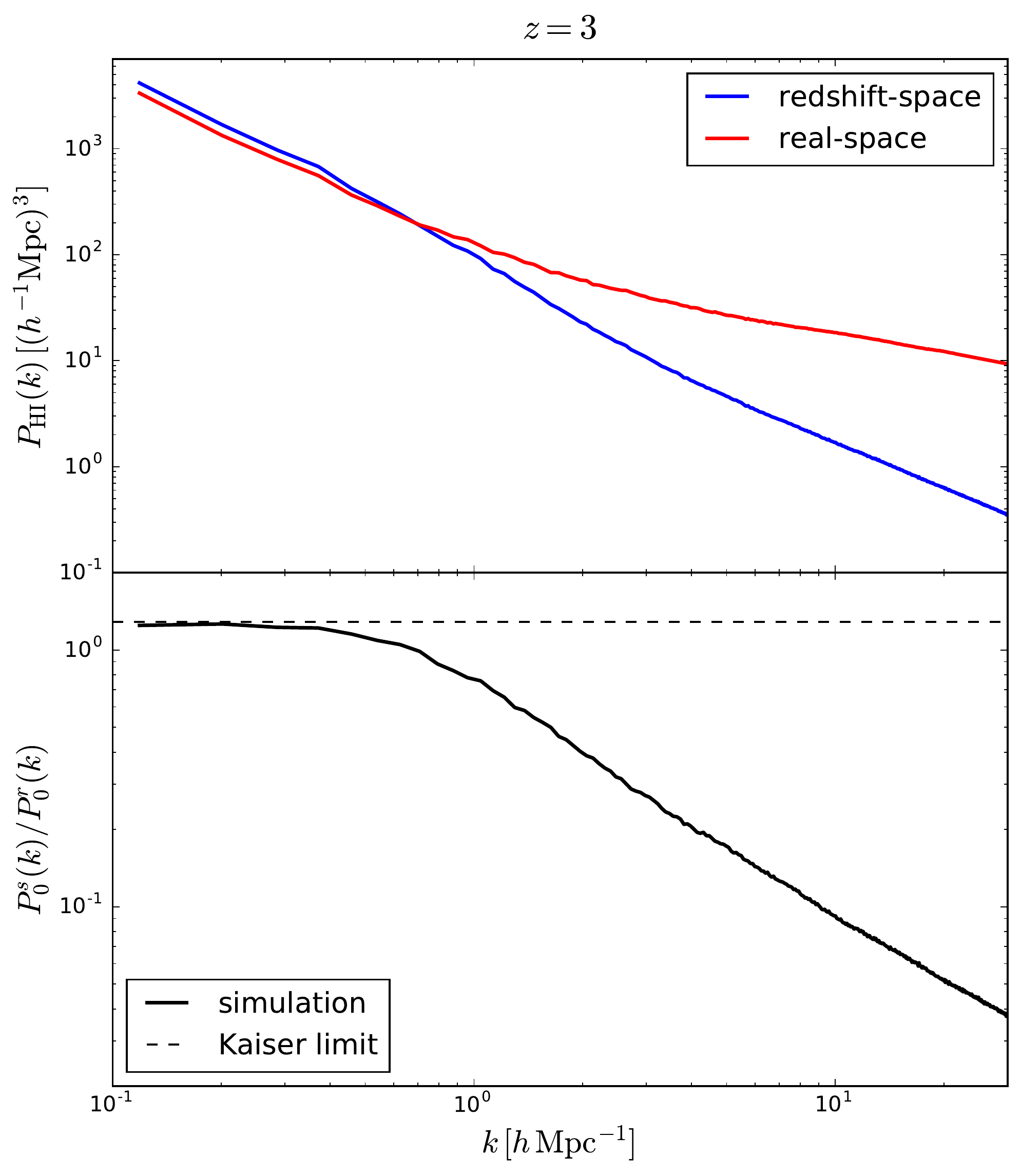}
\includegraphics[width=0.33\textwidth]{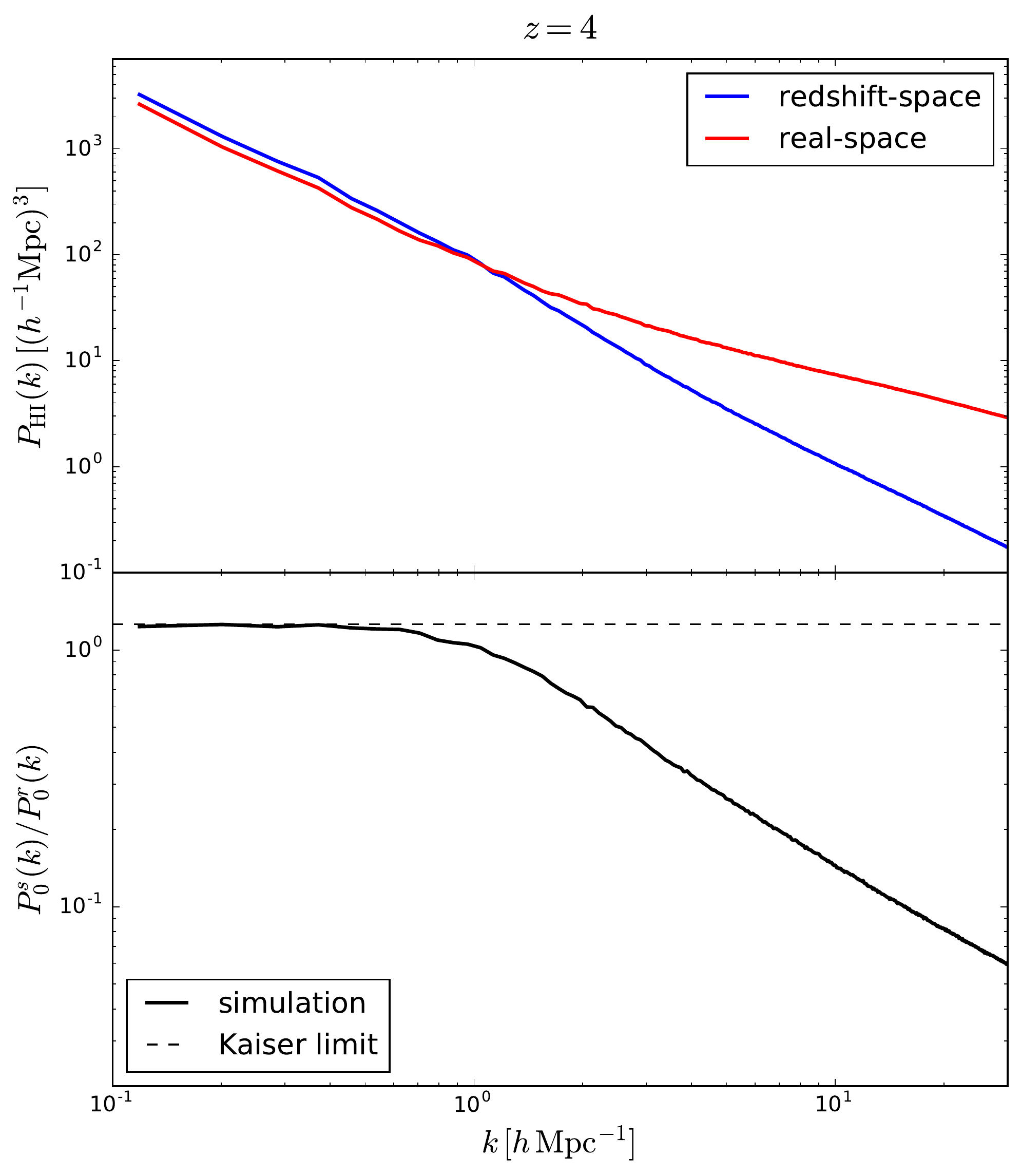}
\includegraphics[width=0.33\textwidth]{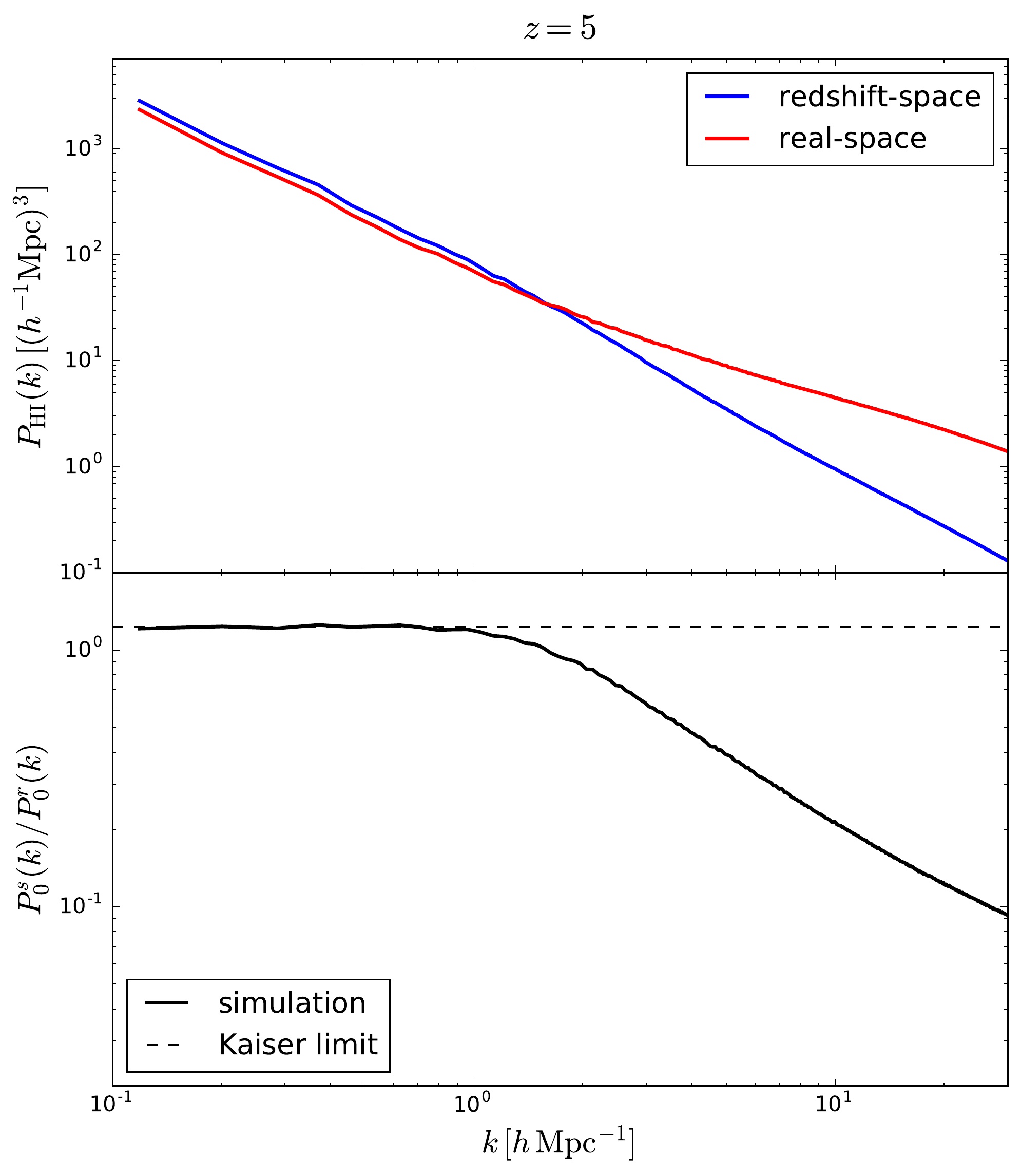}
\caption{Impact of redshift-space distortions on the HI power spectrum at redshifts 0 (top-left), 1 (top-middle), 2 (top-right), 3 (bottom-left), 4 (bottom-middle) and 5 (bottom-right).  The upper part of each panel shows the HI power spectrum (monopole) in real-space (red) and redshift-space (blue). The bottom part displays the ratio between the monopoles in redshift- and real-space (solid black) and the prediction of linear theory (dashed black). Redshift-space distortions enhance/suppress power on large/small scales. Linear theory can explain the HI clustering in redshift-space down to very small scales at high-redshift, while it cannot 
at low redshift on the scales we probe in the simulations.}
\label{fig:HI_RS}
\end{center}
\end{figure*}

The low values of shot-noise we find have important implications. First,
21 cm intensity mapping experiments that aim at
measuring the BAO peak \citep{Obuljen_2016} position such as CHIME, OOTY, BINGO, HIRAX, or
SKA will barely be affected by shot-noise. We illustrate this in
Fig. \ref{fig:nP}, where we plot, in blue,
\be
nP_{\rm 0.2}(z)=\frac{P_{\rm HI}(k=0.2~h{\rm Mpc}^{-1},z)}{P_{\rm SN}(z)}
\ee
as a function of redshift. We obtain values of $nP_{\rm 0.2}$ above
$\simeq15$ at all redshifts, showing that shot-noise contamination on
BAO scales is minimal. This is a great advantage over traditional
methods such as galaxy surveys. 

Second, the shot-noise levels are very low at high redshift. This
means that shot-noise does not erase the cosmological information on small scales. The red line shows the value of $nP$ at $k = 0.5 h/\text{Mpc}^{-1}$, which is still much larger than one and implies these modes can be in principle measured in the cosmic variance limit. With accurate theory predictions as the one described in the previous section, this information
can be retrieved and used to reduce errors in the values of the
cosmological parameters.

\section{Redshift-space distortions}
\label{subsec:RSD}

21 cm intensity mapping observations probe the spatial distribution of
cosmic neutral hydrogen in redshift-space, not in real-space. Thus, it
is of utmost importance to understand the impact of peculiar
velocities on the clustering of HI. In this section we study the
clustering of neutral hydrogen in redshift-space.

Here, we make use of the plane-parallel approximation to
displace particles and Voronoi cells positions from real ($\vec{x}$)
to redshift-space ($\vec{s}$) through
\be
\vec{s}=\vec{x}+\frac{1+z}{H(z)}\vec{v}_{||}(\vec{r})
\ee
where $\vec{v}_{||}(\vec{r})$ is the peculiar velocity of the
particle/cell along the line-of-sight. We use the three cartesian axes
as different lines-of-sight. Our results represent the mean over the
three axes.

We have computed the clustering of HI in redshift-space and show
the results in Fig. \ref{fig:HI_RS}. While we have computed the HI
monopoles, quadrupoles and hexadecapoles, the latter two are too noisy
so we restrict our analysis to the monopoles.

We show with red/blue lines the monopoles in real/redshift space in
this figure.  The two main physical processes governing the effect of
peculiar velocities can clearly be seen. On large-scales the
clustering of HI in redshift-space is enhanced due to the Kaiser
effect \citep{Kaiser}.  On small scales, peculiar velocities of HI
inside halos give rise to Fingers-of-God, suppressing the amplitude of
the HI power spectrum.

The bottom part of each panel of Fig. \ref{fig:HI_RS} shows the ratio
between the monopoles in redshift- and real-space as a solid black
line. The black dashed line displays the prediction of linear theory
for that ratio, i.e.~
\be
\frac{P_0^s(k)}{P_0^r(k)}=1+\frac{2}{3}\beta+\frac{1}{5}\beta^2
\ee
where $\beta=f/b_{\rm HI}$, with $f\simeq\Omega_{\rm m}^{0.545}(z)$
being the linear growth rate. We have estimated $\beta$ using the
values of the linear HI bias from Table \ref{table:SN}.

As expected, linear theory is not able to describe our results at low
redshift. This is because the scales we probe in TNG100 are too small
for linear theory to hold. On the other hand, we find that the Kaiser
factor can reasonably well explain the monopoles ratio down to
$k\simeq0.3$, 0.5 and 1.0 $h{\rm Mpc}^{-1}$ at redshifts, 3, 4 and 5,
respectively.

Given the results in Section~\ref{subsec:HI_bias} our findings are not surprising, and we plan to compare the measurements in TNG100 to redshift space LPT analytical predictions in upcoming work.

\begin{figure*}
\begin{center}
\includegraphics[width=0.33\textwidth]{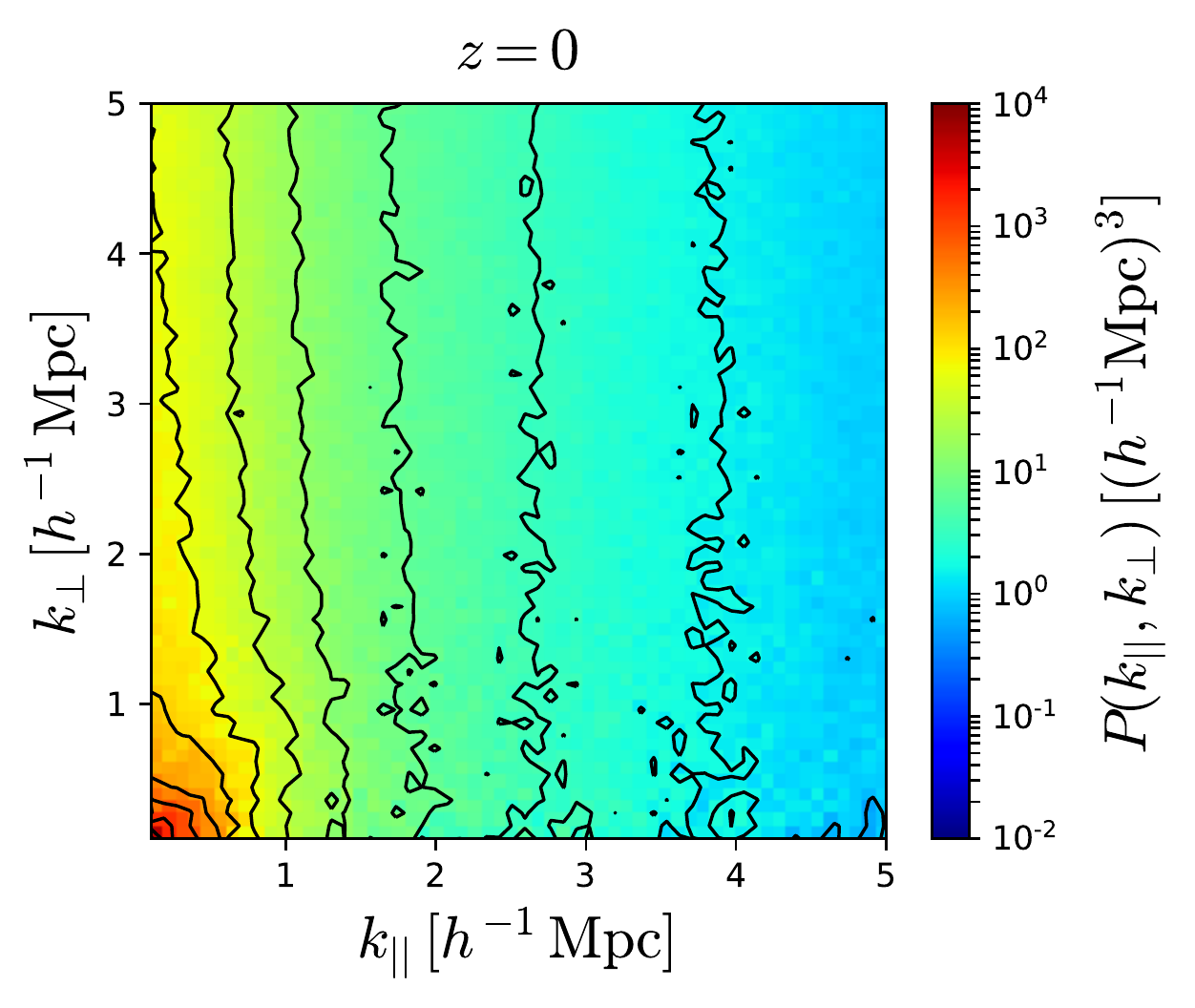}
\includegraphics[width=0.33\textwidth]{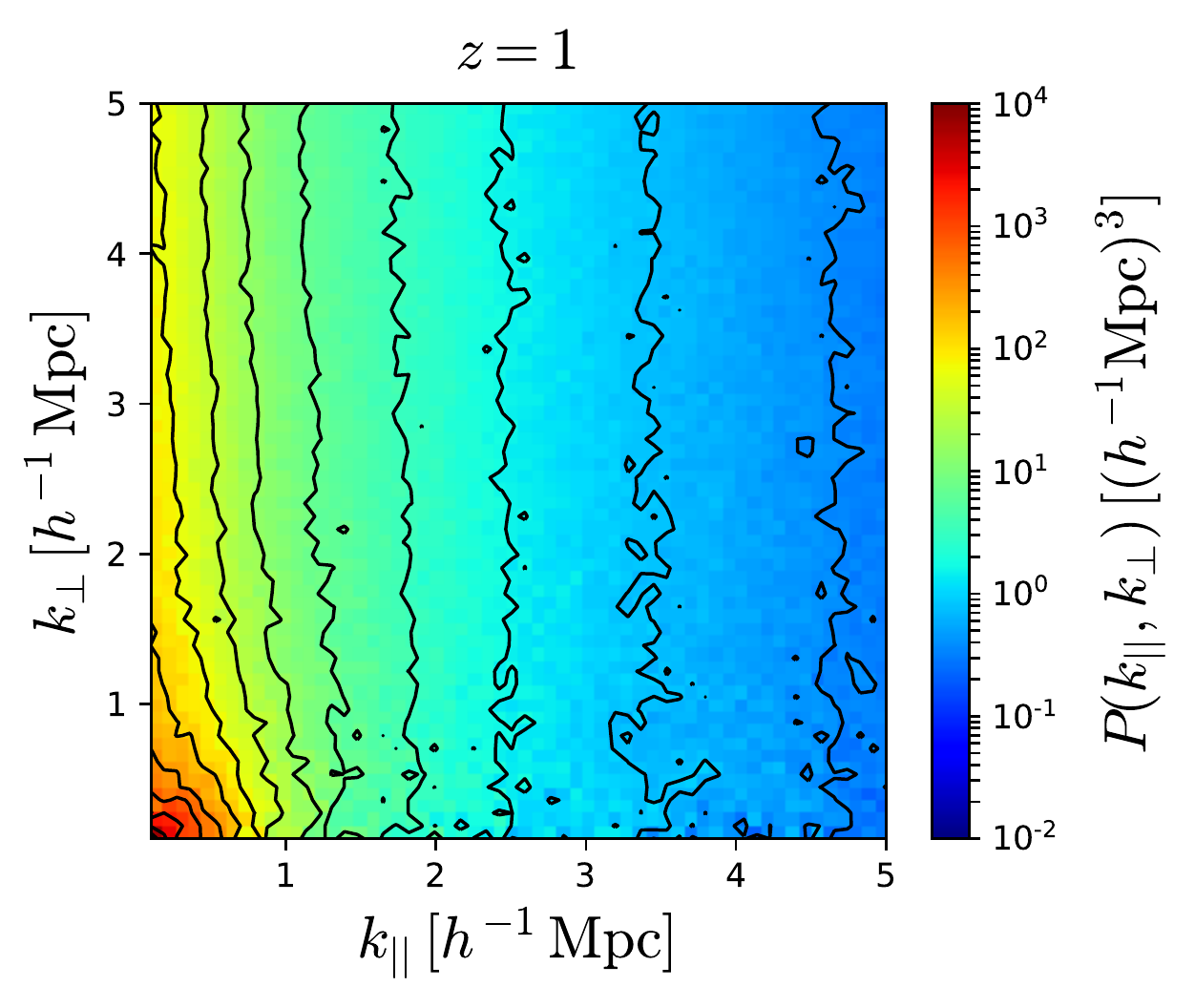}
\includegraphics[width=0.33\textwidth]{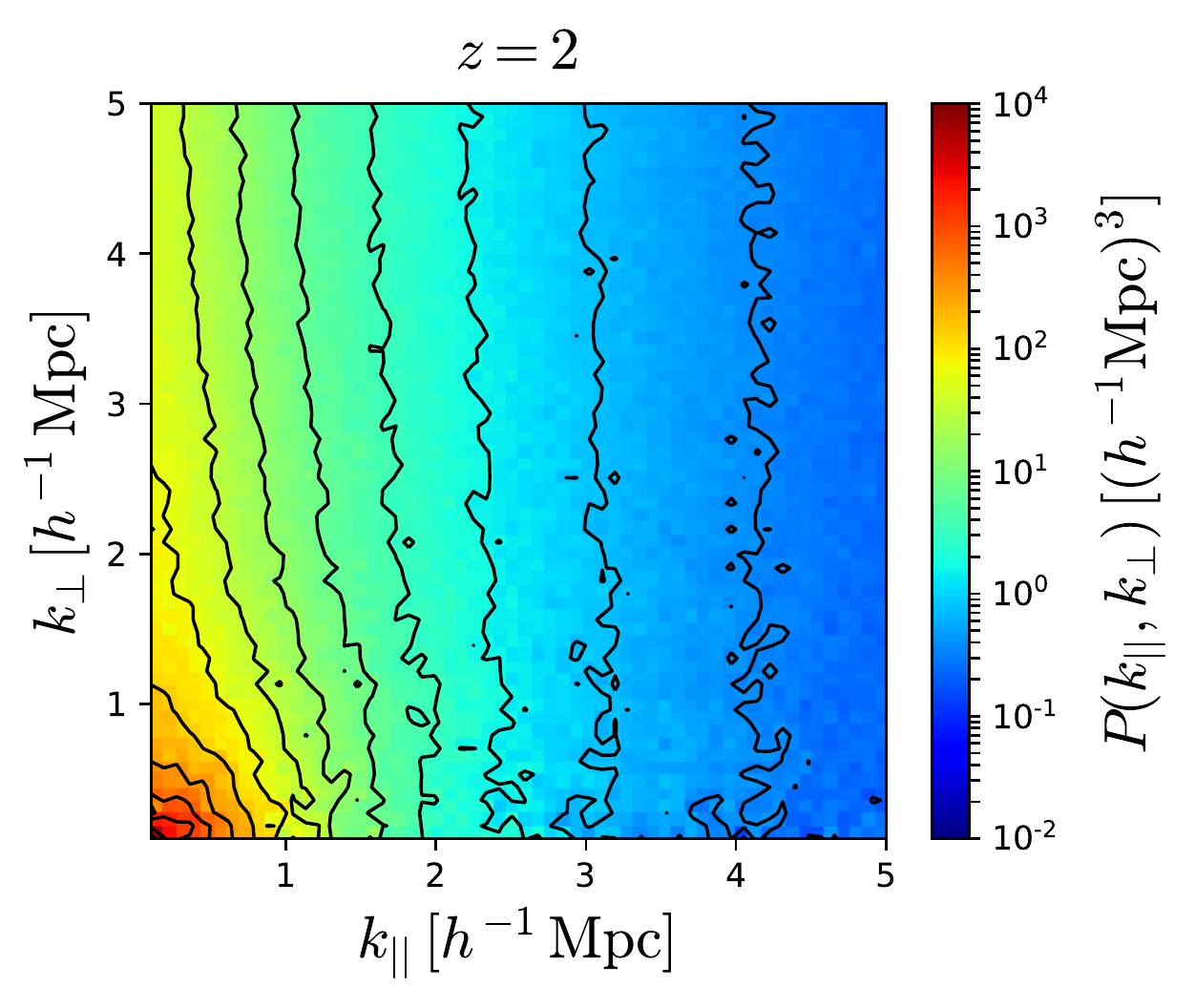}\\
\includegraphics[width=0.33\textwidth]{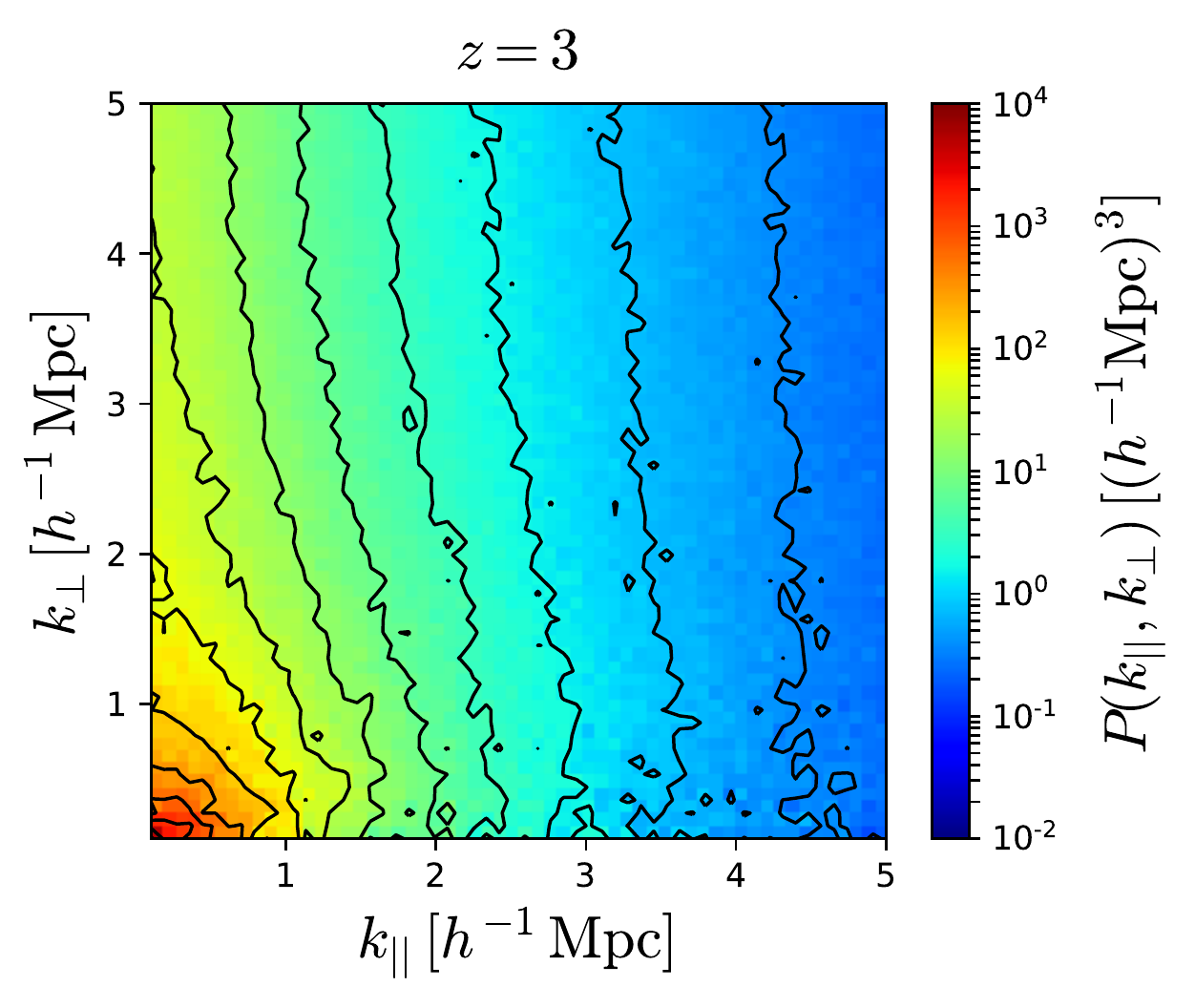}
\includegraphics[width=0.33\textwidth]{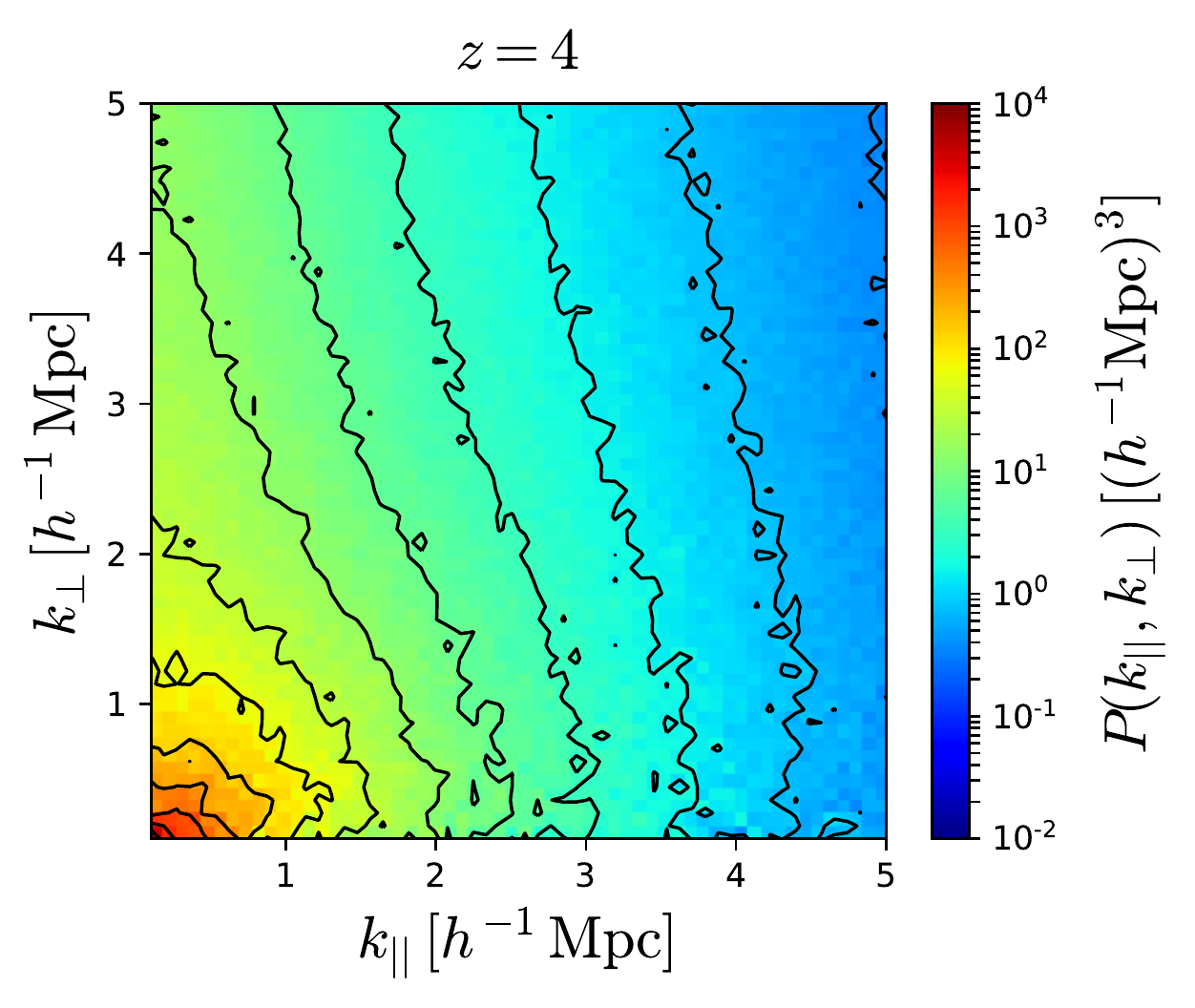}
\includegraphics[width=0.33\textwidth]{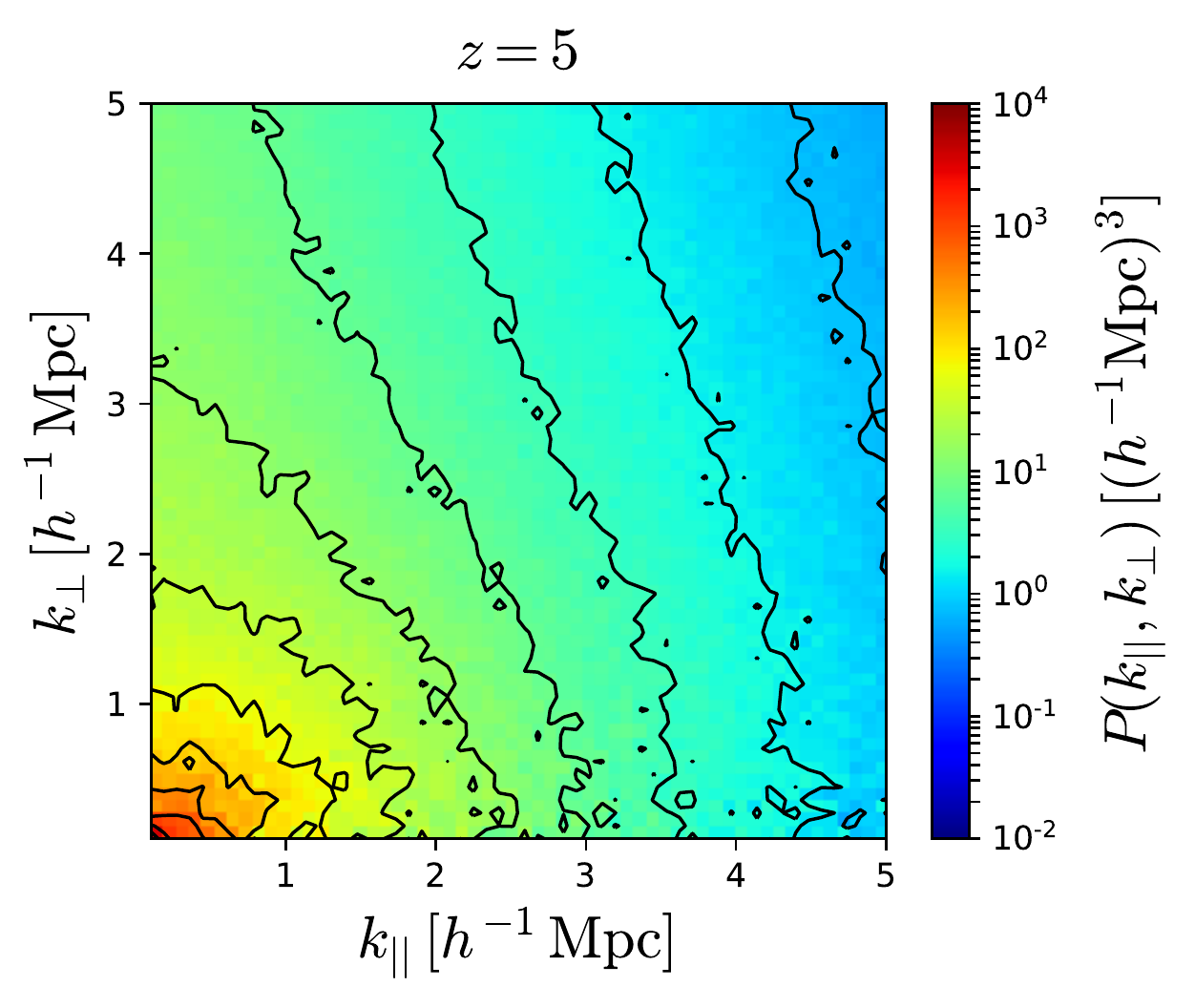}\\[7ex]

\includegraphics[width=0.33\textwidth]{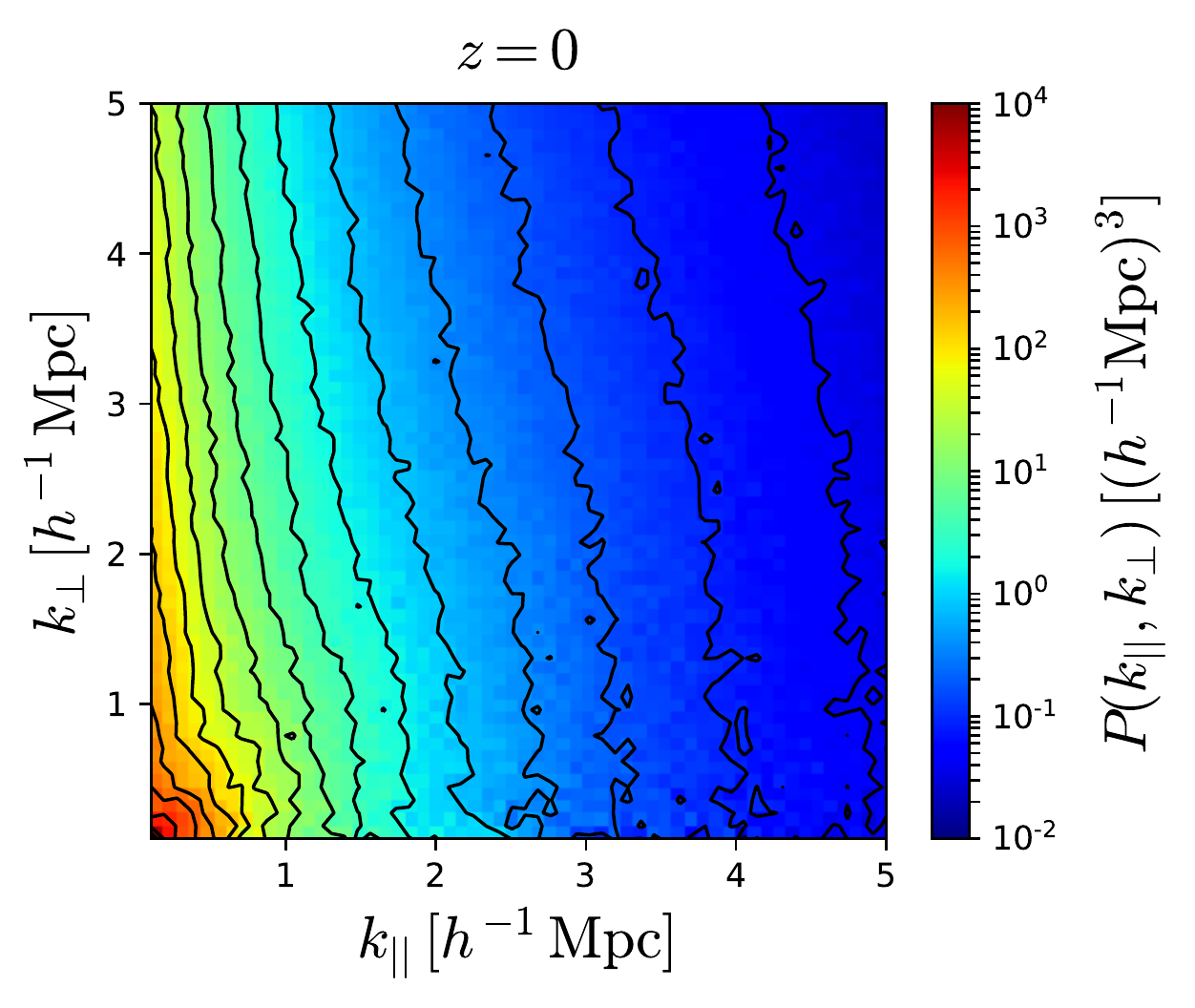}
\includegraphics[width=0.33\textwidth]{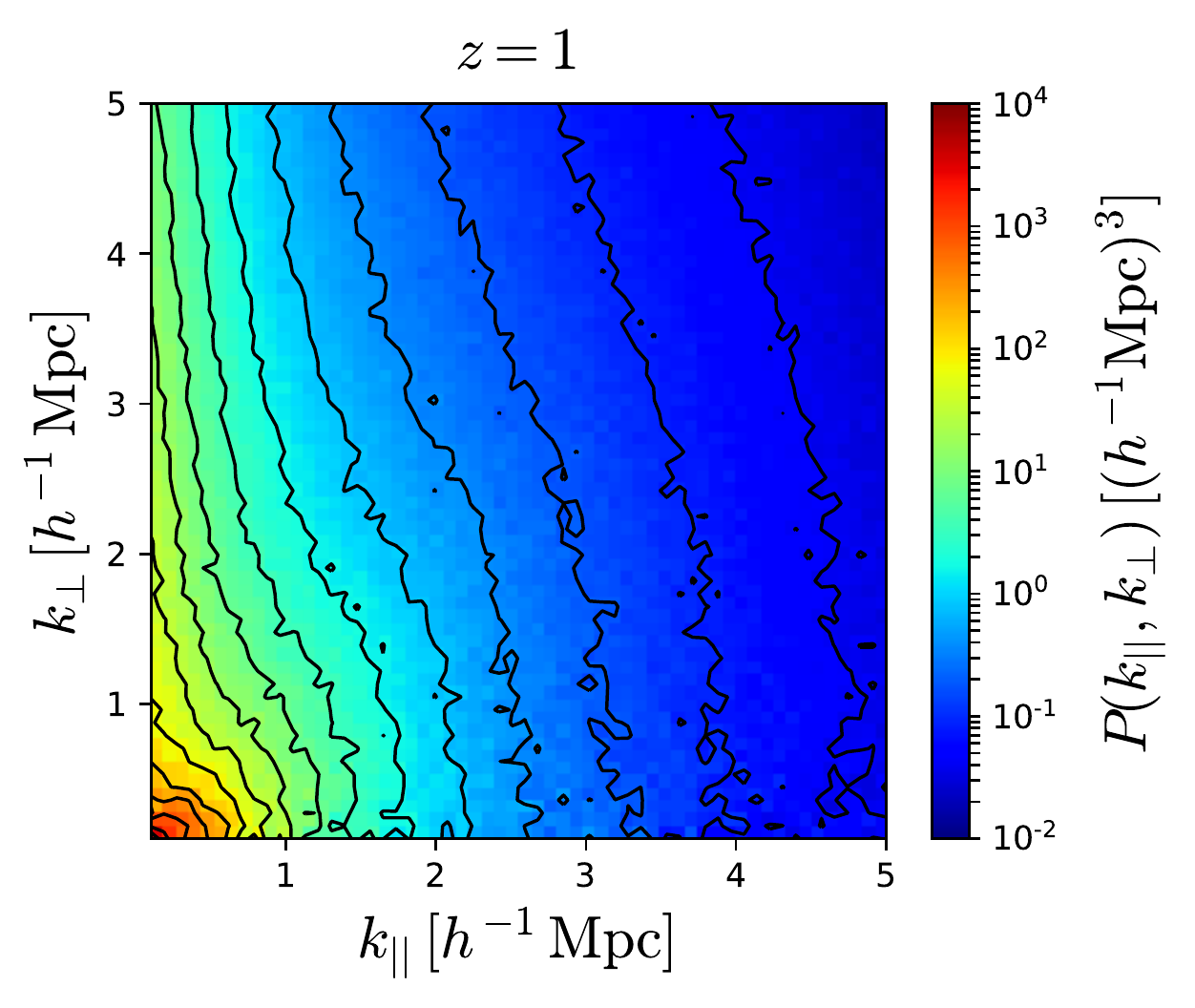}
\includegraphics[width=0.33\textwidth]{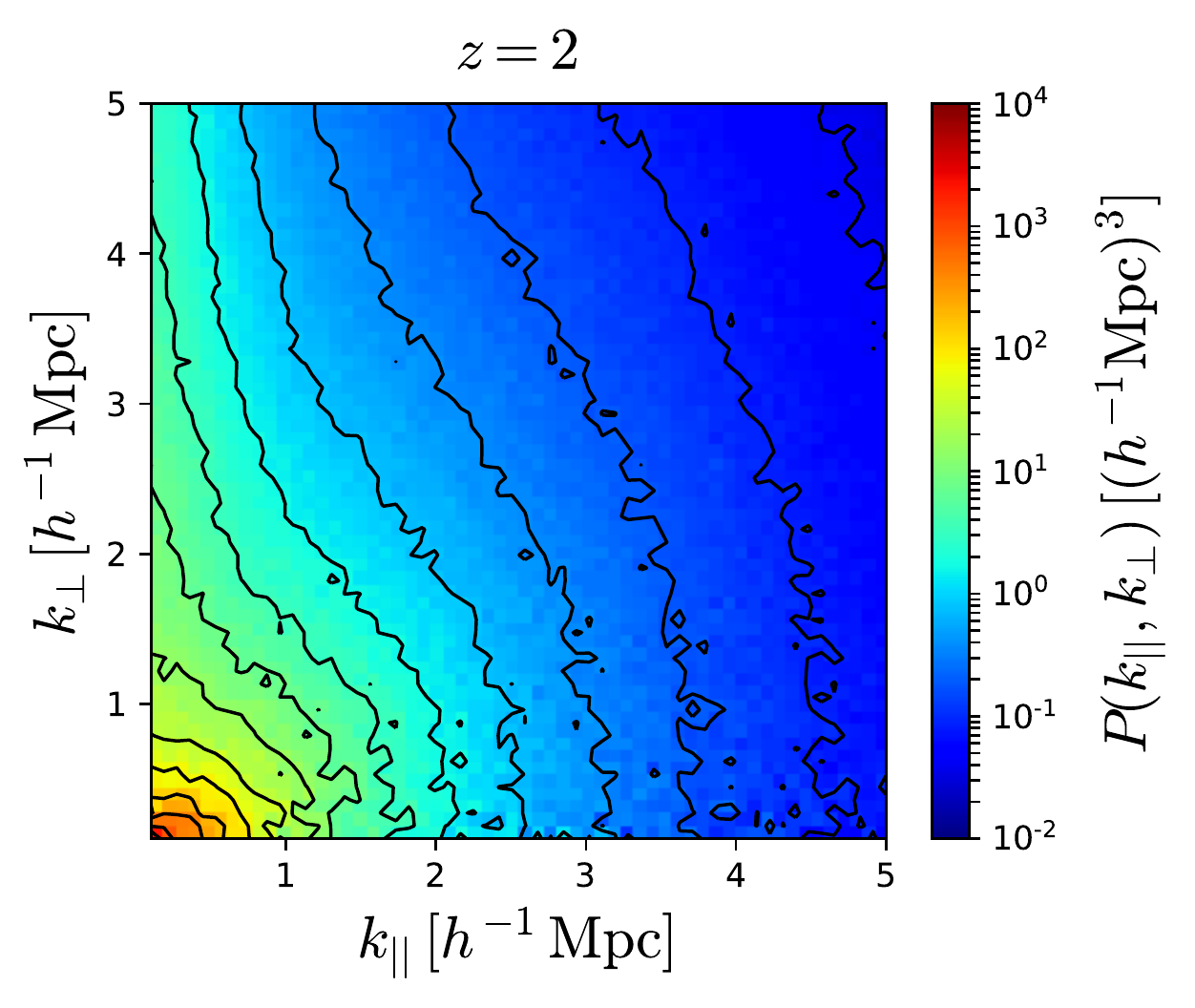}\\
\includegraphics[width=0.33\textwidth]{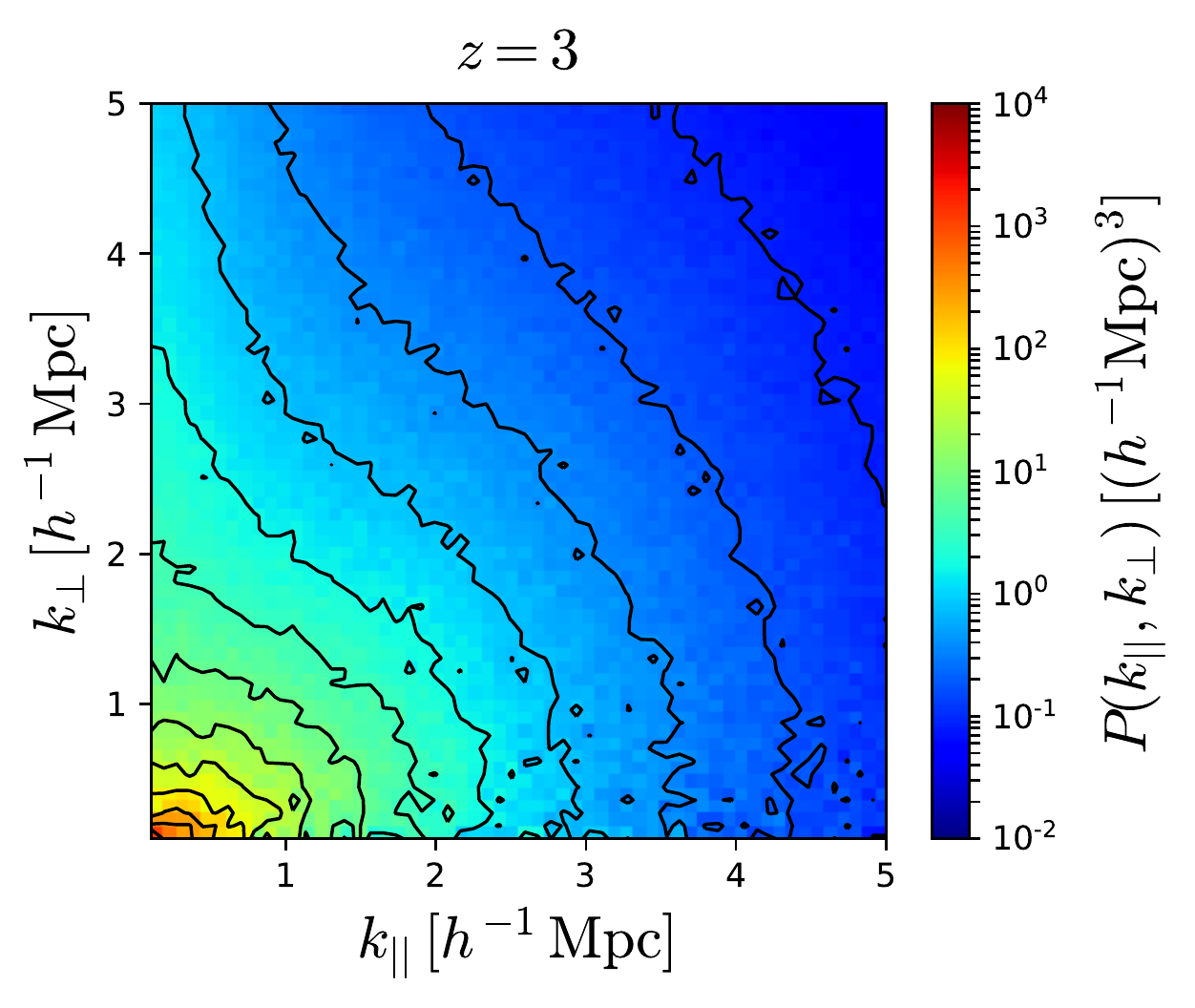}
\includegraphics[width=0.33\textwidth]{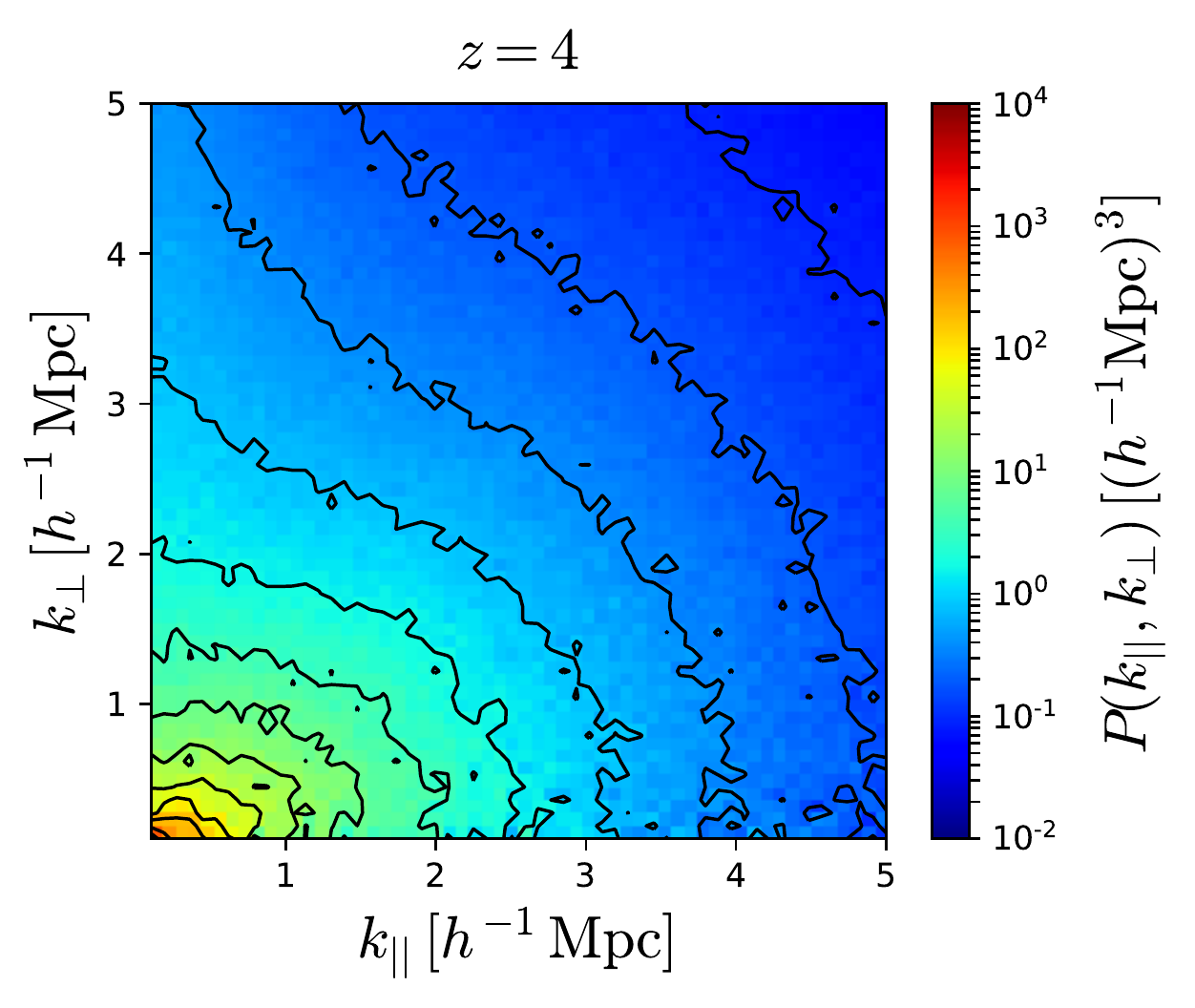}
\includegraphics[width=0.33\textwidth]{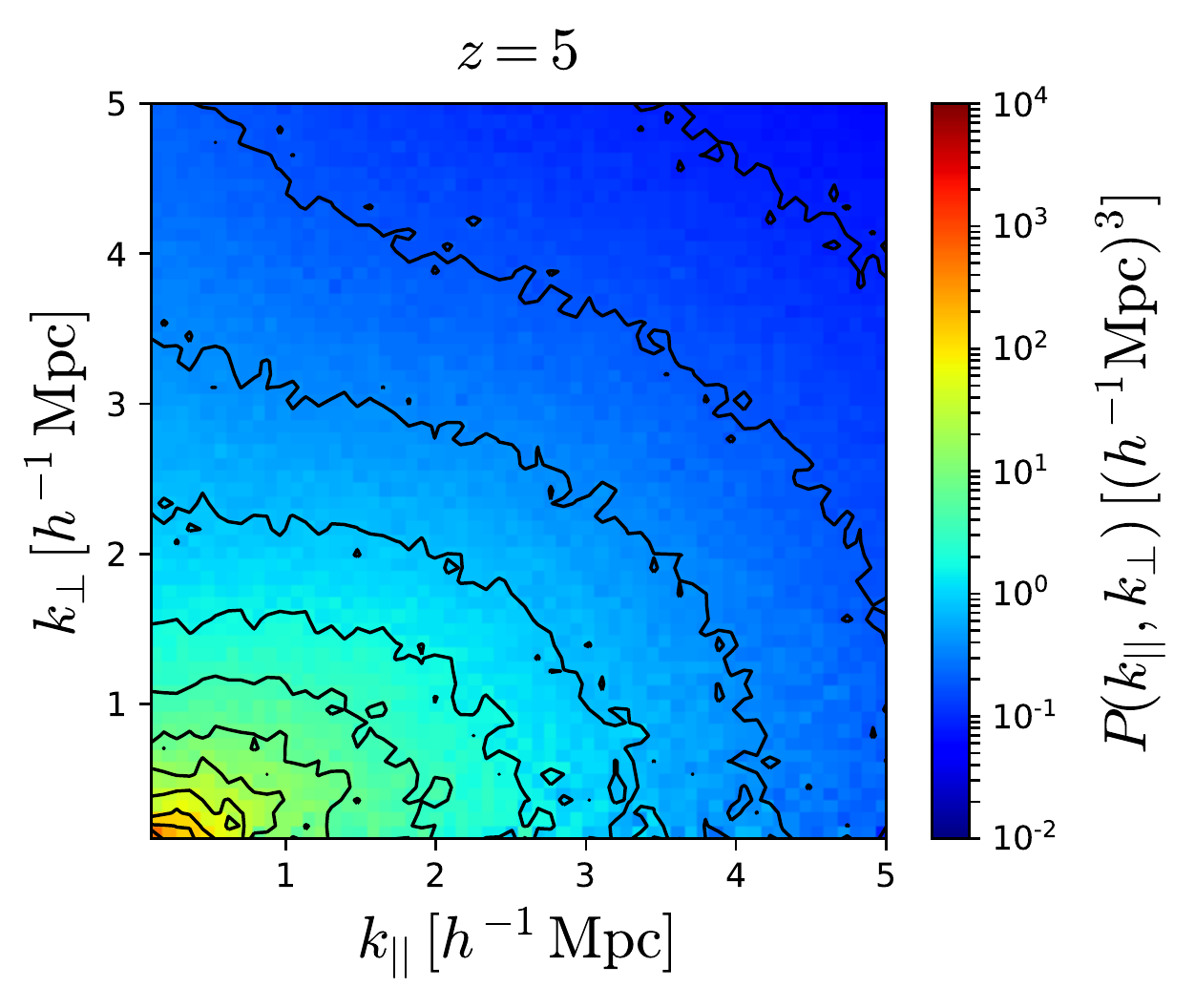}\\
\caption{{\bf Top 2 rows:} 2D power spectrum of HI at redshifts 0 (upper-left), 1 (upper-middle), 2 (upper-right), 3 (bottom-left), 4 (bottom-middle) and 5 (bottom-right). Even though the small volume of our simulation make the 2D power spectrum noisy on large-scales, the Kaiser effect can be seen, particularly at high-redshifts, as isopower contours squeezing in the $k_\bot$ direction. On small scales the 2D power spectrum is dominated by Fingers-of-God, whose amplitude is larger than in the matter field and is more important at low-redshift. {\bf Bottom 2 rows:} same as above but for matter instead of HI.}
\label{fig:HI_2D_1}
\end{center}
\end{figure*}

As an alternate way to visualize the consequences of redshift-space
distortions, we show the 2-dimensional power spectrum of cosmic HI in
the top 2 rows of Fig. \ref{fig:HI_2D_1}, at redshifts from 0 to 5. The Kaiser effect
manifests itself as a squeezing of isopower contours along the
perpendicular direction on large scales. It is apparent down to
relatively small scales at high-redshift, as expected.

On small scales, Fingers-of-God arise as isopower contours propagating
further in the perpendicular than the radial direction. At low
redshift, and on small scales, we find that isopower contours exhibit
a very weak dependence on the perpendicular direction, i.e.~$P_{\rm
  HI}^s(k_{||},k_\bot)\simeq P_{\rm HI}^s(k_{||})$.

We show the results for the 2D matter power spectrum in the bottom rows of
Fig. \ref{fig:HI_2D_1}. It can be seen how: 1) the Kaiser effect is
visible down to smaller scales than in the HI field at high-redshift,
and 2) the magnitude of the Fingers-of-God is lower in the matter
field than in the HI field.

The halo model can be used to understand why the magnitude of
Fingers-of-God is higher in HI than in matter. Each dark matter halo
will show up in redshift-space not as a sphere, but an ellipsoid, due
to internal peculiar velocities. The eccentricity of those ellipsoids
will depend primarily on halo mass: the velocity dispersion in large
halos will be higher than in small halos, so their eccentricity will
be larger. The amplitude of the matter power spectrum on small scales
will be dominated by small halos, whose velocity dispersions are not
large.

On the other hand, we know for HI that since there is a cutoff in the
halo HI mass function, halos below a certain mass will not contribute
to the HI power spectrum. Thus, it is expected that the amplitude of
the Fingers-of-God will be larger in the HI field than in the matter
field because: 1) the velocity dispersions of HI and matter/CDM are
similar (see section \ref{sec:sigma_HI}), and 2) in the HI field we do
not have the contribution of small halos that dominate the amplitude
of the power on small scales.

In order to corroborate this hypothesis we have taken all halos with
masses above $10^{10}~h^{-1}M_\odot$ and we have computed the power
spectrum of the matter inside them. In that case, we observe very
similar results as those from the HI; i.e. the amplitude of the
Fingers-of-God of that particle distribution is much larger than the
one for all matter.

The impact of redshift-space distortions on the HI power spectrum has
recently been studied in \cite{Sarkar_2018} using N-body
simulations. Following their work we have tried to model the 2D HI
power spectrum using the following phenomenological expression
\be
P_{\rm HI}^s(k,\mu)=(1+\beta\mu^2)^2P_{\rm HI}^r(k)\frac{1}{\left(1+\frac{1}{2}k^2\mu^2\sigma_p^2\right)^2}~,
\ee
where $\beta=f/b_{\rm HI}$, $P_{\rm HI}^r(k)$ is the fully non-linear
HI power spectrum and $\sigma_p$ is a phenomenological parameter that
accounts for the Fingers-of-God effect. The reason for using this
expression is that even if the HI bias is non-linear,
we will recover the Kaiser factor
when computing the monopole ratio (see Fig. \ref{fig:HI_RS}).
at high-redshift, where the amplitude of
the Fingers-of-God is small.

By fitting the above expression to our 2D HI power spectrum down to
$k=1~h{\rm Mpc}^{-1}$ and assuming Gaussian errors, we find that this
approach works reasonably well $\chi^2/{\rm d.o.f}\simeq[1.5-2.2]$,
with $\sigma_p={1.73, 2.09, 1.37, 0.93, 0.34, 0}$ at redshifts 0, 1,
2, 3, 4 and 5, respectively. However, by computing the monopole from
the above expression
\be
P_{\rm HI}^0(k)=\frac{1}{2}\int_{-1}^1P_{\rm HI}^s(k,\mu)d\mu
\ee
and comparing with our measurements we do not find good agreement. 

We have also repeated the above exercise for the models considered in
\cite{Sarkar_2018}, concluding that none of those represent a good
description of our results. We leave a more detailed
analysis of this issue for future work.

\section{21cm maps}
\label{sec:21cm}

The HI properties studied in this paper can be used to generate mock
21 cm maps. In this section we study: 1) whether less computationally
expensive simulations can be used to create 21 cm maps, and 2) the
importance of accounting for the 1-halo term when making mocks.

\begin{figure*}
\begin{center}
\includegraphics[width=0.39\textwidth]{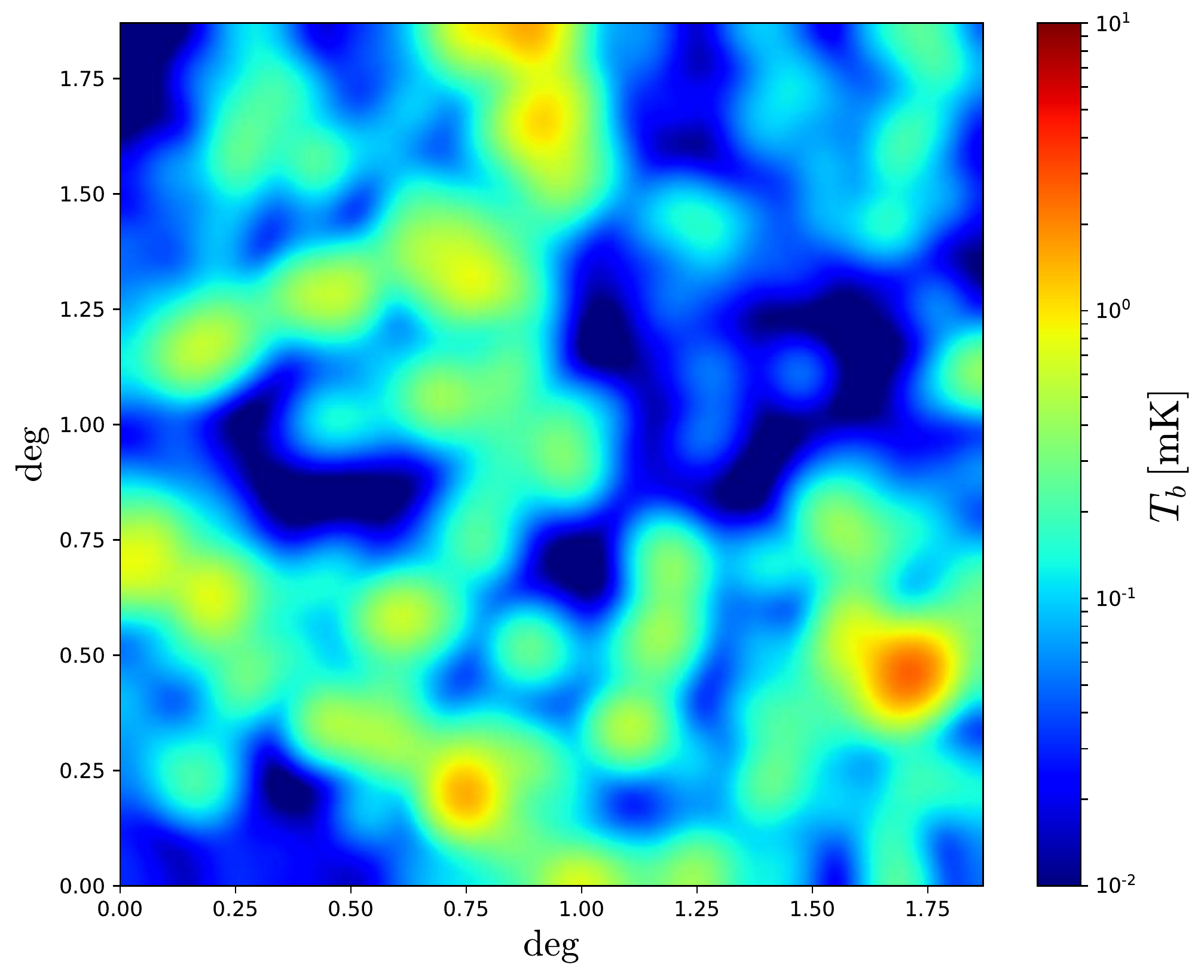}
\includegraphics[width=0.39\textwidth]{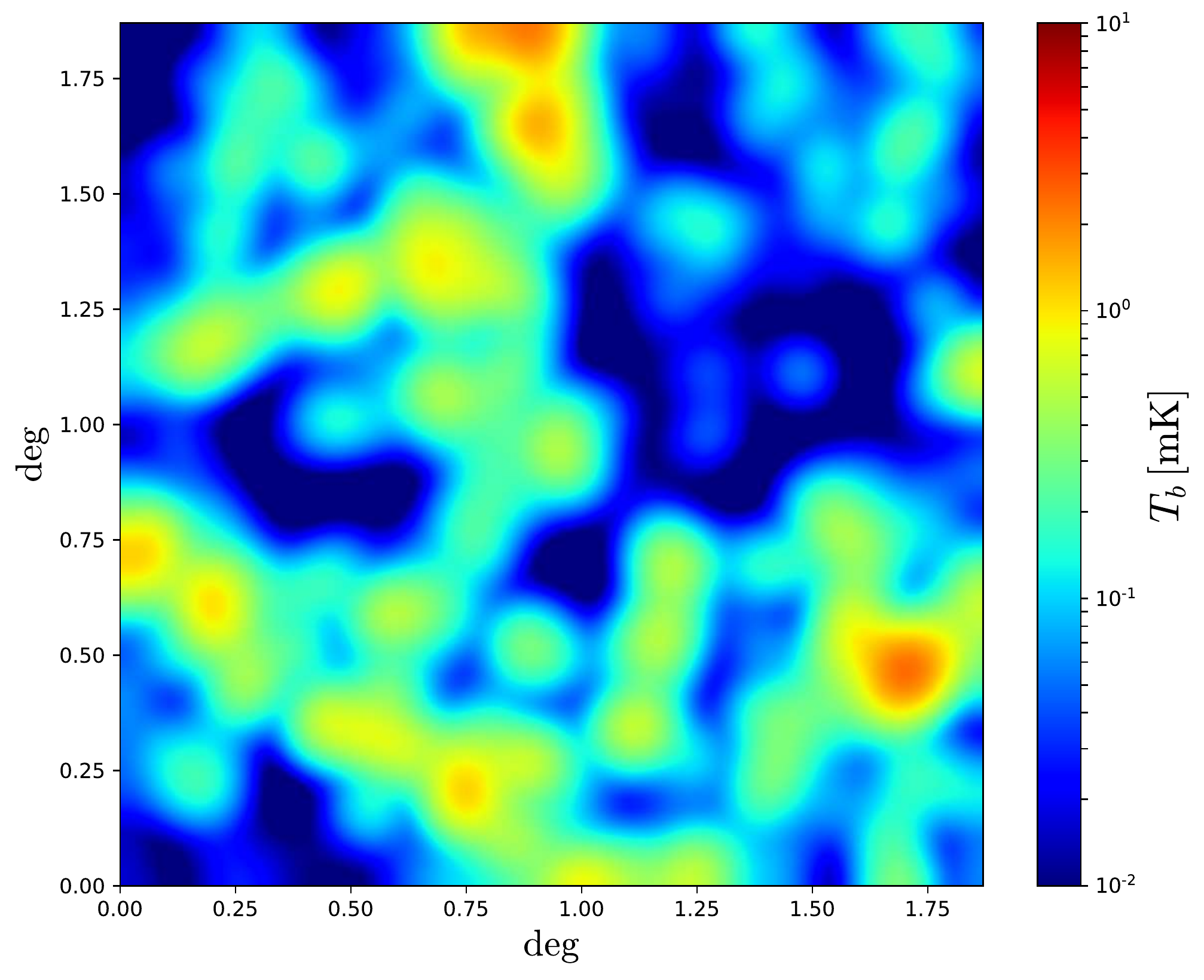}\\
\includegraphics[width=0.39\textwidth]{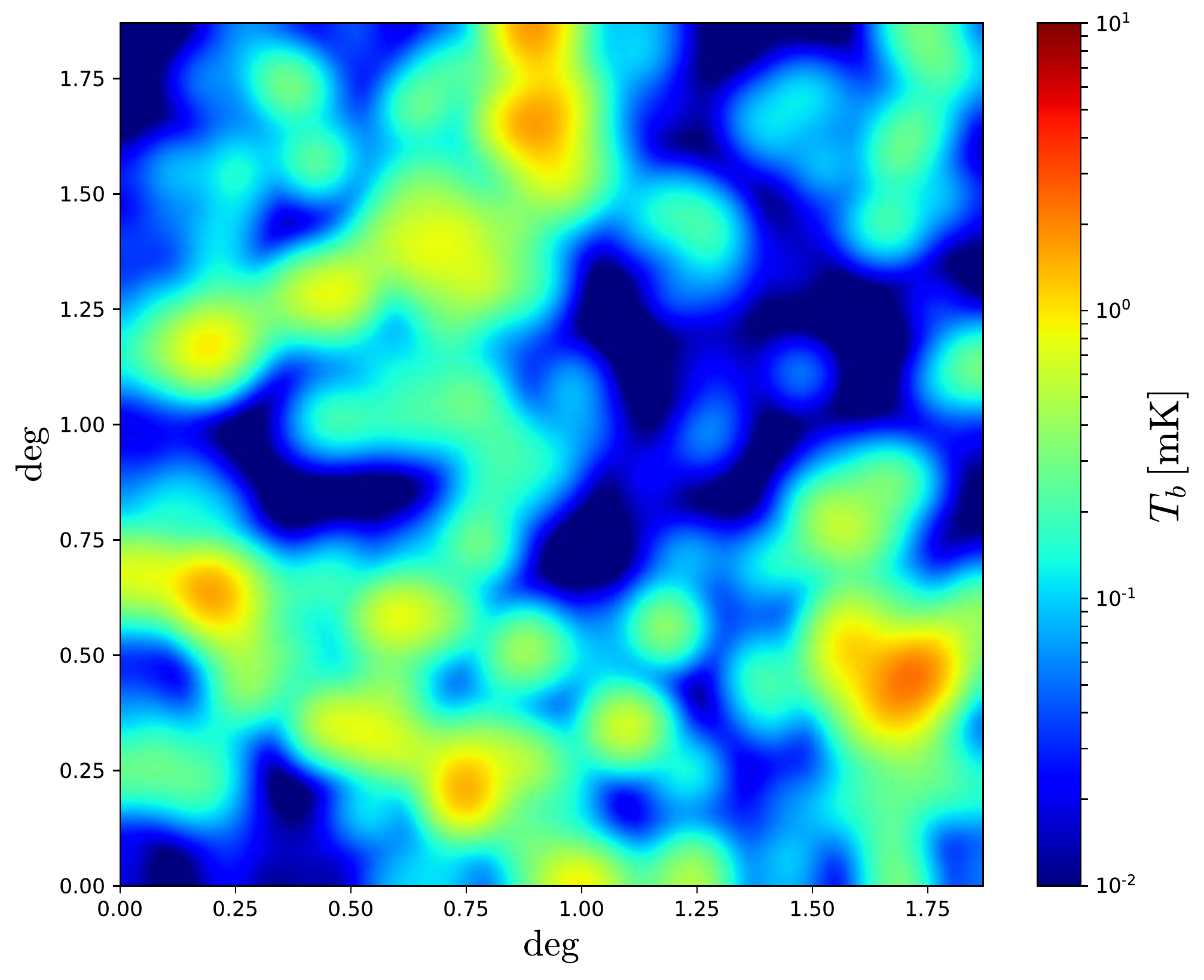}
\includegraphics[width=0.39\textwidth]{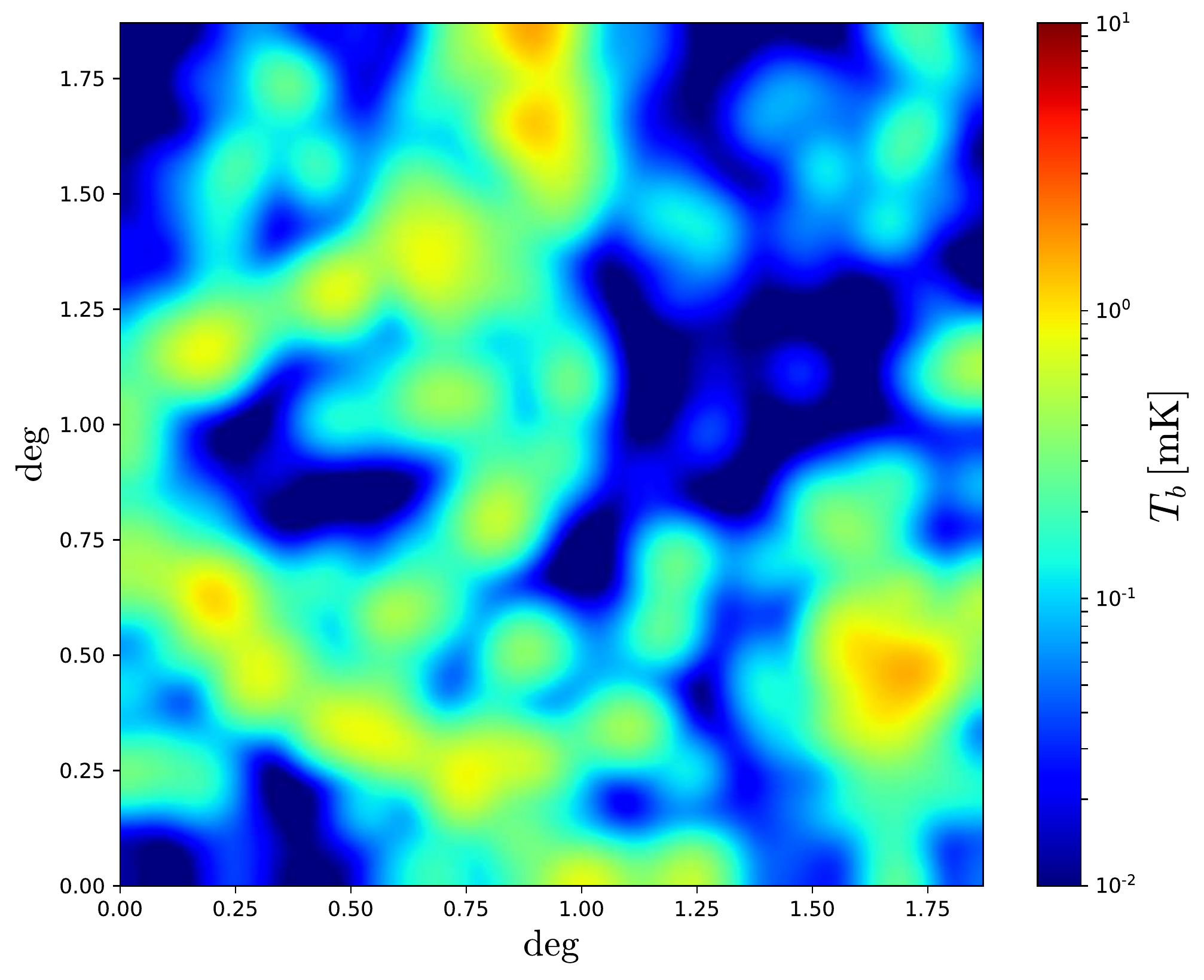}\\
\includegraphics[width=0.39\textwidth]{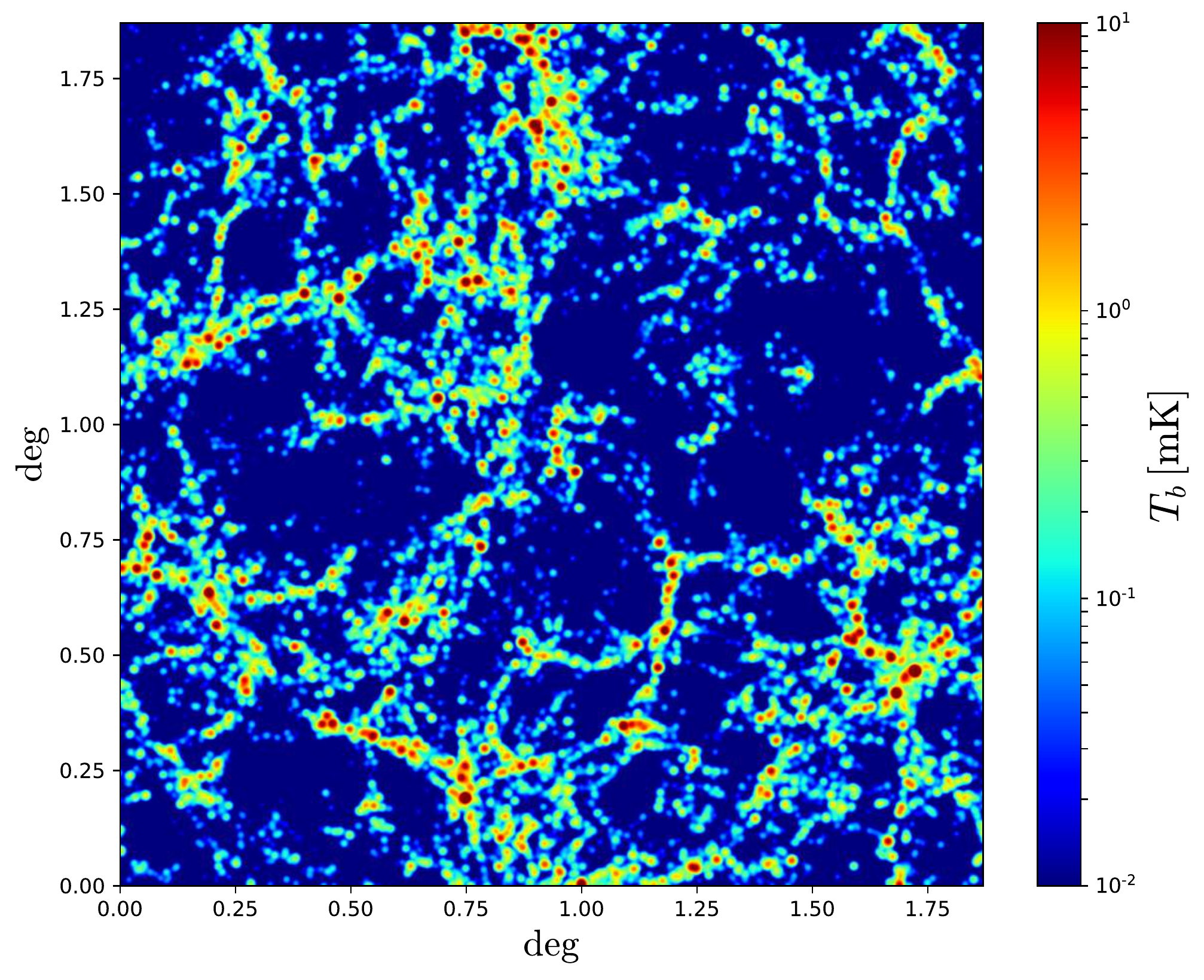}
\includegraphics[width=0.39\textwidth]{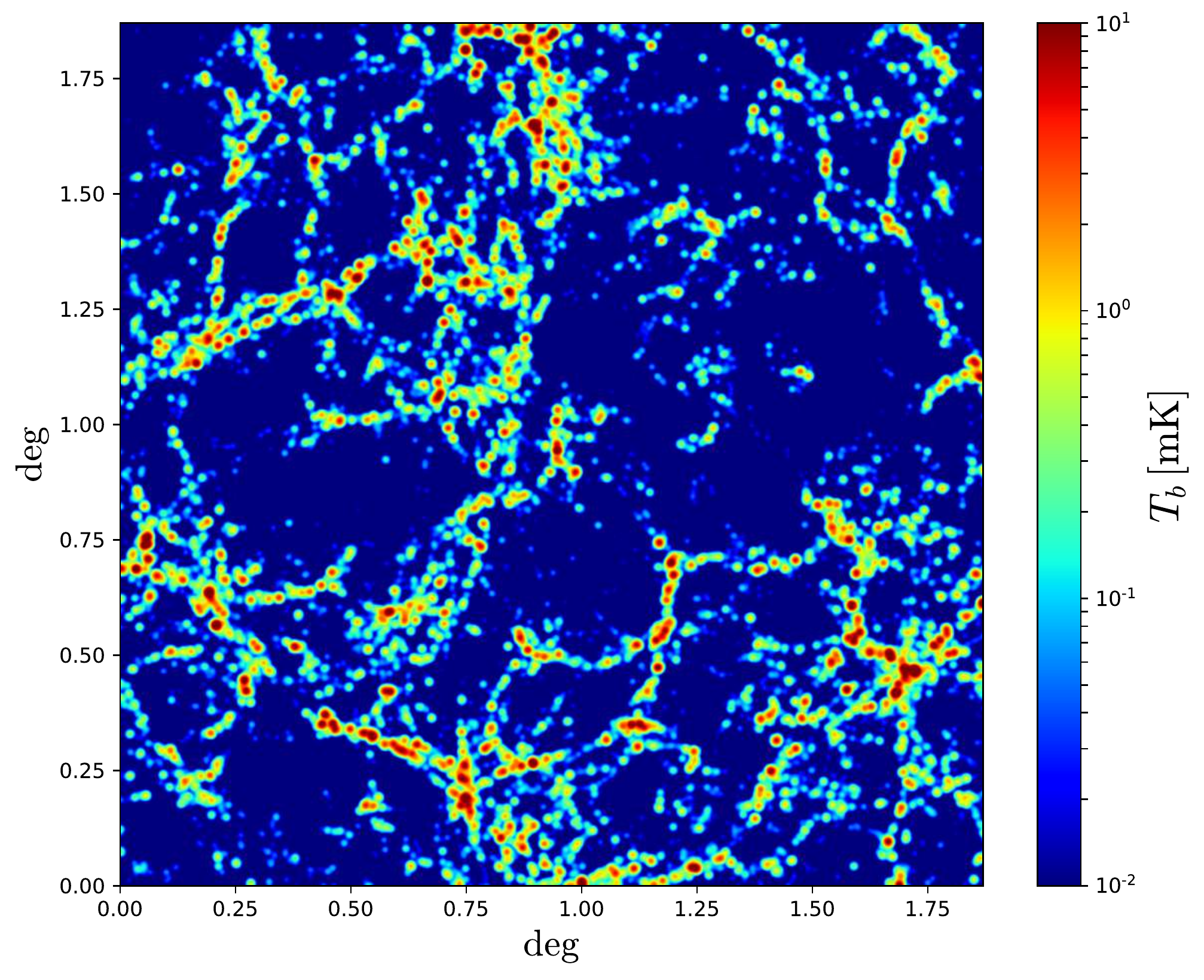}\\
\includegraphics[width=0.39\textwidth]{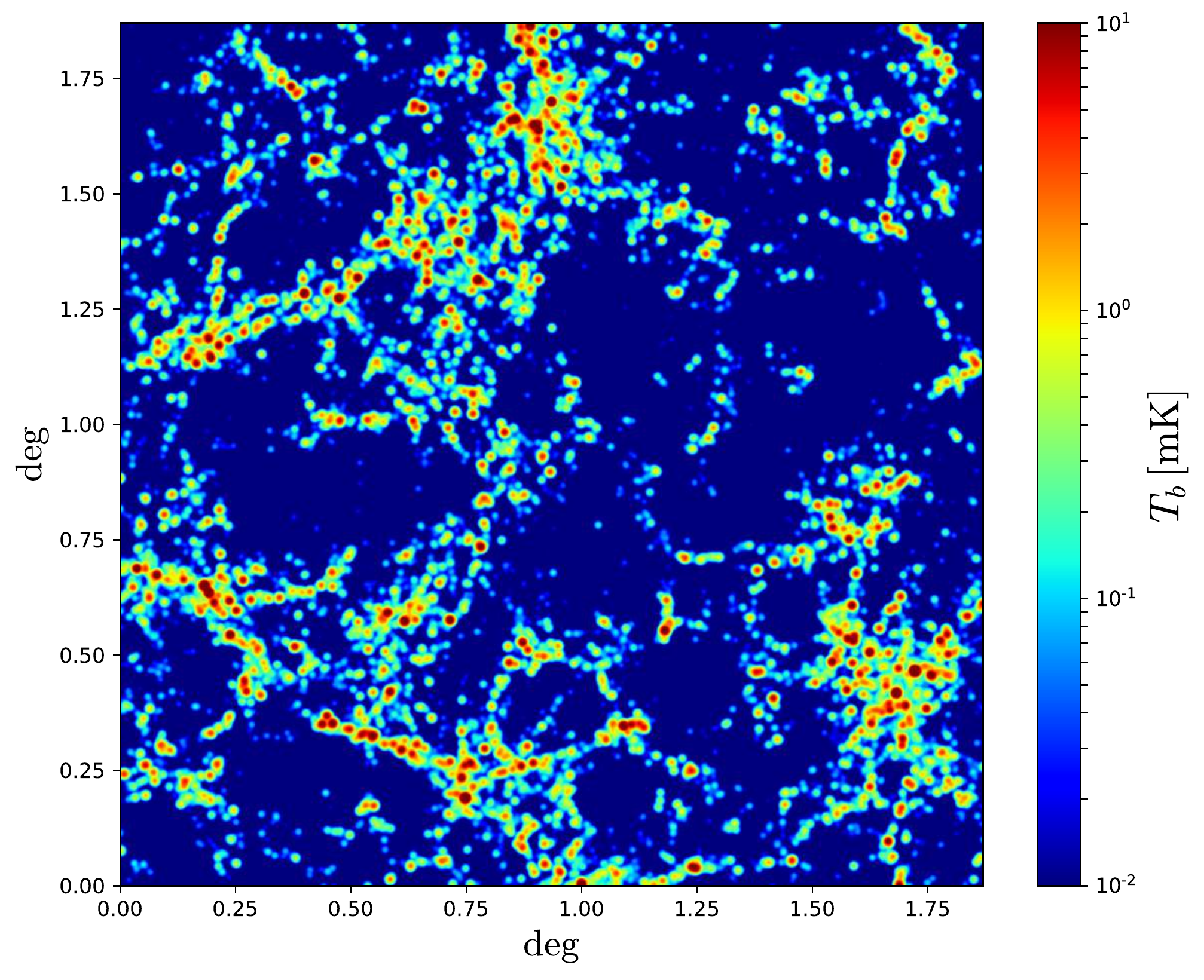}
\includegraphics[width=0.39\textwidth]{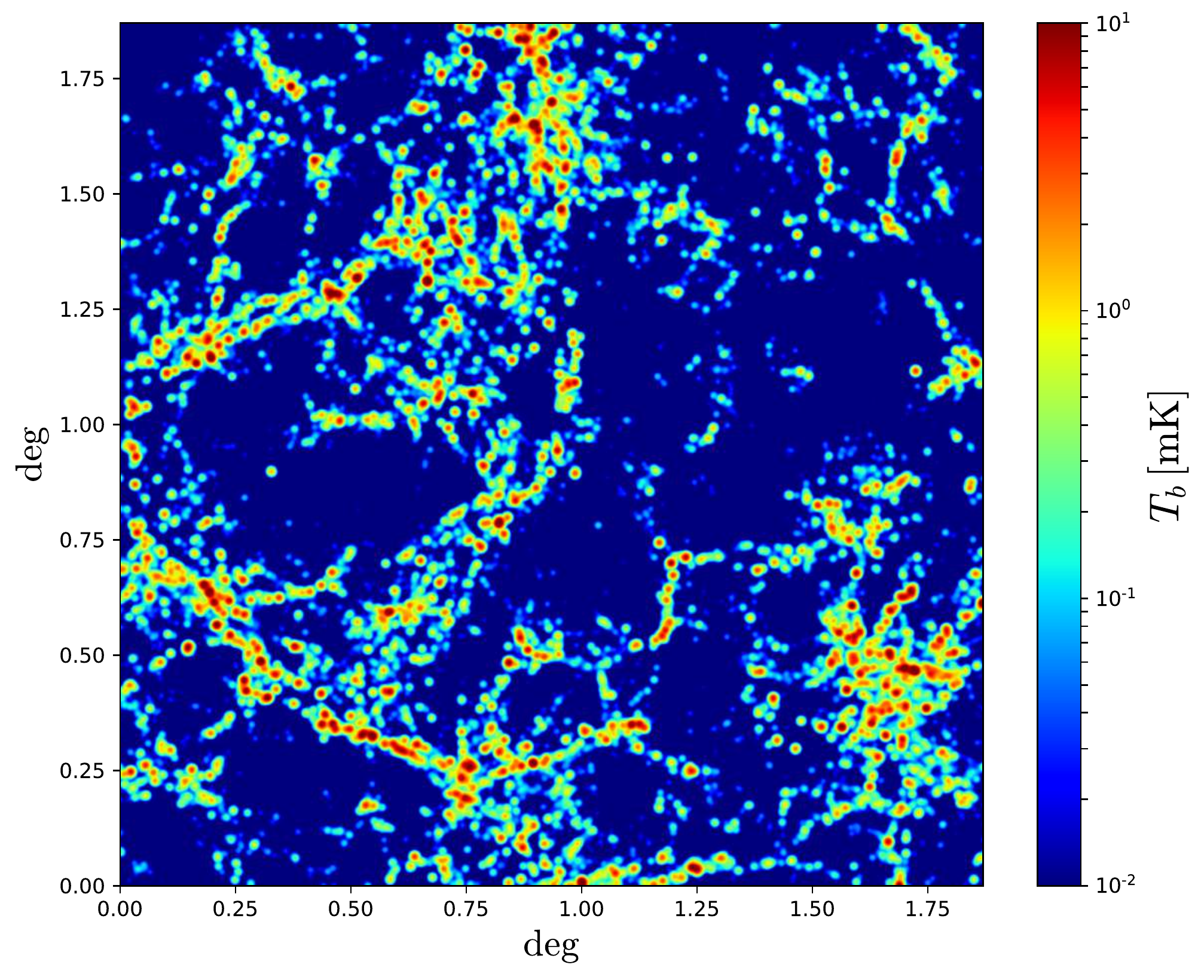}\\
\caption{21 cm maps at $710$ MHz with 1 MHz bandwidth over an area of $\sim4\,{\rm deg}^2$. The upper and bottom pairs of panels show maps with angular resolutions of 3' and 0.3', respectively. Within each pair, the top one was made in real-space and the bottom in redshift-space. The maps on the right column have been generated from the computationally expensive IllustrisTNG simulations, while the maps on the left by painting HI on top of dark matter halos from computationally cheap N-body simulations using the ingredients studied in this paper.}
\label{fig:21cm_maps}
\end{center}
\end{figure*}

The right column of Fig. \ref{fig:21cm_maps} shows 21 cm maps created
from the spatial distribution of HI in the TNG100 simulation. From
top to bottom these maps show: 21 cm map in real-space with 3' angular
resolution (top), 21 cm map in redshift-space with 3' angular
resolution (top-middle), 21 cm map in real-space with 0.3' angular
resolution (bottom-middle), and 21 cm map in redshift-space with 0.3'
angular resolution (bottom). All those maps are centered at a
frequency of 710 MHz ($z=1$) and have a bandwidth of 1 MHz
($\simeq5~h^{-1}$Mpc).

The procedure used to create these maps is as follows. First,
the HI density field is computed by assigning HI masses of gas cells
(either in real- or redshift-space) to a grid of $2048^3$ cells using
the nearest-grid-point (NGP) mass assignment scheme. Then, we select a
slice of the HI density field grid whose width is taken to reproduce
the desired frequency bandwidth. Next, that slice is projected onto a
2-dimensional grid and HI densities are transformed to brightness
temperatures through
\be
T_b(\vec{x}) = 189h\left(\frac{H_0(1.0+z)^2}{H(z)}\right)\frac{\rho_{\rm HI}(\vec{x})}{\rho_{\rm c}}~{\rm mK}~.
\ee
Finally, we convolve that grid with a Gaussian filter of radius
$R=\theta\chi$, where $\theta$ is the desired angular resolution and
$\chi$ is the comoving distance to redshift $z$ (or frequency $\nu$).

The 21 cm maps on the left column of Fig. \ref{fig:21cm_maps} have been
created from an N-body simulation that shares the same initial
conditions as TNG100; i.e. TNG100-1-DM, but whose computational cost
is over an order of magnitude lower. In that simulation, we have
placed an HI mass given by $M_{\rm HI}(M,z)$ from table
\ref{table:M_HI_fit}, on the center of each dark matter halo. Then, we
have followed the procedure outlined above to create the 21 cm maps. To
make 21 cm maps in redshift-space we displace the HI mass that we put
in the center of each halo according to the peculiar velocity of that
halo.

\begin{figure*}
\begin{center}
\includegraphics[width=0.33\textwidth]{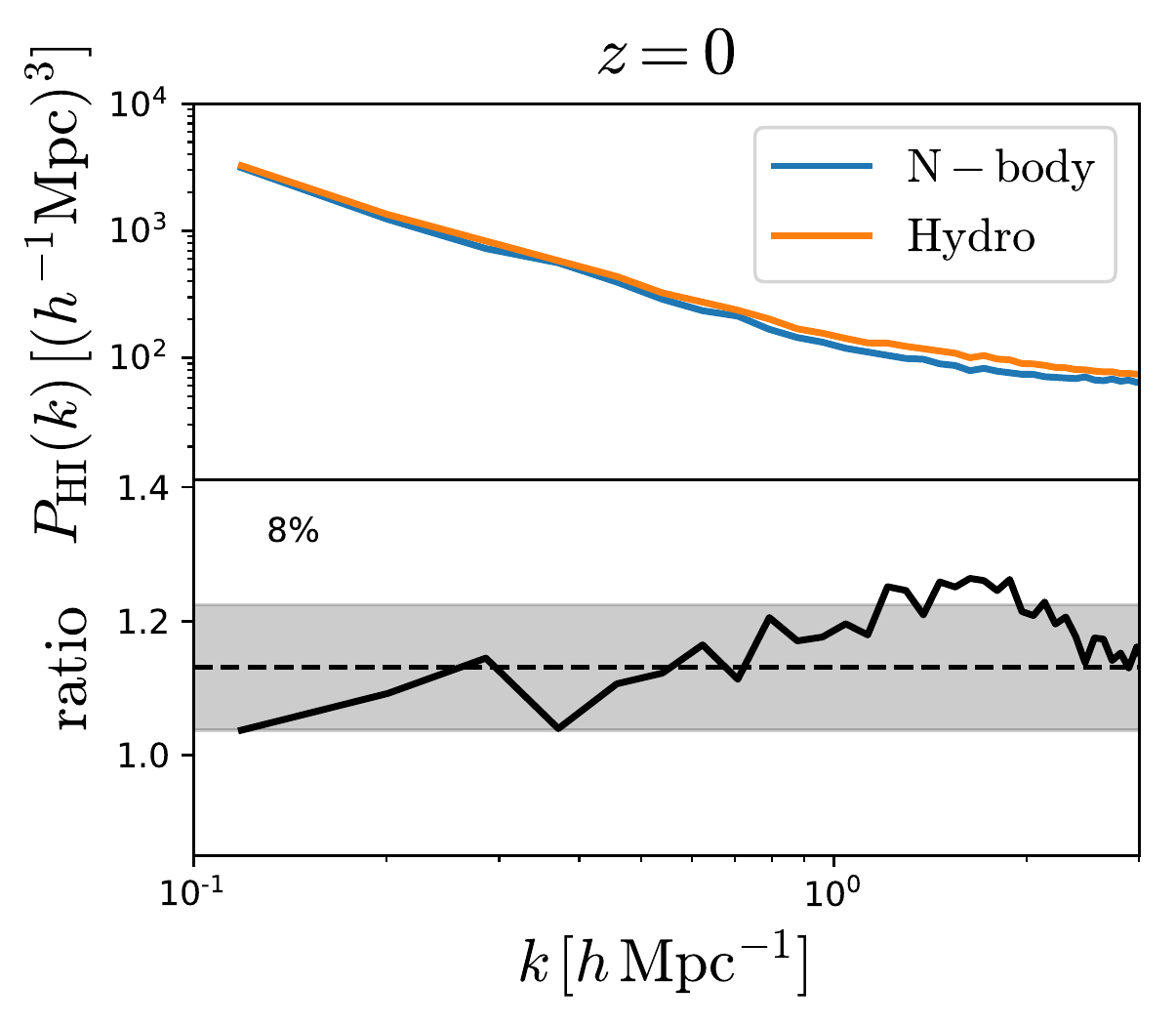}
\includegraphics[width=0.33\textwidth]{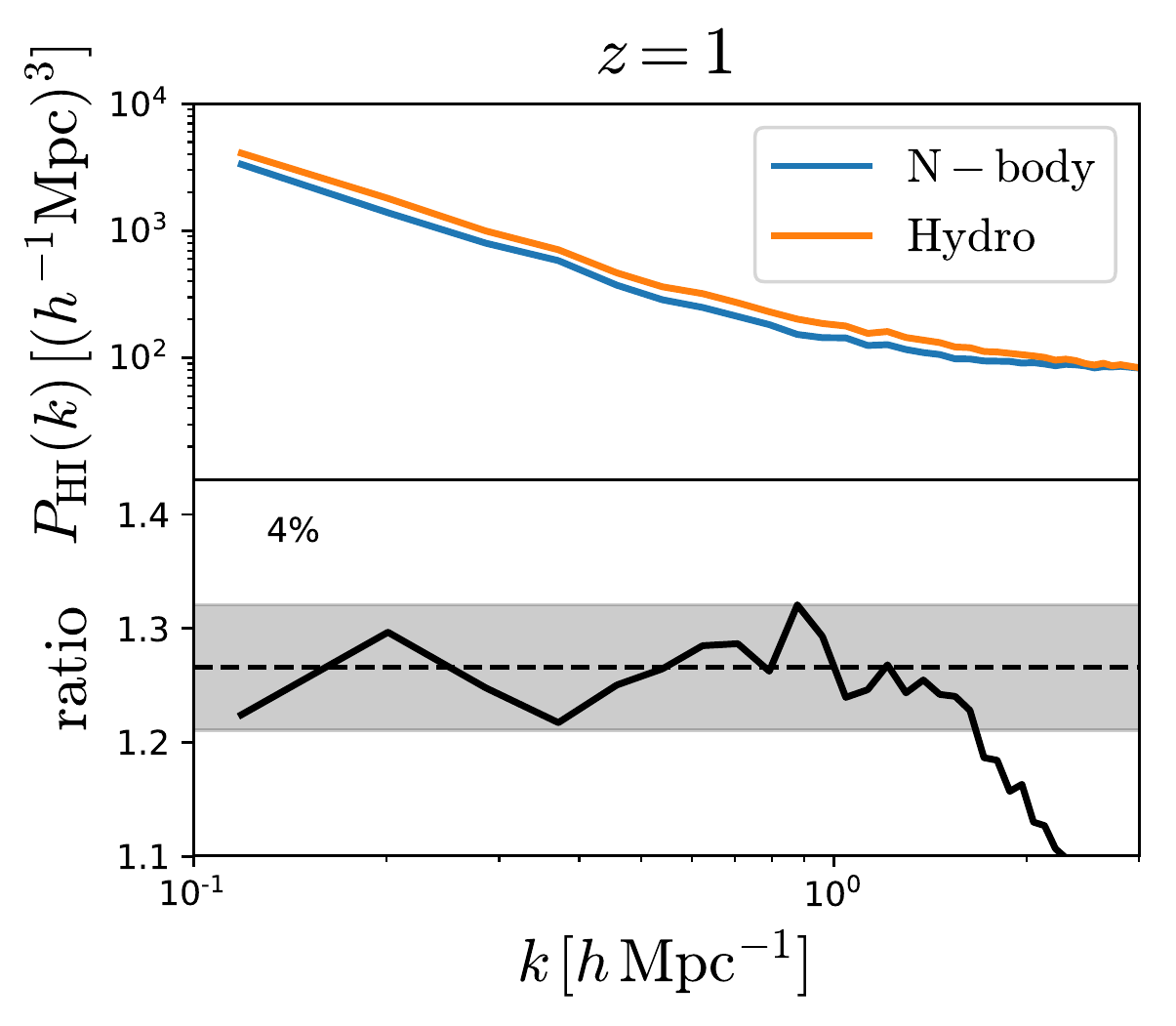}
\includegraphics[width=0.33\textwidth]{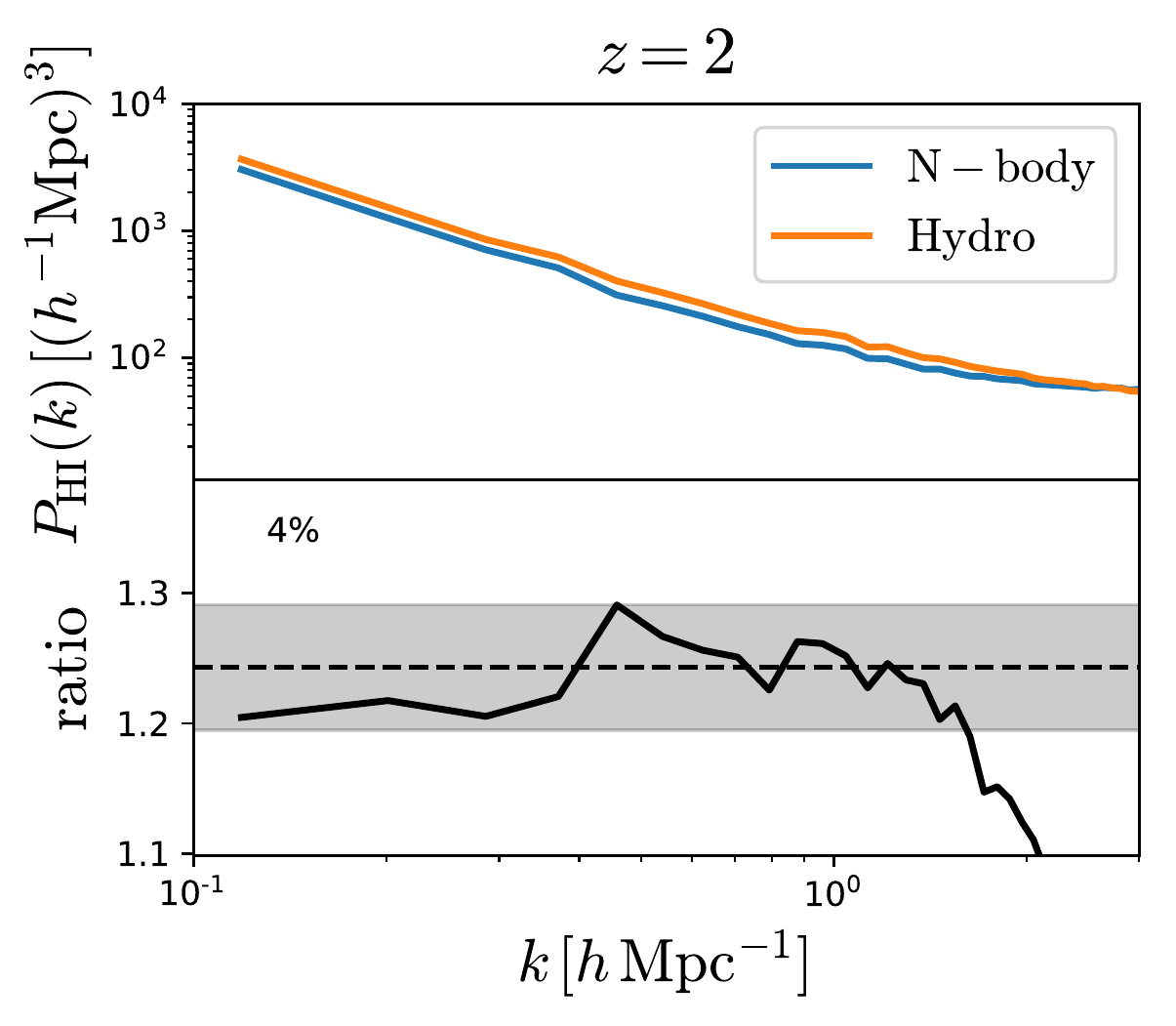}\\
\includegraphics[width=0.33\textwidth]{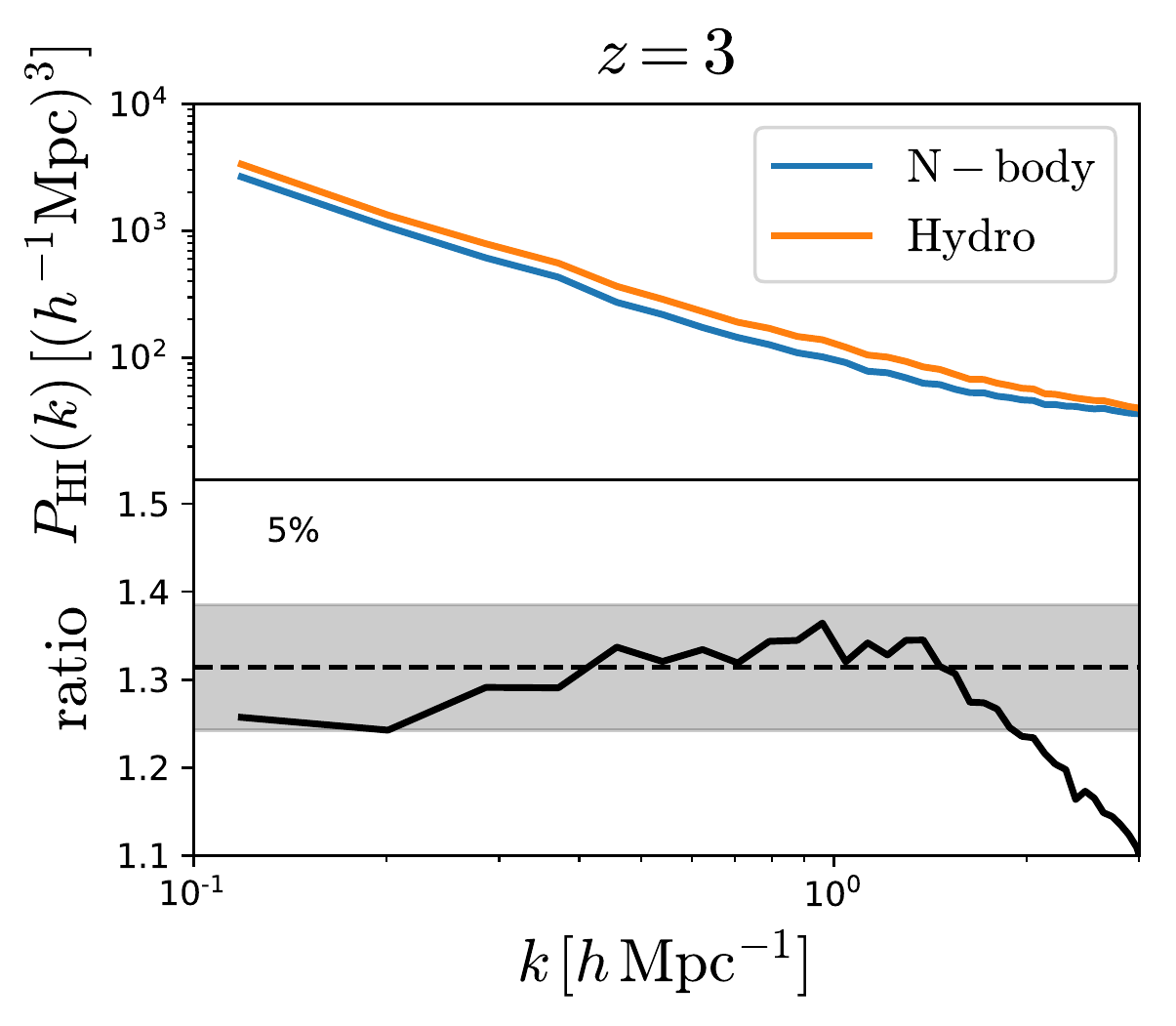}
\includegraphics[width=0.33\textwidth]{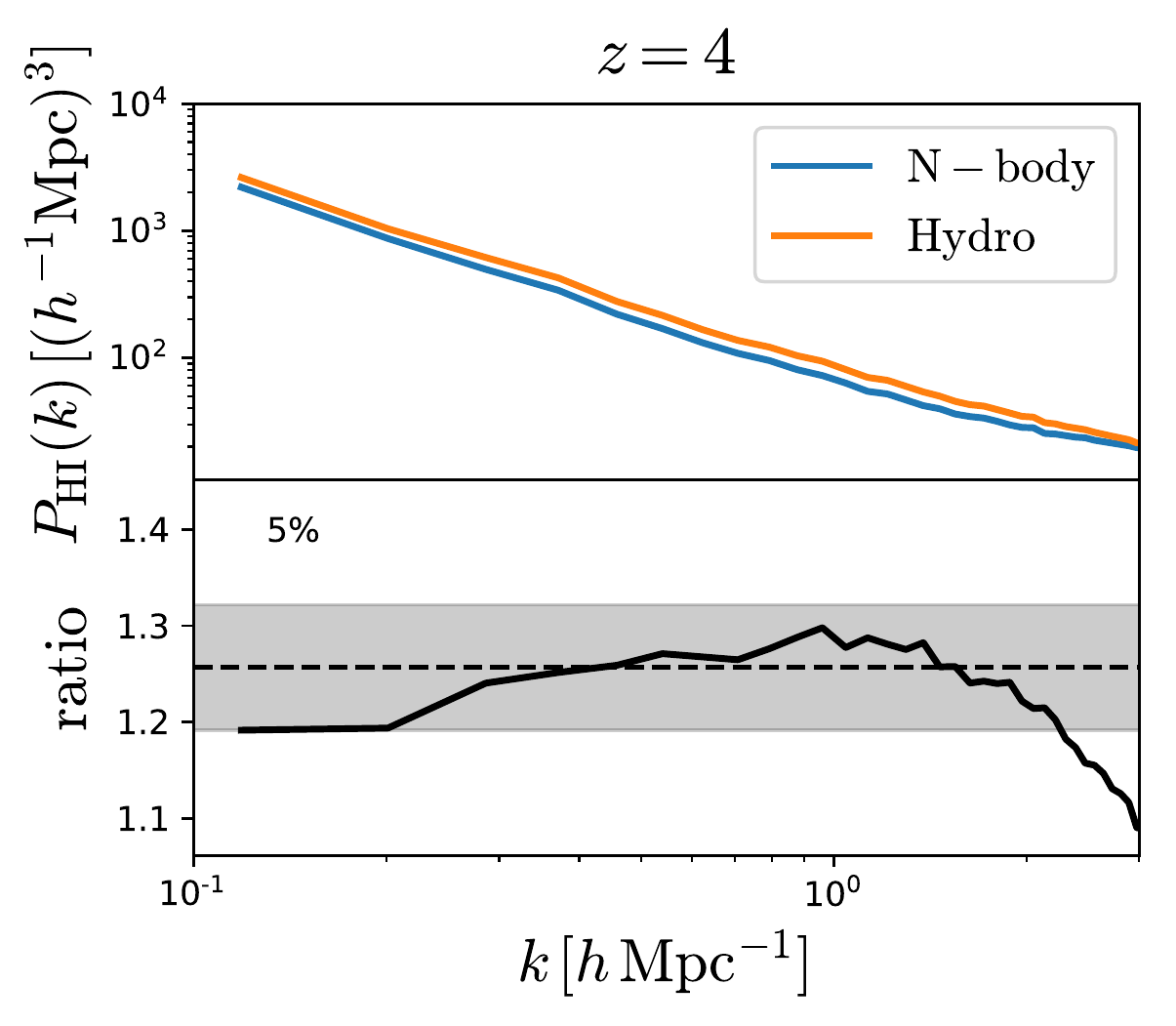}
\includegraphics[width=0.33\textwidth]{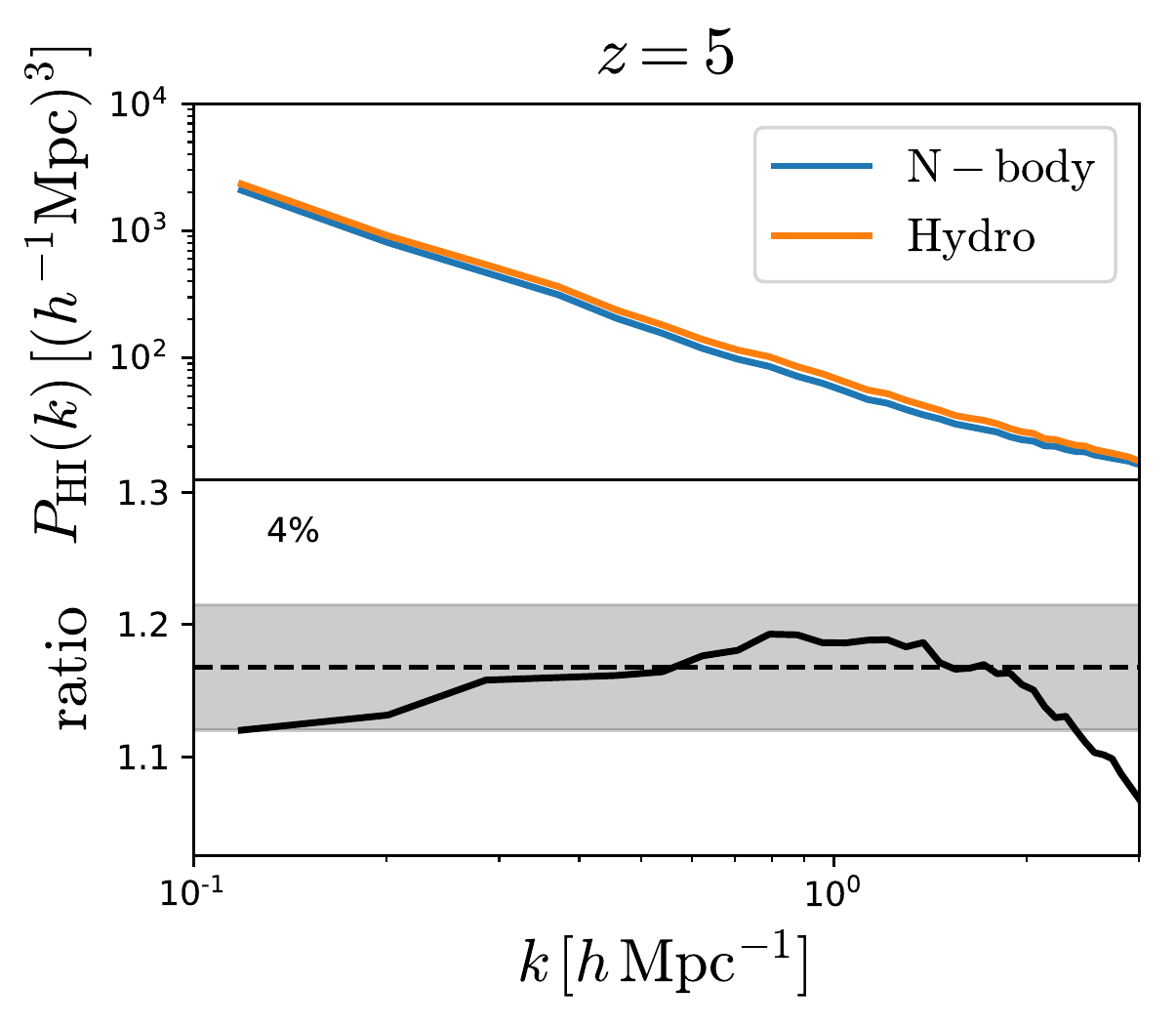}\\
\caption{A comparison of the spatial distribution of HI from IllustrisTNG versus the one obtained by placing HI in the centers of halos in an N-body simulation at redshift 0 (upper-left), 1 (upper-middle), 2 (upper-right), 3 (bottom-left), 4 (bottom-middle) and 5 (bottom-right). The HI mass assigned to each halo in the N-body run is taken from our tabulated $M_{\rm HI}(M,z)$ of Table \ref{table:M_HI_fit}. We compute the HI power spectrum in real-space for both configurations: N-body (blue) and hydro (green). The black line in the bottom part of each panel shows the ratio between the power spectra. Although the overall normalization can be different (see text for more details), it is more important to reproduce the shape. The black shaded region shows the variation in shape from the largest scales to $k=1~h{\rm Mpc}^{-1}$, and quoted with a number in the bottom part of each panel. This simple procedure allows us to generate mock 21 cm maps whose underlying power spectrum is accurate at $\simeq5\%$ down to $k=1~h{\rm Mpc}^{-1}$.}
\label{fig:21cm_real_space}
\end{center}
\end{figure*}

It can be seen that the qualitative agreement between the maps from
the full hydrodynamic simulation and the cheaper N-body one is
very good. In Fig. \ref{fig:21cm_real_space} we quantify this visual
agreement by computing the HI power spectrum of the hydrodynamic and
N-body simulations in real-space at different redshifts.

\begin{figure*}
\begin{center}
\includegraphics[width=0.33\textwidth]{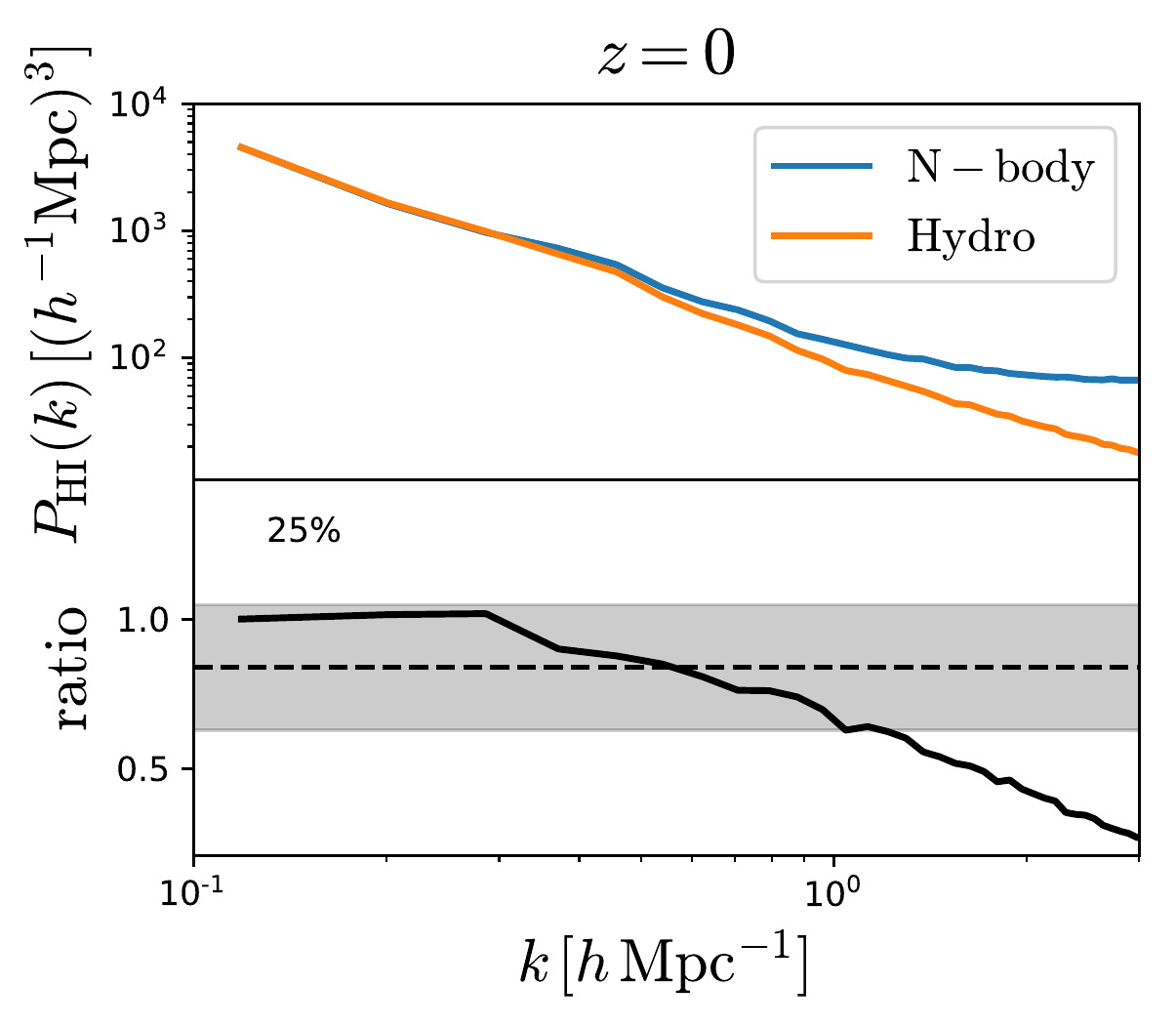}
\includegraphics[width=0.33\textwidth]{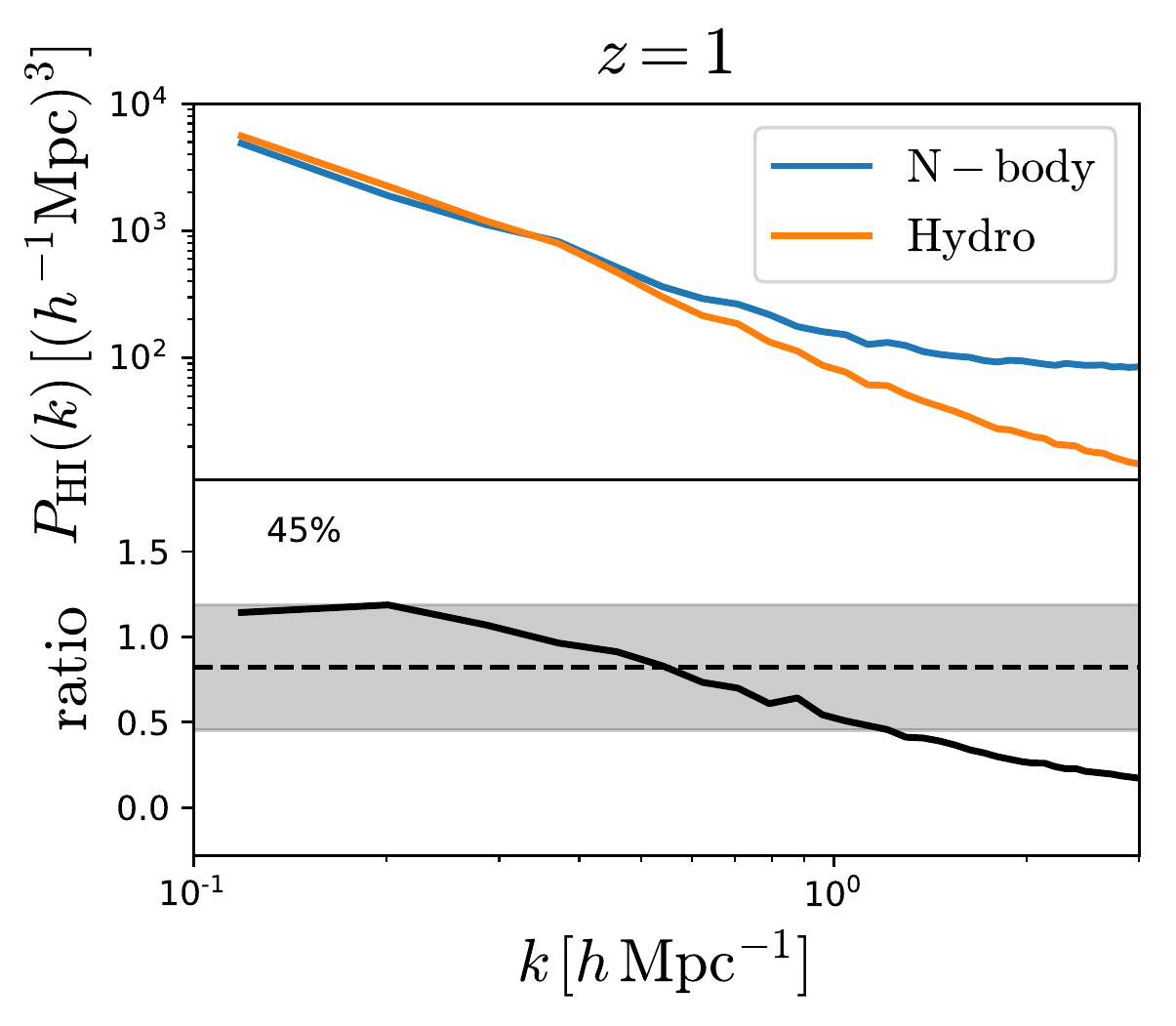}
\includegraphics[width=0.33\textwidth]{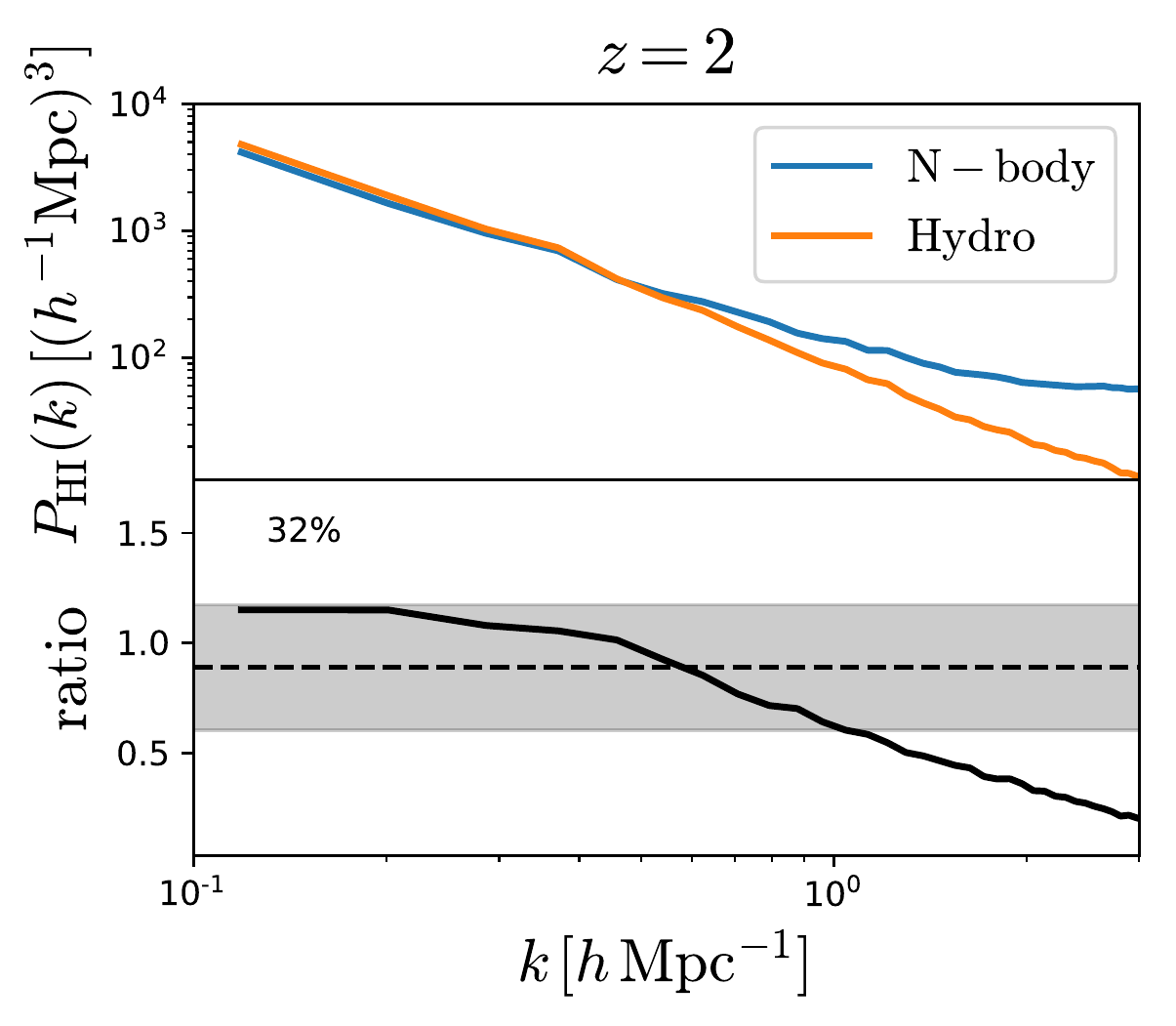}\\
\includegraphics[width=0.33\textwidth]{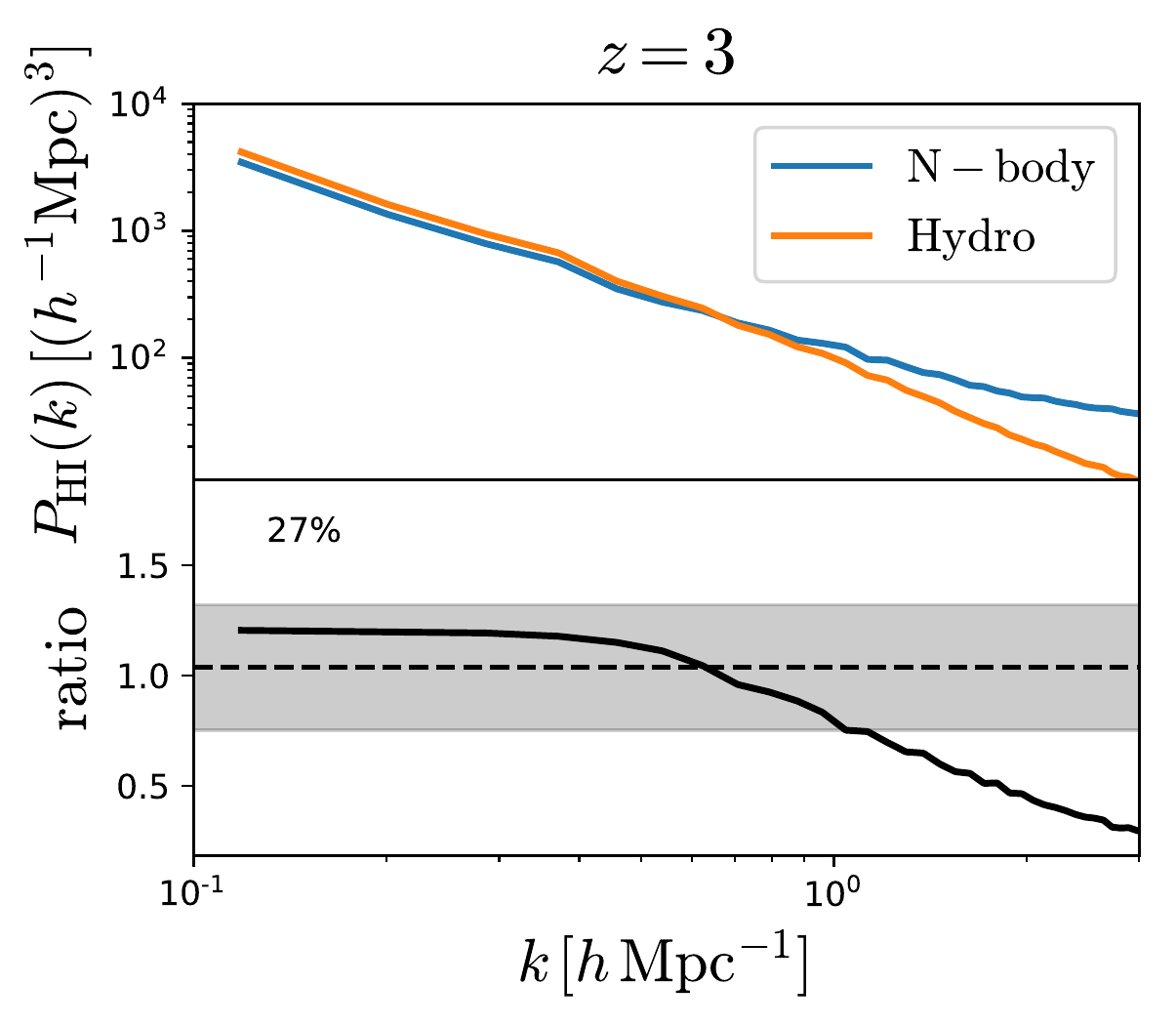}
\includegraphics[width=0.33\textwidth]{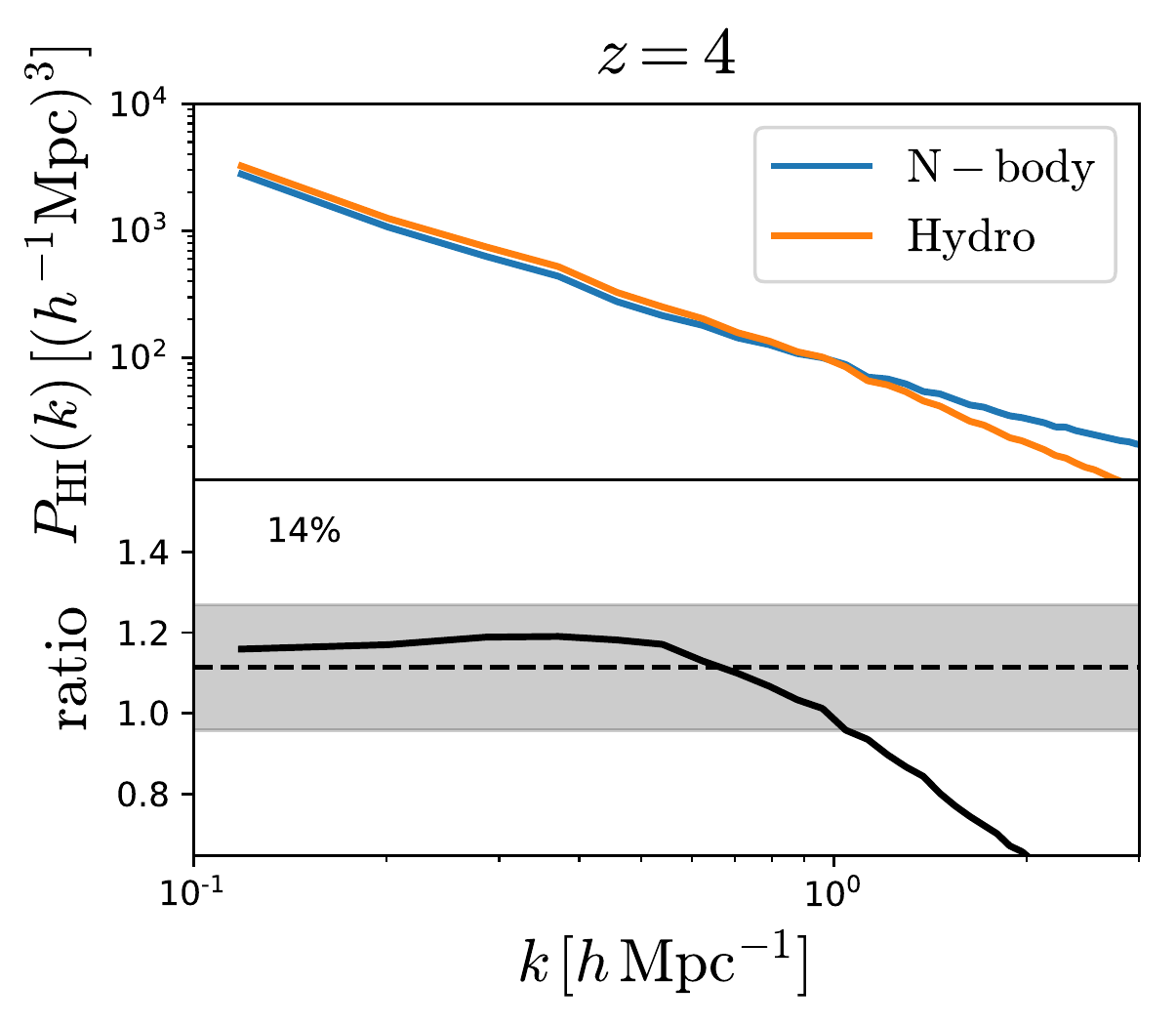}
\includegraphics[width=0.33\textwidth]{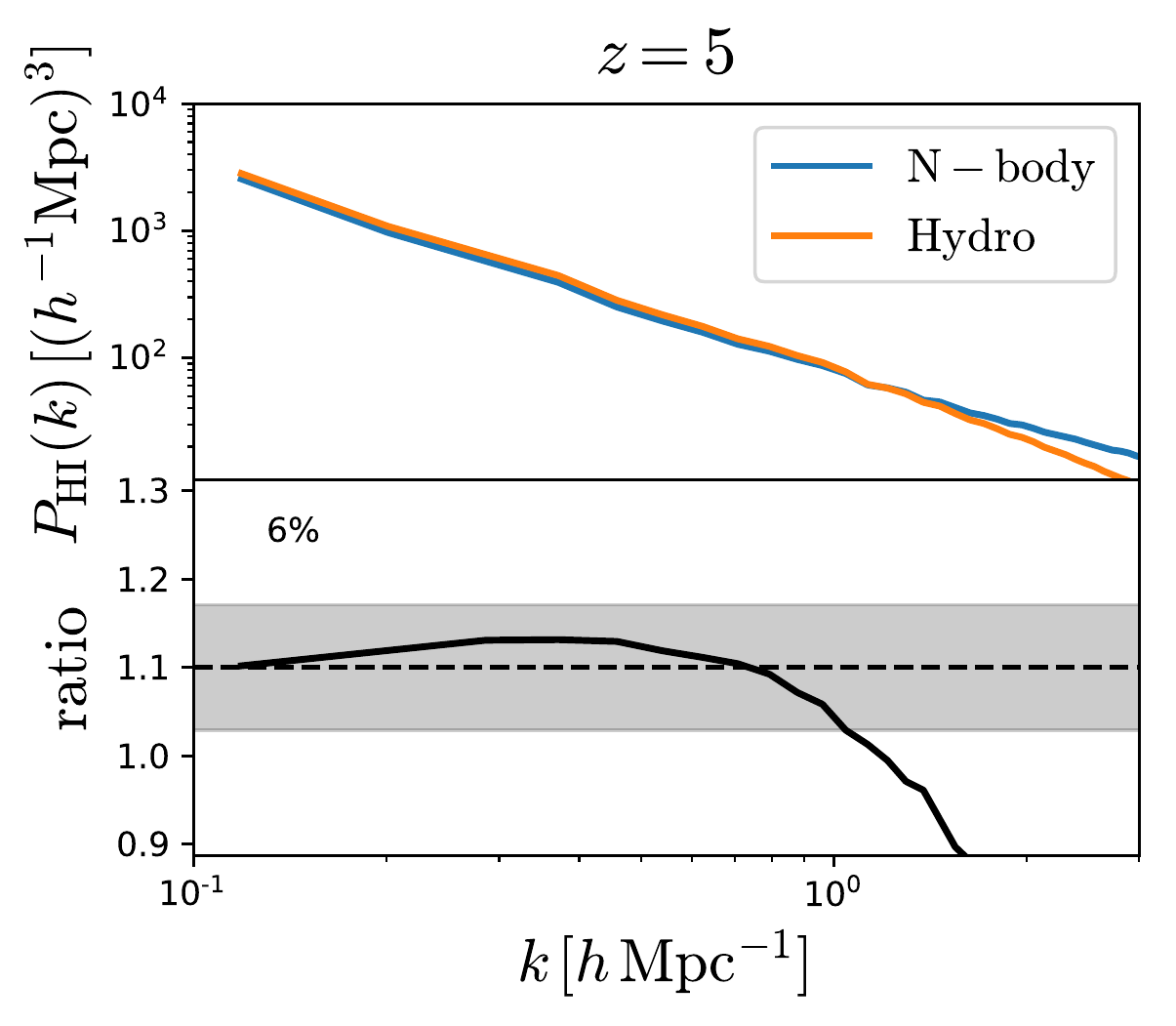}\\
\caption{Same as Fig. \ref{fig:21cm_real_space} but for HI monopole in redshift-space. It can be seen that the distributions of HI in the two configurations differ more significantly than in real-space. The origin of this large discrepancy, more prominent at low-redshift, is the lack of Fingers-of-God in the modeling of HI with the N-body simulations, since we place all HI in the halo center. More realistic 21 cm maps need to account for the HI velocity dispersion inside halos. }
\label{fig:21cm_redshift_space}
\end{center}
\end{figure*}

We find that the amplitude of the two HI power spectra can be
significantly different, between 10\% and 40\%. We attribute these
differences to the inaccuracies of our $M_{\rm HI}(M,z)$ function
(e.g.~we do not require our fit to reproduce $\Omega_{\rm HI}(z)$), to
the effects of baryons on the halo mass function and halo clustering,
which are neglected in this exercise, and to the omission of the
1-halo term at very high $k$.  For the latter, we emphasize that in this section we do
not consider the spatial distribution of HI inside halos, which we
leave for future work. However, once the difference in amplitude is
taken into account, we find that shapes differ by only $\simeq5\%$
from the largest scales down to $k=1~h{\rm Mpc}^{-1}$ at all
redshifts. This is rather remarkable considering that the 1-halo term
was not accounted for. This exercise demonstrates that populating dark
matter halos from a computationally cheap N-body simulation with HI
can yield results that are reasonable, at least in shape, at a few
percent level in the fully non-linear regime at all relevant
redshifts.

In Fig. \ref{fig:21cm_redshift_space} we show the comparison between
the HI power spectra from IllustrisTNG and the N-body simulation in
redshift-space at several redshifts. As for the comparison in
real-space, the amplitude of the two power spectra can be quite
different for the same reasons outlined above. However, even when the
normalization offset is taken into account, there are larger shape
differences than in real-space. For instance, at $z=1$ the
discrepancies up to $k=1~h{\rm Mpc}^{-1}$ can be as large as
45\%. That difference declines with redshift, so that at $z=5$ it is
only $6\%$, very similar to the differences we find in real-space.

The larger differences in redshift-space can be attributed to
Fingers-of-God. Here, we did not attempt to model the 1-halo term and,
therefore, the HI Fingers-of-God are not present in our mock maps,
while they are in those created from IllustrisTNG. As we saw in
section \ref{subsec:RSD}, the HI FoG can propagate to relatively large
scales and affect the amplitude and shape of the HI power spectrum. At
higher redshift, the magnitude of the FoG is lower, so the agreement
between IllustrisTNG and the N-body mocks is expected to improve, as
we observe. Finally, the agreement between the different maps also
depends on bandwidth. For larger bandwidths the FoG effects will have
a smaller impact.
We could also argue that beam smoothing in IM surveys will further reduce the effect of FoG in the final measured power spectrum.

We conclude that 21cm intensity mapping maps can be created via less
computationally expensive simulations like N-body, or fast
simulations, e.g.~COLA \citep{COLA} or Pinocchio \citep{Pinocchio},
instead of expensive hydrodynamic simulations. It is however very
important to account for the 1-halo term, as expected, i.e.~the
FoG, when modeling the distribution of HI in redshift-space.

\section{Summary and Conclusions}
\label{sec:conclusions}

A goal of current and upcoming radio telescopes is to map the
spatial distribution of matter by detecting 21 cm emission from
cosmic neutral hydrogen. The very large volumes that can be sampled
through 21 cm intensity mapping observations will place
tight constraints on the values of the cosmological parameters
\citep{Bull2015,Villaescusa-Navarro_2015a,Obuljen_17,Sprenger_2018}. In
order to extract the maximum information from these surveys, accurate
theory predictions are needed.

Theory predictions to linear order are well known. For instance, the
amplitude and shape of the 21 cm power spectrum is given by
\be
P_{\rm 21cm}(k,\mu)=\bar{T}_b^2(b_{\rm HI}+f\mu^2)^2P_{\rm m}(k)+P_{\rm SN}~,
\ee
where $\bar{T}_b\propto\Omega_{\rm HI}$ is the mean brightness
temperature, $b_{\rm HI}$ is the HI bias, $f$ is the linear growth
rate, $\mu=k_z/k$, $P_{\rm m}(k)$ is the linear matter power spectrum,
and $P_{\rm SN}$ is the HI shot-noise. While the value of $\bar{T}_b$
is relatively well known from several observations across the redshift
range $z\in[0,5]$, little is known about $b_{\rm HI}$ and $P_{\rm
  SN}$. In this work we have quantified them
and studied the scales where linear theory holds, e.g.~when the HI
bias becomes non-linear.

Accurate theory predictions in the mildly or fully-non linear regimes
will allow us to recover the large amount of information embedded in
the spatial distribution of HI on small scales. There are several
techniques for accomplishing this, such as perturbation theory, HI halo
models, or numerical simulations. The purpose of our work has been to
study the ingredients that these techniques employ. For example, HI
halo models require the halo HI mass function and the HI
density profile as inputs.

Our purpose here is not limited to analytic approaches, but is
also to understand how HI is distributed across the Universe and how
it evolves with time. We have shown that even with a subset of the
ingredients studied in this work, one can model the spatial
distribution of HI in the fully non-linear regime without the use of
computationally expensive hydrodynamic simulations, but using HI HOD
models.

We have carried out our analysis using IllustrisTNG
\citep{SpringelV_17a,PillepichA_17a,NelsonD_17a,NaimanJ_17a,MarinacciF_17a},
a sophisticated series of cosmological hydrodynamical simulations
that have shown to be in broad agreement with many basic
observables, run at an unprecedented combination of large volume and
high resolution, therefore providing an excellent testbed for
accurately investigating the distribution of HI from the disks of
spiral galaxies to cosmological scales.

We outline the main conclusions of our work below:

\begin{itemize}

\item We find that almost all HI in the Universe is inside halos: from
  more than $99\%$ at $z=0$ to around $88\%$ at $z=5$. The fraction of
  HI outside halos increases with redshift because the gas in the IGM
  is denser and the amplitude of the UV background decreases with
  redshift (at $z>2$). This justifies the use of halo models
  to model the distribution of HI in the Universe, but quantifies their
  limitations at high redshifts. The fraction of HI inside galaxies is
  slightly lower than in halos. At $z=0$ $\simeq97\%$ of all HI is
  inside galaxies while at $z=5$ this number declines to $\simeq80\%$.

\item We find that the halo HI mass function, i.e.~the average HI mass hosted by a halo of mass $M$ at redshift $z$, is well reproduced by a function like
\be
M_{\rm HI}(M,z)=M_0\left(\frac{M}{M_{\rm min}}\right)^\alpha\exp\left(-(M_{\rm min}/M)^{0.35}\right)\nonumber
\ee
where $M_0$, $M_{\rm min}$ and $\alpha$ are free parameters. The
best-fit values are given in Table \ref{table:M_HI_fit} for both FoF and
FoF-SO halos. The value of $\alpha$ increases with redshift, likely 
indicating that at low redshift processes such as ram-pressure, tidal
stripping, and AGN feedback make galaxies in clusters HI poor. We find
that $M_{\rm min}$ decreases with redshift. On the other
hand, only halos with circular velocities above around $30~{\rm km/s}$ host
a significant HI mass fraction. Although the fit is slightly worse,
our halo HI mass function can also be well-reproduced by the
function
\be
M_{\rm HI}(M,z)=M_0\left(\frac{M}{M_{\rm min}}\right)^\alpha\exp\left(-M_{\rm min}/M\right)\nonumber
\ee
whose best-fit values are given in Table \ref{table:M_HI_fit2}
to facilitate comparison with previous works.

\item We find that the HI density profiles inside halos exhibit a large halo-to-halo variation. 
   HI profiles are sensitive to the physical
  processes that occur in and around halos, such as AGN feedback and
  tidal stripping. The HI profile of small halos
  ($M\lesssim10^{12}~h^{-1}M_\odot$) is not cuspy, but its amplitude saturates. 
  This is expected as HI at high densities
  will turn into molecular hydrogen and then stars in short time
  periods. More massive halos exhibit holes in their centers. For
  galaxy groups this is mostly due to AGN feedback, while in galaxy
  clusters the holes are large and generated by a combination of AGN
  feedback, ram-pressure and tidal stripping. We find that the average HI density profiles are universal and can be reproduced by an expression like 
\be
\rho_{\rm HI}(r)=\frac{\rho_0}{r^{\alpha_\star}}\exp(-r_0/r),
\ee
or
\be
\rho_{\rm HI}(r)=\frac{\rho_0 r_s^3}{(r+3/4r_s)(r+r_s)^2}\exp(-r_0/r),
\ee
where $\rho_0$, $\alpha_*$, $r_0$ and $r_s$ are free parameters. We fix value of $\rho_0$ by requiring that $M_{\rm HI}(M)=\int_0^{R_v}4\pi r^2\rho_{\rm HI}(r)dr$, where $R_v$ is the halo virial radius. The best-fit values for $\alpha_*$, $r_0$ and $r_s$ can be found in table \ref{table:HI_profiles}.

\item We find that the HI mass in small/big halos is mostly located in
  its central/satellites galaxies. The fraction of the total HI mass
  in halos that is within the central galaxy decreases with halo mass,
  while the opposite trend takes place for the satellites.
  For halos of masses $\sim5\times10^{12}~h^{-1}M_\odot$ the
  HI mass in the central galaxy is similar to that of the satellites,
  almost independent of redshift. The HI mass fraction in the
  central galaxy of clusters is negligible at $z=0$. At high-redshift,
  $z\geqslant2$, the fraction of the halo HI mass in satellites is
  roughly $20\%$ for small halos $M\in[10^{10}-10^{11}]~h^{-1}M_\odot$.

\item We find that the pdf of the HI density field is quite different
  from that of the matter field. In general, the HI pdf is broader,
  indicating that the HI is more clustered than matter. The amplitude
  of the pdf for low overdensities is higher for HI than for matter,
  indicating that HI voids are more empty than matter voids. At
  high-redshift the HI and matter density pdf can be well-reproduced
  by a log-normal, while at low-redshift the log-normal is not a good
  description of our results.  

\item We find that the HI column density distribution function is
  nearly constant across redshifts, in agreement with previous studies
  and with observations. In the redshift range $z\in[2,4]$ we find
  that the DLA cross-section depends on both halo mass and HI column
  density, and its mean value can be well reproduced by
\be
\sigma_{\rm DLAs}(M|N_{\rm HI},z)=AM^\alpha\left(1-e^{-(M/M_0)^\beta}\right)
\ee
where $\alpha=0.82$, $\beta=0.85\log_{10}(N_{\rm HI}/{\rm
  cm^{-2}})-16.35$, $A\cdot M_0=0.0141~~h^{-2}{\rm kpc}M_\odot$ and
the best-fit values of $M_0$ are given in Table \ref{table:M0}.
We argued that the small dependence of the above relation on column density implies the bias of different absorbers will be very similar. We estimate the DLAs bias using two methods and find agreement with observations in both.

\item We find that for small halos,
  $M\lesssim10^{12}~h^{-1}~h^{-1}M_\odot$, the bulk velocities of HI
  inside halos trace very well, in modulus and direction, the peculiar
  velocity of the halos they reside in. On the other hand, for bigger
  halos, we observe departures, in modulus and direction, between the
  HI and halo peculiar velocities. This happens because while for
  small halos most of the HI is in the central galaxy, for larger
  halos a significant HI mass is in satellites, whose peculiar
  velocities do not trace that of the halo.

\item We find that the velocity dispersion of HI inside halos can be well reproduced by a simple power-law
\be
\sigma(M)=\sigma_{10}\left(\frac{M}{10^{10}~h^{-1}M_\odot}\right)^\alpha
\ee
where $\sigma_{10}$ and $\alpha$ are free parameters whose best-values
are given in Table \ref{table:HI_sigma}. While at $z=0$ the mean
velocity dispersions of CDM and HI are similar (for halos above
$\simeq5\times10^{10}~h^{-1}M_\odot$), at higher redshifts they depart
for small halos, with HI having a lower amplitude than CDM. The mass
where they diverge increases with redshift, but is typically
around $10^{12}~h^{-1}M_\odot$. In general, for fixed mass and
redshift the variance in the velocity dispersion of HI is larger than
that of CDM, reflecting the larger variation in HI profiles than CDM
profiles inside halos.

\item We find that the values of the HI bias on the largest scales we
  can probe is equal to 0.84, 1.49, 2.03, 2.56, 2.82 and 3.18 at
  redshifts 0, 1, 2, 3, 4 and 5, respectively. While the HI bias is
  relatively flat down to $k\simeq1~h{\rm Mpc}^{-1}$ at $z=1$, it is
  already non-linear at $k\simeq0.3~h{\rm Mpc}^{-1}$ at
  $z\geqslant3$. Our results suggest that the HI bias becomes more
  non-linear with redshift. This is expected as the value of the
  linear bias increases with redshift.
  We have shown the perturbative approaches based on LPT are able to reproduce the clustering measurement up to $k = 1 h/\text{Mpc}^{-1}$, therefore making possible, at least in principle, to extract cosmological information from such small scales.
  
\item We identify a new secondary halo bias. Halos of the same mass are
  clustered differently depending on their HI mass. At low
  redshift HI-poor halos are more clustered than HI-rich
  halos. However, at high redshift the situation is reversed and HI
  rich-halos cluster more strongly than HI-poor halos. We believe that this is
  mainly driven by environment. At low redshift, small halos may lose
  their gas due to stripping by a larger neighboring halo, so
  HI-poor halos will be more clustered than field HI-rich halos. On
  the other hand, at high redshift, gas stripping is likely less
  effective, so HI-rich halos will be found around larger halos and
  therefore their clustering will be higher.

\item We quantify the amplitude of the HI shot-noise to be 104, 124,
  65, 39, 14 and 7 $(h^{-1}{\rm Mpc})^3$ at redshifts 0, 1, 2, 3, 4
  and 5, respectively. These low levels imply that BAO measurements
  through 21 cm intensity mapping are hardly affected by
  shot-noise. Furthermore, the very low shot-noise levels at high
  redshift suggest that a large amount of cosmological information can
  be extracted from the clustering of HI on small scales.

\item We find that the relation between $\rho_{\rm m}$ and $\rho_{\rm
  HI}$ cannot be explained with linear theory for smoothing scales
  $\leqslant5~h^{-1}{\rm Mpc}$ at any redshift. The scatter in that
  relation decreases with redshift, and much larger HI overdensities
  can be found for the same matter overdensities.

\item We find that the Kaiser factor alone cannot explain clustering
  of HI in redshift-space at low redshift, as expected, given the
  small volume of our simulations. But, at high redshift the ratio
  between the monopoles in redshift- and real-space can be explained
  with linear theory down to $0.3$, $0.5$ and $1$ $h{\rm Mpc}^{-1}$ at
  redshifts 3, 4 and 5 respectively. This is rather surprising taking
  into account that the HI bias becomes non-linear already at
  $k=0.3~h{\rm Mpc}^{-1}$ at those redshifts.

\item We find that the 2-dimensional HI power spectrum in
  redshift-space exhibits large differences with respect to the matter
  field. Those differences arise mainly because the amplitude of
  Fingers-of-God is higher for HI than for matter. This can be
  understood taking into account that HI resides only in relatively
  massive halos. While the amplitude of the matter power spectrum on
  small scales is dominated by small halos with low velocity
  dispersion, for HI halos only above $\simeq M_{\rm min}$, i.e.~with
  larger velocity dispersion, can contribute. We find that standard
  phenomenological models to describe the clustering in 2D in
  redshift-space are not adequate for reproducing our results.

\item We show that accurate 21 cm maps can be created from N-body
  simulations, rather than full hydrodynamic simulations, by using the
  ingredients studied in our work. In real-space and without modeling
  the 1-halo term, the agreement in the shape of the 21 cm power
  spectrum from N-body and IllustrisTNG is around $5\%$ down to 1
  $h{\rm Mpc}^{-1}$ at all redshifts. In redshift-space however, the
  lack of the 1-halo term, i.e.~the HI Fingers-of-God, induces much
  larger errors in the 21cm power spectrum from N-body versus hydro at
  low redshift, e.g.~$45\%$ at $z=1$. Modeling the 1-halo term is thus
  crucial for creating mock 21 cm maps.

\end{itemize}

The HI properties investigated in this work will help to improve our
knowledge of the way neutral hydrogen is distributed across the
Universe. The different quantities we have studied can be used as
input to analytic approaches like HI halo models or to create very
accurate mock 21 cm maps.

The python/cython scripts written to carry out the analysis performed
in our work can be found in
\url{https://github.com/franciscovillaescusa/Pylians/tree/master/HI_Illustris}. 
Our scripts made use of the {\sc Pylians} python routines, publicly available at \url{https://github.com/franciscovillaescusa/Pylians}.

\section*{ACKNOWLEDGEMENTS} 
We thank Uros Seljak for detailed comments on shot-noise in discrete tracers. We also thank Sandrine Codis, Philippe Berger, Andreu Font-Ribera, Ariyeh Maller, Anze Slosar, Amiel Sternberg, Martin White and Cora Uhlemann
for useful discussions. The work of FVN, SG and DNS is supported by
the Simons Foundation. The analysis of the simulations has been
carried out on the Odyssey cluster at Harvard University. The
IllustrisTNG simulations were run on the HazelHen Cray XC40
supercomputer at the High-Performance Computing Center Stuttgart
(HLRS) as part of project GCS-ILLU of the Gauss Centre for
Supercomputing (GCS).

\begin{appendix}

\section{Time evolution of HI in the intergalactic medium}
\label{sec:HI_Lya}

\begin{figure}
\begin{center}
\includegraphics[width=0.495\textwidth]{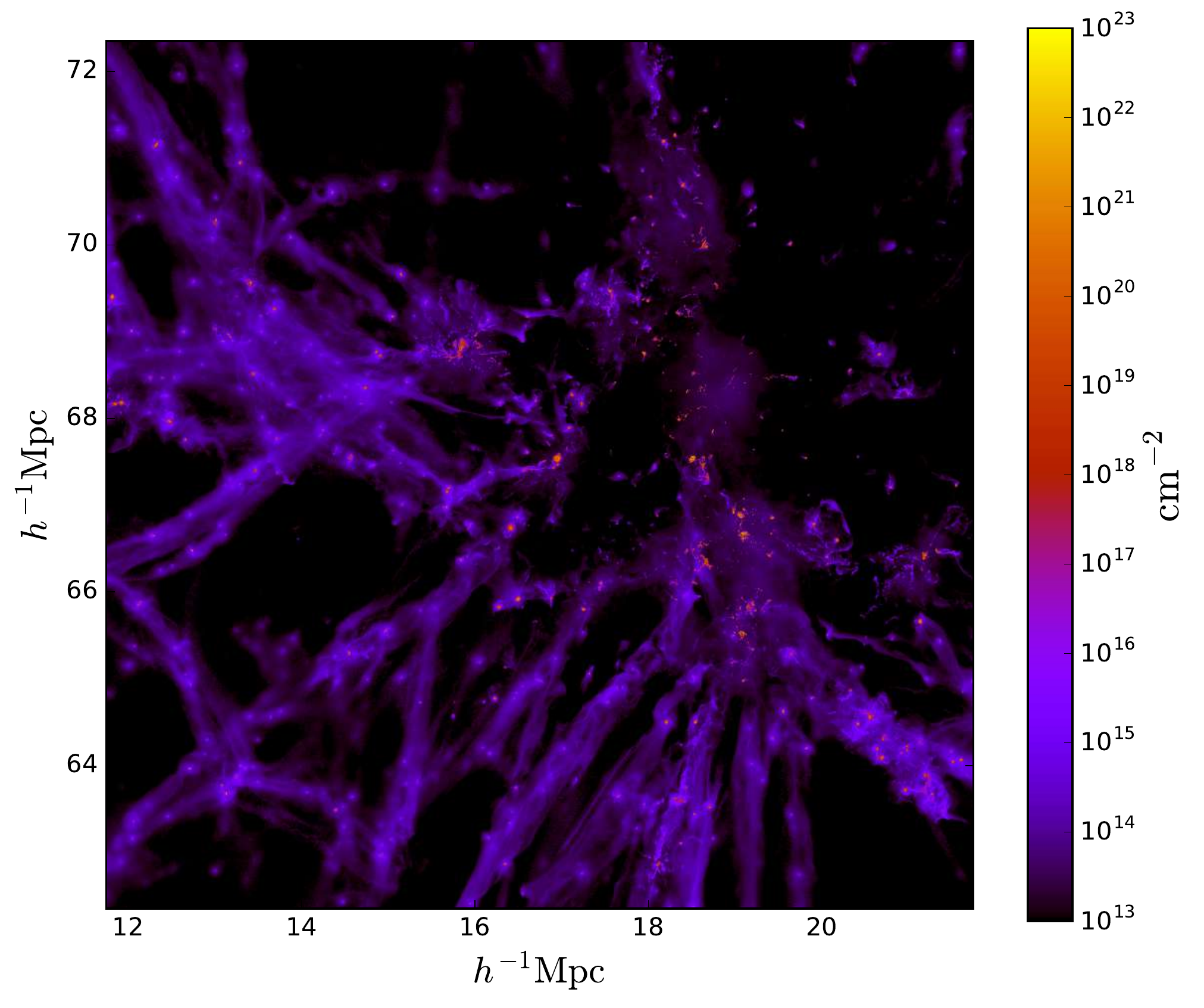}
\includegraphics[width=0.495\textwidth]{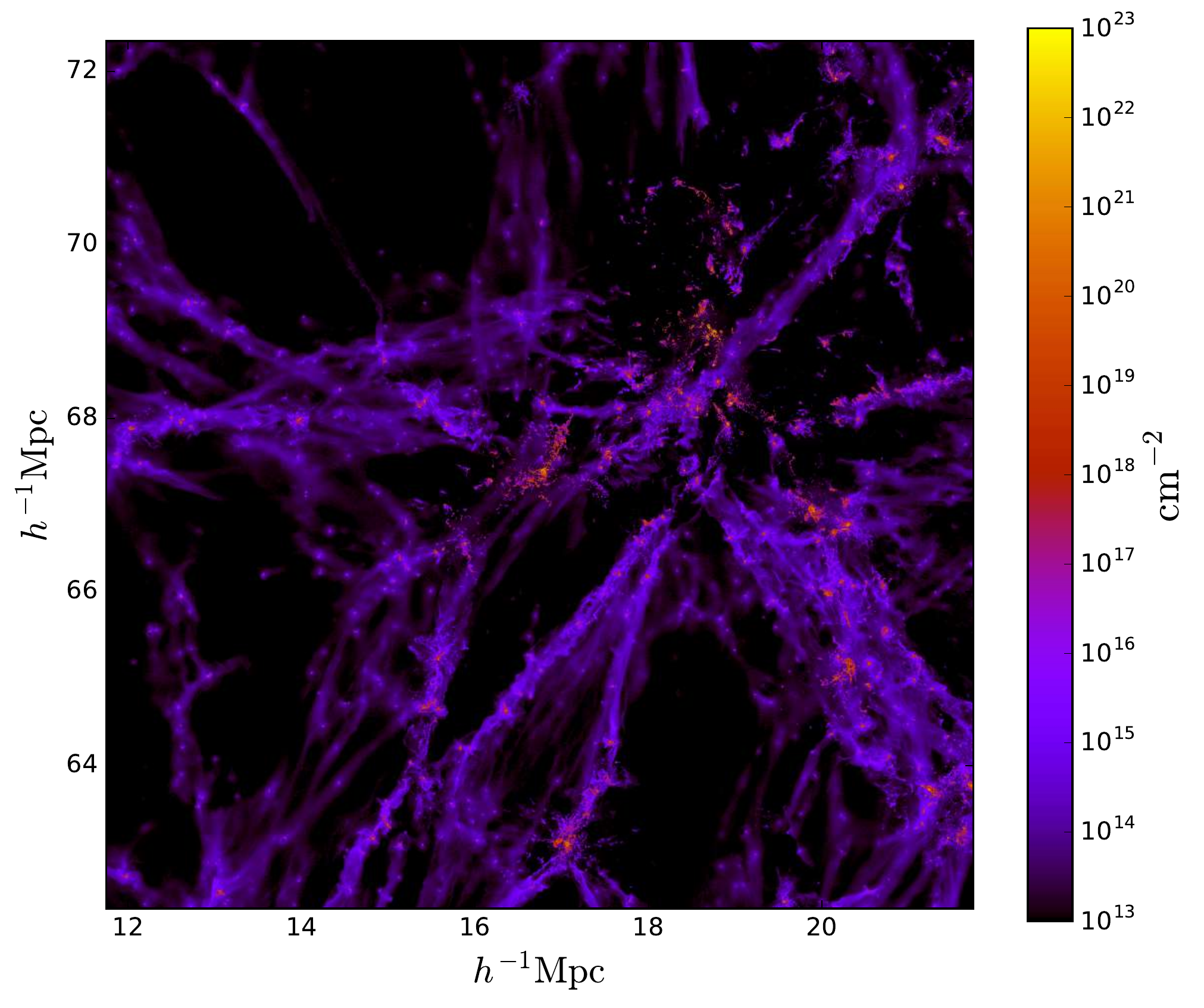}\\
\includegraphics[width=0.495\textwidth]{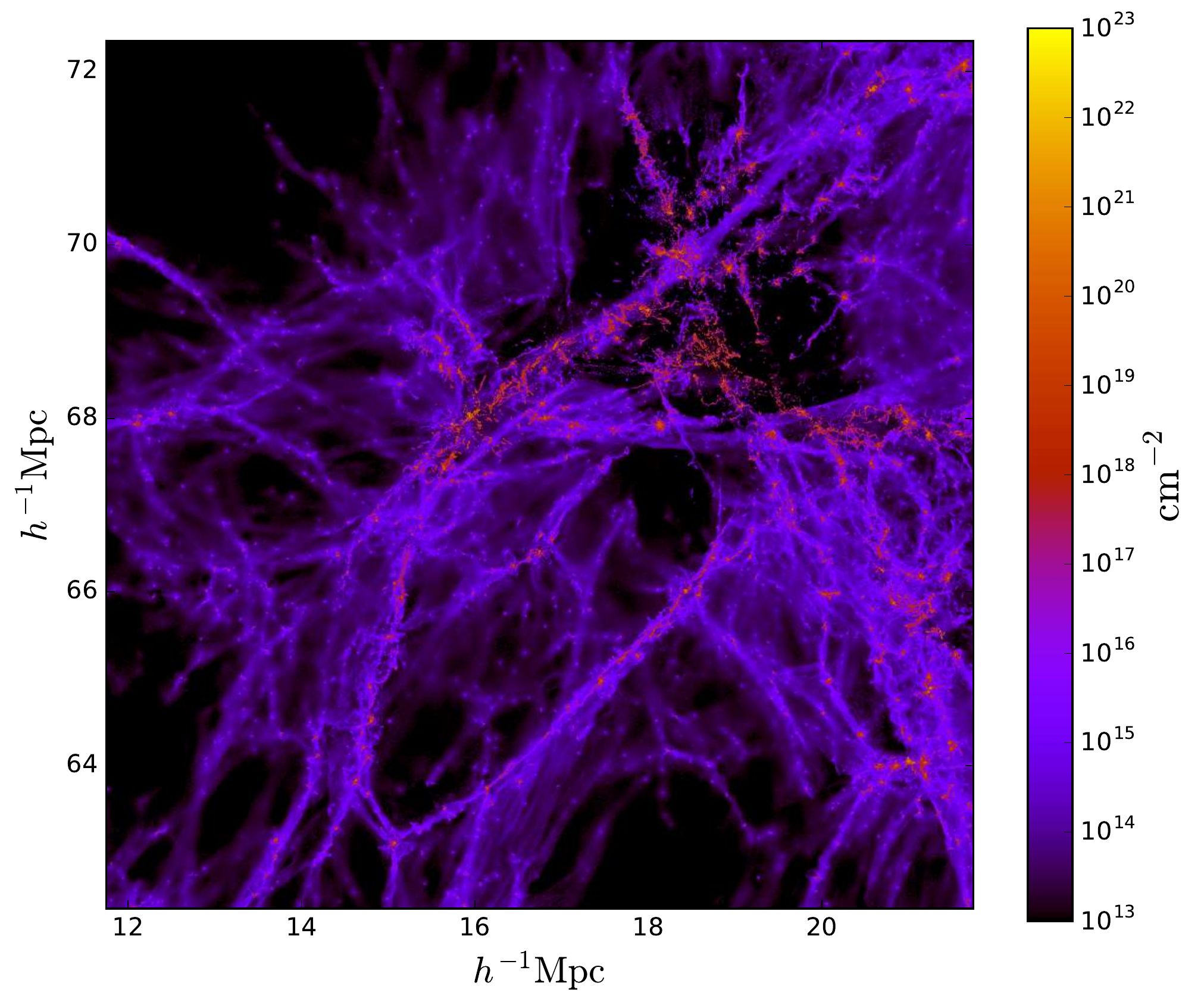}
\includegraphics[width=0.495\textwidth]{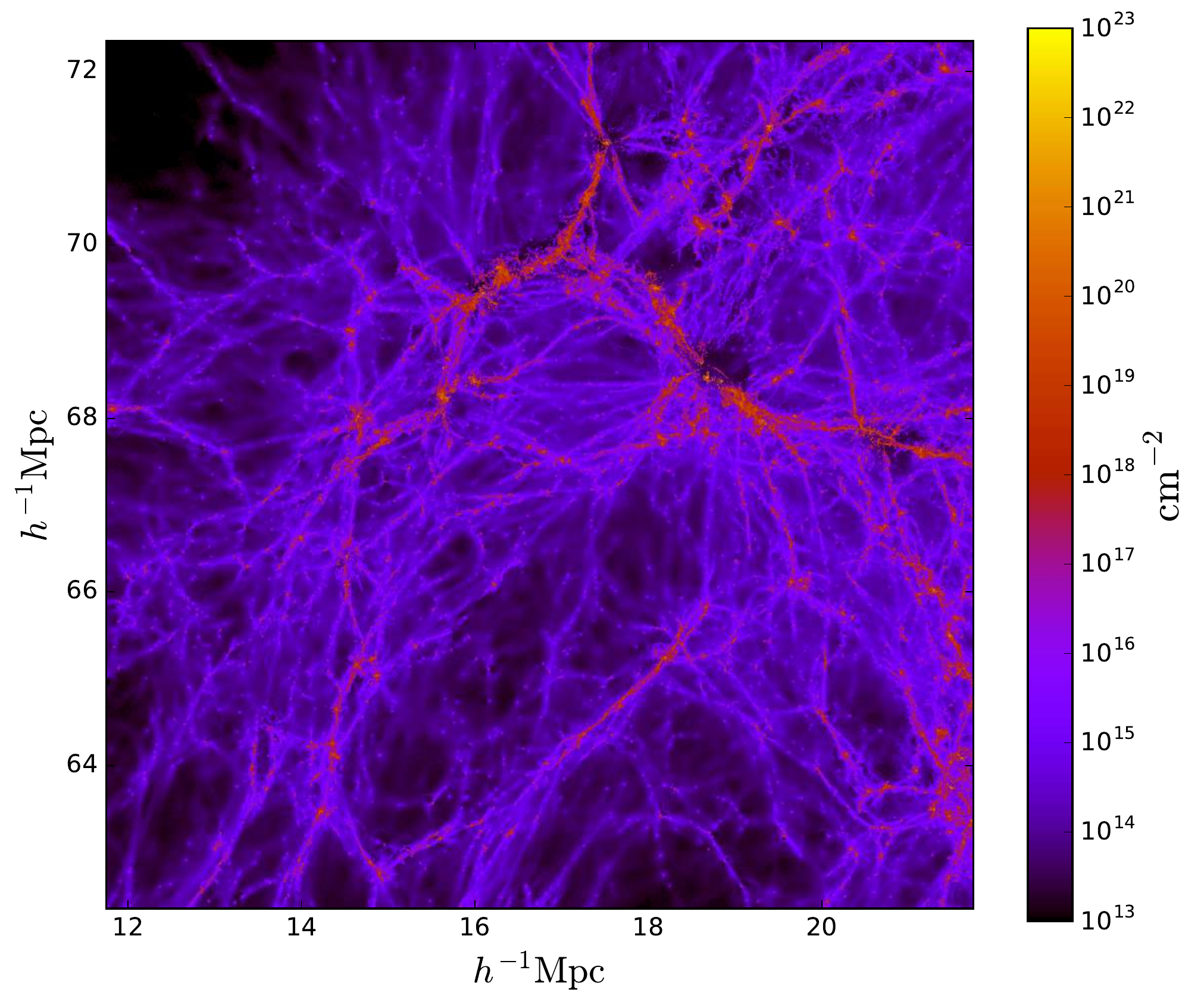}\\
\includegraphics[width=0.495\textwidth]{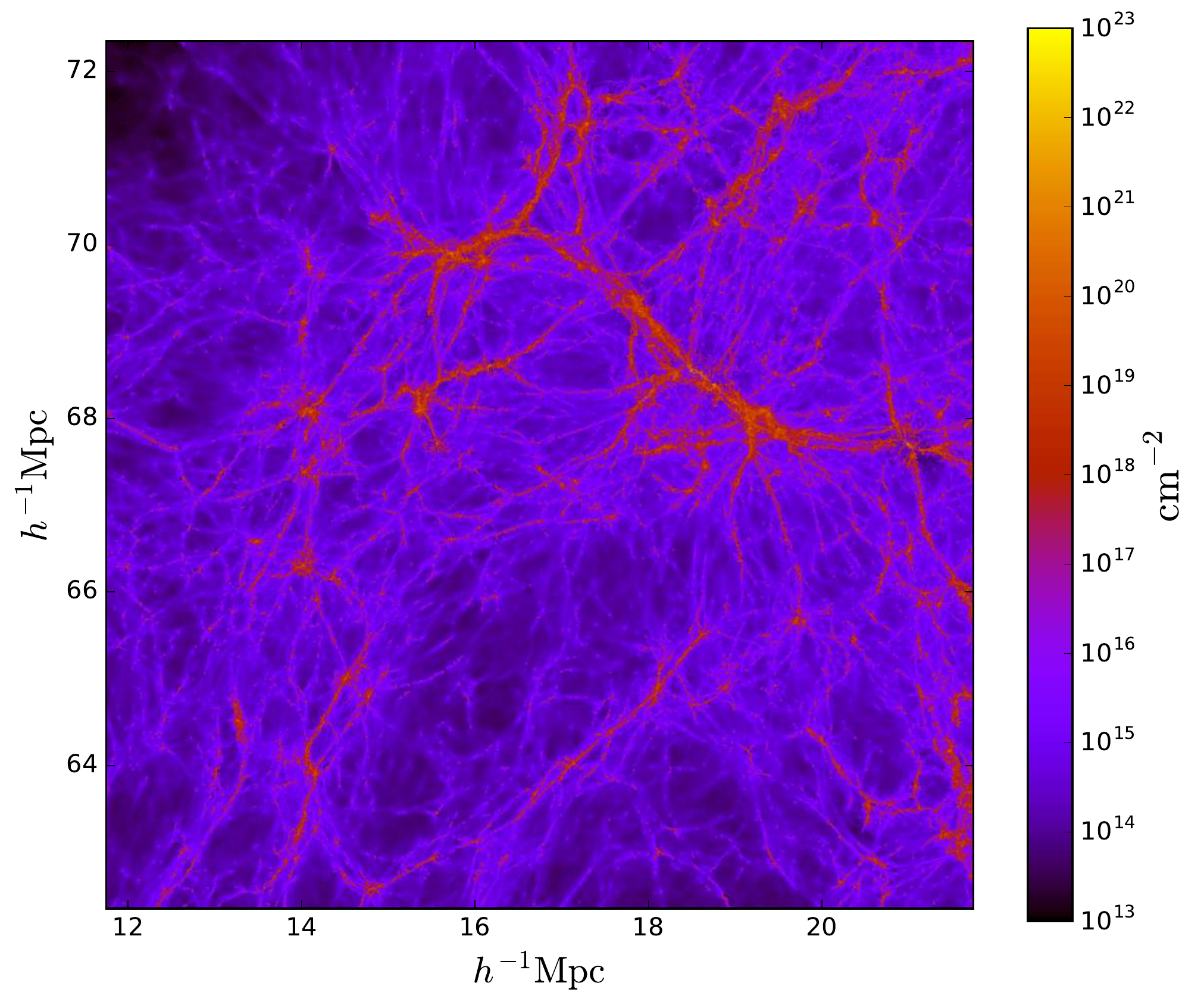}
\includegraphics[width=0.495\textwidth]{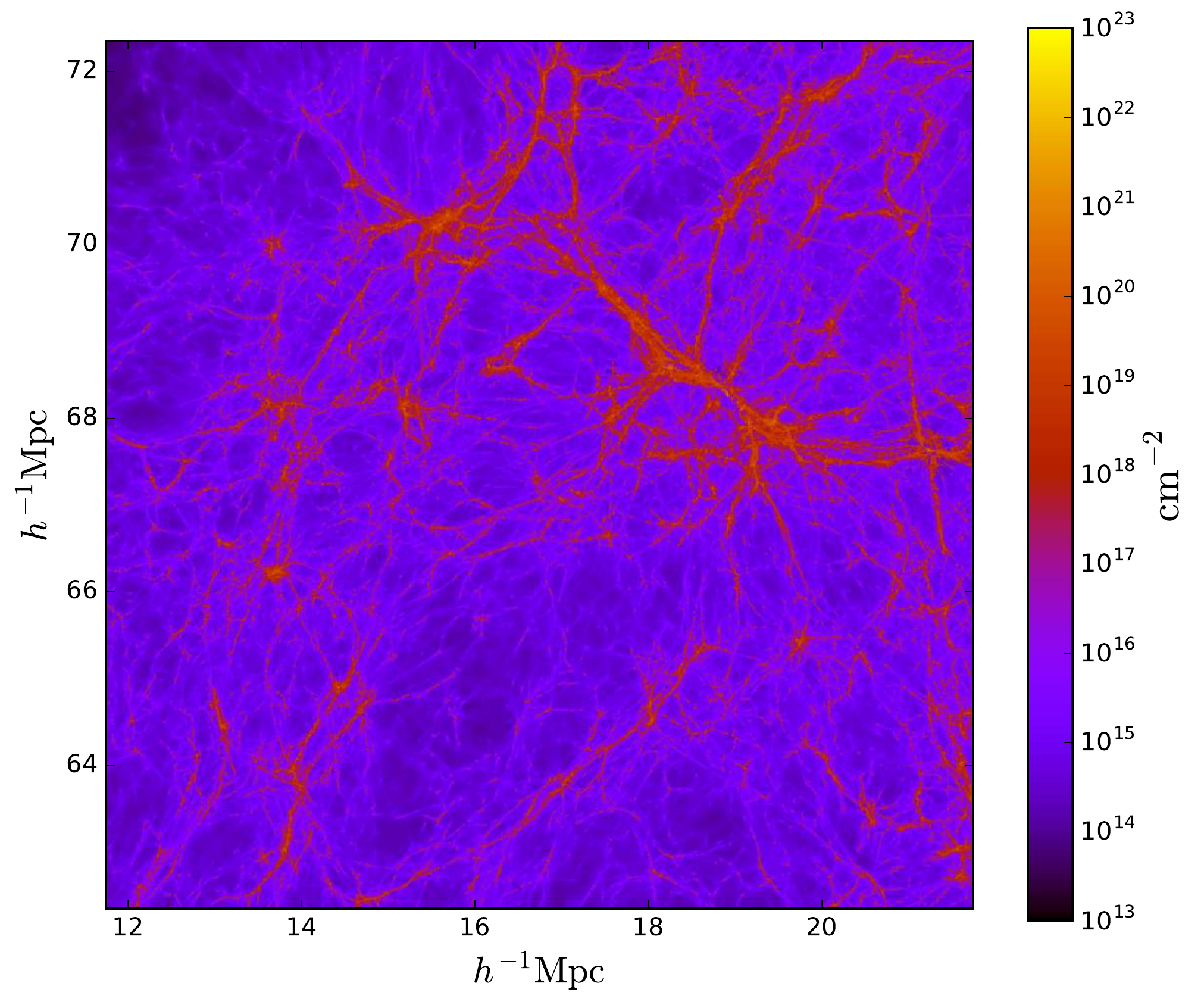}
\caption{Spatial distribution of HI in a slice of 10x10x5 $(h^{-1}{\rm Mpc})^3$ at redshifts 0 (top-left), 1 (top-right), 2 (middle-left), 3 (middle-right), 4 (bottom-left) and 5 (bottom-right). The color indicates the HI column density in comoving ${\rm cm}^{-2}$ units, to facilitate the comparison across redshifts. The region shown is the same as that in Fig. \ref{fig:HI_image} (top and middle panels). At higher gas densities the column densities of HI are higher in the intergalactic medium. It can be seen that some filaments host a significant HI mass at high-redshift, while at low redshift the HI is mostly locked inside galaxies. This explains why the fraction of HI outside halos increases with redshift.}
\label{fig:HI_Lya}
\end{center}
\end{figure}

In section \ref{subsec:HI_in_halos_galaxies} we found that while at
$z\leqslant2$ most of the Universe HI mass is inside halos, at higher
redshift an increasing fraction of it is located outside halos. In
order to visualize this effect we show in Fig. \ref{fig:HI_Lya} the
spatial distribution of HI in a slice of 5 $h^{-1}{\rm Mpc}$ width
across 10x10 $(h^{-1}{\rm Mpc})^2$. We show in that figure HI column
density in comoving units to facilitate the comparison across
redshifts. We note that our color palette may produce the incorrect
impression that at $z=5$ there is much more HI than at $z=0$. We have
explicitly checked that the sum of all column densities across all
pixels in our figures give a similar value across redshift, indicating
that $\Omega_{\rm HI}$ is very similar in all panels.

It can be seen that at low redshift, most of the HI mass is inside
galaxies, while the hydrogen in the filaments is highly ionized. At
higher redshifts, on the other hand, the gas in the intergalactic
medium becomes denser and the filaments contain a larger amount of
HI. The lower amplitude of the UV background at those redshifts
facilitates gas self-shielding. Given these effects it is thus
natural that the fraction of HI outside dark matter halos increases
with redshift.

\section{Origin of the cutoff in the halo HI mass function}
\label{sec:UVB_cutoff}

We saw in section \ref{subsec:M_HI} that the halo HI mass function
exhibits a cutoff at low masses. In this appendix we shed light
on the physical origin of that feature.

\begin{figure}
\begin{center}
\includegraphics[width=0.49\textwidth]{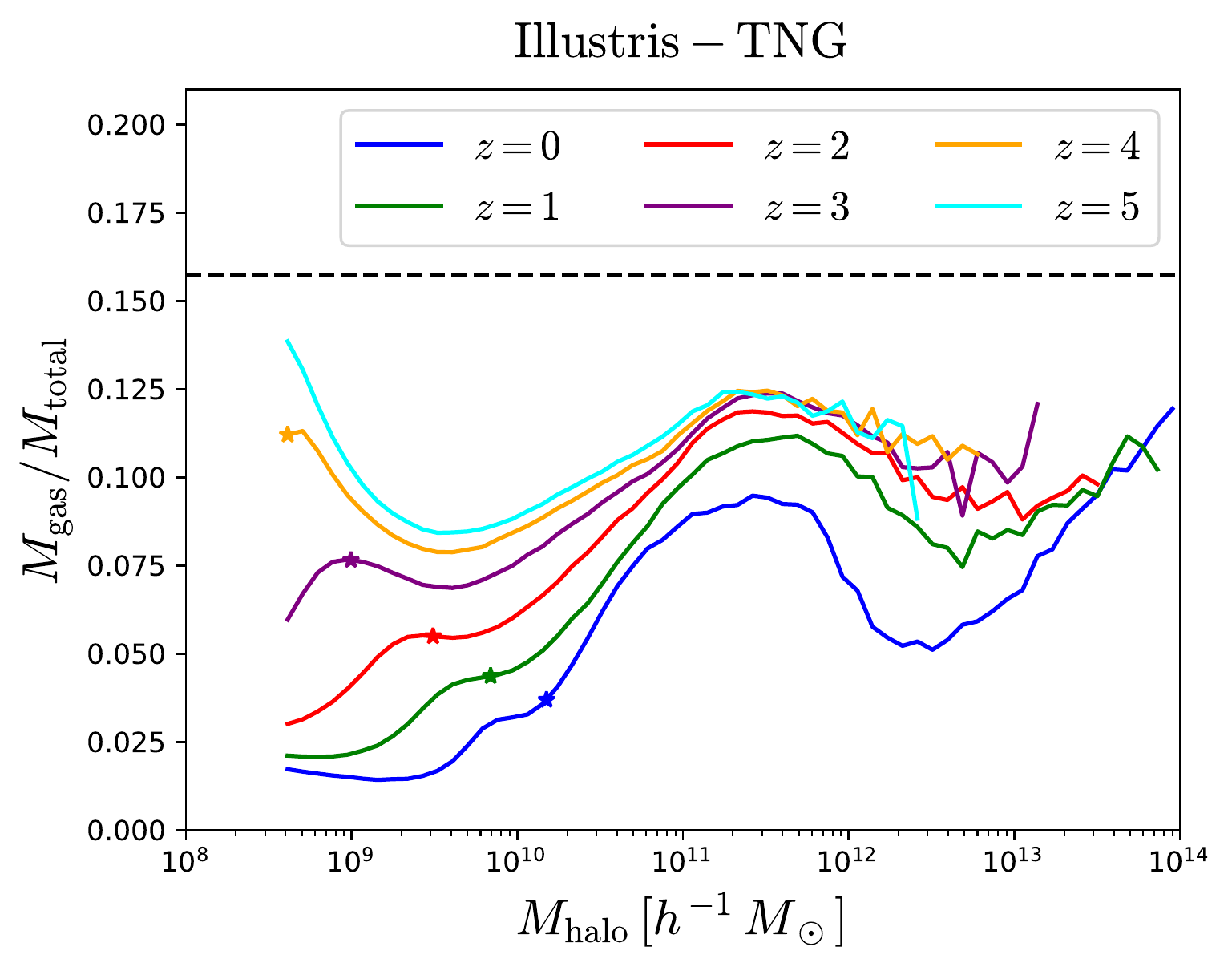}
\includegraphics[width=0.49\textwidth]{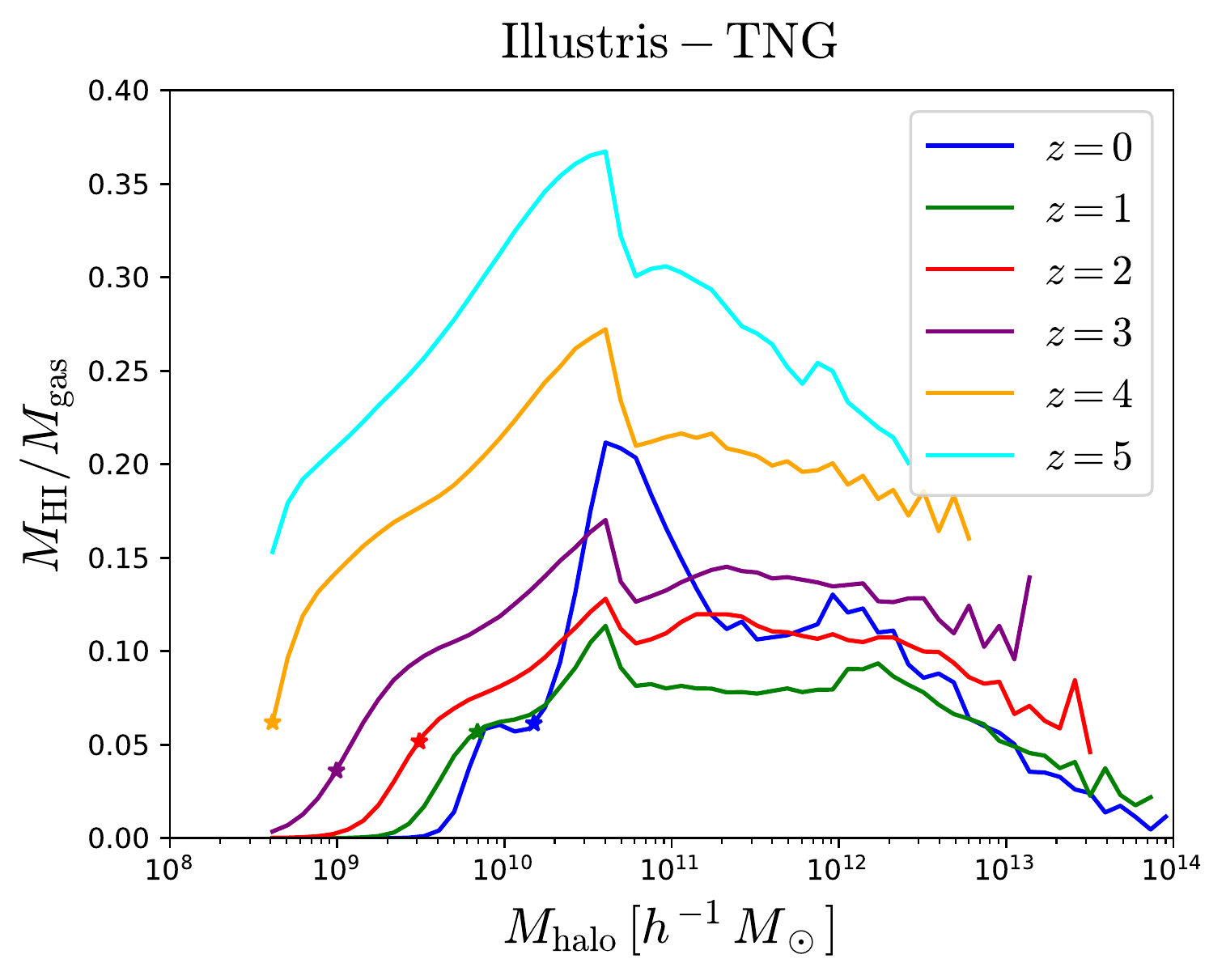}
\caption{Average gas fraction (left) and average HI mass to gas mass ratio (right) in halos as a function of their mass at redshifts 0 (blue), 1 (green), 2 (red), 3 (purple), 4 (yellow) and 5 (cyan) in IllustrisTNG. We show results only for halos with masses larger than 50 CDM particles. The stars indicate the hard cutoff mass (i.e. halos below that mass contain $only 2\%$ of all HI inside halos). The gas content of small halos declines with time. The features we observe in the plot are due to: 1) supernova feedback (removes gas of small halos), 2) AGN feedback (removes gas from large halos) and 3) the UV background (reduces the gas content of small halos). The dashed black line shows the cosmological baryon fraction, $\Omega_{\rm b}/\Omega_{\rm m}$. It can be seen that halos at the hard cutoff mass have a very different gas fraction, while almost all of them have the same HI to gas mass ratio. Thus, the lack of HI gas in small halos is more related to gas being ionized than a lack of gas.}
\label{fig:TNG_gas_fraction}
\end{center}
\end{figure}

The lack of HI gas in small halos may be due to: 1) a deficit in the
abundance of gas in those halos, 2) gas being present but highly
ionized, or a mixture of 1) and 2). We quantify the gas content of
dark matter halos in TNG100 by computing their gas fraction, i.e. the
ratio between the gas mass to the total mass. In the left panel of
Fig. \ref{fig:TNG_gas_fraction} we show the average gas fraction as a
function of halo mass at different redshifts. In this plot we
show results only for halos whose masses are above the mass of $50$ CDM
particles.

We find that the gas fraction of the smallest halos shown at high
redshifts is around $\sim0.12$. As we go to higher halo masses, some
stars form and supernova feedback expel the gas of these halos. In
even more massive halos the gravitational potential is deeper and
hence the mass-loading factors of galactic winds in the TNG model are
lower, rendering it more difficult for supernova feedback to eject gas
from halos. This explains the dip we observe around
$5\times10^9~h^{-1}M_\odot$ halos at high-redshift. In halos with
masses above $\sim3\times10^{11}~h^{-1}M_\odot$, AGN feedback becomes
effective at expelling their gas, explaining the peak at around that
mass. Finally, as we go to even higher mass halos, a smaller fraction
of the gas can be expelled by AGN feedback since the gravitational
potential becomes deeper, explaining the dip around
$3\times10^{12}~h^{-1}M_\odot$ halos.

It is interesting to note that the gas fraction of small halos
decreases quickly with redshift. We will see below that this is caused
by the UV background. At low redshift, the gas fraction of small halos
is small, so it is reasonable to expect that very little HI is
found in those halos. However, as we go to higher redshifts, the gas
fraction of halos at the hard cutoff mass $M_{\rm hard}$ (defined such
that halos with masses below $M_{\rm hard}$ host only $2\%$ of all HI
that is in halos, see Eq. \ref{Eq:Hard_cutoff}) is rather large. We
show this with colored stars in that figure. Thus, the cutoff in the
halo HI mass function cannot be explained, at high-redshift, by the
lack of gas in halos. A better explanation is that the gas
in these halos is highly ionized.

The average HI mass to gas mass ratio is shown in the right panel of
Fig. \ref{fig:TNG_gas_fraction}. We find that halos at the hard cutoff
mass $M_{\rm hard}$ exhibit similarly low HI to gas mass fractions
across redshifts: $\simeq5\%$. Thus, while there is a significant
amount of gas in these halos at high-redshift, HI formation is
impeded, likely due to the gas being at high temperature and diffuse.
The low HI to gas mass ratio shows up at all redshifts for low
mass halos. However, this is the case particularly at low redshift,
where the HI to gas mass fraction of small halos is practically zero,
showing how difficult it is to form HI in these halos. We speculate
that this is related to the fact that for fixed halo mass, the density
of gas and CDM increases with redshift. Thus, for example, while for
halos with masses $\simeq10^9~h^{-1}M_\odot$ the gas fraction 
increases by only $\simeq60\%$ from $z=3$ to $z=5$, the HI to gas mass
ratio changes by more than $400\%$, showing how denser gas enables the
formation of HI.

We thus conclude that the reason why there is almost no HI in small
halos is because the gas in these halos is highly ionized, presumably
because its low density and high-temperature prevents HI formation.

It is interesting to understand why the gas fraction of small halos
decreases with redshift. We argue that this effect is due to the UV
background. In order to demonstrate our claim we have run three
hydrodynamic simulations with radiative cooling and star formation but
without feedback (neither galactic winds nor AGN). In two of them, no
heating by the UV background is included, one with $2\times256^3$
CDM+baryon resolution elements and another with $2\times512^3$
CDM+baryon resolution elements. In the third one, we have heating by
the UV background with $2\times512^3$ CDM+baryon resolution
elements. In all cases the simulation box is 25 $h^{-1}{\rm Mpc}$
across.

\begin{figure}
\begin{center}
\includegraphics[width=0.45\textwidth]{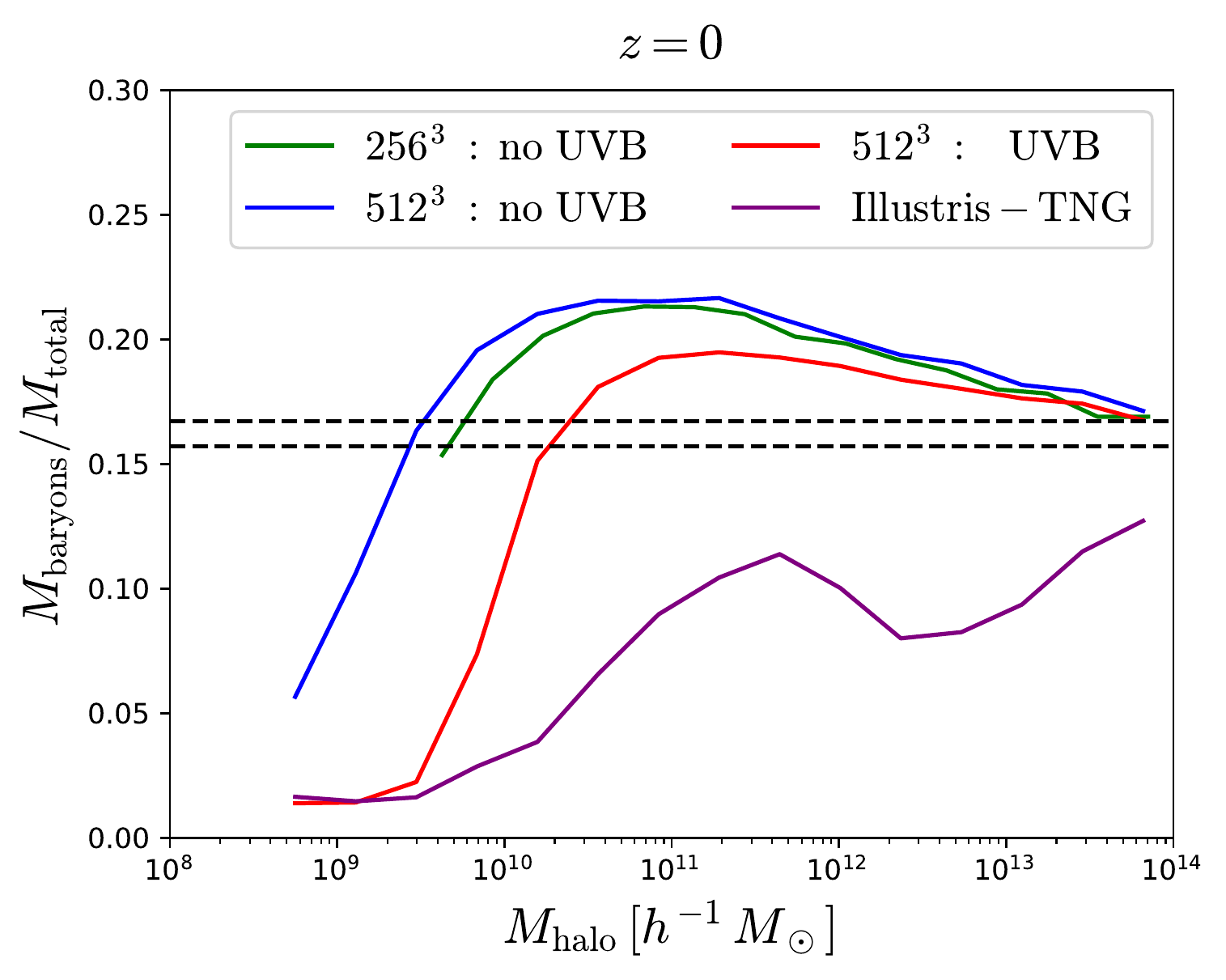}\\
\includegraphics[width=0.45\textwidth]{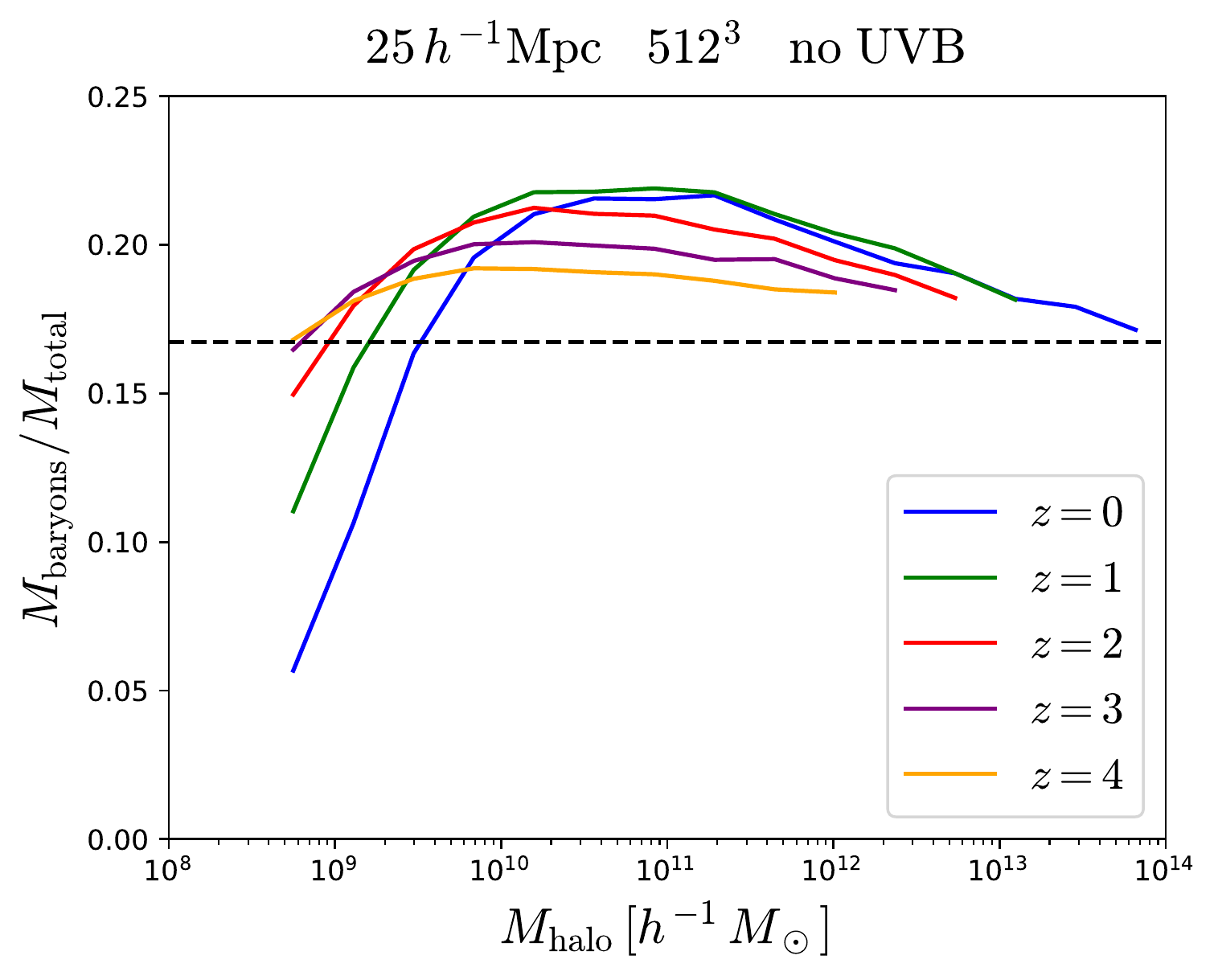}
\includegraphics[width=0.45\textwidth]{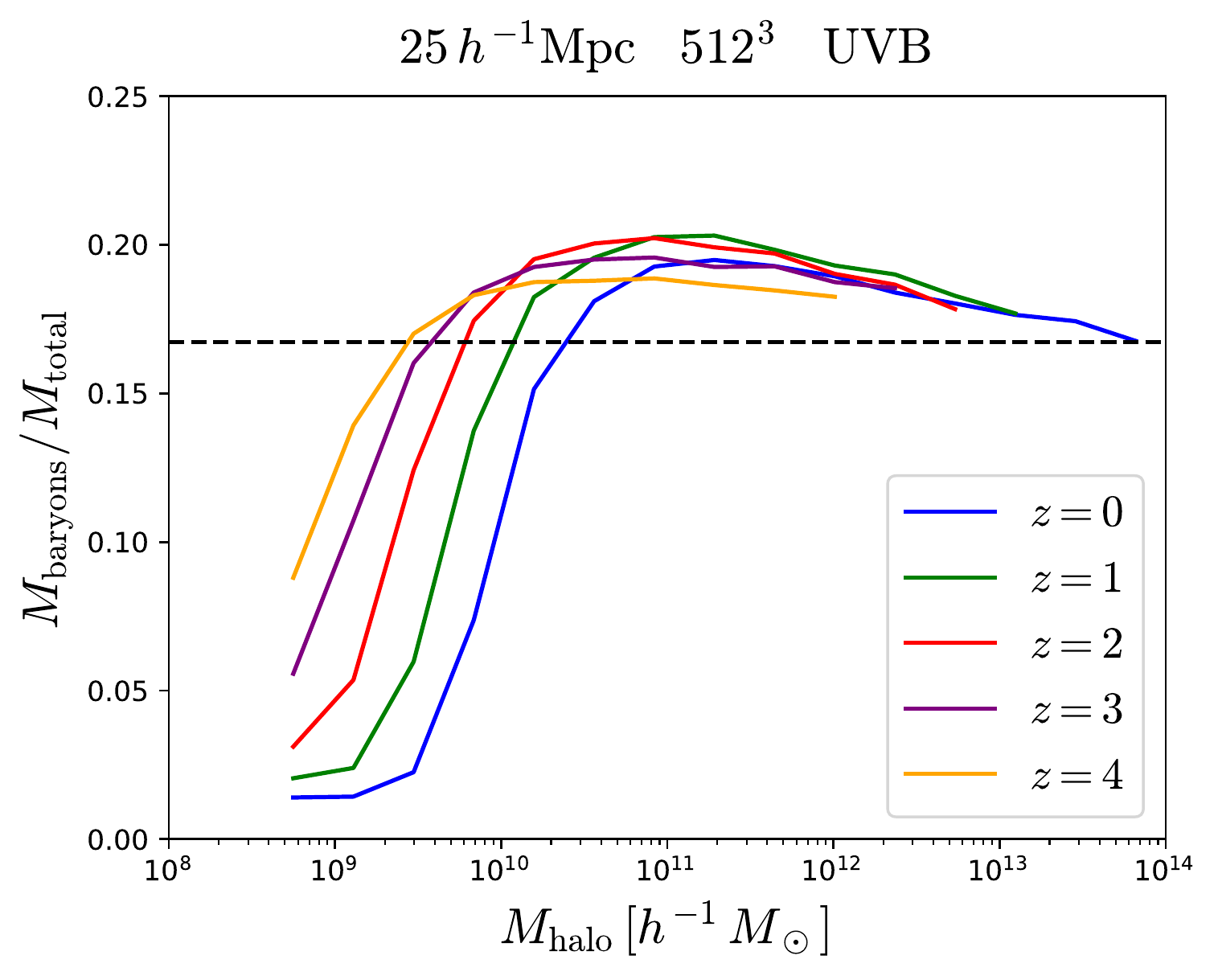}
\caption{We show the average baryon fraction, i.e. the fraction of the mass in baryons (gas+stars) over the total mass in halos, as a function of halo mass. In the upper panel we show results at $z=0$ for three different simulations: 1) a simulation with $2\times256^3$ resolution elements and no UV background (green), 2) a simulation with $2\times512^3$ resolution elements and no UV background (blue), and 3) a simulation with $2\times512^3$ resolution elements and UV background (blue). All simulations are in a box of $25~h^{-1}{\rm Mpc}$ size and no feedback is incorporated in any of them. The bottom panels display the results for the simulations with $2\times512^3$ with (right) and without (left) at different redshifts. The dashed black line indicates the cosmic baryon fraction, $\Omega_{\rm b}/\Omega_{\rm m}$. Since the values of the cosmological parameters are slightly different between IllustrisTNG and the other simulations, we show two horizontal lines in the upper panel. It can be seen that the presence of the UV background removes the baryonic, and therefore also the HI, content of small halos. It is interesting that even if the UV background is not present, very small halos exhibit a deficit in their baryon fraction.}
\label{fig:UVB_cutoff}
\end{center}
\end{figure}

We have computed the average baryon fraction, i.e. the mass in gas and
stars over the total mass, in these simulations and show the results
in Fig. \ref{fig:UVB_cutoff}. The top panel shows the baryon fraction
for these simulations and TNG100 at $z=0$. We find that in simulations
without feedback, the gas content of small halos exhibits a cutoff
that occurs at higher masses when the UVB is present.

The bottom panels of Fig. \ref{fig:UVB_cutoff} show the time evolution
of the average baryon fraction as a function of halo mass. It can be
seen that in both types of simulations, with and without UV
background, the gas fraction of small halos decreases with
redshift. We speculate that in the case of no UV background, many
small halos lie in the vicinity of massive halos, whose presence can
strip the gas from these small halos, and further, that the same
mechanism may explain the dependence of halo clustering on HI mass we
described in section \ref{sec:assembly_bias}.

The effect of the UV background on the baryon fraction of small halos
is more pronounced, as can be seen in the bottom-right panel of
Fig. \ref{fig:UVB_cutoff}. We believe that the reason for this behavior
is that the hot intergalactic gas cannot cluster in small halos since
their gravitational potential is not deep enough \citep{Okamoto_2008,
  Bose_2018}.

We thus conclude that at low redshift, small halos have a low gas
fraction. There are several mechanisms that can remove the gas from
such halos, such as tidal stripping by neighbors, the heating of the
intergalactic medium by the UV background and supernova feedback. Our
results suggest the most effective one for the lowest masses is the
presence of the UV background. The little gas inside those halos
is highly ionized, so no HI is found within them.

\section{HI content in FoF versus FoF-SO halos}
\label{sec:FoF_vs_SO}

We found in section \ref{subsec:HI_in_halos_galaxies} that the HI mass
inside FoF and FoF-SO halos is quite different. Here, we
determine the reason for this difference.

We have computed the total mass inside FoF and FoF-SO halos, and we
find that the former host $\sim9\%$ more mass than the latter at
$z=0$. This difference is almost equal to the deficit in HI mass we
find between FoF-SO and FoF halos at that redshift (see
Fig. \ref{fig:HI_halos_galaxies}). Similar results hold at higher
redshift, where the differences in total mass and HI are slightly
larger ($\simeq25\%$ at $z=5$). This indicates that the deficit in HI
mass we find in FoF-SO halos with respect to FoF is simply due to the
fact that the latter host a larger total mass, and therefore more HI.

\begin{figure}
\begin{center}
\includegraphics[width=0.49\textwidth]{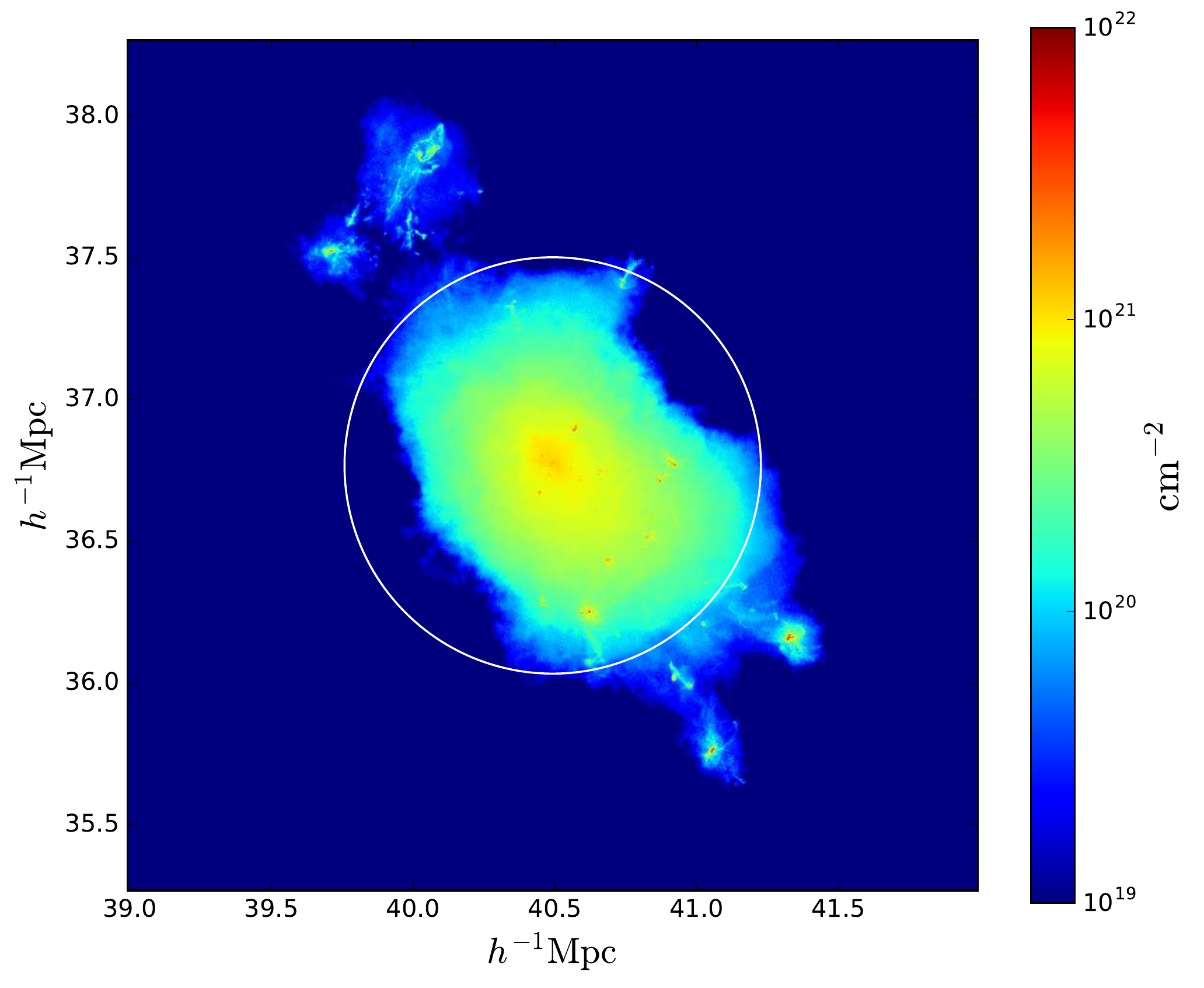}
\includegraphics[width=0.49\textwidth]{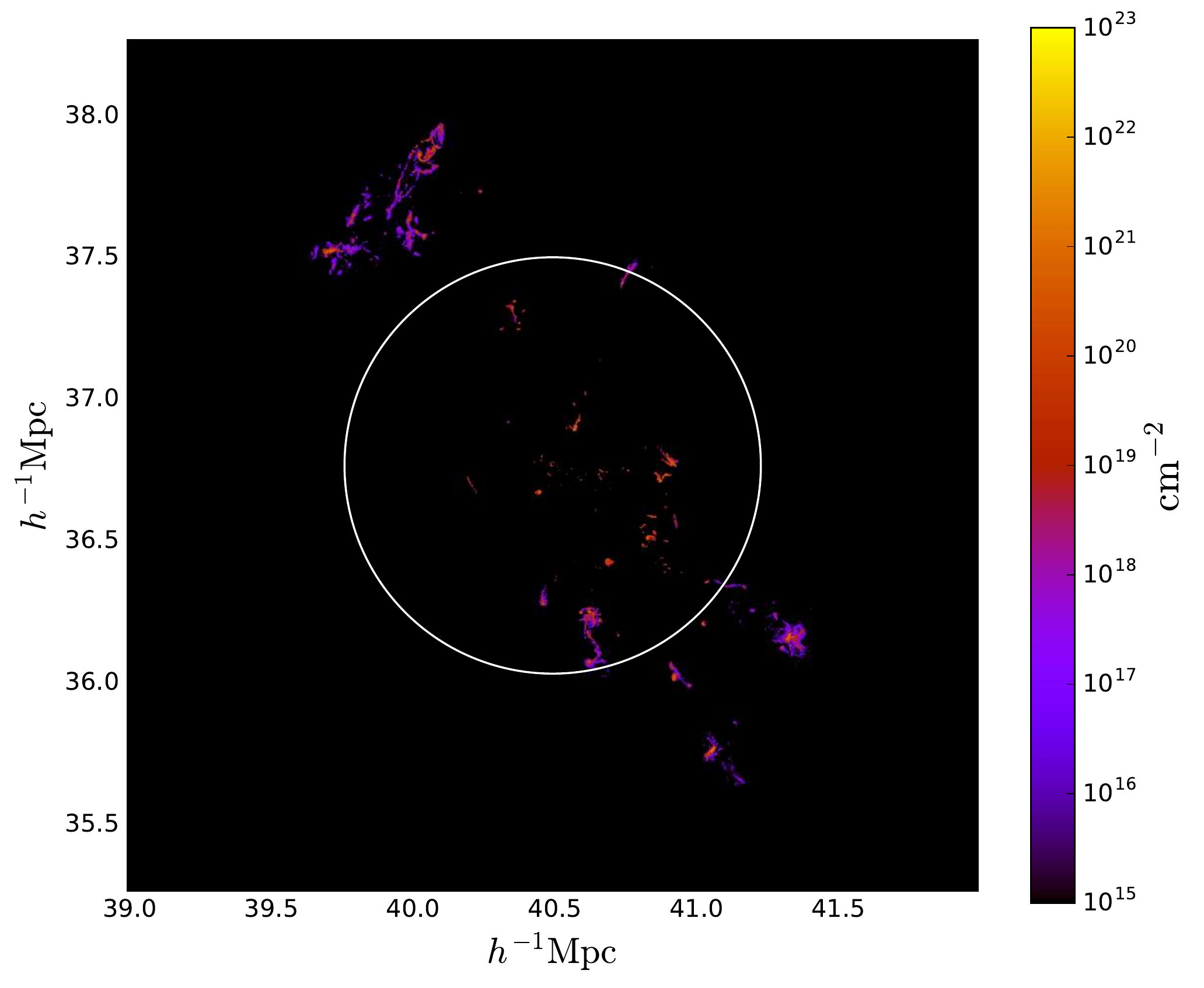}
\caption{Comparison between the gas and HI content in FoF versus FoF-SO halos. The images show the column density of gas (left) and HI (right) from the cells belonging to a FoF halo of mass $5\times10^{13}~h^{-1}M_\odot$ at $z=0$. The white circle shows the position of the halo center and its SO radius. While the gas content in the FoF and SO halo are similar, the FoF halo has almost a factor of 2 more HI than the SO halo.}
\label{fig:FoF_vs_SO}
\end{center}
\end{figure}

We note however that there are some situations where the difference in
HI mass can be much larger than the difference in total mass. We
illustrate one of these situations in Fig. \ref{fig:FoF_vs_SO}. We
have selected the gas cells belonging to a FoF halo of mass
$M=5\times10^{13}~h^{-1}M_\odot$ at redshift $z=0$. The total mass
inside the FoF and FoF-SO halos are nearly the same, while the HI
masses vary by a factor of 2. In Fig. \ref{fig:FoF_vs_SO} we plot the
column density of gas and HI from the gas cells belonging to that
halo. In the same plot, we mark the radius of the corresponding FoF-SO
halo with a white line. It can be seen that, while most of the gas in
the FoF halo is inside the virial radius of the SO halo, the situation
is quite different for HI. The HI mass outside the SO radius is
almost equal to the one inside. The reason is that the FoF algorithm
links the external galaxies to the main halo, and these galaxies are
rich in HI. We have found that this situation is usual for the most
massive halos at each redshift.

\section{Halo HI mass function}
\label{sec:HHIMF}

In section \ref{subsec:M_HI} we found that the halo HI mass function,
$M_{\rm HI}(M,z)$, from IllustrisTNG can be well reproduced by a
fitting formula like $M_0x^\alpha\exp(-1/x^{0.35})$, where $x=M/M_{\rm
  min}$. In the literature, however, an expression of the type
$M_0x^\alpha\exp(-1/x)$ has been widely used,
\citep[e.g.][]{Bagla_2010, EmaPaco,
  Obuljen_17,Paco_Alonso,Penin_17,Padmanabhan_2017}. We have fit our
results to that function, and while the reduced $\chi^2$ is larger
than with our fiducial function, it is still a good fit to the
underlying data. We thus provide in Table \ref{table:M_HI_fit2} the
best-fit values of the parameters $\alpha$, $M_0$ and $M_{\rm min}$ of
the latter expression in order to help in the comparison with previous
works. We note that the value of $M_{\rm hard}$, i.e. the hard cutoff
mass (see section \ref{subsec:M_HI}), is not affected by using a
different parametrization for the halo HI mass function. Thus, the
values we quote in Table \ref{table:M_HI_fit} are valid also here.

\begin{table*}
\begin{center}
\resizebox{1.0\textwidth}{!}{\begin{tabular}{|c|| c| c| c|| c|c|c|} 
 \hline
 & \multicolumn{3}{|c||}{FoF} & \multicolumn{3}{|c|}{FoF-SO}\\[0.5ex]
 \hline
 $z$ & $\alpha$ & $M_0$ & $M_{\rm min}$ & $\alpha$ & $M_0$ & $M_{\rm min}$\\ [0.5ex] 
 & & $[h^{-1}M_\odot]$ & $[h^{-1}M_\odot]$ & & $[h^{-1}M_\odot]$ & $[h^{-1}M_\odot]$ \\[0.5ex] 
 \hline\hline
 
 0 & $0.49\pm0.03$ & $(2.1\pm0.7)\times10^{9}$ & $(5.2\pm1.3)\times10^{10}$ & 
        $0.42\pm0.03$ & $(2.4\pm0.8)\times10^{9}$ & $(5.6\pm1.4)\times10^{10}$\\ 
 
 \hline
1 & $0.76\pm0.03$ & $(4.6\pm2.1)\times10^{8}$ & $(2.6\pm1.0)\times10^{10}$ & 
      $0.67\pm0.04$ & $(6.5\pm3.5)\times10^{8}$ & $(3.3\pm1.5)\times10^{10}$ \\
 
 \hline
2 & $0.80\pm0.03$ & $(4.9\pm2.1)\times10^{8}$ & $(2.1\pm0.7)\times10^{10}$ & 
       $0.72\pm0.03$ & $(5.9\pm2.7)\times10^{8}$ & $(2.4\pm0.9)\times10^{10}$ \\
 
 \hline
 3 & $0.95\pm0.03$ & $(9.2\pm4.7)\times10^7$ &  $(4.8\pm1.9)\times10^{9}$ & 
        $0.90\pm0.03$ & $(1.0\pm0.6)\times10^8$ &  $(5.5\pm2.3)\times10^{9}$\\
 
\hline
4 & $0.94\pm0.02$ & $(6.4\pm3.7)\times10^7$ &  $(2.1\pm1.0)\times10^{9}$ & 
      $0.82\pm0.03$ & $(1.6\pm0.8)\times10^8$ &  $(4.5\pm1.9)\times10^{9}$\\ 
 
 \hline
5 & $0.90\pm0.02$ & $(9.5\pm5.8)\times10^7$ &  $(1.9\pm1.0)\times10^{9}$ & 
      $0.84\pm0.04$ & $(1.1\pm0.9)\times10^8$ &  $(2.6\pm1.7)\times10^{9}$\\
 
 \hline
\end{tabular}}
\caption{We find that an expression like $M_0x^\alpha\exp(-1/x^{0.35})$, where $x=M/M_{\rm min}$, reproduces our results well for the halo HI mass function, $M_{\rm HI}(M,z)$. In the past, expressions like $M_0x^\alpha\exp(-1/x)$ have been widely used to model that quantity. In this work we find that the latter reproduces well the high-mass end of $M_{\rm HI}(M,z)$ but it underestimates the low-mass end. However, that expression still provides a good $\chi^2$ when fitting our results. In order to help in comparisons with previous works, we provide here the best-fit values for $\alpha$, $M_0$ and $M_{\rm min}$ when fitting our halo HI mass function with $M_0x^\alpha\exp(-1/x)$. The left/right part shows the results for the FoF and FoF-SO halos.}
\label{table:M_HI_fit2}
\end{center}
\end{table*}

\section{Fit to HI profiles}
\label{sec:rho_HI_fit}

\begin{figure}
\begin{center}
\includegraphics[width=1\textwidth]{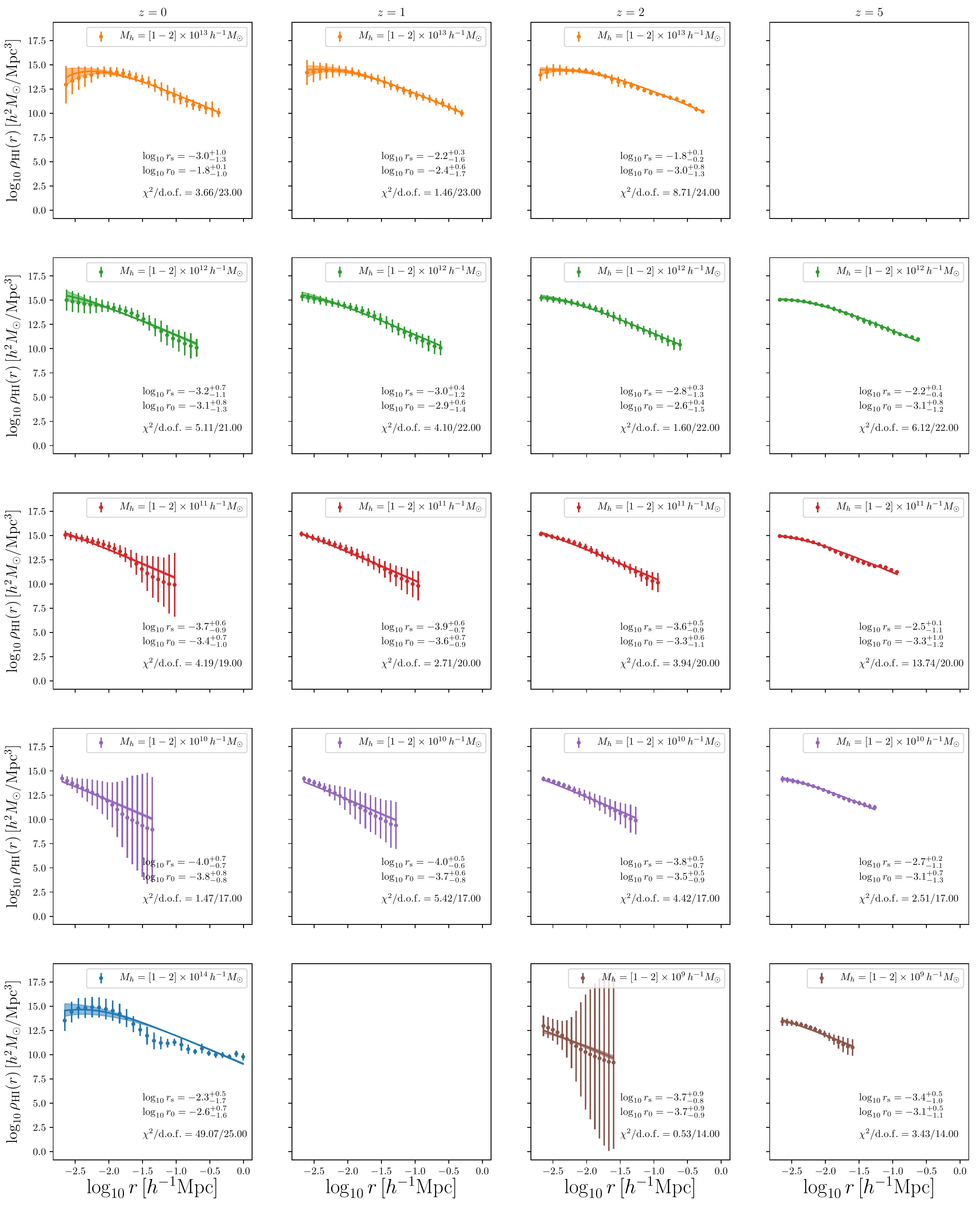}
\caption{Each panel shows the mean and standard deviation of the HI profiles for halos in the mass range indicated in the upper-left part. We fit the results using the form $\rho_{\rm HI}(r)=\exp(-r_0/r)\rho_0r_s^3/[(r+3/4r_s)(r+r_s)^2]$, where $\rho_0$, $r_s$ and $r_0$ are free parameters. The best-fit is shown with a solid line. The dashed region represents the error on the fit. The value of $\rho_0$ is fixed by requiring that $M_{\rm HI}=4\pi\int_0^{R_v} r^2\rho_{\rm HI}(r)dr$, where $R_v$ is the halo virial radius. Each panels show the best-fit values of $r_0$ and $r_s$ and the value of $\chi^2$.}
\label{fig:rho_HI_fit}
\end{center}
\end{figure}

The points with error bars in Fig. \ref{fig:rho_HI_fit} show the mean HI density profiles within halos of different halo masses at different redshifts (the are the same as the blue lines in Fig. \ref{sec:HI_profile}). The solid lines represent the best-fit obtained when we fit those results with a HI density profile as
\be
\rho_{\rm HI}(r)=\frac{\rho_0 r_s^3}{(r+3/4r_s)(r+r_s)^2}\exp(-r_0/r).
\ee
In each panel of the plot we show the best-fit value of $r_s$, $r_0$ and the value of the reduced $\chi^2$. The value of $\rho_0$ is fixed by requiring that $M_{\rm HI}=4\pi\int_0^{R_v} r^2\rho_{\rm HI}(r)dr$, where $R_v$ is the halo virial radius.

\end{appendix}

\bibliography{references}{}
\bibliographystyle{hapj}

\end{document}